\documentclass{article}
\usepackage[utf8]{inputenc}
\usepackage[hscale=.7, vscale=.8]{geometry}
\usepackage{bbm, amsmath, amssymb, amsthm, bm}
\usepackage{natbib}
\usepackage[colorlinks=true, citecolor={blue}]{hyperref}
\usepackage{mathtools}

\usepackage{crossreftools}

\usepackage{subcaption}
\usepackage{graphicx, algorithm, algpseudocode}
\graphicspath{ {./figures/} }

\newcommand{\giv}{\,|\,}

\newcommand{\Pb}{\mathbb{P}}
\mathchardef\mhyphen="2D

\newcommand{\tbeta}{\tilde{\beta}}
\newcommand{\talpha}{\tilde{\alpha}}
\newcommand{\tbm}[1]{\tilde{\bm{#1}}}

\newtheorem{theorem}{Theorem}
\newtheorem{lemma}[theorem]{Lemma}
\newtheorem{proposition}[theorem]{Proposition}

\newtheorem{conjecture}[theorem]{Conjecture}

\theoremstyle{definition}
\newtheorem{definition}[theorem]{Definition}

\theoremstyle{remark}
\newtheorem{remark}{Remark}

\title{ 
Root and community inference on the latent growth process of a network}
\author{Harry Crane, Min Xu\thanks{Corresponding author: 	
			Min Xu, Department of Statistics, Rutgers University, New Brunswick, NJ 08854, USA (E-mail: \textit{mx76@stat.rutgers.edu}). 
			}\\
Department of Statistics \\
Rutgers University, New Brunswick, NJ, USA}
\date{February, 2023}

\begin{document}



\maketitle

\begin{abstract}
Many existing statistical models for networks overlook the fact that most real world networks are formed through a growth process. To address this, we introduce the PAPER (Preferential Attachment Plus Erd\H{o}s--R\'{e}nyi)
model for random networks, where we let a random network G be the
union of a preferential attachment (PA) tree T and additional
Erd\H{o}s--R\'{e}nyi (ER) random edges. The PA tree component captures the underlying growth/recruitment
process of a network where vertices and edges are added sequentially, while the ER
component can be regarded as random noise. Given only a single
snapshot of the final network G, we study the problem of constructing
confidence sets for the early history, in particular the root node, of
the unobserved growth process; the root node can be patient zero in a
disease infection network or the source of fake news in a social media
network. We propose an inference algorithm based on Gibbs sampling
that scales to networks with millions of nodes and provide theoretical
analysis showing that the expected size of the confidence set is small
so long as the noise level of the ER edges is not too large. We also
propose variations of the model in which multiple growth processes
occur simultaneously, reflecting the growth of multiple communities,
and we use these models to provide a new approach to community detection.
\end{abstract}

\section{Introduction}

Network data is ubiquitous. To analyze networks, there are a
variety of statistical models such as Erd\H{o}s--R\'{e}nyi,
stochastic block model (SBM)~\citep{abbe2017community,
  karrer2011stochastic, amini2013, xu2018optimal}, graphon~\citep{diaconis2007graph,
  gao2015rate}, random dot product
graphs~\citep{athreya2017statistical, xie2019optimal}, latent space
models~\citep{hoff2002latent}, configuration graphs~\citep{aiello2000random},
and more. These models usually operate by specifying some
structure, such as community structure in the case of SBM, 
and then adding independent random edges in a way that reflects the
structure. The order in which the edges are added is of no importance
to these models. 

In contrast, real world networks are often formed from growth processes
where vertices and edges are added sequentially. This motivates the
development of Markovian preferential attachment (PA)
models for networks \citep{barabasi1999emergence, barabasi2016network} which produce a
sequence of networks $\bm{G}_1, \bm{G}_2, \ldots, \bm{G}_n$ where
$\bm{G}_1$ starts as a single node which we call the root node and, at each iteration, we add
a new
node and new edges. PA models
naturally produce networks with sparse edges, heavy-tailed degree
distributions, and strands of chains as well as pendants (several
degree 1 vertices linked to a single vertex), which are important features of real world
networks that are difficult to reproduce under a non-Markovian 
model, as observed by~\cite{bloem2018random}.

Although Markovian models are often more realistic, they
have not been as widely used in network data analysis as, say
SBM, because, whereas SBM is useful for recovering the community structure of a
network, it is not obvious what structural information
Markovian models could extract from a network. Recently however,
seminal work from a series of applied probability papers (e.g.
\cite{bubeck2017finding, bubeck2015influence}) demonstrate that Markovian models can indeed recover useful structure: these papers show that, surprisingly, when $\bm{G}_n$ is
a random PA tree, one can infer the early history of $\bm{G}_n$, such
as the root node, even as the size of the tree tends to
infinity. Although these results are elegant, they are theoretical;
their confidence set construction involves large constants that render
the result too conservative. Moreover, most algorithms apply only to
tree-shaped networks, which prohibitively limits their application since trees are rarely encountered in practice.

\begin{figure}
  \centering
  \includegraphics[scale=.3]{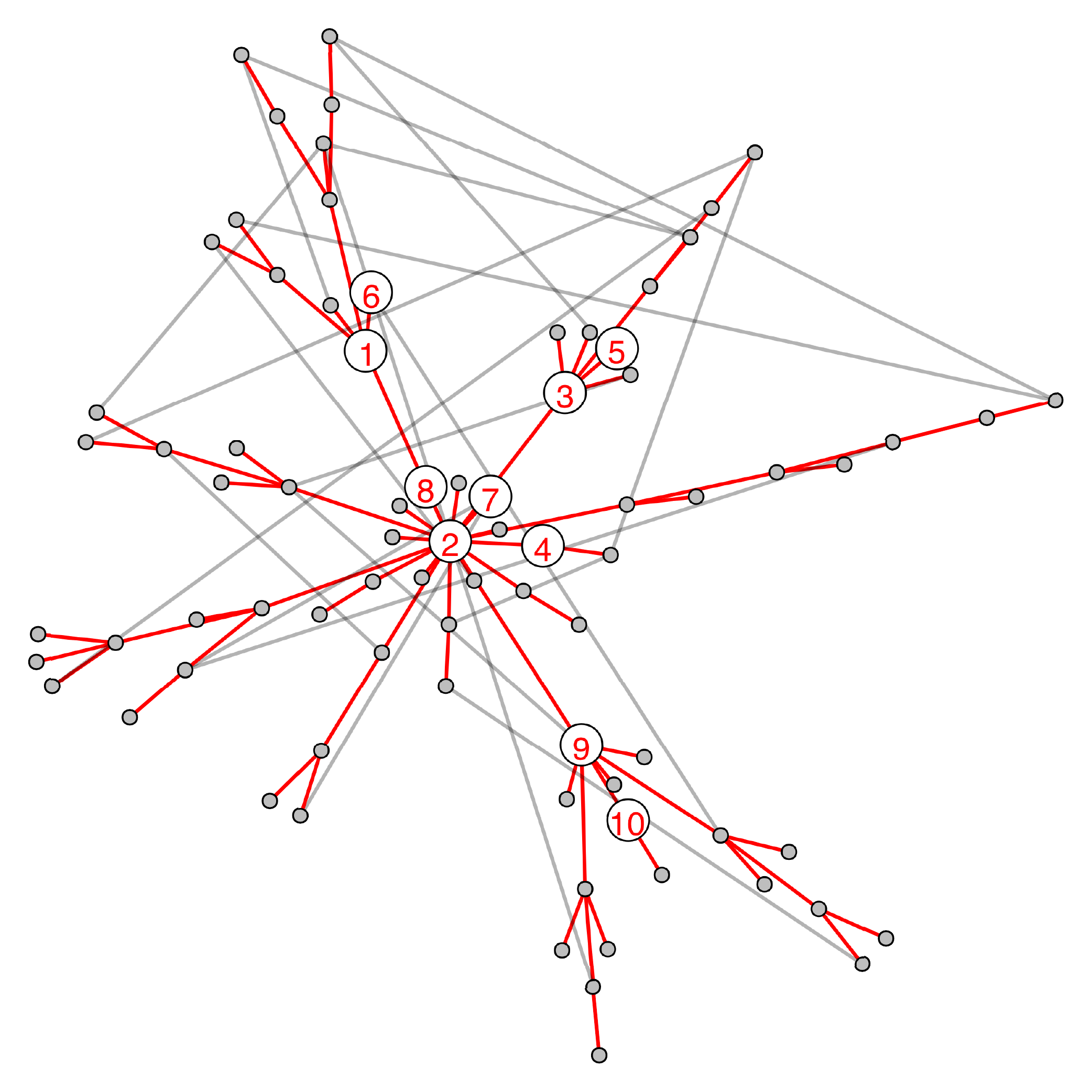}
  \includegraphics[scale=.3]{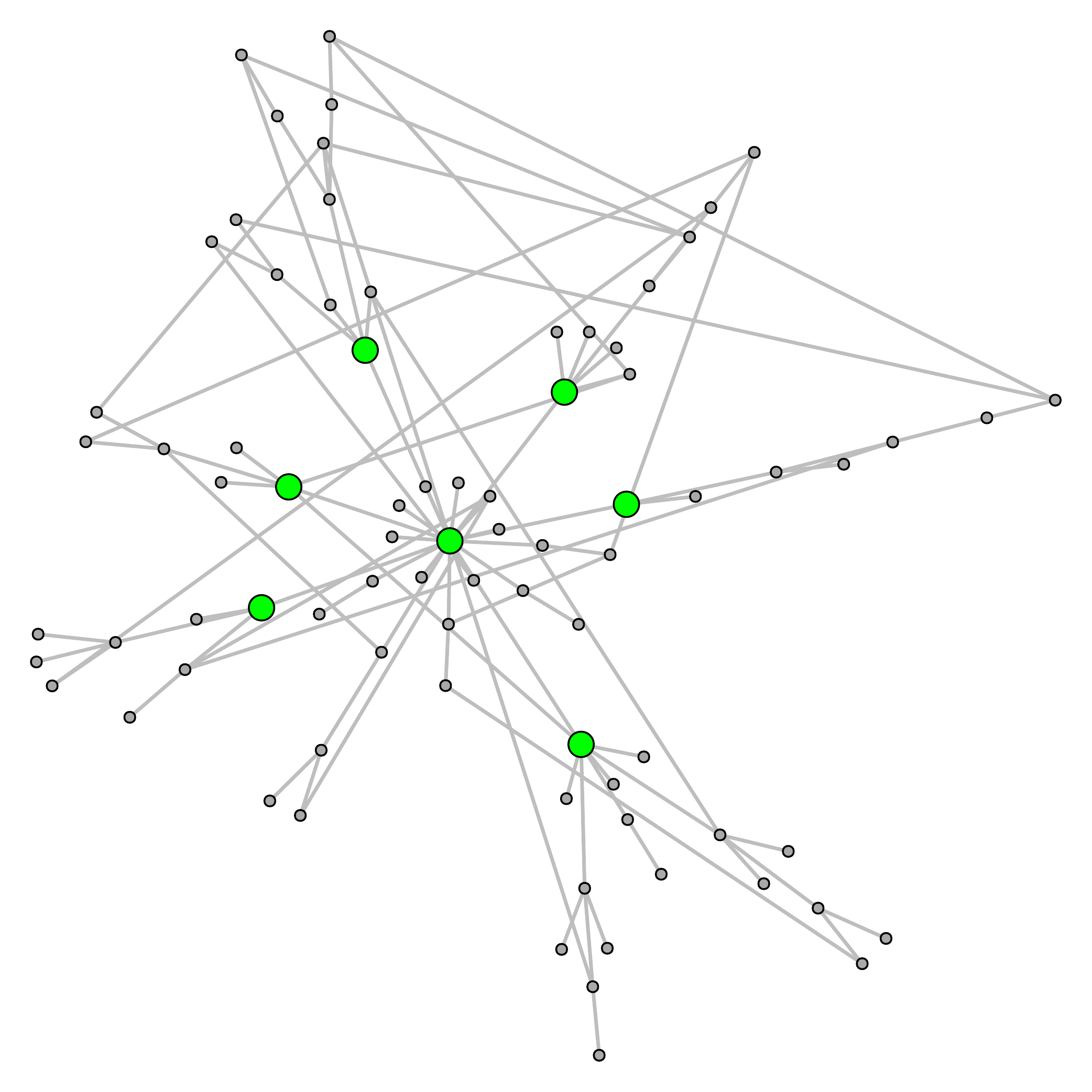}
  \caption{\textbf{Left}: illustration of PAPER model; nodes have latent time
    ordering (only first 10 orderings shown); the red edges form the
    latent tree while gray edges are Erd\H{o}s--R\'{e}nyi. \textbf{Right}:
    80\% confidence set for the root node (node number 1) constructed from the unlabeled graph.}
  \label{fig:intro}
\end{figure}

To overcome these problems, we propose a Markovian model for networks which we call
Preferential Attachment Plus Erd\H{o}s--R\'{e}nyi, or PAPER for short. We say
that $\bm{G}_n$ has the PAPER distribution if it is generated by
adding independent random edges to a preferential attachment
tree $\bm{T}$. The latent PA tree captures the growth process of the
network whereas the ER random edges can be interpreted as additional
noise. Given only a single snapshot of the final graph $\bm{G}_n$, we
study how to infer the early history of the
latent tree $\bm{T}$, focusing on the concrete problem of constructing
confidence sets for the root node that can attain the nominal
coverage. We give a visual illustration of the PAPER model and the
inference problem in Figure~\ref{fig:intro}.


Because we do not know which edges of $\bm{G}_n$ correspond to the
tree and which are noise, most existing methods are not directly applicable. We therefore propose a new approach in
which we first give the nodes new random labels which induce, for a given observation of the network $\bm{G}_n$, a 
posterior distribution of both the latent tree and the latent arrival
ordering of the nodes. Then, we sample from the posterior distribution
to construct a credible set for the inferential target, e.g. the
root node. Bayesian inference statements usually do not have frequentist
validity but we prove in our setting that that the level $1-\epsilon$ credible
set for the root node has frequentist coverage at
exactly the same level.


In order to efficiently sample from the posterior distribution of the
latent ordering and the latent tree, we present a scalable Gibbs sampler that alternatingly samples the ordering and the tree. The algorithm to generate the latent ordering is
based on our previous work \citep{crane2021inference} which studies inference in the
tree setting. The algorithm to generate the latent tree operates
by updating the parent of each of the nodes iteratively. The overall runtime
complexity of one iteration of the outer loop is generally $O(m +
n\log n)$ (where $m$ is the number of edges) and the algorithm can scale to networks of up to a million nodes. 

Since a trivial confidence set for the root node is the set of
all the nodes, it is important to be able to bound the size of a
confidence set. In particular, the presence of noisy Erd\H{o}s--R\'{e}nyi
edges in the PAPER model motivates an interesting question: how
does the size of the confidence set increase with the noise level? In
this paper, we give an initial answer to this question under two
specific settings of the preferential attachment mechanism: 
linear preferential attachment (LPA) and uniform attachment (UA). For
LPA, we prove that the size of our proposed confidence set does not
increase with the number of nodes $n$ so long as the noisy edge
probability is less than $n^{-1/2}$ and for UA, we prove that the size
 is bounded by $n^\gamma$ for some $\gamma < 1$ so long as the noisy edge probability is less than
 $\log(n)/n$. Our analysis shows that the phenomenon discovered by
 \cite{bubeck2017finding}, that there exists confidence sets for the
 root node of $O(1)$ size, is robust to the presence of noise. 

Many real world networks often have community structures. In such
cases, it would be unrealistic to assume that the network originates
from a single root node. We therefore propose variations of the PAPER model in which
$K$ growth processes occur simultaneously from $K$ root nodes. Each of $K$ root nodes can be interpreted as being locally central with respect to a community subgraph. In the
multiple roots model, there is no longer a latent tree but rather a
latent forest (union of disjoint trees), where the
components of the forest can naturally be interpreted as the different
communities of the network. We provide model formulation that allows
$K$ to be either be fixed or random. To analyze networks with multiple
roots, we use essentially the same inferential approach and Gibbs
sampling algorithm that that we develop for the single root setting,
with minimal modifications.

By looking at the posterior probability that a node is in a
particular tree--community, we can estimate the community membership of
each of the nodes. Compared with say
the stochastic block model, the PAPER model approach to community recovery
has the advantage that the inference quality improves with sparsity,
that we can handle heavy-tailed degree distribution without a high-dimensional degree correction parameter
vector, and that the posterior root probabilities also identify the important nodes in the community. Empirically, we show that our approach has
competitive performance on two benchmark datasets and we find that our
community membership estimate is more
accurate for nodes with high posterior root
probability than for the more peripheral nodes. {\color{black} We also use the PAPER model to conduct an extensive analysis of a statistician co-authorship network curated by \cite{ji2016coauthorship} where we recover a large number of communities that accurately reflect actual research communities in statistics.}

We have implemented our inference algorithm in a Python package
called \texttt{paper-network}, which can be installed via command \texttt{pip
  install paper-network}. The code, example scripts, and documentation
are all publicly available at
\texttt{https://github.com/nineisprime/PAPER}. \\

\textbf{Outline for the paper:} In Section~\ref{sec:model}, we define the PAPER model in both the
single root and multiple roots setting. We also formalize the
problem of root
inference and review related work. In
Section~\ref{sec:methodology}, we describe our approach to the root
inference problem, which is to randomize the node labels and analyze
the resulting posterior distribution. We also show that the Bayesian
inferential statements have frequentist validity. In
Section~\ref{sec:algorithm}, we give a sampling algorithm for
computing the posterior probabilities. In Section~\ref{sec:theory}, we
provide theoretical bounds on the size of our proposed confidence sets
and in Section~\ref{sec:empirical}, we provide empirical study on both
simulated and large scale real world networks. 

\subsection{Notation}

\begin{itemize}
\item We take all graphs to be undirected. Given two labeled graphs $\bm{g}$ and $\bm{g}'$ defined on the same set of
nodes, we write $\bm{g} + \bm{g}'$ as the resulting graph if we take
the union of the edges in $\bm{g}$ and $\bm{g}'$ and collapse any
multi-edges. We also write $\bm{g} \subset \bm{g}'$ if $\bm{g}$ is a
subgraph of $\bm{g}'$. 

\item For a labeled graph $\bm{g}$, we write
$D_{\bm{g}}(u)$ as the degree of node $u$ in graph $\bm{g}$ and
$N_{\bm{g}}(u)$ as the set of neighbors  of $u$ (all nodes directly
connected to $u$) with respect to $\bm{g}$; we write $V(\bm{g})$ and $E(\bm{g})$ as
the set of vertices and edges of $\bm{g}$ respectively.

\item For an integer $n$, we write $[n] := \{1, 2, \ldots, n\}$ and
  for a discrete set $A$, we write $|A|$ as the cardinality of
  $A$. For two countable sets $A,B$ of the same cardinality, we write
  $\text{Bi}(A, B)$ as the set of bijections between them. For a vector $\pi$, we write $\pi_{1:K}$ to denote the sub-vector $(\pi_1, \pi_2, \ldots, \pi_K)$.

\item Given a finite
set $V'$ of the same cardinality of $V(\bm{g})$ and given a bijection
$\rho \in \text{Bi}(V(\bm{g}), V')$, we write $\rho \bm{g}$ to denote
a relabeled graph where a pair $(u', v') \in V' \times V'$ is an edge
in $\rho \bm{g}$ if and only if $(u, v) \in V(\bm{g}) \times
V(\bm{g})$ is an edge in $\bm{g}$. 

\item Throughout the paper, we use capital font (e.g. $\bm{G}$) to
denote random objects and lower case font to denote fixed
objects. Graphs are represented via bold font. 
\end{itemize}

\section{Model and Problem}
\label{sec:model}

We first describe the model and inference problem in the single root
setting and then extend the definition to the setting of having fixed
$K$ roots and having random $K$ roots. 

\subsection{PAPER model}
\label{sec:single-root-model}


\begin{definition}
  \label{def:apa1}
The affine preferential attachment tree model, which we denote by
$\text{APA}(\alpha, \beta)$ for parameters $\alpha, \beta \in
\mathbb{R}$, generates an increasing sequence $\bm{T}_1 \subset \bm{T}_2
\subset \ldots \subset \bm{T}_n$ of random trees where $\bm{T}_t$ is a tree with $t$
nodes and where nodes are labeled by their arrival time so that
$V(\bm{T}_t) = [t]$. The first tree $\bm{T}_1 = \{1 \}$ is a singleton and for $t > 2$, we define the transition kernel
$\Pb( \bm{T}_{t} \giv \bm{T}_{t-1})$ in the following way: given
$\bm{T}_{t-1}$, we add a node labeled $t$ and a random edge
$(t, w_t)$ to obtain $\bm{T}_t$, where the existing node $w_t
\in [t-1]$ is chosen with probability 
\begin{align}
  \frac{\beta D_{\bm{T}_{t-1}}(w_t) + \alpha}{ \beta 2 (t-2) + \alpha
  (t-1)}.
\label{eq:pa_tree_prob}
\end{align}
\end{definition}

We may verify that~\eqref{eq:pa_tree_prob} describes a valid
probability distribution by noting that $\bm{T}_{t-1}$ always has $t-2$ edges
and $t-1$ nodes. Before continuing onto the PAPER model, we consider some specific examples
of APA trees:
\begin{enumerate}
  \item setting $\alpha = 1, \beta = 0$ means that we select $w_t$
    uniformly at random from $V(\bm{T}_{t-1})$. This yields the uniform attachment
    (UA) random tree. The resulting degree distribution has
    exponential tail and the maximum degree is of order $\log n$
    \citep{na1970distribution, addario2018high}.
  \item Setting $\alpha = 0, \beta = 1$ means that we select $w_t$ with probability proportional to the degree $D_{\bm{T}_{k-1}}(w_t)$. This yields the linear preferential attachment random (LPA)
tree. LPA has heavy-tailed degree distribution and a maximum degree is of order
    $\sqrt{n}$ \citep{bollobas2001degree, pekoz2014joint}.
    \item We may also set $\beta$ as $-1$ and $\alpha$ as some
      positive integer
      so that the maximum degree of any node is $\alpha$. This
      may be interpreted as an uniform attachment tree growing on top
      of a background infinite $\alpha$-regular tree \citep{khim2017confidence}.
    \end{enumerate}
  
We may generalize Definition~\ref{def:apa1}
by defining a nonparametric function $\phi \,:\, \mathbb{N} \rightarrow [0, \infty)$ and
choose $w_t$ with probability proportional to $\phi(
D_{\bm{T}_{t-1}}(w_t))$. In this paper however, we focus only on the
case where $\phi$ is an affine function. 
  
  \begin{definition}
  \label{def:paper}
To model a general network, we define the $\text{PAPER}(\alpha,
  \beta, \theta)$ (Preferential Attachment Plus Erd\H{o}s--R\'{e}nyi) model
parametrized by $\alpha, \beta \in \mathbb{R}$ and $\theta \in
[0,1]$. We say that a random graph $\bm{G}_n$ distributed according to
the $\text{PAPER}(\alpha, \beta, \theta)$ model if 
\[
\bm{G}_n = \bm{T}_n + \bm{R}_n,
\]
where $\bm{T}_n \sim \text{APA}(\alpha, \beta)$ and $\bm{R}_n \sim
\text{Erd\H{o}s--R\'{e}nyi}(\theta)$ are independent random graphs defined on the same
set of vertices $[n]$. 
  \end{definition}

\begin{figure}
  \centering
  \includegraphics[scale=.2]{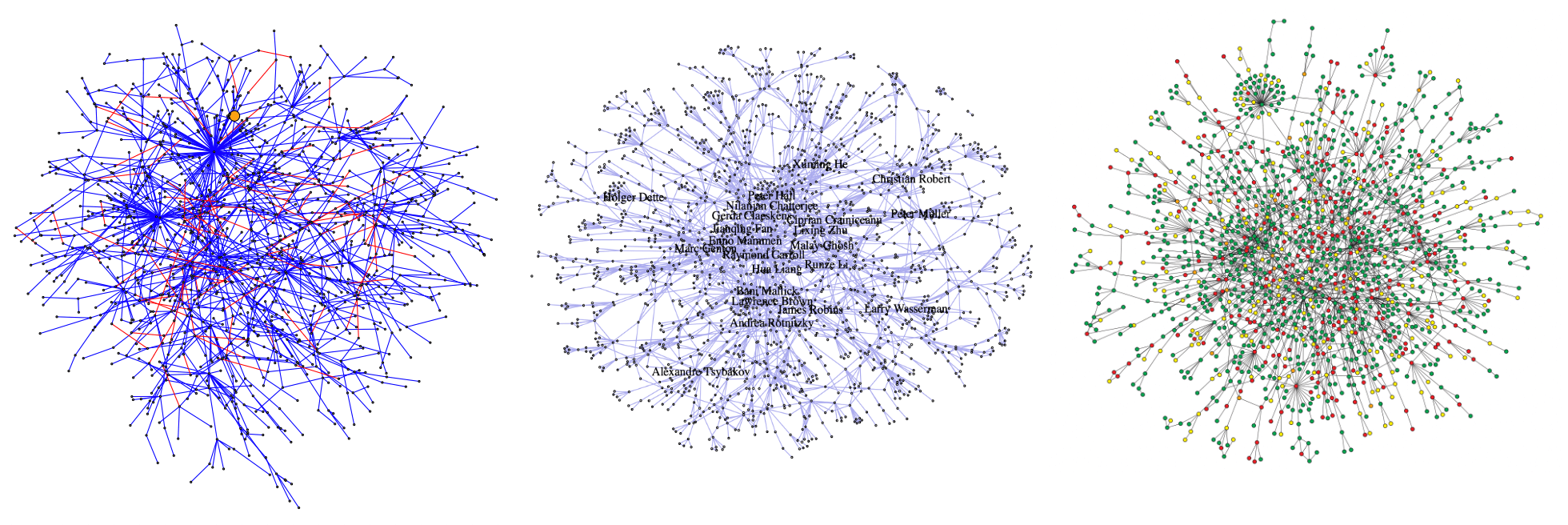}
  \caption{\textbf{Left}: PAPER graph with $\alpha=1,\beta=1$;
    \textbf{Center:} co-authorship graph from~\cite{ji2016coauthorship}; \textbf{Right:} protein-protein
    interaction graph from~\cite{jeong2001lethality}.}
  \label{fig:paperexample}
\end{figure}

Since we collapse any multi-edges that occur when we add $\bm{R}_n$ to
$\bm{T}_n$, we may view $\bm{R}_n$ equivalently as an ER
random graph defined on potential edges excluding those already in the
tree $\bm{T}_n$. The PAPER model can produce networks with either light tailed
or heavy tailed degree distribution depending on the choice of the
parameters $\alpha$ and $\beta$. It produces features that are
commonly seen in real world networks but absent from non-sequential
models like SBM, such as pendants (a node with several degree-1 node
attached to it) and chains of nodes; see
Figure~\ref{fig:paperexample}. It also assigns a non-zero
probability to any connected graph, in contrast to the general
preferential attachment graph model where a fixed $m > 1$ edges are
added at every iteration \citep{barabasi1999emergence}. {\color{black} In computer science terminology, $\bm{G}_n$ is a \emph{planted tree model} where the signal $\bm{T}_n$ is planted in an ER random graph $\bm{R}_n$ in the same sense that stochastic block model is often referred to as the planted partition model. 

An alternative way to define the PAPER model is to specify the total number of edges $m$ in the final graph and generate $\bm{R}_n$ as a uniformly random graph with $m - (n-1)$ edges (since a tree with $n$ nodes always has $n-1$ edges). This is equivalent to the $\text{PAPER}(\alpha, \beta, \theta)$ model where we \emph{condition} on the event that the final graph $\bm{G}_n$ has $m$ edges. To simplify exposition, we use PAPER to refer to this conditional model as well.
}

\begin{remark} 
\label{rem:variations} 
We may view the $\text{PAPER}(\alpha, \beta, \theta)$ model as a Markovian process over a sequence of networks $\bm{G}_1, \bm{G}_2,
\ldots, \bm{G}_n$. We define the transition kernel $\mathbb{P}( \bm{G}_t \,|\,\bm{G}_{t-1})$ for $t \geq 3$ by first adding a new node labeled $t$, then adding a new tree edge $(t, w_t)$ where $w_t$ is chosen with probability~\eqref{eq:pa_tree_prob}, and then, for each existing node $j \in [t-1]$ not equal to $w_t$, we independently add a noise edge $(t, j)$ with probability $\theta$. In Section~\ref{sec:seq_noise}, we extend the PAPER model so that the noise edge probability depends on time $t$ and the state of the tree at time $t$.
\end{remark}

\begin{remark}
\label{rem:paper_prob}
Under $\text{APA}(\alpha, \beta)$ model, the probability of generating a given tree has a closed form expression: $\mathbb{P}(\bm{T}_n = \bm{t}_n) = \frac{ \prod_{v \in [n]} \prod_{j=1}^{D_{\bm{t}_n}(v)-1} (\beta j + \alpha)}{\prod_{t=3}^n 2(t-2)\beta + (t-1)\alpha}$. The important consequence is that the likelihood depends on the tree
  $\bm{t}_n$ only through its degree distribution
  $D_{\bm{t}_n}(\cdot)$. Hence, any two trees with the same degree
  distribution has the same likelihood; \cite{crane2021inference} refers to this property as \emph{shape-exchangeability}. We give the likelihood expression for the multiple roots models and the PAPER model in Section~\ref{sec:model-likelihood} of the Appendix.
\end{remark}

\begin{remark}
  \label{rem:apadegree}
In many settings, it is known that the degree distribution of an
$\text{APA}(\alpha, \beta)$ tree has an asymptotic limit. For example,
if $\beta = 1$ and $\alpha > 0$, then we have by \citet[][Theorem 8.2]{van2016random} that $\frac{1}{n}
\sum_{t=1}^n \mathbbm{1}\{ D_{\bm{T}_n}(t) = k \} \rightarrow
\frac{2+\alpha}{3+2\alpha} \prod_{j=1}^{k-1}
\frac{j+\alpha}{j+3+2\alpha}$ as $n \rightarrow \infty$ uniformly over
all $k$. The limiting distribution is approximately a
power law where the number of nodes with degree $k$ is proportional to
$k^{-(3 + \alpha)}$ (see \citet[][Section 8.4]{van2016random}). Since the ER graph $\bm{R}_n$ only adds an expected
additional degree of at most $n \theta$ to every node, we see that, when
$\theta$ is small, the PAPER
graph can have heavy-tailed degree distribution without any additional degree correction parameters. 
\end{remark}

\noindent \textbf{Single root inference problem: } Let $\bm{G}_n \sim \text{PAPER}(\alpha, \beta, \theta)$ be a random
graph. Since the nodes of $\bm{G}_n$ are labeled by their arrival
time, we observe only the unlabeled shape of $\bm{G}_n$. Equivalently, we may take our observation to be a labeled graph $\bm{G}^*_n$
whose nodes are labeled by an arbitrary alphabet $\mathcal{U}_n$ of
$n$ elements, i.e., $V(\bm{G}^*_n) = \mathcal{U}_n$. Then, there exists a label bijection $\rho \in
\text{Bi}([n], \mathcal{U}_n)$ such that $\rho \bm{G}_n =
\bm{G}^*_n$. 

The unobserved label bijection $\rho$ captures precisely the arrival
time of the nodes in that for
any $t \in [n]$, the node with label $\rho_t$ in $\bm{G}^*_n$
correspond precisely to node $t$ in $\bm{G}_n$. Therefore, we call any
label bijection in $\text{Bi}([n], \mathcal{U}_n)$ an
\emph{ordering} of the nodes. 

\begin{figure}[htp]
  \centering
  \includegraphics[scale=0.43]{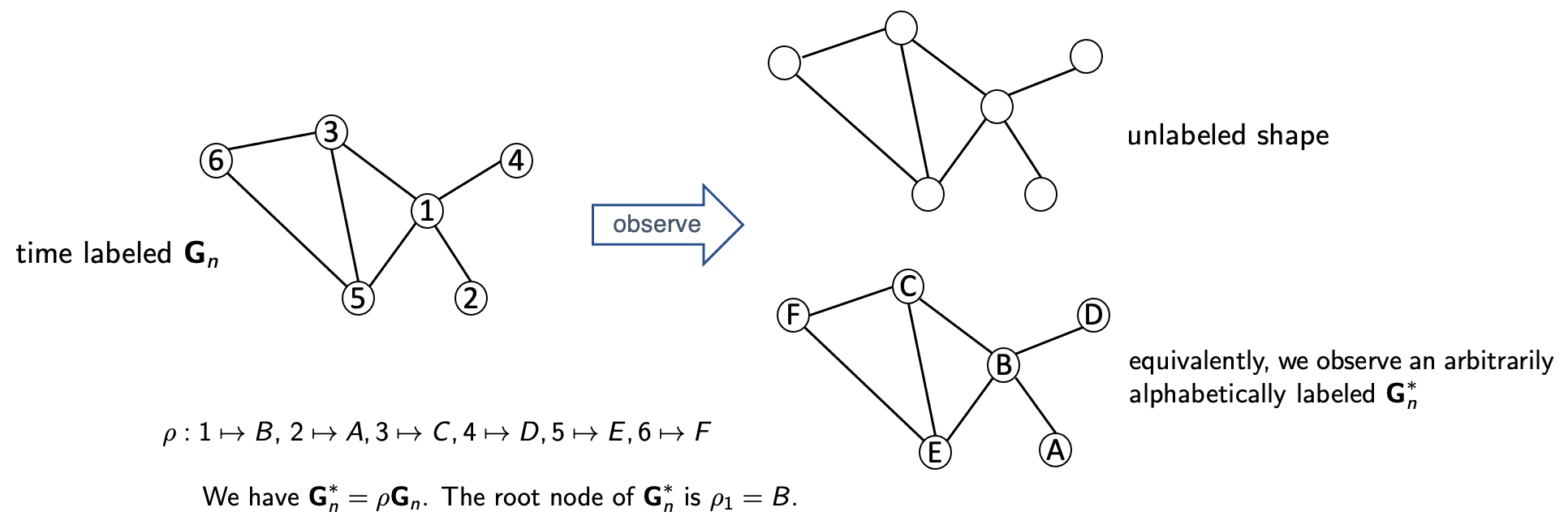}
  \caption{Our observation is the unlabeled shape or alphabetically labeled
    $\bm{G}^*_n$ instead of time labeled $\bm{G}_n$. There exists an unobserved
    ordering $\rho \in \text{Bi}([n], \mathcal{U}_n)$ such that
    $\bm{G}^*_n = \rho \bm{G}_n$.}
  \label{fig:observation1}
\end{figure}

Our goal is to infer various aspects of the latent ordering
$\rho$. We focus on the simpliest example of root inference. Since
the node labeled 1 is the root node in $\bm{G}_n$, it corresponds to
the node labeled $\rho_1$ in $\bm{G}^*_n$. Therefore, the node
$\rho_1 \in \mathcal{U}_n$ is the root node, that is, the first node in the
growth process. To illustrate the setting clearly, we provide a
specfic example in Figure~\ref{fig:observation1}.

\begin{definition}
For $\epsilon \in (0, 1)$, we say that a set $C_\epsilon(\bm{G}_n^*)
\subset \mathcal{U}_n$ is a level $1-\epsilon$ confidence set for the
root node if
\begin{align}
\mathbb{P}\bigl( \rho_1 \in C_\epsilon(\bm{G}_n^*) \bigr)
\geq 1 - \epsilon.  \label{eq:single_root_coverage}
\end{align}
\end{definition}

One may construct a trivial confidence set for the root nodes by taking the set
of all the nodes. We aim therefore to make the confidence set
$C_\epsilon(\cdot)$ as small as possible. Although we focus on the problem of root inference, the approach that
we develop is applicable to more general problems such as inferring
the first two or three nodes or inferring the arrival time of a particular node. 

\begin{figure}[htp]
  \centering
  \includegraphics[scale=.4]{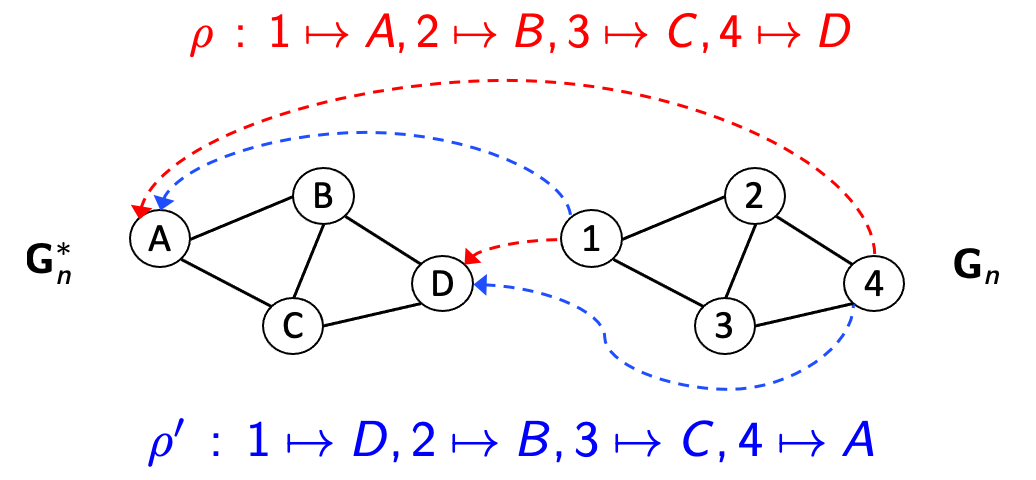}
  \caption{Both $\rho$ (red) and $\rho'$ (blue) are distinct
    bijections in $\text{Bi}([n],
    \mathcal{U}_n)$ but they both satisfy $\bm{G}^*_n = \rho \bm{G}_n
    = \rho' \bm{G}_n$. The root node is $D$ according to $\rho$ but
    $A$ according to $\rho'$. Note that nodes $A$ and $D$ are
    indistinguishable if the labels are removed.}
  \label{fig:remark1}
\end{figure}

\begin{remark} 
  \label{rem:label-equivariance}
  It is important to note that $\bm{G}^*_n$ may have multiple nodes that are indistinguishable once the node labels are removed, which may lead to the paradoxical scenario that which node of $\bm{G}^*_n$ correspond to the true root node depends on the choice of the label bijection $\rho$. Luckily, this is a technical issue that does not pose a problem so long as we restrict ourselves to confidence sets $C_{\epsilon}(\cdot)$ that are labeling equivariant in that they do not depend on the specific node labeling. Labeling equivariance is a very weak condition that only rules out confidence sets that can access side information about the nodes somehow.
  
  Formally, we note that there may exist $\rho, \rho' \in
\text{Bi}([n], \mathcal{U}_n)$ where $\rho_1 \neq \rho'_1$ but both
satisfy $\bm{G}^*_n = \rho \bm{G}_n = \rho' \bm{G}_n$; in other words, root node can only be well-defined up to an automorphism. We illustrate a concrete
example in Figure~\ref{fig:remark1}. We define $C_\epsilon(\cdot)$ to be \emph{labeling equivariant} if, for all $\tau \in \text{Bi}(\mathcal{U}_n, \mathcal{U}_n)$, we
have $\tau C_\epsilon( \bm{G}^*_n) = C_{\epsilon} ( \tau
\bm{G}^*_n)$; if the confidence set algorithm contains randomization
(to break ties for example), then we say it is labeling equivariant if
$\tau C_\epsilon(\bm{G}^*_n) \stackrel{d}{=} C_{\epsilon}( \tau
\bm{G}^*_n)$ for all $\tau \in \text{Bi}(\mathcal{U}_n, \mathcal{U}_n)$. If a confidence set $C_\epsilon(\cdot)$ is labeling equivariant, then for
any $\rho, \rho' \in \text{Bi}([n], \mathcal{U}_n)$ such that
$\bm{G}^*_n = \rho
\bm{G}_n = \rho' \bm{G}_n$, we have that $(\rho' \circ \rho^{-1})
\bm{G}^*_n = \bm{G}^*_n$ and hence, 
\[
  \rho_1 \in C_\epsilon(\bm{G}^*_n) \Leftrightarrow (\rho' \circ
  \rho^{-1}) \rho_1 \in (\rho' \circ \rho^{-1}) C_{\epsilon}(
  \bm{G}^*_n) \Leftrightarrow \rho'_1 \in C_{\epsilon}( (\rho' \circ
  \rho^{-1}) \bm{G}^*_n) \Leftrightarrow \rho'_1 \in 
  C_\epsilon(\bm{G}^*_n).
\]
Therefore, the coverage
probability~\eqref{eq:single_root_coverage} does not depend on the
choice of $\rho$. 
\end{remark}

\subsection{Multiple roots models}

Many real world networks have multiple communities that grow
simultaneously form multiple sources. The APA model
allows for only one root node in the graph but we can augment the
model to describe networks that grow from multiple roots. When there
are $K$ roots, we start the growth process with an initial network of
$K$ singleton nodes and attach each new node to an existing node $w_t$ with
probability proportional to $\beta \cdot \text{(degree of $w_t$)} + \alpha$ as
before.

However, one complication is that when $\alpha = 0$, the probability
of attaching to a singleton node is 0. Thus, for convenience, we give each
root node an unobserved imaginary self-loop edge for the purpose of computing the attachment
probabilities. 

\begin{definition}
\label{def:apak}
  We first define the $\text{APA}(\alpha, \beta, K)$
  model for a random forest of $K$ disjoint component trees: let $K
  \in \mathbb{N}$ and for $t \in S := \{1, 2, \ldots, K\}$ (the set
  $S$ is the set of root nodes), let $\bm{F}_{t}$ be the set of
singleton nodes $1, 2, \ldots, t$. For $t > K$, we define the transition kernel
$\Pb( \bm{F}_t \,|\, \bm{F}_{t-1})$ in the following way: given
$\bm{F}_{t-1}$, we add a new node $t$ and a new random edge $(t, w_t)$ where
the existing node $w_t \in [t-1]$ is chosen with probability
\begin{align}
   \frac{\beta D_{\bm{F}_{t-1}}(w_t) 
  + 2\beta \mathbbm{1}\{ w_t \in S\} + \alpha}{(2\beta + \alpha)(t-1) }. \label{eq:apak}
  \end{align}

We then say that a random graph $\bm{G}_n \sim \text{PAPER}(\alpha, \beta, K, \theta)$
if $\bm{G}_n = \bm{F}_n + \bm{R}_n$ where $\bm{F}_n \sim \text{APA}(\alpha, \beta, K)$ and $\bm{R}_n \sim \text{ER}_\theta$ is an
Erd\H{o}s--R\'{e}nyi random graph independent of $\bm{F}_n$ defined on the
same set of nodes $[n]$. We refer to this setting as the \emph{fixed
  $K$ setting}. In contrast, we refer to the $\text{PAPER}(\alpha,
\beta, \theta)$ model in Section~\ref{sec:single-root-model} as the
\emph{single root setting}.
\end{definition}

We can verify the normalization term~\eqref{eq:apak} by noting that
each root node starts with one imaginary self-loop and that we add one
node and one edge at every iteration. The theory of Polya's urn
immediately implies that the number of nodes in each of the $K$
component trees, divided by $n$, has the asymptotic distribution of
$\text{Dirichlet}(\frac{1}{K}, \ldots, \frac{1}{K})$. \\

To deal with networks in which the number of roots $K$ is unknown, we propose a variation of the
PAPER model with random $K$ number of roots. We can express the model as a sequential growth
process where every newly arrived node has some probability of
becoming a new root. Similar to the fixed $K$ setting, we give each
new root node an imaginary self-loop edge for the purpose of
determining the attachment probabilities. 

\begin{definition}
\label{def:apa3}
  We first define the $\text{APA}(\alpha, \beta,
  \alpha_0)$ model for a random forest graph: let $\bm{F}_1$ be a singleton node and let $S = \{ 1 \}$. For $k > 1$,
we define the transition kernel
$\Pb( \bm{F}_t \,|\, \bm{F}_{t-1})$ in the following way: given
$\bm{F}_{t-1}$, we add a new node $t$. With probability
\[
\frac{\alpha_0}{ (2\beta + \alpha)(t-1) + \alpha_0 },
\]
we let $t$ be a new root node to form $\bm{F}_t$ and add $t$ to set
$S$. Or, we add a new edge $(t, w_t)$ to $\bm{F}_{t-1}$ to obtain
$\bm{F}_t$ where the existing node $w_t \in [t-1]$ is chosen with probability
\[
  \frac{ \beta D_{\bm{F}_{t-1}}(w_t) + \alpha  + 2 \beta \mathbbm{1}\{ w_t \in S \}}
{ (2\beta + \alpha)(t-1) + \alpha_0 }.
\]
Note that the resulting set of root nodes $S \subset [n]$ of
$\bm{F}_n$ is a random set. 

We then say that a random graph $\bm{G}_n$ has the
$\text{PAPER}(\alpha, \beta, \alpha_0, \theta)$ distribution 
if $\bm{G}_n = \bm{F}_n + \bm{R}_n$ where $\bm{F}_n \sim \text{APA}(\alpha, \beta, \alpha_0)$ and $\bm{R}_n \sim \text{ER}(\theta)$ is an
Erd\H{o}s--R\'{e}nyi random graph independent of $\bm{F}_n$ defined on the
same set of nodes $[n]$. We refer to this setting as the \emph{random
  $K$ setting}.
\end{definition}

In the random $K$ setting, each node has some probability of becoming
a new root node and creating a new component tree in the same way as
the Dirichlet process mixture model, which is often called the Chinese
restaurant process. Therefore, the expected number of component trees
is $ (1 + o(1)) \frac{\alpha_0}{(2\beta + \alpha)} \log n $
\citep[][Section 2.2]{crane2016ESF}. \\

\noindent \textbf{Local roots inference problem:} We observe $\bm{G}^*_n = \rho \bm{G}_n$ for an unknown label bijection
$\rho \in \text{Bi}([n], \mathcal{U}_n)$. In both the $\text{APA}(\alpha, \beta,
K)$ and the $\text{APA}(\alpha, \beta, \alpha_0)$ models, the root nodes is a set
$S$ which is fixed to be $[K]$ in the first model and random in the
second model. Intuitively, we interpret $S$ as a set of \emph{local} roots, where each root is central with respect to a specific community or sub-network represented by a component tree in the forest $\bm{F}_n$ in Definition~\ref{def:apak} or~\ref{def:apa3}. The root inference problem is then, for a given
$\epsilon \in (0, 1)$, to construct a confidence set
$C_\epsilon(\bm{G}^*_n)$ such that
\begin{align*}
\mathbb{P}\bigl( \rho S \subseteq C_{\epsilon}( \bm{G}^*_n) \bigr)
  \geq 1 - \epsilon.
\end{align*}
We illustrate this notion of local roots in a synthetic example in Figure~\ref{fig:multipleroots}.

{\color{black}

\begin{remark}
\label{rem:imbalance}
{\bf (Interpretation of community under the PAPER model)} 

The disjoint component trees of $\bm{F}_n$ induce a community structure on the graph $\bm{G}_n$. This way of modeling community by adding Erd\H{o}s--R\'{e}nyi noise to disjoint subgraphs follows the same spirit as stochastic block model (SBM): a SBM with $K$ communities, $p$ as the within-community edge probability, and $q < p$ as the between-community edge probability can be similarly defined as first generating $K$ disjoint $\text{ER}(\frac{p-q}{1-q})$ graphs on each of the communities and then taking the union of that with $\text{ER}(q)$ noisy edges on all the nodes, collapsing multi-edges. 

The PAPER notion of community is however different from that described by SBM. The PAPER notion of community is based on Markovian growth process and intuitively characterized by the imbalance of spanning trees on a network, that is, we believe a network to contain multiple communities if the spanning trees of the network tend to be highly imbalanced (see Figure~\ref{fig:imbalance}), which would suggest that the network is very unlikely to have been formed from a single homogeneous growth process.

\begin{figure}[htp]
\centering
\includegraphics[scale=0.30]{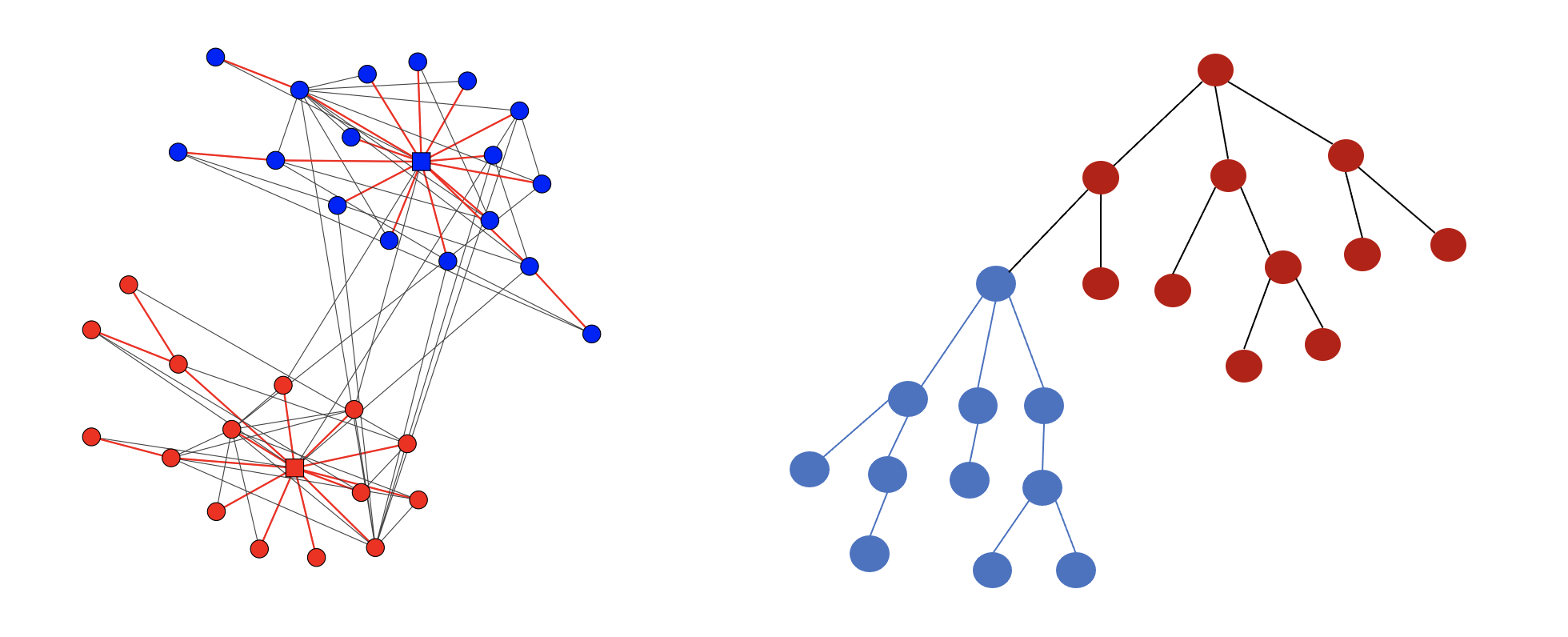}
\caption{The karate club network (\textbf{left}) has two true communities. Most spanning trees of the whole karate club network would be imbalanced (such as the tree on the \textbf{right}), showing that the karate club network is very unlikely to have been formed from a single homogeneous growth process and hence very likely to contain multiple communities.}
\label{fig:imbalance}
\end{figure}

The PAPER model also produces more within-community edges than between-community edges because each community has a spanning tree. However, since a tree on $n$ nodes only has $n-1$ edges, the difference in the within-community edge density and the between-community edge density is diminishingly small when the noise level $\theta$ is of an order larger than $\omega(\frac{1}{n})$. In this case, the peripheral leaf nodes of a community-tree become impossible to cluster but it is still possible to recover the root node of each of the community-trees, as our experimental results show. One disadvantage of the PAPER notion of community is that it is not able to capture non-assortative clusters where nodes in the same clusters are unlikely to form edges.

The PAPER notion of community is appropriate in many application. For example, for a co-authorship network where there exists an underlying growth process, our empirical analysis in Section~\ref{sec:coauthor} shows that the PAPER model captures clusters that accurately reflect salient research communities. We can also combine both notions by a PAPER-SBM mixture model, where we generate a preferential attachment forest $\bm{F}_n$ via the mechanism described in Definition~\ref{def:apak} or~\ref{def:apa3}, then, for every pair of nodes $u$ and $v$, we add a noisy edge $(u,v)$ with probability $\theta_1$ if $u$ and $v$ belong to the same tree in $\bm{F}_n$ and with a different probability $\theta_2$ if $u$ and $v$ belong to different trees. The inference method and algorithm that we develop in this manuscript can extend to such a PAPER-SBM mixture model, but the computational run-time would be substantially slower. We relegate a detailed study of a PAPER-SBM mixture model to a future work. 
\end{remark}
}


\begin{figure}[htp]
  \centering
  \includegraphics[scale=.27]{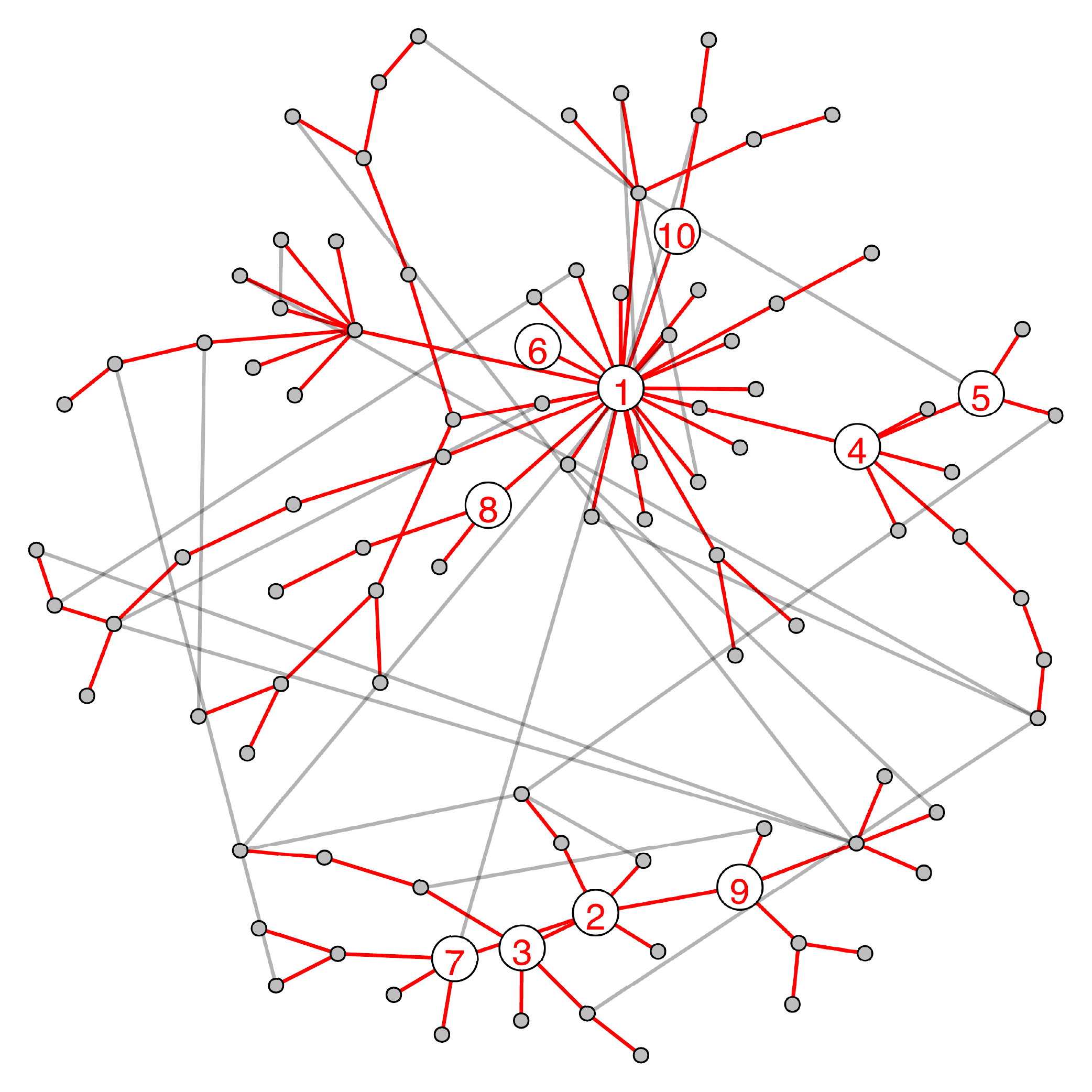}
  \includegraphics[scale=.27]{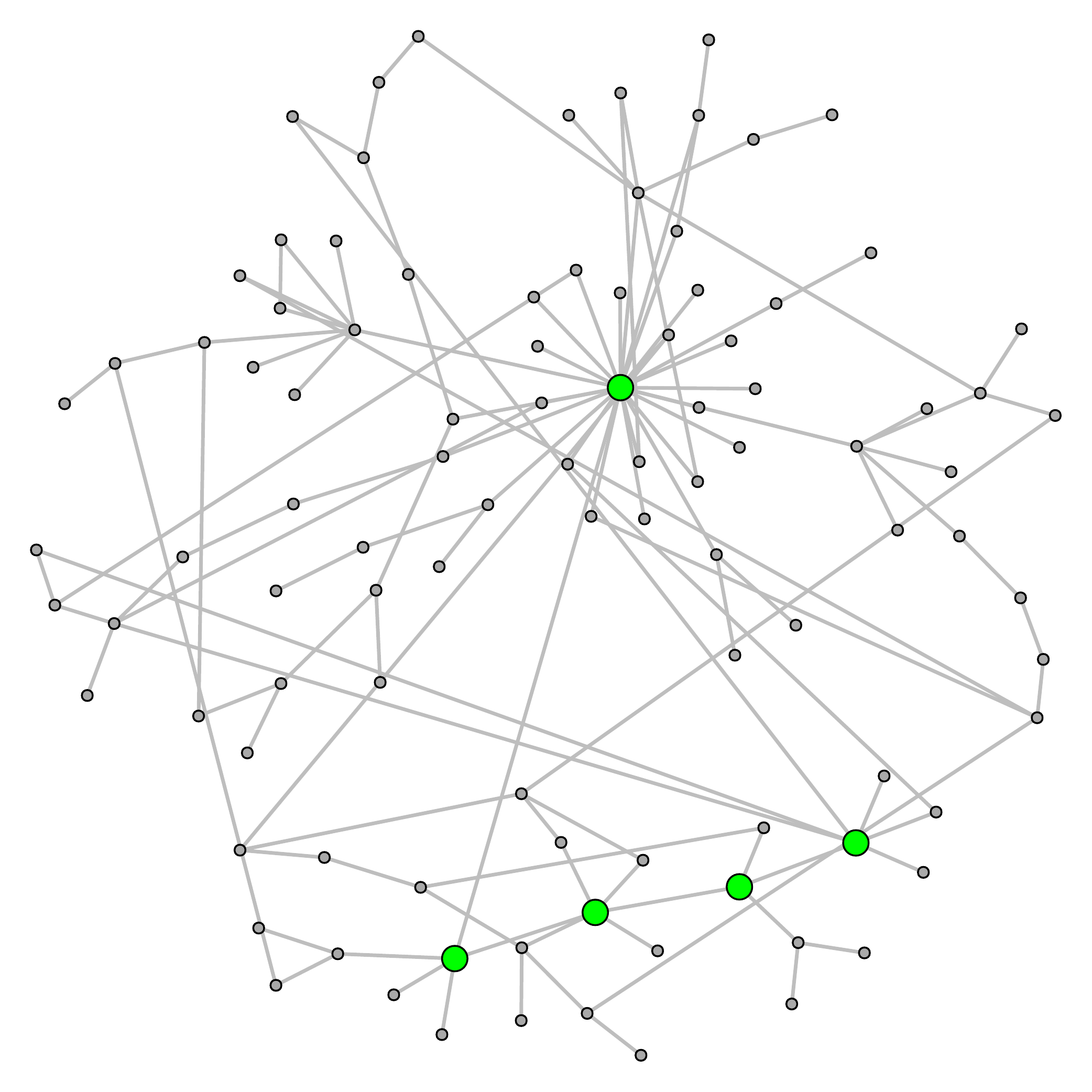}
  \caption{\textbf{Left}: illustration of PAPER model with $K=2$
    underlying trees; nodes have latent time
    ordering (only first 10 orderings shown); the red edges form the
    latent tree while gray edges are Erd\H{o}s--R\'{e}nyi. \textbf{Right}:
    80\% confidence set for the set of root nodes (node number 1 for
    tree 1 and node number 2 for tree 2) constructed from the unlabeled graph.}
  \label{fig:multipleroots}
\end{figure}

{\color{black}
\subsection{Sequential noise models}
\label{sec:seq_noise}

As suggested in Remark~\ref{rem:variations}, PAPER model is a special case of a general Markovian process over a sequence of networks $\bm{G}_1, \bm{G}_2,\ldots, \bm{G}_n$ based on a latent sequence of trees $\bm{T}_1, \bm{T}_2, \ldots, \bm{T}_n$. In the general framework, we specify the
transition kernel
$\mathbb{P}( \bm{G}_t \,|\,\bm{G}_{t-1})$ by specifying two stages:
\begin{enumerate}
\item (tree stage) $\Pb( \bm{T}_t \,|\, \bm{T}_{t-1}, \bm{G}_{t-1})$ which adds one
  node $t$ and one tree edge and
\item (noise stage) $\Pb( \bm{G}_t \,|\, \bm{T}_t, \bm{G}_{t-1})$ which adds
  more random edges to obtain $\bm{G}_t$.
\end{enumerate}
We can of course define $\mathbb{P}(\bm{G}_t \,|\, \bm{G}_{t-1})$
without having an underlying tree but the key insight of our approach is
that augmenting the model with the latent tree $\bm{T}_n$ greatly
facilitates the design of tractable models and inference algorithms
because calculations on trees are easy and fast. In addition, the
latent tree has a real world interpretation as the recruitment
history – a tree edge between nodes $(u, v)$ implies that node $u$ recruited node $v$ into the network.

In the noise stage, if we independently adds noise edges between the new node $t$ and the existing nodes with the same probability $\theta$, then we get back the single root PAPER model. More generally, we can let the noise edge probability depend on the time $t$ and the state of the graph at time $t$. We define the following extension which we refer to as the \textbf{seq-PAPER} model with parameters $(\alpha, \beta, \theta, \talpha, \tbeta)$: 

\begin{definition}
We start with a singleton root node $\bm{T}_1 = \bm{G}_1 = \{ 1 \}$. At time $t=2$, we add node $2$ and attach it to node $1$. At time $t \geq 3:$
\begin{enumerate}
\item (tree stage) We add new node $t$; we select node an existing node $w_t \in [t-1]$ with probability $\frac{\beta D_{\bm{T}_{t-1}}(w_t) + \alpha}{2(t-2)\beta + (t-1)\alpha}$
  and add edge $(t, w_t)$ to $\bm{T}_{t-1}$ to form $\bm{T}_t$;
\item (noise stage) for each existing node $j \in [t-1]$, we add edge $(t, j)$ independently with probability
  \begin{align}
    q_j := \theta \frac{ \tbeta D_{\bm{T}_{t-1}}(j) + \talpha}{2(t-2)\tbeta + (t-1)\talpha} \wedge 1. \label{eq:qj_defn}
  \end{align}
  It is possible that we add the tree edge $(j, w_t)$ in the noise stage in which case we collapse the multi-edge.
\end{enumerate}
\end{definition}
\noindent In general, we may take $\tbeta = \beta$ and $\talpha = \alpha$ but we allow them to be distinct in the model definition for greater flexibility. We discuss parameter estimation in Section~\ref{sec:seq_param} of the Appendix.

When $t$ is large, the independent Bernoulli generative process approximates a Poisson growth model (see e.g. \cite{sheridan2008preferential}) where we first generate $M \sim \text{Poisson}(\theta)$, and then repeat $M$ times the procedure where we draw an existing node $j \in [t-1]$ with probability $q_j$ (also with replacement) and then add the edge $(t, j)$ to the random network, collapsing multi-edges if any are formed. We thus add an average of approximately $\theta$ noise edges at each time step. In contrast, under the PAPER model where the noise edge probability is $\theta$, we add on average $(t-2) \cdot \theta$ noise edges at time $t$. 

The approximation error between the Bernoulli mechanism and the Poisson mechanism, in each iteration $t$, converges to 0 in total variation distance as $t$ increases; see rigorous statement and proof in Proposition~\ref{prop:poisson_approx} of Section~\ref{sec:poisson_approx} in the Appendix.  
{\color{black} However, it is important to note that the two mechanisms could still produce final random graphs whose overall distributions have total variation distance bounded away from 0. For example, UA or LPA trees are known to be sensitive to initialization so that different initial seeds could lead to very different distributions over the final observed graph, see e.g.  \cite{bubeck2015influence} and \cite{curien2015scaling}.} In this work, we prefer the Bernoulli generative process in order to simplify the inference algorithm. Even with the Bernoulli approximation however, inference under the sequential setting is much more computationally intensive than the vanilla PAPER model. 

A more realistic extension of the seq-PAPER model is to replace the tree degree $D_{\bm{T}_{t-1}}(j)$ with the graph degree $D_{\bm{G}_{t-1}}(j)$ in the noise probability~\ref{eq:qj_defn}. This small change unfortunately leads to additional significant slowdown in the resulting inference algorithm; see Remark~\ref{rem:swap} for more detail. We note that an even more sophisticated model of sequential noise is one where the additional noise edges are generated by a random walk mechanism~\citep{bloem2018random}; \cite{bloem2018random} proposes a sequential Monte Carlo inference method which may not scale well to large networks. 

We have so far considered additive noise where new edges are added to the network. We can also model deletion noise where each tree edge is removed from the observed network independently with some probability $\eta > 0$. Having deletion noise under the vanilla PAPER model can adversely increase the size of the confidence set for the root node. However, the seq-PAPER model is much more resilient to deletion noise, especially when $\tbeta = \beta$ and $\talpha = \alpha$ since the noise edges also contain sequential information. To be precisely, we define the $\text{seq-PAPER}^*(\alpha, \beta, \theta, \talpha, \tbeta, \eta)$ as the model where we first generate $\bm{G}_n$ according to the $\text{seq-PAPER}(\alpha, \beta, \theta, \talpha, \tbeta)$ model with latent spanning tree $\bm{T}_n$; we then remove each edge of $\bm{T}_n$ from the final graph $\bm{G}_n$ independently with probability $\eta$.
}

\subsection{Related Work}

Many researchers in statistics~\citep{kolaczyk2009statistical}, computer
science~\citep{bollobas2001degree}, engineering, and
physics~\citep{callaway2000network} have been interested in the probabilistic
properties of various random growth processes of networks, including the preferential attachment
model~\citep{barabasi1999emergence}. Recently however, the specific
problem of root inference on trees has received increased
attention.

These efforts began with the ground-breaking work of \cite{bubeck2017finding,
  bubeck2015influence, bubeck2017trees}, which shows that, given an
observation of an LPA or UA tree of size $n$, for any $\epsilon \in (0, 1]$, one can construct asymptotically valid
confidence sets for the root node with size $K_{LPA}(\epsilon)$ and
$K_{UA}(\epsilon)$ for LPA or UA trees respectively. Importantly and
surprisingly, $K_{LPA}(\epsilon)$ and $K_{UA}(\epsilon)$ do not depend
on $n$ so that the confidence set have size that is
$O(1)$. To construct the confidence sets, \cite{bubeck2017finding}
computes a centrality value for every node, which can for instance be
based on inverse of the size of the maximum subtree of a node (a
concepted sometimes called Jordan centrality on trees, different from
the notion of a Jordan center, which is the node with the minimum
farthest distance to the other nodes); they then sort the nodes by centrality and take the top
$K(\epsilon)$ nodes where the size $K(\epsilon)$ is determined by
probabilistic bounds.

\cite{khim2017confidence} further extends these results
to the setting of uniform attachment over an infinite regular
tree. \cite{banerjee2020root} improves the analysis of Jordan
centrality on trees and derives tight upper and lower bounds on the confidence
set size. \cite{devroye2018discovery, lugosi2019finding} study the
more general problem of seed-tree inference instead of root node
inference. The aforementioned results apply only to tree shaped
networks but very recently, \cite{banerjee2021degree} studies
confidence sets constructed from the degrees of the nodes which
applies to preferential attachment models in which a fixed $m$ edges
are added at every iteration. After the completion of this paper, \cite{briend2022archaeology} propose confidence sets for the root node on a class of uniform-attachment-based general Markovian graphs by detecting anchors of double-cycle subgraphs within the network; they show the confidence set sizes to be $O(1)$ and give explicit bounds in terms of confidence level $\epsilon$.

A line of work in the physics literature also explores the problem of full
or partial recovery of a tree network history~\citep{young2019phase,
  cantwell2019recovering, sreedharan2019inferring}. In computer science and engineering, researchers
have studied the related problem of estimating the source of an
infection spreading over a background network \cite{shah2011rumors,
  fioriti2014predicting, shelke2019source}, with approaches that range
from using Jordan centers, eigenvector centrality, and belief
propagation (see survey in \cite{jiang2016identifying}).

\section{Methodology}
\label{sec:methodology}

Our approach to root inference and related problems is to randomize
the node labels, which induces a posterior distribution over the
latent ordering. 

\subsection{Label randomization}
\label{sec:label-randomization}

\begin{figure}
  \centering
  \includegraphics[scale=.35]{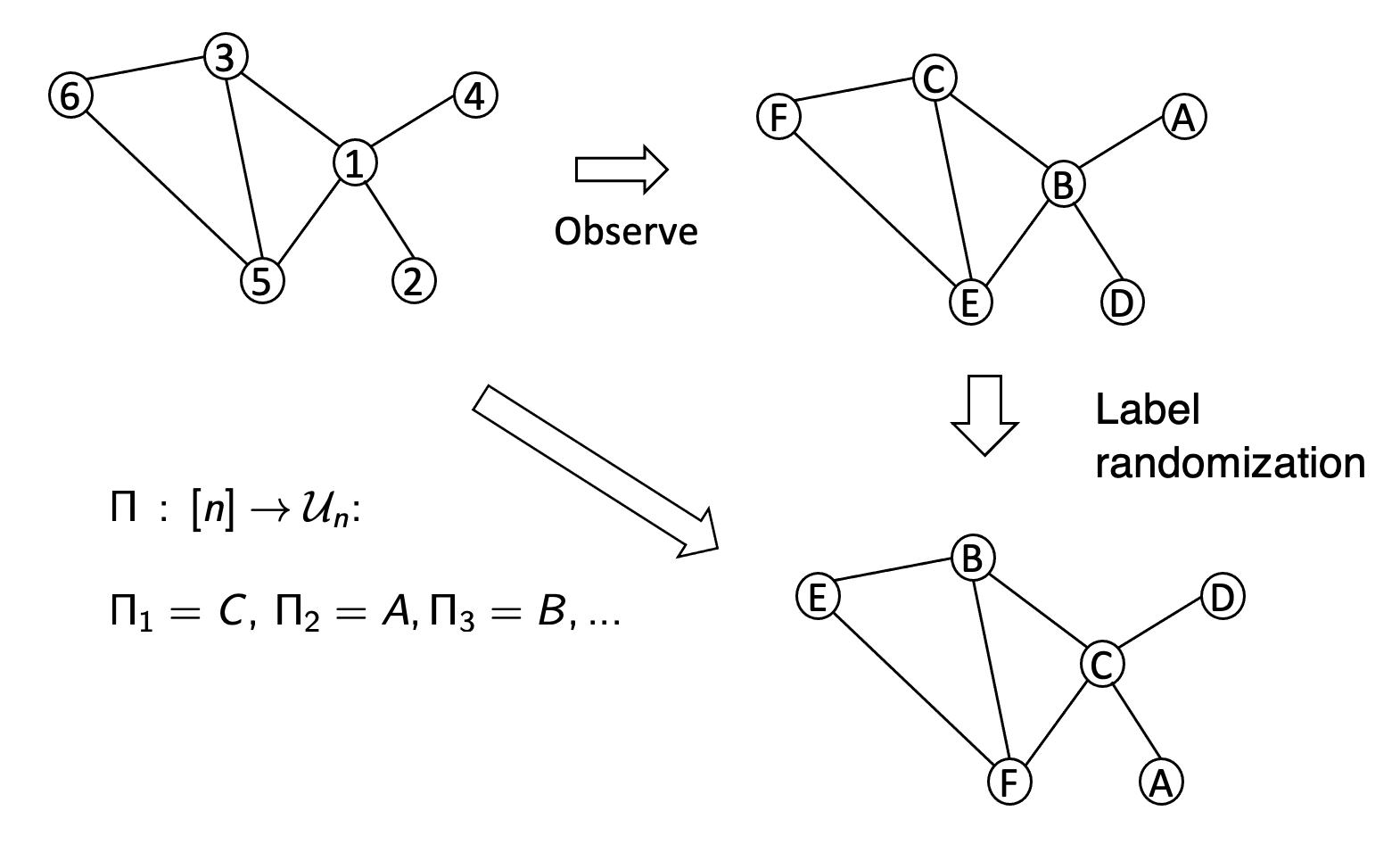}
  \caption{Label randomization induces a random latent arrival
    ordering $\Pi$.}
  \label{fig:randomlabel}
\end{figure}

Suppose $\bm{G}_n$ is a time labeled graph distributed according to a
$\text{PAPER}$ model and $\bm{G}^*_n$ is the alphabetically labeled
observation where $\bm{G}^*_n = \rho \bm{G}_n$ for some label
bijection $\rho \in \text{Bi}([n], \mathcal{U}_n)$. We may
independently generate a random bijection $\Lambda \in
\text{Bi}(\mathcal{U}_n, \mathcal{U}_n)$ and apply it to $\bm{G}^*_n$
to obtain a randomly labeled graph
\[
\tilde{\bm{G}}_n := \Lambda \bm{G}^*_n = \underbrace{(\Lambda \circ
  \rho)}_{\text{$\Pi$}}
\bm{G}_n. 
  \]
By defining $\Pi = \Lambda \circ \rho$, we see that $\tilde{\bm{G}}_n
= \Pi \bm{G}_n$ where $\Pi$ is a random bijection drawn uniformly in
$\text{Bi}([n], \mathcal{U}_n)$ independently of
$\bm{G}_n$ (see Figure~\ref{fig:randomlabel}). We define the randomly labeled latent forest
$\tilde{\bm{F}}_n = \Pi \bm{F}_n$. We may view label randomization as
an augmentation of the probability space. An outcome of a 
$\text{PAPER}$ model is a time labeled graph $\bm{g}_n$ whereas an
outcome after label randomization is a pair $(\tilde{\bm{g}}_n,
\pi)$ where $\tilde{\bm{g}}_n$ is an alphabetically labeled graph and
$\pi$ is an ordering of the nodes. We now make two simple but important observations regarding label randomization.

Our first key observation is that, with respect to $\tilde{\bm{G}}_n$, the
random labeling $\Pi$ describes the arrival time of the nodes in the
sense that if $\Pi_t = u$, then the node with alphabetical label $u$
in $\tilde{\bm{G}}_n$ has the true arrival time $t$. Therefore, in the
single root setting, we may infer the root
node if we can infer $\Pi_1$; in the multiple roots setting, we may
infer the set of root nodes if we can infer $\Pi S$. 

Our second key observation is that label randomization allows us to
define the posterior distribution
\begin{align}
  \Pb( \Pi = \pi \,|\, \tilde{\bm{G}}_n = \tilde{\bm{g}}_n)
  &= \frac{ \Pb( \tilde{\bm{G}}_n = \tilde{\bm{g}}_n \,|\, \Pi = \pi) }{
    \sum_{\pi' \in \text{Bi}([n], \mathcal{U}_n)}
    \Pb( \tilde{\bm{G}}_n = \tilde{\bm{g}}_n \,|\, \Pi = \pi')}
    \label{eq:posterior-history}
\end{align}
which follows because $\Pb(\Pi = \pi) = \frac{1}{n!}$. This posterior
distribution is
supported on the subset of bijection $\pi$ such that $\pi^{-1}
\tilde{\bm{g}}_n$ has non-zero probability under the PAPER model. In the case of the single root PAPER or seq-PAPER model, the support of~\eqref{eq:posterior-history} has a simple
characterization: for every time point $t \in
[n]$, define $\pi_{1:t} \cap \tilde{\bm{g}}_n$ as the
subgraph of $\tilde{\bm{g}}_n$ restricted to nodes in
$\pi_{1:t}$. Then, $\Pb( \Pi = \pi \,|\, \tilde{\bm{G}}_n =
\tilde{\bm{g}}_n) > 0$ if and only if $\pi_{1:t} \cap
\tilde{\bm{g}}_n$ is connected for all $t \in [n]$.

From a Bayesian perspective, label randomization adds a uniform prior
distribution on the arrival ordering of the nodes in the observed 
alphabetically labeled graph $\bm{G}^*_n$; {\color{black} this is sometimes used in Bayesian parameter inference on network models \citep{sheridan2012measuring, bloem2018sampling}}. This prior however is not
subjective. Indeed, we will see in
Theorem~\ref{thm:frequentist-coverage1} that Bayesian
inference statements in our setting directly have frequentist validity
as well and, from Section~\ref{sec:equivalence-mle}, that the
posterior root probability of a node is equal to the 
likelihood of that node being the root node up to normalization.

We describe how to compute~\eqref{eq:posterior-history} tractably in
Section~\ref{sec:algorithm}. For computation, we will also be interested in the posterior
probability over both the ordering $\Pi$ as well as the latent forest
$\tilde{\bm{F}}_n$:
\begin{align}
\Pb( \Pi = \pi, \tilde{\bm{F}} = \tilde{\bm{f}}_n \,|\,
  \tilde{\bm{G}}_n = \tilde{\bm{g}}_n). \label{eq:posterior_history_forest}
\end{align}

In the single root setting, $\tilde{\bm{f}}_n$ is actually a tree,
which we may write as $\tilde{\bm{t}}_n$. It is then clear
that~\eqref{eq:posterior_history_forest} is non-zero only if
$\tilde{\bm{t}}_n$ is a \emph{spanning tree} of $\tilde{\bm{g}}_n$,
i.e., $\tilde{\bm{t}}_n$ is a connected subtree of $\tilde{\bm{g}}_n$
that contains all the vertices.

{\renewcommand{\arraystretch}{1.1}
\begin{table}
  \begin{center}
    \begin{tabular}{|c|c|c|c|}
      \hline
      $\bm{G}_n$ & time labeled graph (unobserved)
      & $\bm{F}_n$ & latent time labeled forest \\
      \hline
      $\bm{G}^*_n$ & observed alpha. labeled graph 
      & $\bm{F}^*_n$ & latent alpha. labeled forest \\
      \hline
      $\tilde{\bm{G}}_n$ & randomly alpha. labeled graph
      & $\tilde{\bm{F}}_n$ & latent randomly alpha. labeled forest \\
      \hline
      $\rho$ & fixed unobserved ordering; $\bm{G}_n^* = \rho \bm{G}_n$
      & $\Pi$ & latent random ordering; $\tilde{\bm{G}}_n = \Pi \bm{G}_n$
      \\
      \hline
      $S$ & time labeled root nodes of $\bm{G}_n$
      & $\tilde{S}$ & latent alpha. labeled root nodes; $\tilde{S}
                      = \Pi S$ \\
      \hline
    \end{tabular}
  \end{center}
  \caption{Quick reference of important notation and definitions.}
  \label{tab:def-ref}
\end{table}}

\subsection{Confidence set for the single root}
\label{sec:confset-root}

To make the ideas clear, we first consider the single root model. Since the root node is the node labeled $\Pi_1$ after label randomization, a
natural approach is to first construct a level $1 - \epsilon$ Bayesian
\emph{credible
  set} for the node $\Pi_1$ by using its posterior distribution, which
we call the posterior root distribution.

More concretely, let $\tilde{\bm{g}}_n$ be an alphabetically labeled graph. For each
node $u \in \mathcal{U}_n$ of $\tilde{\bm{g}}_n$, we define the posterior root probability as $\Pb(
\Pi_1 = u \,|\, \tilde{\bm{G}}_n = \tilde{\bm{g}}_n)$. We sort
the nodes $u_1, \ldots, u_n$ so that
\[
\Pb( \Pi_1 = u_1 \,|\, \tilde{\bm{G}}_n = \tilde{\bm{g}}_n) \geq \Pb(
\Pi_1 = u_2 \,|\, \tilde{\bm{G}}_n = \tilde{\bm{g}}_n) \ldots \geq
\Pb( \Pi_1 = u_n \,|\, \tilde{\bm{G}}_n = \tilde{\bm{g}}_n),
\]
and define
\begin{align}
  L_\epsilon( \tilde{\bm{g}}_n) = \min \biggl\{ k \in [n] \,:\,
  \sum_{i=1}^k \Pb( \Pi_1 = u_i \,|\, \tilde{\bm{G}}_n =
    \tilde{\bm{g}}_n ) \geq 1 - \epsilon \biggr\} \label{eq:Lepsilon}
\end{align}
We then define the
$\epsilon$-credible set as
\begin{align}
B_\epsilon(\tilde{\bm{g}}_n) = \bigl\{ u_1, u_2, \ldots,
u_{L_\epsilon(\tilde{\bm{g}}_n)} \bigr\}, \qquad \text{(breaking ties
  at random).} \label{eq:Bepsilon}
\end{align}
By definition, $B_\epsilon(\tilde{\bm{g}})$ is the smallest set of
nodes with Bayesian coverage at
level $1-\epsilon$ in
that $\Pb( \Pi_1 \in B_{\epsilon}(\tilde{\bm{g}}_n) \,|\,
\tilde{\bm{G}}_n = \tilde{\bm{g}}_n) \geq 1 - \epsilon$. In general, credible sets do not have valid frequentist confidence coverage. However, our next theorem shows that in our setting, the credible set $B_\epsilon$ is in fact an honest confidence set in that $\mathbb{P}\{ \text{root node} \in B_\epsilon(\bm{G}_n^*)\} \geq 1 - \epsilon$.

\begin{theorem}
\label{thm:frequentist-coverage1}
Let $\bm{G}_n \sim \text{PAPER}(\alpha, \beta, \theta)$ or $\text{seq-PAPER}(\alpha,\beta, \theta, \talpha, \tbeta)$ and let
$\bm{G}^*_n$ be the alphabetically labeled observation. Let $\rho \in
\text{Bi}([n], \mathcal{U}_n)$ be any label bijection such that $\rho \mathbf{G}_n = \mathbf{G}^*_n$. We have that, for any $\epsilon \in (0, 1)$,
\[
\mathbb{P}\bigl\{ \rho_1 \in B_\epsilon(\mathbf{G}^*_n) \bigr\} \geq 1 - \epsilon.
\]
\end{theorem}

The proof is very similar to that of \citet[][Theorem~1]{crane2021inference}. Since the
proof is short, we provide it here for readers' convenience. 

\begin{proof}
We first claim that $B_\epsilon(\cdot)$ is labeling-equivariant
(cf. Remark~\ref{rem:label-equivariance}) in the sense that for any
$\tau \in \text{Bi}(\mathcal{U}_n, \mathcal{U}_n)$ and any
alphabetically labeled graph $\tilde{\bm{g}}_n$, we have that $\tau
B_{\epsilon}(\tilde{\bm{g}}_n) \stackrel{d}{=} B_\epsilon(\tau
\tilde{\bm{g}}_n)$ (note that $B_\epsilon(\cdot)$ uses randomization
to break ties). Indeed, since $(\Pi, \tilde{\bm{G}}_n)
\stackrel{d}{=} (\tau^{-1} \circ \Pi, \tau^{-1} \tilde{\bm{G}}_n)$, we have that, for any $u \in \mathcal{U}_n$,
\begin{align*}
\mathbb{P}( \Pi_1 = u \,|\, \tilde{\bm{G}}_n = \tilde{\bm{g}}_n) = 
\mathbb{P}( \Pi_1 = \tau(u) \,|\,
\tilde{\mathbf{G}}_n = \tau\tilde{\mathbf{g}}_n). 
\end{align*}
Therefore, for any $u, v \in \mathcal{U}_n$, we have that $\mathbf{P}( \Pi_1 = u \,|\, \tilde{\mathbf{G}}_n = \tilde{\mathbf{g}}_n) \geq \mathbf{P}( \Pi_1 = v \,|\, \tilde{\mathbf{G}}_n = \tilde{\mathbf{g}}_n)$ if and only if 
$\mathbf{P}( \Pi_1 = \tau(u) \,|\, \tilde{\mathbf{G}}_n = \tau \tilde{\mathbf{g}}_n) \geq \mathbf{P}( \Pi_1 = \tau(v) \,|\, \tilde{\mathbf{G}}_n = \tau \tilde{\mathbf{g}}_n)$.
Since $B_\epsilon(\mathbf{G}^*_n)$ is constructed by taking the top elements of $\mathcal{U}_n$ that maximize the cumulative posterior root probability, the claim follows. 

Now, let $\rho \in \text{Bi}([n], \mathcal{U}_n)$ be such that $\rho
\mathbf{G}_n = \mathbf{G}^*_n$ and let $\Lambda$ be a random bijection
drawn uniformly in $\text{Bi}(\mathcal{U}_n, \mathcal{U}_n)$ and let
$\Pi = \Lambda \circ \rho$. Then,
\begin{align*}
\mathbb{P}( \rho_1 \in B_\epsilon(\mathbf{G}^*_n)) &=
\mathbb{P}( \rho_1 \in B_\epsilon(\rho \mathbf{G}_n) ) \\
&= \mathbb{P}\bigl\{ (\Lambda \circ \rho)_1 \in B_\epsilon( (\Lambda \circ \rho) \mathbf{G}_n) \,|\, \Lambda = \text{Id} \bigr\}\\
&= \mathbb{P}\bigl\{ (\Lambda \circ \rho)_1 \in B_\epsilon( (\Lambda \circ \rho) \mathbf{G}_n) \bigr\} \\
&= \mathbb{P}( \Pi_1 \in B_\epsilon(\tilde{\mathbf{G}}_n)) \geq 1 - \epsilon,
\end{align*}
where the penultimate equality follows from the labeling-equivariance of $B_\epsilon$ and where the last inequality follows because $\mathbf{P}( \Pi_1 \in B_\epsilon(\tilde{\mathbf{G}}_n) \,|\, \tilde{\mathbf{G}}_n = \tilde{\mathbf{g}}_n) \geq 1 - \epsilon$ for any labeled tree $\tilde{\mathbf{g}}_n$ (with labels in $\mathcal{U}_n$) by the definition of $B_{\epsilon}$.
\end{proof}

\begin{remark}
  \label{rem:likelihood}
We show in Theorem~\ref{thm:equivalence-mle} of the appendix that the
posterior root probability $\Pb( \Pi_1 = u \,|\, \tilde{\bm{G}}_n
= \tilde{\bm{g}}_n)$ is equal to the likelihood of node $u$ being the
root node on observing the unlabeled shape of
$\tilde{\bm{g}}_n$. Therefore, the set
$B_{\epsilon}(\tilde{\bm{g}}_n)$ is in fact the maximum likelihood confidence
set. Because the likelihood in this setting is complicated to even write
down, we leave all the details to
Section~\ref{sec:equivalence-mle} of the appendix. 
\end{remark}

\begin{remark}
One may see from the proof that
Theorem~\ref{thm:frequentist-coverage1} applies more broadly then just PAPER models. It in fact applies to any random graph $\bm{G}_n$ whose nodes are labeled by $\{1,2,\ldots, n\}$. For the PAPER model, the integer labels encode arrival time and thus contain information about the graph. In a model where the integer labels are uninformative of the graph connectivity structure, Theorem~\ref{thm:frequentist-coverage1} is still valid although the posterior probability $\mathbb{P}(\Pi_1 = \cdot \,|\, \tbm{G}_n = \tbm{g}_n)$ would be uniform. {\color{black} A reviewer of this paper also pointed out that Theorem~\ref{thm:frequentist-coverage1} is related to the classical literature on invariant/equivariant estimation where credible sets constructed from uniform (Haar) priors may also be valid confidence sets; see e.g. \citet[][Theorem 6.78]{schervish2012theory}.}
\end{remark}

\subsection{Confidence set for local roots}
\label{sec:confset-rootset}

First consider the fixed $K$ setting where $\bm{G}_n \sim
\text{PAPER}(\alpha, \beta, \theta, K)$; let $\Pi$ be a uniformly random
ordering in $\text{Bi}([n], \mathcal{U}_n)$ and let
$\tilde{\bm{G}}_n = \Pi \bm{G}_n$. The latent
set of root nodes of $\tilde{\bm{G}}_n$ in this case is $\tilde{S} :=
\Pi S = \{ \Pi_1, \ldots, \Pi_K\}$. We then define the posterior root
probability for any node $u \in \mathcal{U}_n$ as
\begin{align*}
\Pb( u \in \tilde{S} \,|\, \tilde{\bm{G}}_n = \tilde{\bm{g}}_n),
\end{align*}
that is, the probability that node $u$ is an element of the latent
root set $\tilde{S}$. 

To form the credible set $B_\epsilon(\tilde{\bm{g}}_n) \subseteq \mathcal{U}_n$, we sort the
nodes by the posterior root probabilities
\begin{align}
  \Pb( u_1 \in \tilde{S} \,|\, \tilde{\bm{G}}_n = \tilde{\bm{g}}_n) \geq
  \Pb( u_2 \in \tilde{S} \,|\, \tilde{\bm{G}}_n = \tilde{\bm{g}}_n) \geq
  \ldots \geq \Pb( u_n \in \tilde{S} \,|\, \tilde{\bm{G}}_n =
  \tilde{\bm{g}}_n). \label{eq:posterior_rootset_sort}
\end{align}

We may then take $B_\epsilon(\tilde{\bm{g}}_n)$ to be the smallest set
of nodes such that $P\bigl( \tilde{S} \subsetneq B_\epsilon(\tilde{\bm{g}}_n)  \,|\, \tilde{\bm{G}}_n =
\tilde{\bm{g}}_n\bigr) \leq \epsilon$. More precisely, define the integer
\begin{align}
L_\epsilon( \tilde{\bm{g}}_n) = \min \biggl\{ k \in [n] \,:\,
  &
  \sum_{i=k+1}^n \Pb( u_i \in \tilde{S} \,|\, \tilde{\bm{G}}_n =
    \tilde{\bm{g}}_n ) \leq \epsilon \biggr\} \label{eq:posterior_rootset_size}
\end{align}
and then define the credible set as
\begin{align}
B_\epsilon(\tilde{\bm{g}}_n) = \bigl\{ u_1, u_2, \ldots,
u_{L_\epsilon(\tilde{\bm{g}}_n)} \bigr\} \qquad \text{(breaking ties
  at random).}\label{eq:posterior_rootset_credset}
\end{align}

In the $\text{PAPER}(\alpha, \beta, \alpha_0,
\theta)$ model where the number of roots $K$ is random, the set of root nodes is $\tilde{S} = \Pi S$ which comprises,
according to the ordering $\Pi$, of the node that is first to arrive in each of
the component trees of $\tilde{\bm{F}}_n$. We may then sort the nodes
as in~\eqref{eq:posterior_rootset_sort}, compute
$L_\epsilon(\tilde{\bm{g}}_n)$ as
in~\eqref{eq:posterior_rootset_size} and
$B_\epsilon(\tilde{\bm{g}}_n)$ as
in~\eqref{eq:posterior_rootset_credset}.

Similar to Theorem~\ref{thm:frequentist-coverage1}, we may show that
$B_\epsilon(\cdot)$ in fact also has frequentist coverage at the same
level $1 - \epsilon$. 

\begin{theorem}
\label{thm:frequentist-coverage2}
Let $\bm{G}_n \sim \text{PAPER}(\alpha, \beta, K, \theta)$ or
$\text{PAPER}(\alpha, \beta, \alpha_0, \theta)$ and let
$\bm{G}^*_n$ be the alphabetically labeled observation. Let $\rho \in
\text{Bi}([n], \mathcal{U}_n)$ be any label bijection such that $\rho
\mathbf{G}_n = \mathbf{G}^*_n$ and let $S \subset [n]$ be the time
labels of the root
nodes (see Definitions~\ref{def:apak} and~\ref{def:apa3}). We have that, for any $\epsilon \in (0, 1)$,
\[
  \mathbb{P}\bigl\{ \rho S \subseteq B_\epsilon(\mathbf{G}^*_n) \bigr\} \geq 1 - \epsilon.
\]
\end{theorem}

\begin{proof}
The proof is very similar to that of
Theorem~\ref{thm:frequentist-coverage1}. First, since the random set $\tilde{S}$ is a
function of the random ordering $\Pi$ in the fixed $K$ setting and a
function of both the random ordering $\Pi$ and the random forest
$\tilde{\bm{F}}_n$, we write $\tilde{S}(\Pi)$ or $\tilde{S}(\Pi,
\tilde{\bm{F}}_n)$ to be precise. 

We then observe that $\tilde{S}(\Pi)$ in the fixed $K$ setting or $\tilde{S}(\Pi,
\tilde{\bm{F}}_n)$ in the random $K$ setting, are labeling equivariant
in that for any $\tau \in \text{Bi}(\mathcal{U}_n, \mathcal{U}_n)$, we
have that $\tilde{S}( \tau^{-1} \Pi) = \tau^{-1} \tilde{S}(\Pi)$ or, in the random
$K$ setting, $\tilde{S}(\tau^{-1} \Pi, \tau^{-1} \tilde{\bm{F}}_n) = \tau^{-1}
\tilde{S}(\Pi, \tilde{\bm{F}}_n)$. Therefore, since $(\Pi, \tilde{\bm{G}}_n)
\stackrel{d}{=} (\tau^{-1} \Pi, \tau^{-1} \tilde{\bm{G}}_n)$ for any
$\tau \in \text{Bi}(\mathcal{U}_n, \mathcal{U}_n)$, we have
$\tilde{S}(\Pi, \tilde{\bm{F}}_n) \stackrel{d}{=} \tau^{-1} \tilde{S}(\Pi,
\tilde{\bm{F}}_n)$ and thus, for any $u \in \mathcal{U}_n$,
\[
\Pb( u \in \tilde{S} \,|\, \tilde{\bm{G}}_n = \tilde{\bm{g}}_n) = \Pb( \tau(u) \in \tilde{S} \,|\, \tilde{\bm{G}}_n = \tau \tilde{\bm{g}}_n).
\]
The rest the proof proceeds in an identical manner to that of
Theorem~\ref{thm:frequentist-coverage1}.
\end{proof}

When there are multiple roots, an alternative way of inferring the root set is to
construct the confidence set $B_\epsilon(\cdot)$ as a set of subsets of
the nodes and then require that $\tilde{S} \in B_\epsilon$ with probability at
least $1 - \epsilon$. We can take the same approach to construct such
confidence set over sets but it becomes much more computationally
intensive to compute them in practice. 

\subsection{Combinatorial interpretation}
\label{sec:combinatorial}

Before we describe the Gibbs sampling algorithm for computing the
posterior root probabilities $\Pb( \Pi_1 = u \,|\, \tilde{\bm{G}}_n =
\tilde{\bm{g}}_n)$, we provide an intuitive combinatorial interpretation of the
posterior root probability in the single root PAPER model (Definition~\ref{def:paper}). The definitions and calculations here are also important for
deriving the algorithm in Section~\ref{sec:algorithm}.\\

\textbf{The noiseless case:} We first consider the simpler setting in
which we can observe the tree $\tilde{\bm{T}}_n$ (with a single root) distributed according to
the APA model. In this case, we have
\[
\Pb( \Pi_1 = \cdot \,|\, \tilde{\bm{T}}_n =
\tilde{\bm{t}}_n) = \sum_{\pi \,:\, \pi_1 = u} \Pb( \Pi = \pi \,|\, \tilde{\bm{T}}_n =
\tilde{\bm{t}}_n).
\]

\begin{figure}
  \centering
  \includegraphics[scale=0.4]{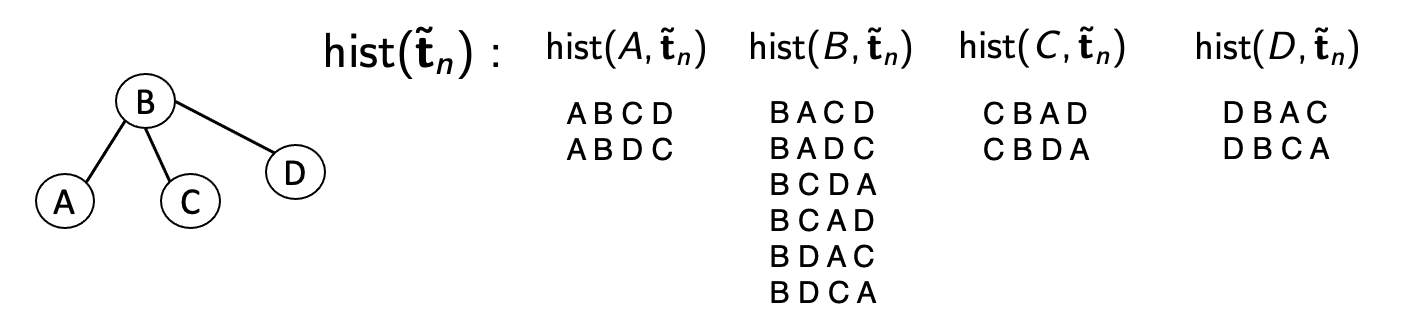}
  \caption{All histories of a tree with 4 nodes.}
  \label{fig:treehist}
\end{figure}

Recall that $\tilde{\bm{T}}_n = \Pi
\bm{T}_n$ where $\bm{T}_n$ is a random time labeled tree with
$\text{APA}(\alpha, \beta)$ distribution and $\Pi$ is an independent
uniformly random ordering in $\text{Bi}([n], \mathcal{U}_n)$. The distribution $\Pb( \Pi = \pi \,|\,
\tilde{\bm{T}}_n = \tilde{\bm{t}}_n)$ is supported on a subset of the
the bijections $\text{Bi}([n], \mathcal{U}_n)$ because $\pi^{-1}
\tilde{\bm{T}}_n$ must be a valid time labeled tree (also called
\emph{recursive tree} in discrete mathematics). To be precise, we define the histories of $\tilde{\bm{t}}_n$ as
\begin{align*}
  \text{hist}(\tilde{\bm{t}}_n)
  &:= \bigl\{ \pi \in \text{Bi}([n],
    \mathcal{U}_n) \,:\, \Pb( \bm{T}_n = \pi^{-1} \tilde{\bm{t}}_n ) > 0
    \bigr \}, \text{ and } \\
    h(\tilde{\bm{t}}_n) &:= | \text{hist}(\tilde{\bm{t}}_n)|
\end{align*}
as the number of distinct histories. Since the APA tree distribution
assigns a non-zero probability to any valid time labeled trees, we see
that $\text{hist}(\tilde{\bm{t}}_n)$ contains the elements $\pi$ of
$\text{Bi}([n], \mathcal{U}_n)$ such that for all $t \in [n]$, the subtree restricted only to nodes in $\pi_{1:t}$,
i.e. $\tilde{\bm{t}}_n \cap \pi_{1:t}$, is connected. Thus,
$\text{hist}(\tilde{\bm{t}}_n)$ is the set of bijections $\pi$ which represent a
valid arrival ordering for the nodes of the given tree
$\tilde{\bm{t}}_n$. Similarly, we define, for any node $u \in \mathcal{U}_n$,
\begin{align*}
\text{hist}(u, \tilde{\bm{t}}_n) := \bigl\{ \pi \in
  \text{hist}(\tilde{\bm{t}}_n) \,:\, \pi_1 = u \bigr\} \\
  h(u, \tilde{\bm{t}}_n) := | \text{hist}(u, \tilde{\bm{t}}_n) |,
\end{align*}
as histories of $\tilde{\bm{t}}_n$ that start at node $u$. We
illustrate an example of the set of histories for a simple tree in Figure~\ref{fig:treehist}.

By definition, $\Pb( \Pi = \cdot \,|\,
\tilde{\bm{T}}_n = \tilde{\bm{t}}_n)$ is supported on
$\text{hist}(\tilde{\bm{t}}_n)$. For most values of $\alpha$ and
$\beta$, the posterior distribution is in fact uniform over
$\text{hist}(\tilde{\bm{t}}_n)$:
\begin{proposition} \label{prop:apa-posterior}
  \citep[][Theorem 4 and Proposition
  3]{crane2021inference} 
  Let $\alpha, \beta$ be two real numbers such that either (1) $\beta \geq 0$ and $\alpha \geq -\beta$ or (2) $\beta < 0$ and $\alpha = -D \beta$ for some integer $D \geq 2$. 
  Suppose $\bm{T}_n \sim \text{APA}(\alpha, \beta)$. Let $\Pi$ be a
  uniformly random ordering taking value in $\text{Bi}([n],
  \mathcal{U}_n)$ and let $\tilde{\bm{T}}_n = \Pi \bm{T}_n$. Then,
\begin{align}
\Pb( \Pi = \pi \,|\, \tilde{\bm{T}}_n = \tilde{\bm{t}}_n ) =
  \frac{1}{h(\tilde{\bm{t}}_n)} \mathbbm{1}\{ \pi \in
  \text{hist}(\tilde{\bm{t}}_n) \}. \label{eq:single_tree_uniform}
\end{align}
\end{proposition}
The full proof of Proposition~\ref{prop:apa-posterior} is in \cite{crane2021inference} but we
give a short justification here: the posterior is uniform because
$\Pb( \Pi = \pi \,|\, \tilde{\bm{T}}_n =
\tilde{\bm{t}}_n ) = \frac{ \Pb( \tilde{\bm{T}}_n = \tilde{\bm{t}}_n
  \,|\, \Pi = \pi) \frac{1}{n!}}{ \Pb(\tilde{\bm{T}}_n =
  \tilde{\bm{t}}_n)} = \frac{\Pb( \bm{T}_n = \pi^{-1}
  \tilde{\bm{t}}_n) \frac{1}{n!}}{\Pb(\tilde{\bm{T}}_n =
  \tilde{\bm{t}}_n)}$. Moreover, the probability $\Pb( \bm{T}_n = \pi^{-1}
  \tilde{\bm{t}}_n)$ is actually the same for any $\pi \in
  \text{hist}(\tilde{\bm{t}}_n)$ by Proposition~\ref{prop:apa-prob1}.

  By Proposition~\ref{prop:apa-posterior}, we have that
  \begin{align*}
    \Pb( \Pi_1 = u \,|\, \tilde{\bm{T}}_n = \tilde{\bm{t}}_n)
    &= \frac{h(u, \tilde{\bm{t}}_n)}{h(\tilde{\bm{t}}_n)}.
  \end{align*}

\begin{figure}
  \centering
  \includegraphics[scale=0.5]{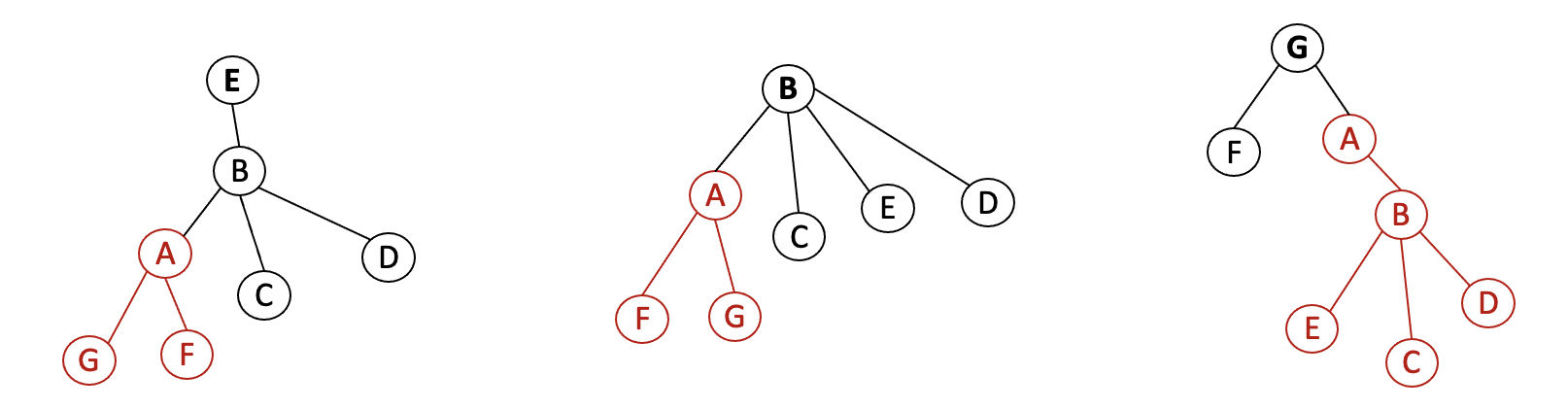}
  \caption{Same tree $\tilde{\bm{t}}_n$ in three rooted
    orientations. Left: $\tilde{\bm{t}}_n^{(E)}$ rooted at $E$; the
    subtree of $A$ (denoted $\tilde{\bm{t}}^{(E)}_A$) contains nodes
    $A, F, G$; node $A$ is the parent of $F, G$. Center:
    $\tilde{\bm{t}}_n^{(B)}$ rooted at $B$; the subtree of $A$
    (denoted $\tilde{\bm{t}}^{(B}_A$) contains nodes $A, F, G$; node
    $A$ is the parent of $F,G$. Right:
    $\tilde{\bm{t}}_n^{(G)}$ rooted at $G$; the subtree of $A$
    (denoted $\tilde{\bm{t}}^{(G)}_A$) contains nodes $A, B, E, C, D$;
    node $A$ is the parent of $B$.}
  \label{fig:rootedtree}
\end{figure}

Therefore, we need only count the histories $h(u, \tilde{\bm{t}}_n)$
for every node $u \in \mathcal{U}_n$. We give a well-known
characterization of $h(u, \tilde{\bm{t}}_n)$ that leads to a linear
time algorithm for counting the size of the histories: define, for any node $u, v \in \mathcal{U}_n$, the tree
  $\tilde{\bm{t}}_v^{(u)}$ as the subtree of node $v$ where we view
  the whole tree as being rooted (hanging from) node $u$;
  $\tilde{\bm{t}}^{(u)}_u$ is thus the entire tree rooted at $u$. See
  Figure~\ref{fig:rootedtree} for an example. We then have that, by
  \cite{knuth1997art} or \cite{shah2011rumors}, 
  \begin{align}
    h(u, \tilde{\bm{t}}_n) = n! \prod_{v \in \mathcal{U}_n} \frac{1}{|
    \tilde{\bm{t}}^{(u)}_v|}. \label{eq:subtreeprod}
  \end{align}
Therefore, we can compute $h(u, \tilde{\bm{t}}_n)$ by viewing
$\tilde{\bm{t}}_n$ as being rooted at $u$ and taking the product of
the inverse of the sizes of all the subtrees. By using the fact that
$h(u, \tilde{\bm{t}}_n)$ can be directly computed from $h(u',
\tilde{\bm{t}}_n)$ for any neighbor $u'$ of $u$, \cite{shah2011rumors}
derive an $O(n)$ algorithm for computing the size of the histories
over all roots $\{ h(u, \tilde{\bm{t}}_n)\}_{u \in \mathcal{U}_n}$,
which we give in Section~\ref{sec:methodology-supp} of the appendix for readers' convenience.\\

\begin{figure}
  \centering
  \includegraphics[scale=0.4]{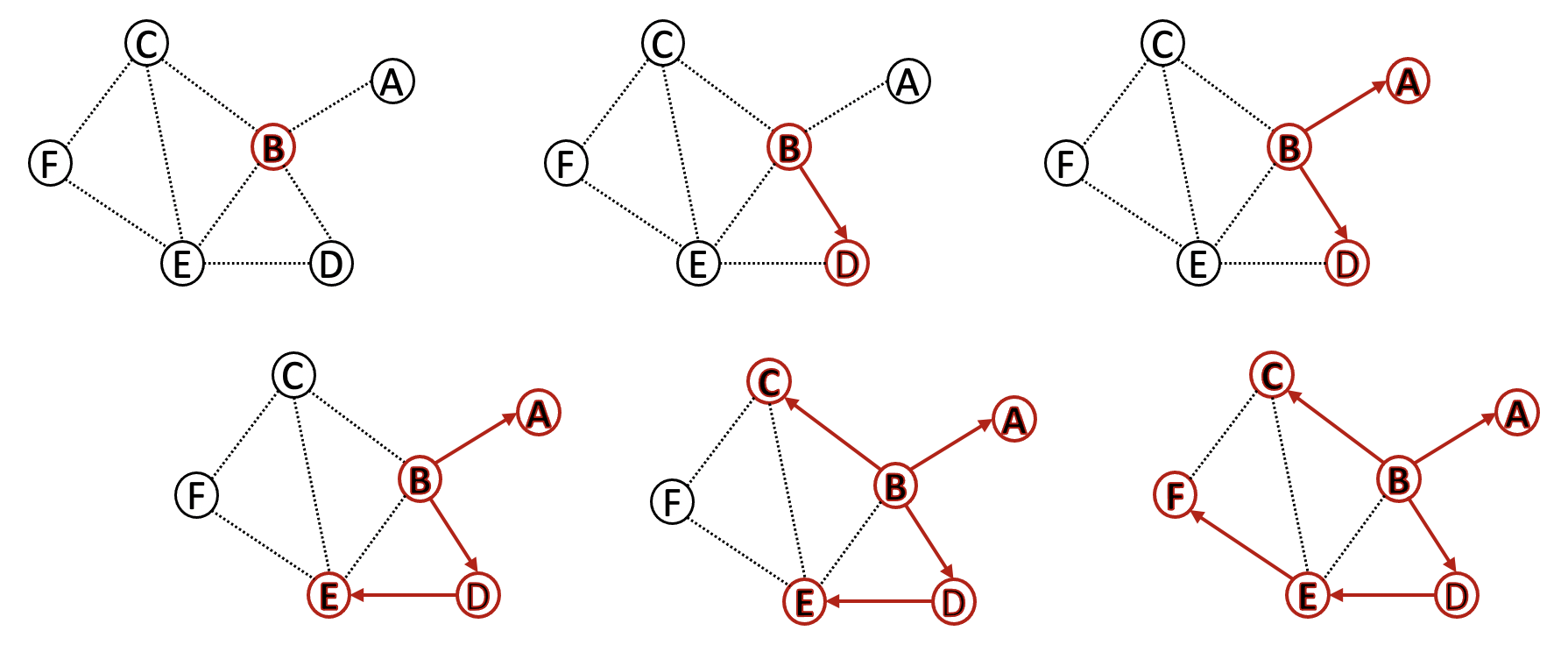}
  \caption{One possible growth realization starting from node B.}
  \label{fig:realization}
\end{figure}

\textbf{The general case:} Now suppose we have the label randomized graph $\tilde{\bm{G}}_n$ from
the PAPER model. We then have that
\begin{align}
  \Pb( \Pi_1 = u \,|\, \tilde{\bm{G}}_n = \tilde{\bm{g}}_n)
  &= \sum_{\tilde{\bm{t}}_n \subseteq \tilde{\bm{g}}_n} \sum_{\pi \in \text{hist}(u,
    \tilde{\bm{t}}_n) } \Pb( \Pi = \pi, \tilde{\bm{T}}_n =
    \tilde{\bm{t}}_n \,|\, \tilde{\bm{G}}_n = \tilde{\bm{g}}_n)
    \nonumber \\
  & \propto \sum_{\tilde{\bm{t}}_n \subseteq \tilde{\bm{g}}_n} \sum_{\pi \in \text{hist}(u,
    \tilde{\bm{t}}_n) } \Pb( \Pi = \pi, \tilde{\bm{T}}_n =
    \tilde{\bm{t}}_n) \underbrace{\Pb( \tilde{\bm{G}}_n = \tilde{\bm{g}}_n \,|\,
    \tilde{\bm{T}}_n = \tilde{\bm{t}}_n, \Pi = \pi)}_{\binom{n(n-1)/2
    - (n-1)}{m-(n-1)}^{-1}}.  \nonumber \\
  &\propto \sum_{\tilde{\bm{t}}_n \subseteq \tilde{\bm{g}}_n} \sum_{\pi \in \text{hist}(u,
    \tilde{\bm{t}}_n) } \Pb( \tilde{\bm{T}}_n =
    \tilde{\bm{t}}_n \,|\, \Pi = \pi) = \sum_{\tilde{\bm{t}} \subseteq \tilde{\bm{g}}_n} \sum_{\pi \in \text{hist}(u,
    \tilde{\bm{t}}) } \Pb( \bm{T}_n = \pi^{-1} \tilde{\bm{t}}_n ), \label{eq:finalroot}
\end{align}
where, in the outer summation, we require $\tilde{\bm{t}}_n$ to be a
subtree of $\tilde{\bm{g}}_n$ with $n$ nodes, that is, we require
$\tilde{\bm{t}}_n$ to be a spanning tree of $\tilde{\bm{g}}_n$ (see~\eqref{eq:spanningtree}). If $\bm{T}_n$
has the uniform attachment distribution ($\alpha=1, \beta=0$), then we
have that $\Pb( \bm{T}_n = \pi^{-1} \tilde{\bm{t}}_n ) =
\frac{1}{(n-1)!}$ by Proposition~\ref{prop:apa-prob1} and hence,
\[
  \Pb(
\Pi_1 = u \,|\, \tilde{\bm{G}}_n = \tilde{\bm{g}}_n) \propto \sum_{
  \tilde{\bm{t}}_n \subseteq \tilde{\bm{g}}_n} h(u,
\tilde{\bm{t}}_n).
\]
Thus, the posterior root probability of $u$ is
simply proportional to the number of all possible realizations of
growth process that start from node $u$ and end up with graph
$\tilde{\bm{g}}_n$; see Figure~\ref{fig:realization}. When $\bm{T}_n$
has the LPA distribution $(\alpha=0, \beta=1)$, then $\Pb( \bm{T}_n =
\pi^{-1} \tilde{\bm{t}}_n )$ depends on the degree sequence of the
tree $\tilde{\bm{t}}_n$ so that the posterior root probability is proportional to a
weighted count of all possible growth realizations.

\section{Algorithm}
\label{sec:algorithm}

The inference approach that we described in
Sections~\ref{sec:confset-root} and~\ref{sec:confset-rootset} requires
computing posterior probabilities such as the posterior root probability $P( \Pi_1 = u \,|\,
\tilde{\bm{G}}_n = \tilde{\bm{g}}_n)$ for a fixed alphabetically
labeled graph $\tilde{\bm{g}}_n$. In this section, we derive a Gibbs
sampling algorithm to generate an ordering
$\pi \in \text{Bi}([n], \mathcal{U}_n)$ and a forest $\tilde{\bm{f}}_n$ according to
the posterior probability
\begin{align}
\Pb( \Pi = \pi, \tilde{\bm{F}}_n = \tilde{\bm{f}}_n \,|\,
  \tilde{\bm{G}}_n = \tilde{\bm{g}}_n). \label{eq:posterior_history_tree2}
\end{align}

As discussed towards the end of Section~\ref{sec:label-randomization},
in the single root setting, the posterior
probability~\eqref{eq:posterior_history_tree2} over $\Pi, \tilde{\bm{F}}_n$ is non-zero only if
$\tilde{\bm{f}}_n$ is a spanning tree of the graph
$\tilde{\bm{g}}_n$. We formally define the set of spanning trees of a
connected graph $\tilde{\bm{g}}_n$ as 
\begin{align}
  \mathcal{T}(\tilde{\bm{g}}_n) := 
  \bigl\{ \tilde{\bm{f}}_n \,:\, \text{$\tilde{\bm{f}}_n$ is connected
  subtree of $\tilde{\bm{g}}_n$ and $V(\tilde{\bm{f}}_n) =
  V(\tilde{\bm{g}}_n)$} \bigr\}. \label{eq:spanningtree}
\end{align}

We note that $\mathcal{T}(\tilde{\bm{g}}_n)$ is non-empty if and only
if $\tilde{\bm{g}}_n$ is connected. For the multiple roots setting, we define the spanning forest of
$\tilde{\bm{g}}_n$ with $K$ components as
\begin{align*}
  \mathcal{F}_K( \tilde{\bm{g}}_n) :=
  \bigl\{ \tilde{\bm{f}}_n \,:\, \text{$\tilde{\bm{f}}_n$ is sub-forest
  of $\tilde{\bm{g}}_n$ with $K$ disjoint component trees and $V(\tilde{\bm{f}}_n) =
  V(\tilde{\bm{g}}_n)$} \bigr\}
\end{align*}
so that $\mathcal{F}_1( \tilde{\bm{g}}_n) =
\mathcal{T}(\tilde{\bm{g}}_n)$. Then, for the fixed $K$ roots model,
the posterior probability~\eqref{eq:posterior_history_tree2} is
non-zero only if $\tilde{\bm{f}}_n \in
\mathcal{F}_K(\tilde{\bm{g}}_n)$ and for the random $K$ roots model,
probability~\eqref{eq:posterior_history_tree2} is non-zero only if
$\tilde{\bm{f}}_n \in \mathcal{F}( \tilde{\bm{g}}_n) := \cup_{K=1}^n \mathcal{F}_K(\tilde{\bm{g}}_n)$.

The value of the posterior probability~\eqref{eq:posterior_history_tree2} depends on the
parameters of the model, e.g. $\alpha, \beta, \theta$ in the single
root setting. We provide an estimation procedure for these parameters
in Section~\ref{sec:parameter-estimation} but for now, to keep the presentation simple, we assume
that all parameters are known.

Our Gibbs sampler alternates between two stages:
\begin{enumerate}
\item[(A)] We fix the forest $\tilde{\bm{f}}_n$ and generate an ordering
  $\pi$ with probability $\mathbb{P}( \Pi = \pi \,|\, \tilde{\bm{G}}_n =
    \tilde{\bm{g}}_n, \tilde{\bm{F}}_n = \tilde{\bm{f}}_n)$.
\item[(B)] We fix the ordering $\pi$ and generate a new forest $\tilde{\bm{f}}_n$
  by iteratively sampling a new parent for each of the nodes. 
\end{enumerate}

We give the details for stage A in the next section and for stage B in
Section~\ref{sec:sample-forest}. 

\begin{remark}
In Section~\ref{sec:collapsed-gibbs}, we give an alternative collapsed
Gibbs sampling algorithm in which we collapse stage (A) so that we
only sample the roots instead of the whole history $\pi$. The
collapsed Gibbs sampler requires fewer iterations to converge but each
iteration is more computationally intensive. Practically, the sampling
algorithm that we present in Section~\ref{sec:sample-ordering}
and~\ref{sec:sample-forest} appears to be faster except for the random
$K$ roots model on some data sets. 
\end{remark}

\subsection{Sampling the ordering}
\label{sec:sample-ordering}
In this section, we provide an algorithm for the first stage of the
Gibbs sampler where we sample an ordering. We fix a spanning forest $\tilde{\bm{f}}_n$ of the
observed graph $\tilde{\bm{g}}_n$, let $K$ be the number of component
trees of $\tilde{\bm{f}}_n$, and let $m = |E(\tilde{\bm{g}}_n)|$ be
the number of edges of $\bm{g}_n$. We have that
\begin{align}
  \mathbb{P}( \Pi = \pi \,|\, \tilde{\bm{G}}_n =
    \tilde{\bm{g}}_n, \tilde{\bm{F}}_n = \tilde{\bm{f}}_n) 
  \propto \Pb( \Pi = \pi \,|\, \tilde{\bm{F}}_n = \tilde{\bm{f}}_n)
    \Pb( \tilde{\bm{G}}_n = \tilde{\bm{g}}_n \,|\, \tilde{\bm{F}}_n =
    \tilde{\bm{f}}_n, \Pi = \pi). \label{eq:cond_order_prob}
\end{align}
Under the non-sequential noise PAPER models, since the non-forest edges of $\tilde{\bm{G}}_n$ are
    independent Erd\H{o}s--R\'{e}nyi random edges, we have $ \Pb( \tilde{\bm{G}}_n = \tilde{\bm{g}}_n \,|\, \tilde{\bm{F}}_n =
    \tilde{\bm{f}}_n, \Pi = \pi) = \binom{ \binom{n}{2} - (n-K)}{m -
      (n-K)}^{-1}$ and may thus ignore the non-forest edges and consider only on the
    posterior probability $\Pb( \Pi = \pi \,|\, \tilde{\bm{F}}_n =
      \tilde{\bm{f}}_n)$ when sampling $\pi$. In the sequential noise $\text{seq-PAPER}$ model, the $\Pb( \tilde{\bm{G}}_n = \tilde{\bm{g}}_n \,|\, \tilde{\bm{F}}_n =
    \tilde{\bm{f}}_n, \Pi = \pi)$ term must be taken into account but can be computed efficiently. We give the detailed algorithms for each of the settings.\\

\noindent \textbf{Single root setting:} 
In the single root setting, $\tilde{\bm{f}}_n$ is
connected and hence a tree; we thus change to the notation
$\tilde{\bm{t}}_n := \tilde{\bm{f}}_n$ to be consistent with the
notation used in Definition~\ref{def:apa1}.

Hence, by our discussion in Section~\ref{sec:combinatorial}, sampling $\pi$ according to $\Pb( \Pi = \cdot \,|\,
\tilde{\bm{T}}_n = \tilde{\bm{t}}_n)$ is equivalent to sampling $\pi$ uniformly from
$\text{hist}(\tilde{\bm{t}}_n)$. \cite{crane2021inference} and also
\cite{cantwell2021inference} derive a
procedure to sample uniformly from $\text{hist}(\tilde{\bm{t}}_n)$ and
we provide a concise description of the procedure here for the
readers' convenience. 

To generate $\pi$ uniformly from $\text{hist}(\tilde{\bm{t}}_n)$, we
generate the first node $\pi_1$ by taking the set of all nodes and drawing a node $u$ with
probability 
\begin{align}
  \Pb( \Pi_1 = u \,|\,
\tilde{\bm{T}}_n = \tilde{\bm{t}}_n) = \frac{h(u,
  \tilde{\bm{t}}_n)}{h(\tilde{\bm{t}}_n)}.
  \label{eq:single_tree_root}
\end{align}
The entire collection
$\{ h(u, \tilde{\bm{t}}_n) \}_{u \in \mathcal{U}_n}$ can be computed
in $O(n)$ time (c.f. Section~\ref{sec:combinatorial} and~\ref{sec:methodology-supp}) and thus we require at most
$O(n)$ time to generate the first node $\pi_1$.

To generate the subsequent ordering $\pi_{2:n}$, we view the tree
$\tilde{\bm{t}}_n$ as being rooted at $\pi_1$ and use the notation
$\tilde{\bm{t}}_n^{(\pi_1)}$ make the root explicit. For each node $v \in
\mathcal{U}_n$, we define $\tilde{\bm{t}}_v^{(\pi_1)}$ as the subtree
of the node $v$, viewing the whole tree as being rooted at node
$\pi_1$. We give an example of these definitions in Figure~\ref{fig:rootedtree}.

Then, by \citet[Proposition 9][]{crane2021inference}, for every $t \in [n-1]$,
\begin{align}
\Pb( \Pi_{t+1} = v \,|\, \tilde{\bm{T}}_n = \tilde{\bm{t}}_n, \Pi_{1:t}
  = \pi_{1:t}
  )
  &= \left\{ \begin{array}{cc} \frac{|\tilde{\bm{t}}^{(\pi_1)}_v|}{n -
               t + 1} & \text{ if $v$ is a neighbor of $\pi_{1:t}$ in $\tilde{\bm{t}}_n$} \\
               0 & \text{ else}
                   \end{array} \right. \label{eq:iterative_hist}
\end{align}
One may verify this by showing that the probability of generating a
particular ordering is $\frac{1}{n!} \prod_{v \in \mathcal{U}_n} |
\tilde{\bm{t}}_n^{(u)}| = \frac{1}{h(u, \tilde{\bm{t}}_n)}$
by~\eqref{eq:subtreeprod}. 

\begin{figure}
  \centering
  \includegraphics[scale=0.4]{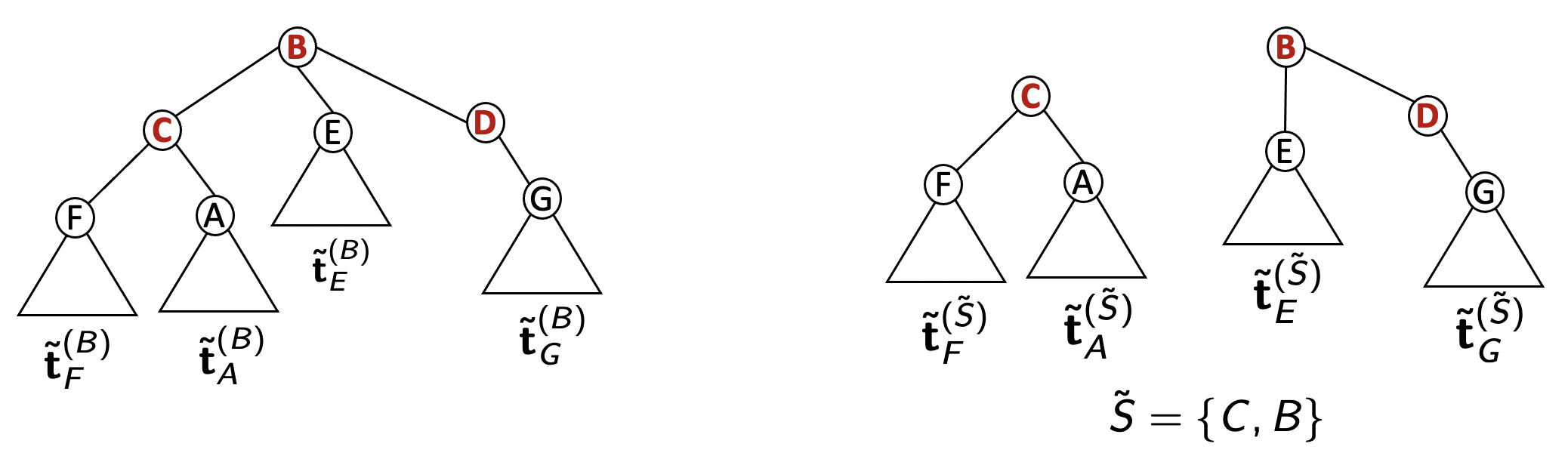}
  \caption{Example of sampling an ordering. In both cases, suppose
    $\pi_{1:3} = \{B, C,
    D\}$, then draw $\pi_4$ from the neighbors $\{F, A, E, G\}$ with
    probability proportional to the size of their subtrees. }
  \label{fig:tree-sampling}
\end{figure}

Thus, we may generate
$\pi_2$ by considering all neighbors of $\pi_1$ in $\tilde{\bm{t}}_n$
and drawing a node $v$ with probability proportional to the size of
its subtree $|
\tilde{\bm{t}}^{(u_1)}_v|$ and similar for $\pi_3$, $\pi_4$, etc. The
entire sampling process can be efficiently done by generating a
permutation uniformly at random and modifying it in place so that it
obeys the $\text{hist}(\tilde{\bm{f}}_n)$ constraint. We summarize
this in Algorithm~\ref{alg:draw-history} with $K=1$ and also give a
visual illustration in Figure~\ref{fig:tree-sampling}. The runtime of
the sampling algorithm is upper bounded by $O( n
\text{diam}(\tilde{\bm{t}}_n))$ \citep[][Proposition
10]{crane2021inference}. Trees generated by the $\text{APA}(\alpha,
\beta)$ model have diameter $O_p(\log n)$ (see
e.g. \citet[][Theorem~6.32]{drmota2009random} and
\citet[][Theorem~18]{bhamidi2007universal}) and the
overall runtime is therefore $O(n\log n)$. The computational complexity is the same under the fixed $K$ setting and the
random $K$ setting. \\

\noindent \textbf{Fixed $K$ roots setting:} For the $\text{PAPER}(\alpha, \beta, K, \theta)$ model, we may generate
from $\Pb( \Pi = \cdot \,|\, \tilde{\bm{F}}_n = \tilde{\bm{f}}_n)$ in
a similar way. In this case, $\tilde{\bm{f}}_n$ is a forest that
contains $K$
disjoint component trees, which we denote by $\tilde{\bm{t}}^1, \ldots,
\tilde{\bm{t}}^K$. We first generate a root for each component tree. For each $k \in
[K]$, we draw $u^k \in V(\tilde{\bm{t}}^k)$ with probability 
\begin{align}
\frac{h(u^k, \tilde{\bm{t}}^k) (\beta D_{\tilde{\bm{t}}^k}(u^k) + \beta +
  \alpha)(\beta D_{\tilde{\bm{t}}^k}(u^k) + \alpha)}
{ \sum_{v \in V(\tilde{\bm{t}}^k)} h(v, \tilde{\bm{t}}^k) (\beta D_{\tilde{\bm{t}}^k}(v) + \beta +
  \alpha)(\beta D_{\tilde{\bm{t}}^k}(v) + \alpha)}. \label{eq:multi-tree-root}
\end{align}
We note that~\eqref{eq:multi-tree-root} is different from the corresponding probability in the single
tree setting~\eqref{eq:single_tree_root} because we give each root
node an imaginary self-loop edge. We leave the detailed derivation
of~\eqref{eq:multi-tree-root} to Section~\ref{sec:rootderivation} of the appendix.  

We let $\tilde{s} = \{u^1, \ldots, u^k\}$ denote the set of roots that we have
generated. By the definition of the $\text{PAPER}(\alpha, \beta, K,
\theta)$ model (Definition~\ref{def:apak}), the root nodes $\tilde{s}$
occupy the first $K$ positions of the ordering $\pi$ and we thus let
$\pi_{1:K}$ be the elements of $\tilde{s}$ placed in a random
ordering. 

Next, we view each component
tree $\tilde{\bm{t}}^k$ as being rooted at $u_k$ and, for every
node $v \in V(\tilde{\bm{f}}_n)$, we denote the subtree of node $v$ by
$\tilde{\bm{t}}_v^{(\tilde{s})}$. We then generate
$\pi_{(K+1):n}$ according to probability~\eqref{eq:iterative_hist} where we use the size of the subtree
$|\tilde{\bm{t}}^{(\tilde{s})}_v|$. This is equivalent to
generating a full history (excluding the root node) for every tree and
then interleaving them at random. We again summarize the whole
procedure in Algorithm~\ref{alg:draw-history}.\\

\noindent \textbf{Random $K$ roots setting:} Now consider the random $K$ roots setting with the $\text{PAPER}(\alpha, \beta, \alpha_0, \theta)$
model and suppose $\tilde{\bm{f}}_n$ comprises of $K$ disjoint
trees $\tilde{\bm{t}}^1, \ldots, \tilde{\bm{t}}^K$. We again generate the
set of roots $\tilde{s} = \{u^1, \ldots, u^K\}$ by drawing $u^k$ from
$\tilde{\bm{t}}^k$ with probability~\eqref{eq:multi-tree-root}. In
contrast with the fixed $K$ roots setting, the root nodes $u^1, \ldots,
u^K$ need not occupy the first $K$ positions of the ordering $\pi$. 

To generate the ordering $\pi$, we first choose $u^k \in \tilde{s}$ with probability
$|\tilde{\bm{t}}^k|$ and set $\pi_1 = u^k$. We then draw $\pi_{2:n}$
iteratively using the conditional distribution
\begin{align}
\Pb( \Pi_{t+1} = v \,|\, \tilde{\bm{F}}_n = \tilde{\bm{f}}, \Pi_{1:t}
  = \pi_{1:t}
  )
  &= \left\{ \begin{array}{cc} \frac{|\tilde{\bm{t}}^{(\tilde{s})}_v|}{n -
               t + 1} & \text{ if $v$ is a neighbor of $\pi_{1:t}$ in
                        $\tilde{\bm{f}}_n$ or if $v \in \tilde{s}$} \\
               0 & \text{ else}
                   \end{array} \right. \label{eq:iterative_hist2}
\end{align}
We note that for a root node $u^k \in \tilde{s}$, the subtree
$\tilde{\bm{t}}^{(\tilde{s})}_{u^k}$ is precisely the whole tree
$\tilde{\bm{t}}^k$. We summarize this procedure in Algorithm~\ref{alg:draw-history}.\\

{\color{black}

\noindent \textbf{Sequential noise setting:} Under the seq-PAPER model described in Section~\ref{sec:seq_noise}, we no longer have a direct sampling algorithm to draw from $\mathbb{P}(\Pi = \cdot \,|\, \tbm{G}_n = \tbm{g}_n, \tbm{T}_n = \tbm{t}_n)$ because we have to take into account the $\mathbb{P}( \tbm{G}_n = \tbm{g}_n \,|\, \tbm{T}_n = \tbm{t}_n, \Pi = \pi)$ term in~\eqref{eq:cond_order_prob}. For seq-PAPER models, we propose instead a Metropolis--Hastings algorithm to update $\pi$ by sampling new transpositions. 

Let $\pi$ be the current sample of arrival ordering. To generate a new proposal $\pi^*$, we randomly choose a pair $j, k \in \{2,\ldots, n\}$ and construct $\pi^*$ by swapping the $j$-th and the $k$-th entries of $\pi$, that is, $\pi^*_j = \pi_k$ and $\pi^*_k = \pi_j$ and all other entries are equal. If $\pi^* \notin \text{hist}(\tbm{t}_n)$, then we reject the proposal; otherwise, we accept it with probability
\begin{align}
1 \wedge \frac{\mathbb{P}( \tbm{G}_n = \tbm{g}_n \,|\, \Pi = \pi^*, \tbm{T}_n = \tbm{t}_n)}{ \mathbb{P}( \tbm{G}_n = \tbm{g}_n \,|\, \Pi = \pi^, \tbm{T}_n = \tbm{t}_n)},
\label{eq:swap_accept_prob}
\end{align}
which follows because $\mathbb{P}(\Pi = \pi \,|\, \tbm{T}_n = \tbm{t}_n) = \mathbb{P}(\Pi = \pi^* \,|\, \tbm{T}_n = \tbm{t}_n)$. The ratio in~\eqref{eq:swap_accept_prob} has a complicated expression but can be computed in time proportional to only the degrees, with respect to $\tbm{g}_n$, of $\pi_j, \pi_k$, and the parent nodes $\text{pa}(\pi_j), \text{pa}(\pi_k)$, where the notion of parent node is defined in~\eqref{eq:pa_node}. We give a detailed description of how to efficiently compute~\eqref{eq:swap_accept_prob} and determine whether $\pi^* \in \text{hist}(\tbm{t}_n)$ in Section~\ref{sec:seq_noise_detail} of the Appendix; in particular, see Section~\ref{sec:seq_noise_transposition} which uses results from Section~\ref{sec:seq_noise_preliminary}. Even with our efficient implementation however, updating $\pi$ by sampling transpositions is considerably slower than sampling $\pi$ directly via~\eqref{eq:iterative_hist}.  

The transposition sampler does not change the root node since $j, k$ are not allowed to take on the value $1$. To sample a new root node, we fix $k_0 \in \mathbb{N}$ and generate a new proposal $\pi^*$ by shuffling the first $k_0$ entries of $\pi$. We then accept $\pi^*$ if it is a valid history and with probability~\eqref{eq:swap_accept_prob}. Finally, we note that under the $\text{seq-PAPER}^*$ model with tree edge removal, our method for sampling $\pi$ is exactly the same. Since we condition on $\tbm{T}_n$, it makes no difference whether we have deletion noise or not.

\begin{remark}
\label{rem:swap}
\cite{sheridan2012measuring} and \cite{bloem2018sampling} use the idea of swapping \emph{adjacent} elements of an ordering $\pi$ for a Poisson growth attachment models and a sequential edge-growth model referred to as Beta NTL respectively. In contrast, under the seq-PAPER model, we can compute non-adjacent swap proposal probabilities efficiently and hence, we can explore the permutation space of $\pi$ faster. This is because the seq-PAPER is a simpler model and also because we restrict ourselves to a spanning tree, which simplifies many parts of the calculations. We note that sampling $\pi$ through non-adjacent pair swaps can also be used for the model $\bm{G}_n = \bm{T}_n + \bm{R}_n$ where $\bm{T}_n$ is not shape-exchangeable, for instance when the attachment probability is $\phi( D_{\bm{T}_{t-1}}(w_t) )$ for some non-affine function $\phi(\cdot)$ instead of the affine expression given in~\eqref{eq:pa_tree_prob}. Finally, We emphasize that inference for the vanilla PAPER model is significantly faster than any form of swapping-based Metropolis samplers since it directly samples the entire ordering. 
\end{remark}

}

\begin{algorithm}[htp]
\caption{Generating $\pi \in \text{hist}(\tilde{\bm{f}}_n)$
  according to $\mathbb{P}(\Pi = \pi \,|\, \tilde{\bm{F}}_n =
  \tilde{\bm{f}}_n)$ in ER noise settings.}
\label{alg:draw-history}
\textbf{Input:} Labeled forest $\tilde{\bm{f}}_n$ with $K$ trees,
denoted $\tilde{\bm{t}}^{1}, \ldots, \tilde{\bm{t}}^{K}$. \\
\textbf{Output:} $\pi \in \text{hist}(\tilde{\bm{f}}_n)$.
\begin{algorithmic}[1]
\For{$k=1,2,\ldots, K$}:
  \State Choose node $u^k \in V(\tilde{\bm{t}}^{(k)})$ with probability
  \eqref{eq:single_tree_root} with $\text{PAPER}(\alpha, \beta,
  \theta)$ model and with probability~\eqref{eq:multi-tree-root}
  under $\text{PAPER}(\alpha, \beta, K, \theta)$ or
  $\text{PAPER}(\alpha, \beta, \alpha_0, \theta)$. 
  \EndFor
  \State Let $\tilde{s} = \{u^1, u^2, \ldots, u^K\}$ be the set of roots, and 
  \begin{itemize}
    \item under $\text{PAPER}(\alpha, \beta,
\theta)$, let $\pi_1 = u^1$ and let
$t_0 = 2$,
\item under $\text{PAPER}(\alpha, \beta, K, 
\theta)$, let $\pi_{1:K} = \tilde{s}$ in a random ordering and let $t_0
= K+1$.
\item under $\text{PAPER}(\alpha, \beta, \alpha_0, 
\theta)$, choose $u^k \in \tilde{s}$ with probability
$|\tilde{\bm{t}}^k|/n$, let $\pi_1 = u^k$, let $t_0 = 2$.
\end{itemize}
\State Generate $\pi_{t_0:n}$ as a uniformly random permutation of
$\mathcal{U}_n \backslash \pi_{1:(t_0 - 1)}$. 
\For{$t = t_0, t_0 +1, \ldots, n$}: 
  \State Let $v_1 = \pi_t$, $v_2 = \text{pa}(v_1),\, \ldots,
  v_k = \text{pa}(v_{k-1})$ where $k$ is the largest integer such that
  $v_1, v_2, \ldots, v_k \notin \pi_{1:(t-1)}$.
  \Comment{$\text{pa}(v)$
    denotes the parent of $v$ with respect to $\tilde{\bm{f}}_n$ rooted at $\tilde{s}$.}
  \State Set $\pi_{t} = v_k$, $t_k = \pi^{-1}(v_k)$, and $\pi_{t_k} = v_1$. 
\EndFor
\end{algorithmic}
\end{algorithm}

\subsection{Sampling the forest}
\label{sec:sample-forest}

In this section, we describe stage B of the Gibbs sampling
algorithm. For a fixed ordering $\pi$ and a spanning forest
$\tilde{\bm{f}}_n$, we may obtain a set of roots $\tilde{s}$ for each of the
component trees of $\tilde{\bm{f}}_n$ by taking the earliest node
(according to $\pi$) of each tree. Viewing $\tilde{\bm{f}}_n$ as being
rooted at $\tilde{s}$ induces parent-child relationships between all
the nodes. 

To define the parent-child relationship formally, let $\tilde{\bm{f}}_n$ be a forest with disjoint component trees
$\tilde{\bm{t}}^1, \ldots, \tilde{\bm{t}}^K$ and let
$\tilde{s} = \{u^1, u^2, \ldots, u^K\}$ be a set of root nodes such that $u^k
\in V(\tilde{\bm{t}}^k)$. Let $u$ be any node not in $\tilde{s}$ and suppose $u \in
V(\tilde{\bm{t}}^k)$. There
exists a unique node $v \in V(\tilde{\bm{t}}^k)$ such that $v$ is a
neighbor of $u$ in $\tilde{\bm{f}}_n$ and that the unique path from
$u$ to the root $u^k$ contains $v$. We say $v$ the \emph{parent node}
of $u$ and write
\begin{align}
pa(u) \equiv pa_{\tilde{\bm{f}}^{(\tilde{s})}_n}(u) = \text{parent of $u$ with respect to
  $\tilde{\bm{f}}^{(\tilde{s})}$}. \label{eq:pa_node}
\end{align}
For a root node $u \in \tilde{s}$, we let $pa(u) :=
\emptyset$ for convenience. Since every edge in $\tilde{\bm{f}}_n$ is between a node
and its parent, the set of parents $\bigl\{ pa(u) \bigr\}_{u
    \in \mathcal{U}_n}$ specifies the $n-K$ edges in $\tilde{\bm{f}}_n$
    and hence uniquely specifies the forest $\tilde{\bm{f}}_n$ and the
    root nodes $\tilde{s}$.

    Our Gibbs sampler updates the forest $\tilde{\bm{f}}_n$ by iteratively updating the parent of
    each of the nodes, which adds and removes a single edge from
    $\tilde{\bm{f}}_n$ (it is possible to add and remove the same edge so that
    the forest does not change) or, in the random $K$ setting, we may remove a
    single edge and add a new root node or remove a root node and add
    a single edge. 

\begin{figure}[htp]
  \centering
  \includegraphics[scale=.3]{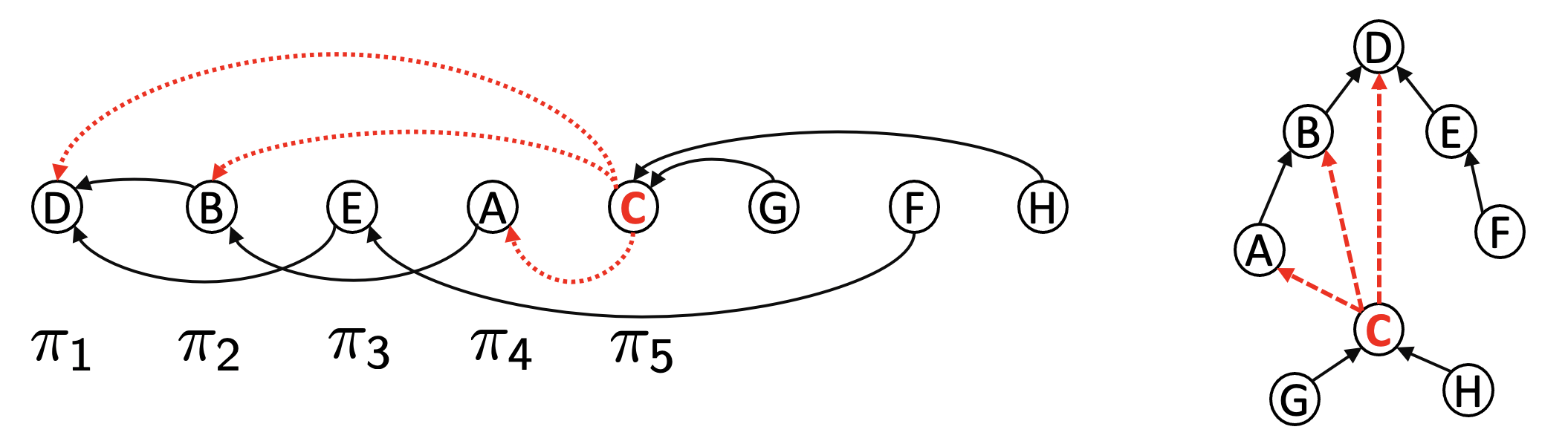}
  \caption{Sampling a parent for $\pi_5$ (node C).}
  \label{fig:sample-parent}
\end{figure}

    To be precise, the latent tree $\tilde{\bm{F}}_n$ and root set
    $\tilde{S}$ induces a latent parent of each node which we denote
    $pa_{\tilde{\bm{F}}_n^{(\tilde{S})}}(\cdot)$. For every node $u$, we
    generate a new parent $u'$ according
    to the conditional distribution
    \begin{align}
Q_u(u') := \Pb\biggl( pa_{\tilde{\bm{F}}^{(\tilde{S})}_n}(u) = u' \,\bigg|\, \Pi = \pi, \tilde{\bm{G}}_n =
      \tilde{\bm{g}}_n,
      \bigl\{ pa_{\tilde{\bm{F}}^{(\tilde{S})}_n}(v) = pa_{\tilde{\bm{f}}^{(\tilde{s})}_n}(v) \bigr\}_{v \neq u} \biggr), \label{eq:cond_pa}
    \end{align}
and then replace the old edge $(u, pa(u))$ with $(u, u')$. Since we condition on the arrival ordering $\Pi$,
probability~\eqref{eq:cond_pa} is non-zero only when $u'$
arrives prior to $u$, i.e. $\pi^{-1} u' < \pi^{-1}u$, and
$(u, u') \in E(\tilde{\bm{g}}_n)$. In other words, if $\pi^{-1}
u = t$, then $Q_u(\cdot)$ is supported on the set of nodes $\pi_{1:(t-1)} \cap
N_{\tilde{\bm{g}}_n}(u)$. In the random $K$ setting, $u'$ is allowed
to be empty in which case $Q_u(\cdot)$ is supported on
$\{\emptyset\} \cup \bigl( \pi_{1:(t-1)} \cap
N_{\tilde{\bm{g}}_n}(u) \bigr)$ where $N_{\tilde{\bm{g}}}(u)$ is the
set of neighbors of $u$ on the graph $\tilde{\bm{g}}_n$. Our sampling
procedure then generate the parents for $\pi_1,
\pi_2, \pi_3, \ldots$ sequentially. In Figure~\ref{fig:sample-parent}, we illustrate
how we may generate a new parent for $\pi_5$ (node C) by choosing one of the
edges that connects $\pi_5$ with one of the earlier nodes
$\pi_{1:4}$.

At iteration $t$, to compute $Q_{\pi_t}(\cdot)$ with respect to
$\pi_t$, for each node $v$ in the support of $Q_{\pi_t}(\cdot)$,
we let $\tilde{\bm{f}}_n^{(v, \pi_t)}$ denote the forest formed by
removing the old edge $(\text{pa}(\pi_t), \pi_t)$ and adding the new edge $(v, \pi_t)$. We
note that $v$ is allowed to be the old parent so that we may
have $\tilde{\bm{f}}_n =
\tilde{\bm{f}}_n^{(v, \pi_t)}$. Then, for any $w_t$ in the support of $Q_{\pi_t}(\cdot)$, we have
\begin{align}
Q_{\pi_t}(w_t) = \frac{ \Pb( \tilde{\bm{F}}_n =
  \tilde{\bm{f}}_n^{(w_t, \pi_t)} \,|\, \Pi = \pi, \tilde{\bm{G}}_n =
  \tilde{\bm{g}}_n) }
  { \sum_{v}
  \Pb( \tilde{\bm{F}}_n =
  \tilde{\bm{f}}_n^{(v, \pi_t)} \,|\, \Pi = \pi, \tilde{\bm{G}}_n =
  \tilde{\bm{g}}_n) }. \label{eq:cond_pa_prob}
\end{align}
In the PAPER models with Erd\H{o}s--R\'{e}nyi edges, We can compute the conditional distribution $\Pb(
\tilde{\bm{F}}_n = \cdot \,|\,
\Pi = \pi, \tilde{\bm{G}}_n = \tilde{\bm{g}}_n)$ by using the fact that once when we condition on $\tilde{\bm{F}}_n =
\tilde{\bm{f}}_n$, the remaining edges of $\tilde{\bm{G}}_n$ are
uniformly random and the fact that $\Pi$ and $\bm{F}_n$ are
independent. Thus,
\begin{align}
&  \Pb( \tilde{\bm{F}}_n = \tilde{\bm{f}}_n \,|\, \Pi = \pi,
    \tilde{\bm{G}}_n = \tilde{\bm{g}}_n) \nonumber \\
  &\qquad \propto \Pb( \tilde{\bm{G}}_n = \tilde{\bm{g}}_n \,|\,
    \tilde{\bm{F}}_n = \tilde{\bm{f}}_n, \Pi = \pi) \Pb(
    \tilde{\bm{F}}_n = \tilde{\bm{f}}_n \,|\, \Pi = \pi) \nonumber \\
  &\qquad = \binom{\binom{n}{2} - (n - K(\tilde{\bm{f}}_n))}{m - (n -
    K(\tilde{\bm{f}}_n))}^{-1} \Pb( \bm{F}_n = \pi^{-1}
    \tilde{\bm{f}}_n) \mathbbm{1}\{ \tilde{\bm{f}}_n \in
    \mathcal{F}(\tilde{\bm{g}}_n) \} \nonumber \\
  &\qquad \propto \biggl\{ \prod_{k=1}^{K(\tilde{\bm{f}}_n)} \frac{ n(n-1)/2 - n + k)
    }{m - n + k} \biggr\}  \Pb( \bm{F}_n = \pi^{-1}
    \tilde{\bm{f}}_n)
    \mathbbm{1}\{ \tilde{\bm{f}}_n \in
    \mathcal{F}(\tilde{\bm{g}}_n) \}. \label{eq:sample-forest-weight1}
\end{align}

We now discuss the sampling procedure in detail in all the settings.\\

\noindent \textbf{Single root setting}: In the single root setting, we again use the notation $\tilde{\bm{t}}_n =
\tilde{\bm{f}}_n$ to be consistent with Definition~\ref{def:apa1}. The first term
of~\eqref{eq:sample-forest-weight1} is a constant since
$K(\tilde{\bm{t}}_n) = 1$ and may thus be ignored. Using the likelihood of APA trees (see Remark~\ref{rem:paper_prob} as well as Proposition~\ref{prop:apa-prob1} from the Appendix) and using the
fact that $\Pb( \bm{T}_n = \pi^{-1}
  \tilde{\bm{t}}_n) > 0$ when $\pi \in
  \text{hist}(\tilde{\bm{t}}_n)$, we have that, for any $w_t \in \pi_{1:(t-1)} \cap N_{\tilde{\bm{g}}_n}(\pi_t)$,
  \begin{align*}
Q_{\pi_t}(w_t) = \frac{\beta D_{\tilde{\bm{t}}^{(\cdot, \pi_t)}_n}( w_t) + \alpha}{ \sum_{v
    \in \pi_{1:(t-1)} \cap N_{\tilde{\bm{g}}_n}(\pi_t)} \beta
    D_{\tilde{\bm{t}}^{(\cdot, \pi_t)}_n}( v) + \alpha},
  \end{align*}
where $\tilde{\bm{t}}^{(\cdot, \pi_t)}_n$ is the disconnected graph obtained by removing the old
edge $(\text{pa}(\pi_t), \pi_t)$ from
$\tilde{\bm{t}}_n$. We summarize the resulting procedure in
Algorithm~\ref{alg:draw-forest1}. Since we visit every node once and,
for a single node $u$, it takes time $O( D_{\tilde{\bm{g}}_n}(u))$ to
generate a new parent, the overall runtime of the second stage of the
algorithm is $O(m)$. The computational complexity is the same under the fixed $K$ setting and the
random $K$ setting. \\

\noindent \textbf{Fixed $K > 1$ setting:} Since the number of trees $K$ is fixed, the first term
of~\eqref{eq:sample-forest-weight1} is again a constant. Using likelihood of APA trees again (see
Proposition~\ref{prop:apa-prob2} from the Appendix), we have that for any $w_t \in
\pi_{1:(t-1)} \cap N_{\tilde{\bm{g}}_n}(\pi_t)$, 
\begin{align*}
Q_{\pi_t}(w_t) = \frac{\beta D_{\tilde{\bm{f}}_n^{(\cdot, \pi_t)}}( w_t) + 2 \beta
  \mathbbm{1}\{ w_t \in \pi_{1:K}\} + \alpha}{ \sum_{v
    \in \pi_{1:(t-1)} \cap N_{\tilde{\bm{g}}_n}(\pi_t)} \beta
    D_{\tilde{\bm{f}}_n^{(\cdot, \pi_t)}}(v) + 2\beta \mathbbm{1}\{ v \in \pi_{1:K} \} + \alpha},
\end{align*}
where, as with the single root setting, $\tilde{\bm{f}}_n^{(\cdot, \pi_t)}$ is the forest obtained by removing the old
edge $(\text{pa}(\pi_t), \pi_t)$ from
$\tilde{\bm{f}}_n$. The only difference from the single root setting is that we have a higher probability to
attach to a root node because of the imaginary self-loop edge. We
summarize the procedure in Algorithm~\ref{alg:draw-forest1}. \\

\begin{algorithm}[htp]
  \caption{Generating spanning forest $\tilde{\bm{f}}_n$ of
    $\tilde{\bm{g}}_n$ under either $\text{PAPER}(\alpha, \beta,
    \theta)$ or $\text{PAPER}(\alpha, \beta, K,
    \theta)$}
  \label{alg:draw-forest1}
  \textbf{Input:} Graph $\tilde{\bm{g}}_n$, ordering
  $\pi \in \text{Bi}([n], \mathcal{U}_n)$, and a spanning
  forest $\tilde{\bm{f}}_n$ with $K$ component trees.  \\
  \textbf{Effect:} Modifies $\tilde{\bm{f}}_n$ in place. 
  \begin{algorithmic}[1]
    \For{$t = K + 1, \ldots, n$}:
    \State Remove old edge $(\pi_t,
    pa(\pi_t))$ from $\tilde{\bm{f}}_n$ to obtain $\tilde{\bm{f}}_n^{(\cdot, \pi_t)}$.
    \State Choose a node $w_t \in \pi_{1:(t-1)}\cap N_{\tilde{\bm{g}}_n}(\pi_t)$ with
    probability proportional to
    \begin{align*}
      \left\{ \begin{array}{cc}
         \beta D_{\tilde{\bm{f}}^{(\cdot, \pi_t)}_n}(w_t) + \alpha 
                & \text{ under $\text{PAPER}(\alpha, \beta, \theta)$} 
                \\
                 \beta D_{\tilde{\bm{f}}^{(\cdot, \pi_t)}_n}(w) +
      2 \beta \mathbbm{1}\{w \in \pi_{1:K}\} + \alpha 
                & \text{ under $\text{PAPER}(\alpha, \beta, K,
                  \theta)$}
              \end{array} \right.
    \end{align*}
    \State Add new edge $(\pi_t, w_t)$ to $\tilde{\bm{f}}_n$.
    \EndFor
  \end{algorithmic}
\end{algorithm}

\noindent \textbf{Random $K$ roots setting:} Under the $\text{PAPER}(\alpha, \beta, \alpha_0, \theta)$ model, a
node may become a new root in the sampling process and thus we
must take into account the first term
of~\eqref{eq:sample-forest-weight1}. Moreover, in this setting,
$Q_{\pi_t}(\cdot)$ for node $\pi_t$ is
supported on $\{\emptyset\} \cup \bigl( \pi_{1:(t-1)} \cap
N_{\tilde{\bm{g}}_n}(\pi_t) \bigr)$ since we may turn the node $\pi_t$ into a
new root node, in which case we set its parent to $\emptyset$ by convention. Define $\tilde{\alpha}_0 :=  
\alpha_0 \frac{m - n + K + \mathbbm{1}\{ \pi_t
  \notin \tilde{s} \} }{n(n-1)/2 - n + K + \mathbbm{1}\{ \pi_t \notin \tilde{s}
  \}}$; we then have that, by Proposition~\ref{prop:apa-prob3}, 
for any $w_t \in \{\emptyset\} \cup \bigl( \pi_{1:(t-1)} \cap
N_{\tilde{\bm{g}}_n}(\pi_t) \bigr)$, 
\begin{align*}
Q_{\pi_t}(w_t) &= \frac{ \tilde{\alpha}_0 }
  { \tilde{\alpha}_0 
  +  \sum_{v \in \pi_{1:(t-1)} \cap N_{\tilde{\bm{g}}_n}(\pi_t)} \beta
  D_{\tilde{\bm{f}}^{(\cdot, \pi_t)}_n}(v) + 2\beta \mathbbm{1}\{ v \in \tilde{s} \} + \alpha
                 } \quad \text{ if $w_t = \emptyset$}\\
\textrm{and } Q_{\pi_t}(w_t) &= \frac{\beta D_{\tilde{\bm{f}}_n^{(\cdot, \pi_t)}}( w_t) + 2 \beta
  \mathbbm{1}\{ w_t \in S\} + \alpha}{ \tilde{\alpha}_0 + \sum_{v
    \in \pi_{1:(t-1)} \cap N_{\tilde{\bm{g}}_n}(\pi_t)} \beta
    D_{\tilde{\bm{f}}_n^{(\cdot, \pi_t)}}(v) + 2\beta \mathbbm{1}\{ v \in \tilde{s} \} +
                               \alpha} \quad \text{ if $w_t \neq \emptyset$},
\end{align*}
where, if $\pi_t$ is not a root node, $\tilde{\bm{f}}_n^{(\cdot, \pi_t)}$ is the forest obtained by removing the old
edge $(\pi_t, \textbf{pa}(\pi_t))$ and if $\pi_t$ is a root node, then
$\tilde{\bm{f}}_n^{(\cdot, \pi_t)} = \tilde{\bm{f}}_n$. We summarize the resulting procedure in
Algorithm~\ref{alg:draw-forest2}.\\

\begin{algorithm}[htp]
  \caption{Generating spanning forest $\tilde{\bm{f}}_n$ of
    $\tilde{\bm{g}}_n$ under $\text{PAPER}(\alpha, \beta, \alpha_0, \theta)$}
  \label{alg:draw-forest2}
  \textbf{Input:} Graph $\tilde{\bm{g}}_n$, ordering
  $\pi \in \text{Bi}([n], \mathcal{U}_n)$, and a spanning
  forest $\tilde{\bm{f}}_n$.  \\
  \textbf{Effect:} Modifies $\tilde{\bm{f}}_n$ in place. 
  \begin{algorithmic}[1]
    \State Let $\tilde{s}$ be the set of root nodes.
    \For{$t = 2, 3, \ldots, n$}:
    \State If $\pi_t \notin \tilde{s}$, remove edge $(\pi_t,
    pa(\pi_t))$ from $\tilde{\bm{f}}_n$ to get
    $\tilde{\bm{f}}^{(\cdot, \pi_t)}_n$. Else, let $\tilde{s} = \tilde{s}
    \backslash \{w_t \}$ and let $\tilde{\bm{f}}^{(\cdot, \pi_t)}_n = \tilde{\bm{f}}_n$.
    \State Choose a node $w_t \in \{\emptyset\} \cup \bigl(
    \pi_{1:(t-1)}\cap N_{\tilde{\bm{g}}_n}(\pi_t) \bigr)$ with
    probability proportional to
    \begin{align*}
      \left\{ \begin{array}{cc}
         \alpha_0 
                & \text{ for $w_t = \emptyset$} 
                \\
                 \beta D_{\tilde{\bm{f}}^{(\cdot, \pi_t)}_n}(w_t) +
      2 \beta \mathbbm{1}\{w_t \in \tilde{s}\} + \alpha 
                & \text{ for $w_t \neq \emptyset$}
              \end{array} \right.
    \end{align*}
    \State If $w_t \neq \emptyset$, let $\tilde{\bm{f}}_n =
    \tilde{\bm{f}}_n^{(\cdot, \pi_t)} \cup (\pi_t, w_t)$. Otherwise, let $\tilde{s} =
    \tilde{s} \cup \{\pi_t\}$ and $\tilde{\bm{f}}_n = \tilde{\bm{f}}^{(\cdot, \pi_t)}_n$.
    \EndFor
  \end{algorithmic}
\end{algorithm}

{\color{black}
\noindent \textbf{Sequential noise setting:} 
Under the seq-PAPER setting, we use the same sampling procedure but the sampling probabilities become more complicated. From~\eqref{eq:cond_pa_prob}, we see that, for $w \in N_{\tbm{g}_n} \cap \pi_{1:(t-1)}$,
\begin{align*}
Q_{\pi_t}(w) &\propto \mathbb{P}( \tbm{T}_n = \tbm{t}_n^{(w, \pi_t)} \,|\, \Pi = \pi, \tbm{G}_n = \tbm{g}_n) \\
&\propto \underbrace{\mathbb{P}( \tbm{G}_n = \tbm{g}_n \,|\, \tbm{T}_n = \tbm{t}_n^{(w, \pi_t)}, \Pi = \pi)}_{\text{noise term}} \mathbb{P}( \tbm{T}_n = \tbm{t}_n^{(w, \pi_t)} \,|\, \Pi = \pi).
\end{align*}
Under the seq-PAPER model, the noise term also depends on $w$ since choosing a new parent for $\pi_t$ would change the tree degrees of some of the nodes. Naively computing $Q_{\pi_t}(w)$ takes time $O(n)$, but in Section~\ref{sec:seq_noise_tree_sampling} of the Appendix (using results from Section~\ref{sec:seq_noise_preliminary}), we give a detailed algorithm to compute $Q_{\pi_t}(w)$ in time $O(D_{\tilde{\bm{g}}_n}(w))$ so that overall, we can sample a new parent for $\pi_t$ in time proportional to the number of neighbors of neighbors of $\pi_t$. 

When we have deletion noise, as the case of the $\text{seq-PAPER}^*$ model, the latent tree $\tbm{T}_n$ need not be a subgraph of $\tbm{G}_n$ and hence, when sampling a new parent for $\pi_t$, we must consider all of $\pi_{1:(t-1)}$ and not just graph neighbors of $\pi_t$. Thus, we draw $w \in \pi_{1:(t-1)}$ with probability $Q_{\pi_t}(w)$ and set $\text{pa}(\pi_t) = w$. We give the detailed algorithm for computing $Q_{\pi_t}(w)$ in Section~\ref{sec:seq_noise_tree_sampling} of the Appendix.\\

}

\subsection{Other aspects of the algorithm}

\noindent \textbf{Parameter estimation:} To estimate $\alpha$ and $\beta$, we derive an EM algorithm in Section~\ref{sec:parameter-estimation} of the Appendix. The noise level $\theta$ is easy to estimate via $\hat{\theta} = \frac{m
  - (n - 1)}{ n(n-1)/2 - (n - 1)}$ in the single root setting. 
The inference algorithm in fact does not require knowledge of $\theta$
since it conditions on the number of edges $m$ of the observed graph. We discuss some ways to select the number of trees $K$ in the
fixed $K$ root setting and ways
to estimate $\alpha_0$ in the random $K$ roots setting in Section~\ref{sec:practical} of the
Appendix. \\

\noindent \textbf{Inference from posterior samples:} The Gibbs sampler described in Section~\ref{sec:sample-ordering} and
Section~\ref{sec:sample-forest} generates a Monte Carlo sequence $\{ (\pi^{(j)},
\tilde{\bm{f}}_n^{(j)} ) \}_{j=1}^J$ where $J$ is the number of Monte
Carlo samples. A straightforward way to approximate the posterior root
probability is to use the empirical distribution based on all the
$\pi^{(j)}$'s. However, we can construct a much more accurate
approximation by taking advantage of the fact that the posterior root
probability is easy to compute on a tree.

Consider the single root setting for simplicity where the posterior
root probability is $\Pb( \Pi_1 = u \,|\, \tilde{\bm{G}}_n =
\tilde{\bm{g}}_n)$ for any node $u$. In this case, we may compute
distributions $Q^{(1)}, Q^{(2)}, \ldots, Q^{(J)}$ over the nodes by
\begin{align*}
Q^{(j)} &= \Pb( \Pi_1 = u \,|\, \tilde{\bm{T}}_n = \tilde{\bm{t}}_n^{(j)},
      \tilde{\bm{G}}_n = \tilde{\bm{g}}_n) 
    = \Pb( \Pi_1 = u \,|\, \tilde{\bm{T}}_n = \tilde{\bm{t}}_n^{(j)})
  = \frac{h(u, \tilde{\bm{t}}_n^{(j)})}{h(\tilde{\bm{t}}_n^{(j)})}.
\end{align*}
Then, we output $\frac{1}{J} \sum_{j=1}^J Q^{(j)}$ as our
approximation of the posterior root distribution. In the multiple
roots setting, we use the same procedure except that we compute
$u \mapsto \Pb( u \in \tilde{S} \,|\, \tilde{\bm{F}}_n = \tilde{\bm{f}}_n^{(j)})$
and then average across $j \in \{1, 2, \ldots, J \}$.

In the multiple roots setting, each Monte Carlo sample of the forest
$\tilde{\bm{f}}_n^{(j)}$ contain either $K$ disjoint trees in the
fixed $K$ setting or a random number of disjoint trees in the random
$K$ setting. These disjoint trees provide a posterior sample of the
communities on the network and using them, we may estimate the community structure
of the network. We provide details on one way of using posterior
samples for community recovery in Section~\ref{sec:fixed-K-real-data}
and~\ref{sec:random-K-real-data}.

The Gibbs sampling algorithm scales to large networks. We are able to
run it on networks of up to a million nodes
(c.f. Section~\ref{sec:single-root-subgraph}) on a single 2020 MacBook
Pro laptop. To give a rough sense of the runtime, it takes about 1
second to perform one outer loop of the Gibbs sampler on a graph of
10,000 nodes and 20,000 edges. In Section~\ref{sec:practical} of the
appendix, we provide more details on practical usage of the Gibbs
sampler such as convergence criterion. \\

{\color{black}
\noindent \textbf{Initialization:} In the single root setting, to initialize the Gibbs sampling algorithm, we recommend generating the initial tree $\tilde{\bm{t}}_n$ uniformly at random from the set of spanning trees $\mathcal{T}(\tilde{\bm{g}}_n)$ of the observed graph, which can be efficiently done via elegant random-walk-based algorithms such as the Aldous--Broder algorithm \citep{broder1989generating, aldous1990random} or Wilson's algorithm \citep{wilson1996generating}. We then initialize $\pi$ by drawing an ordering uniformly from the history of the initial tree. This initialization distribution is guaranteed to be overdispersed and works very well in practice. The same initialization works for the random $K$ setting. For the fixed $K$ setting, we can form the initial forest by constructing uniformly random spanning tree $\tbm{t}_n$ and uniformly random ordering $\pi$ as usual, taking the first $K$ nodes of the $\pi$ as the root nodes, and removing all tree edges between them to obtain an initial $\tbm{f}_n$. We use Wilson's algorithm in our implementation. }

\section{Theoretical Analysis}
\label{sec:theory}

We provide theoretical support for our approach by deriving bounds on the size of our proposed confidence
sets when the observed graph has the PAPER distribution. In particular, we aim to quantify how the quality of inference
deterioriates with the noise level $\theta$, that is, how the size of
the confidence set increases with $\theta$. For simplicity, for
consider only the single root setting and we do not take into account
approximation errors introduced by the Gibbs sampler, that is, we
analyze the confidence set constructed from the exact posterior root probabilities. 

We begin with a type of optimality statement which shows that the
size of the confidence set $B_\epsilon(\cdot)$, as defined
in~\eqref{eq:Bepsilon}, is of no larger order than any other
asymptotically valid confidence set. Intuitively, this is because
$B_\epsilon(\cdot)$ can be interpreted as a ``Bayes estimator'' for
the root node. 

\begin{lemma}
\label{lem:comparison}
Let $\epsilon$ be in $(0, 1)$, let $\bm{G}_n \sim
\text{PAPER}(\alpha, \beta, \theta)$, and let $\bm{G}^*_n = \rho
\bm{G}_n$ be the observed alphabetically labeled graph for some $\rho
\in \text{Bi}([n], \mathcal{U}_n)$. Let $B_\epsilon(\mathbf{G}^*_n)$
be defined as in~\eqref{eq:Lepsilon} and~\eqref{eq:Bepsilon}. Fix any $\delta \in (0,1)$ and let
$C_{\delta \epsilon}(\mathbf{G}_n^*)$ be any confidence set for the
root node that is labeling-equivariant and has asymptotic coverage
level $1 - \delta \epsilon$, that is, $\limsup_{n \rightarrow \infty}
\mathbb{P}( \rho_1 \notin C_{\delta \epsilon}(\mathbf{G}^*_n)) \leq \delta \epsilon$. Then, we have that
\[
\limsup_{n \rightarrow \infty} \mathbb{P}\bigl( |B_{\epsilon}(\mathbf{G}_n^*)| \geq | C_{\delta \epsilon}(\mathbf{G}_n^*)| \bigr) \leq \delta.
\]
\end{lemma}

We provide the proof of Lemma~\ref{lem:comparison} in
Section~\ref{sec:proof} of the appendix. 

Ideally, we would compare the size of $B_{\epsilon}(\cdot)$ with
$C_{\epsilon}(\cdot)$ at the same level. It is however much easier to
compare with the more conservative $C_{\delta \epsilon}(\cdot)$. In
many cases, the size of a confidence set $|C_{\epsilon}(\cdot)|$ has
bounds of the form $f(n) g(\epsilon^{-1})$ for some functions $f$ and $g$ (see e.g.  \cite{banerjee2020root}) so
that comparing with $C_{\delta \epsilon}(\cdot)$ adds only a
multiplicative constant to the bound.

Lemma~\ref{lem:comparison} is useful because it is difficult to directly bound the confidence set
$B_\epsilon(\cdot)$ as a function of $n$ and the parameters;
Lemma~\ref{lem:comparison} shows that we can indirectly upper bound it by
analyzing a simpler asymptotically valid confidence set. Our strategy then is to construct confidence sets based on the degree of
the nodes whose size is much easier to bound through well-understood
probabilistic properties of preferential attachment trees. This leads
to our next result which provides explicit bounds on the size of the
confidence set $B_\epsilon(\cdot)$ when the underlying tree is LPA.

\begin{theorem}
  \label{thm:pa-conf-size}
Let $\bm{G}_n \sim \text{PAPER}(\alpha, \beta, \theta)$ for $\beta=1$,
$\alpha=0$, and $\theta \in [0,1]$. For $t \in [n]$, let
$D_{\bm{G}_n}(t)$ be the degree of node with arrival time $t$
and for $k \in [n]$, let $k\mhyphen\max(D_{\bm{G}_n})$ be the $k$-th largest degree of $\bm{G}_n$. 
Let $\delta > 0$ be arbitrary and suppose $\theta \leq n^{-\frac{1}{2}
  - \delta}$. Then, for any $\epsilon > 0$, there exists $L_\epsilon
\in \mathbb{N}$ (dependent on $\delta$ but not on $n$) such that
\begin{align}
\limsup_{n \rightarrow \infty} \mathbb{P}\bigl\{  D_{\bm{G}_n}(1) \leq
  L_\epsilon\mhyphen\max(D_{\bm{G}_n}) \bigr\} \leq \epsilon.
  \label{eq:lpa-degree-bound}
\end{align}

As a direct consequence, if $\theta = O(n^{-\frac{1}{2} - \delta})$ for any $\delta > 0$, then, for any $\epsilon \in (0,1)$,
\[
|B_\epsilon( \bm{G}^*_n) | = O_p(1).
\]

\end{theorem}

We relegate the proof of Theorem~\ref{thm:pa-conf-size} in
Section~\ref{sec:lpa-proof} of the appendix and provide a short sketch
here: we use results from \cite{pekoz2014joint} which
show that the degree sequence of an LPA tree, when normalized by
$\frac{1}{\sqrt{n}}$, converges to a limiting distribution in the
$\ell_q$ sequential metric sense, which shows
that~\eqref{eq:lpa-degree-bound} holds for the tree degree
$D_{\bm{T}_n}(\cdot)$, that is, the degree of the root node is one of
the highest among all the nodes. Since $D_{\bm{G}_n} = D_{\bm{T}_n} + D_{\bm{R}_n}$, we show that if the noise level $\theta$
is less than $n^{-1/2 - \delta}$ for some $\delta > 0$, then the
degree of the noisy edges $D_{\bm{R}_n}$ has a second order effect
and~\eqref{eq:lpa-degree-bound} remains valid.

We know from existing results (such as~\citet[][Theorem
6]{bubeck2017finding}; see also~\citet[][Corollary
7]{crane2021inference}) that $| B_\epsilon(\bm{T}_n^*)|$ is $O_p(1)$
in the $\theta=0$ case where we observe the LPA tree
$\bm{T}_n^*$. Theorem~\ref{thm:pa-conf-size} shows that this
phenomenon is quite robust to noise. Indeed, when $\theta = n^{-1/2 -
  \delta}$, the observed graph would have approximately $n^{3/2 - \delta}$
noisy edges and only $n-1$ tree edges.

The situation is different when the underlying latent tree has the UA
distribution, where $\alpha = 1$ and $\beta = 0$. In this case, we
have the following result:

\begin{theorem}
  \label{thm:ua-conf-size}
  Let $\bm{G}_n \sim \text{PAPER}(\alpha, \beta, \theta)$ for $\alpha=1$,
$\beta=0$, and $\theta \in [0,1]$. For $t \in [n]$, let
$D_{\bm{G}_n}(t)$ be the degree of node with arrival time $t$
and for $k \in [n]$, let $k\mhyphen\max(D_{\bm{G}_n})$ be the $k$-th
largest degree of $\bm{G}_n$. Suppose $\theta = o\bigl( \frac{\log
  n}{n} \bigr)$ and let $\epsilon \in (0,1)$ be arbitrary. For any
$\eta \in (0,1)$, define $L_{\eta, n, \epsilon} := n^\eta + \epsilon^{-1}
n^{1 - (2-\eta) h\bigl(\frac{\eta}{2 - \eta} \bigr) }$ where $h(x) =
(1+x)\log(1+x) - x$ for $x \geq 0$. Then, we have that
\begin{align}
\limsup_{n \rightarrow \infty} \Pb \bigl\{ D_{\bm{G}_n}(1) \leq
  L_{\eta,n,\epsilon}\mhyphen\max( D_{\bm{G}_n}) \bigr\} \leq \epsilon.
  \label{eq:ua-degree-bound}
\end{align}
As a direct consequence, if $\theta = o( \frac{\log n}{n} )$, then, for some $\gamma \leq 0.8$, we have that
\[
n^{-\gamma} \epsilon^{-1} |B_\epsilon(\bm{G}^*_n)| = O_p(1)\, \quad
\text{for any $\epsilon \in (0, 1)$}.
\]
\end{theorem}

We relegate the proof of Theorem~\ref{thm:ua-conf-size} to
Section~\ref{sec:ua-proof} of the appendix. The proof technique is
similar to that of Theorem~\ref{thm:pa-conf-size} except that we use
concentration inequalities to derive~\eqref{eq:ua-degree-bound}. 

\begin{figure}
  \centering
  \includegraphics[scale=.47]{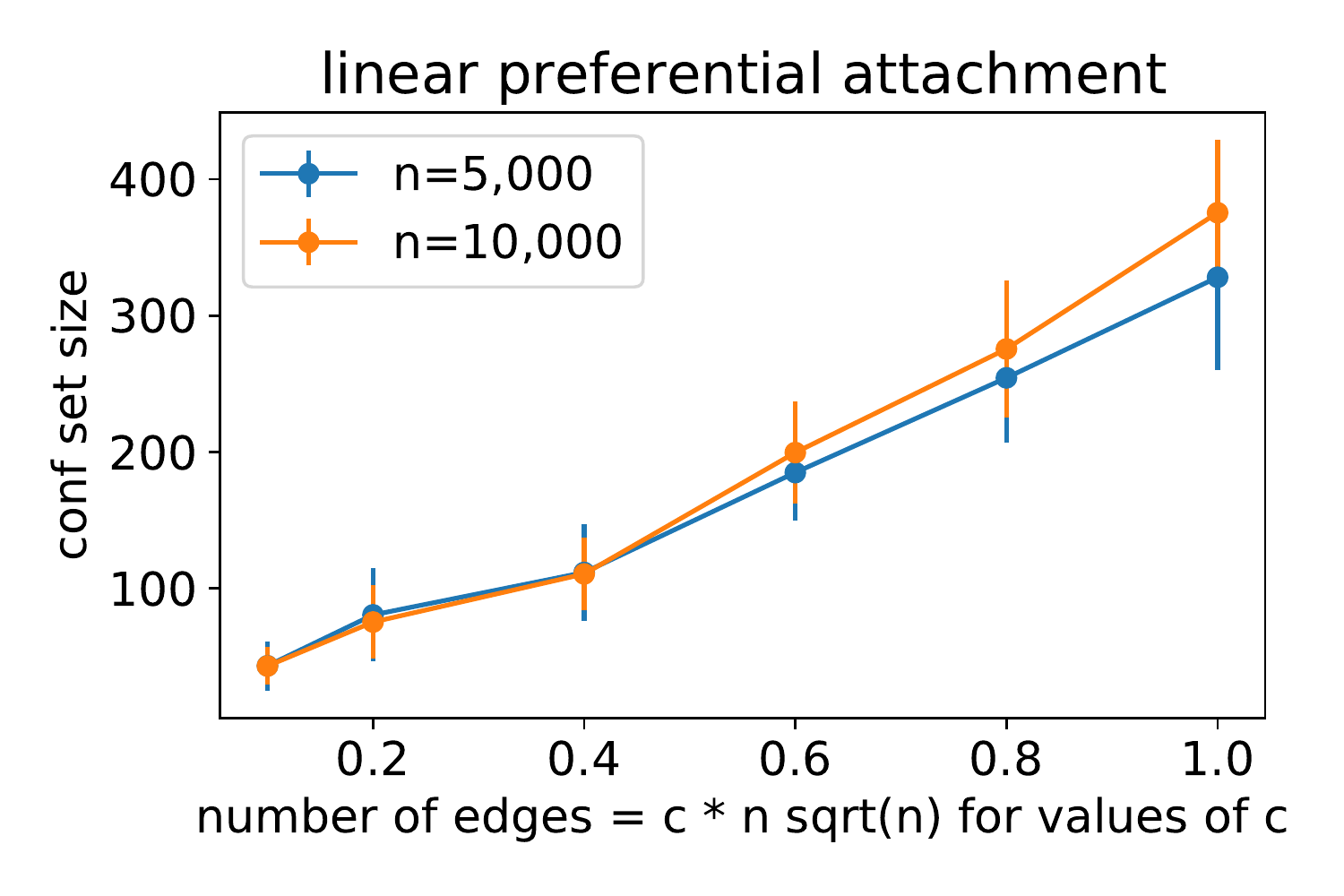}
  \includegraphics[scale=.47]{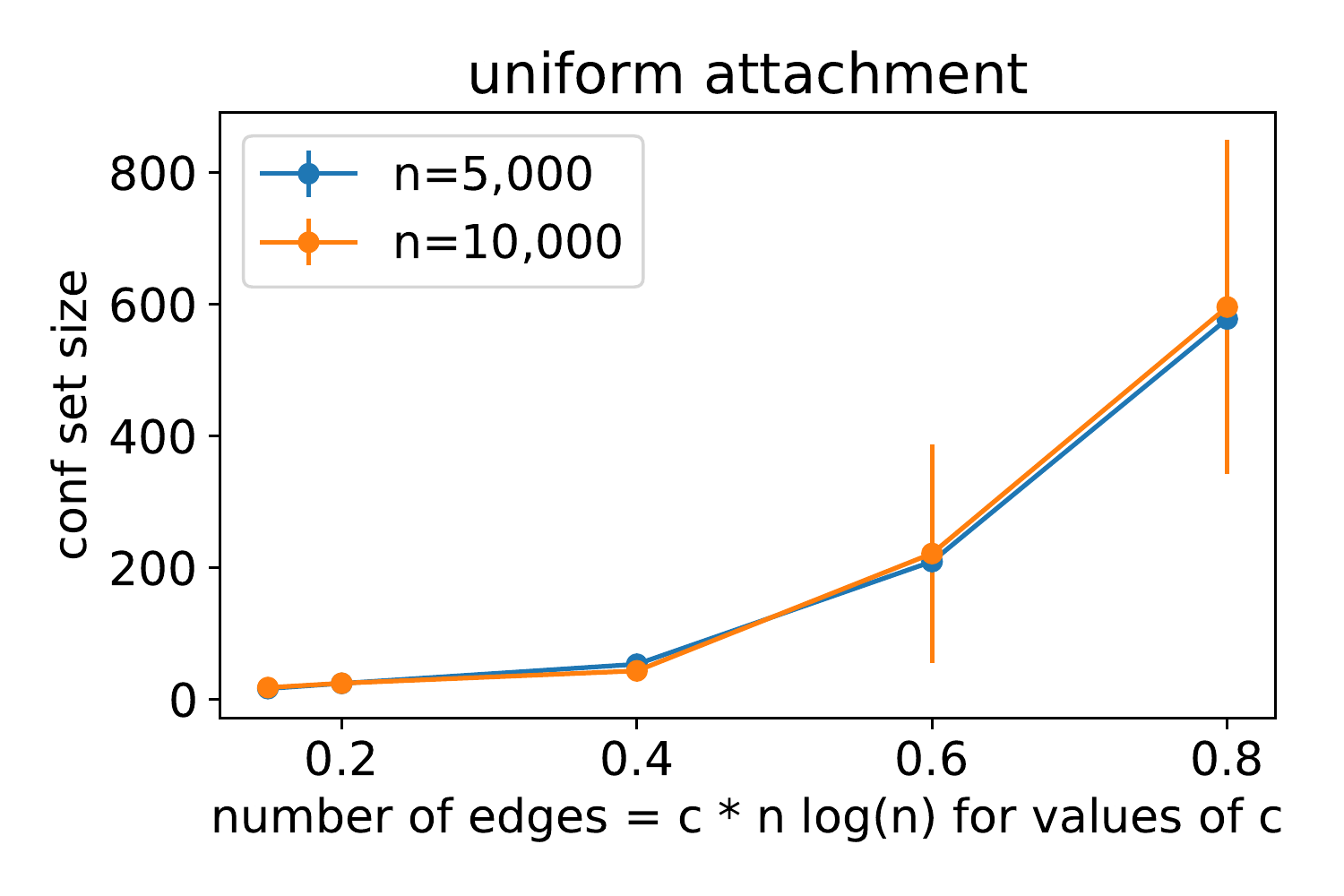}
  \caption{Size of the confidence set vs. the number of edges.}
  \label{fig:size}
\end{figure}

Comparing Theorem~\ref{thm:ua-conf-size} with
Theorem~\ref{thm:pa-conf-size}, we see two important
differences. First, even if the noise level is small, we can no longer
guarantee that $|B_{\epsilon}(\bm{G}_n^*)|$ is bounded even as $n$
increases. Instead, we have the much weaker bound that
$|B_{\epsilon}(\bm{G}^*_n)|$ is less than $O(n^{\gamma})$ for some
$\gamma < 0.8$. We believe this bound is not tight; we observe
from simulations in Section~\ref{sec:simulation} (see Figure~\ref{fig:size}) that the size of the
confidence set $B_{\epsilon}(\cdot)$ is indeed $O_p(1)$ even when the
noise level is of order $\frac{\log n}{n}$. The bound is sub-optimal
because the degree of the nodes is not informative of their latent
ordering when the latent tree has the UA distribution; hence,
$B_\epsilon(\cdot)$ could be much smaller than confidence sets
constructed solely from degree information. Intuitively, this is because largest degree nodes do not persist in uniform attachment as opposed to linear preferential attachment \citep{dereich2009random, galashin2013existence}.

The second difference is that the noise tolerance is much smaller. We
require $\theta$ to be smaller than $\frac{\log n}{n}$ rather than
$n^{-1/2}$. We conjecture that these rates are tight in the following sense:

\begin{conjecture}
  \label{con:size}
Let $\bm{G}_n \sim \text{PAPER}(\alpha, \beta, \theta)$ for $\alpha=1$,
$\beta=0$, and $\theta \in [0,1]$.
\begin{enumerate}
  \item Suppose $\alpha = 0$ and $\beta=1$ (LPA). If $\theta = o(n^{-1/2})$,
    then $|B_{\epsilon}(\bm{G}^*_n)| = O_p(1)$ and if $\theta = \omega(
    n^{-1/2})$, then every asymptotically valid confidence set has
    size that diverges with $n$.
  \item Suppose $\alpha = 1$ and $\beta=0$ (UA). If $\theta =
    o(\frac{\log n}{n})$,
    then $|B_{\epsilon}(\bm{G}^*_n)| = O_p(1)$ and if $\theta = \omega(
    \frac{\log n}{n})$, then every asymptotically valid confidence set has
    size that diverges with $n$.
  \end{enumerate}
\end{conjecture}

We provide empirical support for this conjecture in
Section~\ref{sec:simulation}, particularly Figure~\ref{fig:size}. In
those experiments, we see that, when the latent tree has the LPA
distribution and when $\theta = c
n^{-1/2}$ where $c > 0$ is small, the size of
$B_{\epsilon}$ does not increase with $n$; however, when $c$ (and
hence $\theta$) is large, $B_{\epsilon}$ is larger when the size of
the graph $n$ is larger. The same phenomenon holds when the latent
tree has the UA distribution when $\theta = c \frac{\log n}{n}$. 

\section{Empirical Studies}
\label{sec:empirical}

We have implemented the inference approach in
Section~\ref{sec:methodology} and the sampling algorithm in
Section~\ref{sec:algorithm} in a Python package named
\texttt{paper-network}, which can be installed via command line \texttt{pip
  install paper-network} on the terminal and then imported in Python
via \texttt{import PAPER}. The source code of the package, along with
examples and documentation, are available at the website
\texttt{https://github.com/nineisprime/PAPER}. All the code used in
this Section are also available there under the directory
\texttt{paperexp}. {\color{black} We also give detailed sampler diagnostics information in Section~\ref{sec:diagnostics} of the Appendix.}

\subsection{Simulation}
\label{sec:simulation}
\noindent \textbf{Frequentist coverage in the single root setting: } In our first simulation study, we empirically verify
Theorem~\ref{thm:frequentist-coverage1} by showing that a level
$1-\epsilon$ credible
set for the root node constructed from the posterior root
probabilities has frequentist coverage at exactly the same level
$1-\epsilon$. We consider three different settings of parameters:
$\alpha=0, \beta=1$ (LPA), $\alpha=1, \beta=0$ (UA), and $\alpha=8,
\beta=1$. We generate $\bm{G}_n^*$ according to the
$\text{PAPER}(\alpha, \beta, \theta)$ model with $n=3,000$ nodes and
$m=7,500$ edges. We then estimate $\alpha$ and $\beta$ using the
method given in Section~\ref{sec:parameter-estimation}, compute the level $\epsilon \in \{0.2, 0.05, 0.01\}$
credible sets, and record whether they cover the true root node. We
repeat the experiment over 300 independent trials and report the
results in Table~\ref{tab:coverage1}. We observe that the credible
sets attain the nominal coverage and that the size of the credile sets
are small compared to the number of nodes $n$. \\

\begin{table}[htp]
\begin{center}
\begin{tabular}{c|c|c|c|c|c|c|c|c|c|}
\hline
\text{$(\alpha, \beta)$} & (0, 1) & (1, 0) & (8, 1) & (0, 1) & (1, 0) & (8, 1)
  & (0, 1) & (1, 0) \\
\hline
\text{Theoretical coverage} & 0.8 & 0.8 & 0.8 & 0.95 & 0.95 & 0.95 & 0.99
                            & 0.99 \\
\hline
{\bf \text{Empirical coverage}} & {\bf 0.8} & {\bf 0.823} & {\bf 0.82} & {\bf 0.937} & {\bf 0.943} & {\bf 0.94} & {\bf 0.983} & {\bf 0.993} \\
  \hline
{ \text{Ave. conf. set size}} & 7 & 12 & 9 & 42 & 42 & 31 & 183
          & 115 \\
  \hline
\end{tabular}
\end{center}
\caption{Empirical coverage of our confidence set for the root
  node. We report the average over 300 trials. Graph has $n= 3000$ nodes
  and $m= 7,500$ edges in all cases.}
\label{tab:coverage1}
\end{table}

\noindent \textbf{Size of the confidence set:} In our second simulation study, we study the effect of the sample size
$n$ and the magnitude of the noisy edge probability $\theta$ on the
size of the confidence set. We let $\bm{G}^*_n$ be the observed graph
with $n$ nodes and $m$ edges according to the $\text{PAPER}(\alpha,
\beta, \theta)$ model where we consider $(\alpha, \beta) = (0, 1)$ (LPA) or
$(1, 0)$ (UA). Since a tree with $n$ nodes always
contains $n-1$ edges, $\frac{n^2}{2} \theta + n$ is approximately equal to the 
number of edges $m$ in the observed graph $\bm{G}_n^*$.

We empirically show that the
confidence set size does not depend on $n$ so long as $\theta$ is much
smaller than $n^{-1/2}$ for LPA and much smaller than $\frac{\log n}{n}$ for UA. To that end, we
set $m = c 
n \sqrt{n}$ for $c \in \{0.1, 0.2, 0.4, 0.6, 0.8, 1\}$ for LPA and $m = c n
\log n$ for $c \in \{0.15, 0.2, 0.4, 0.6, 0.8 \}$ for UA. We then plot the
average size of the confidence set with respect to $c$ for $n \in
\{5000, 10000\}$. We plot the curve for $n=5,000$ and for $n=10,000$
on the same figure and observe that, when $c$ is small, the two curves
overlap completely but when $c$ is large, the $n=10,000$ curve lies
above the $n=5,000$ curve. This
provides empirical support to Theorem~\ref{thm:pa-conf-size} and
Theorem~\ref{thm:ua-conf-size}. In fact, this experiment shows that
the bound of $n^\gamma$ on the size of the confidence set in 
Theorem~\ref{thm:ua-conf-size} is loose; the actual size does not
increase with $n$. The fact that
the confidence set size seems to diverge with $n$ when $c$ is larger
supports Conjecture~\ref{con:size} and suggests that the
problem of root inference exhibits a phase transition when 
$\theta \approx \frac{1}{\sqrt{n}}$ under the LPA model and $\theta \approx \frac{\log
  n}{n}$ under the UA model. \\

{\color{black}
\noindent \textbf{Frequentist coverage under sequential noise models:} In our third simulation study, we verify Theorem~\ref{thm:frequentist-coverage1} for the seq-PAPER model with sequential noise described in Section~\ref{sec:seq_noise}. We generate $\bm{G}_n^*$ according to both the $\text{seq-PAPER}(\alpha, \beta, \theta, \talpha, \tbeta)$ model and the $\text{seq-PAPER}^*(\alpha, \beta, \theta, \talpha, \tbeta, \eta)$ model with deletion noise. We then construct the credible sets for the root node from posterior root probabilities computed via the algorithm given in Section~\ref{sec:algorithm}. We repeat the experiment over 200 independent trials and report the results in Tables~\ref{tab:seq1} and~\ref{tab:seq2}. We observe that the credible sets attain the nominal coverage. We also note that Table~\ref{tab:seq2} shows that the $\text{seq-PAPER}^*$ model can tolerate tree deletion probability up to $\eta = 0.08$ without significant increase in the confidence set sizes.  \\ 
}

\begin{table}[htp]
\begin{center}
\begin{tabular}{c|c|c|c|c|c|c|c|c|c|}
\hline
\text{$(\alpha, \beta)$ (with $\talpha=\alpha$, $\tbeta=\beta$)} & (0, 1) & (1, 0) & (0, 1) & (1, 0) 
  & (0, 1) & (1, 0) \\
\hline
\text{Theoretical coverage} & 0.8 & 0.8 & 0.95 & 0.95 & 0.99
                            & 0.99 \\
\hline
{\bf \text{Empirical coverage}} & {\bf 0.795} & {\bf 0.895} & {\bf 0.935} & {\bf 0.965} & {\bf 0.970} & {\bf 0.995} \\
  \hline
{ \text{Ave. conf. set size}} & 7 & 7 & 25 & 16 & 56
          & 28 \\
  \hline
\end{tabular}
\end{center}
\caption{Empirical coverage of our confidence set for the $\text{seq-PAPER}(\alpha, \beta, \theta, \talpha, \tbeta)$ model without deletion noise, with $\theta = 1.5$ and $\talpha = \alpha$ and $\tbeta = \beta$. We report the average over 200 trials. Graph has $n=600$ nodes and around $m \approx 1500$ edges in all cases.}
\label{tab:seq1}
\end{table}

\begin{table}[htp]
\begin{center}
\begin{tabular}{c|c|c|c|c|c|c|c|c|c|}
\hline
\text{$\eta$ (tree edge deletion probability)} & 0 & 0 & 0.04 & 0.04 
  & 0.08 & 0.08 \\
\hline
\text{Theoretical coverage} & 0.8 & 0.95 & 0.8 & 0.95 & 0.8
                            & 0.95 \\
\hline
{\bf \text{Empirical coverage}} & {\bf 0.825} & {\bf 0.96} & {\bf 0.84} & {\bf 0.95} & {\bf 0.85} & {\bf 0.98} \\
  \hline
{ \text{Ave. conf. set size}} & 5.9 & 14.1 & 6.3 & 15.0 & 6.7
          & 15.9 \\
  \hline
\end{tabular}
\end{center}
\caption{Empirical coverage of our confidence set for the $\text{seq-PAPER}^*(\alpha, \beta, \theta, \talpha, \tbeta, \eta)$ model with deletion noise, with $\alpha=0, \beta=1, \talpha=8, \tbeta=1, \theta = 1.5$ in all cases. We report the average over 200 trials. Graph has $n=300$ nodes and around $m \approx 750$ edges in all cases.}
\label{tab:seq2}
\end{table}

\noindent \textbf{Frequentist coverage for multiple roots:} Our next simulation
study is similar to the first except that we generate graphs from the
$\text{PAPER}(\alpha, \beta, K, \theta)$ model with $K=2$. We construct
our credible sets as described in Section~\ref{sec:confset-rootset}
and verify Theorem~\ref{thm:frequentist-coverage2} by showing that the
credible set at level $1-\epsilon$ also has frequentist coverage at
exactly the same level. We consider two different settings of parameters:
$\alpha=0, \beta=1$ (LPA) and $\alpha=1, \beta=0$ (UA). We generate $\bm{G}_n^*$ according to the
$\text{PAPER}(\alpha, \beta, K, \theta)$ model with $n=700$ nodes,
$m=1,000$ edges, and $K=2$. We then estimate $\alpha$ and $\beta$
using the method given in Section~\ref{sec:parameter-estimation}, compute the level $\epsilon \in \{0.2, 0.05, 0.01\}$
credible sets, and record whether they contain the true set of root nodes. We
repeat the experiment over 200 independent trials and report the
results in Table~\ref{tab:coverage2}. We observe that the credible
sets attain the nominal coverage. In the LPA setting, the size of the credible sets
are small but in the UA setting, the sizes of the credible sets become
much larger. We relegate an in-depth analysis of this phenomenon to
future work. \\

\begin{table}[htp]
\begin{center}
\begin{tabular}{c|c|c|c|c|c|c|c|c|c|}
\hline
\text{$(\alpha, \beta)$} & (0, 1) & (1, 0) & (0, 1) & (1, 0) 
  & (0, 1) & (1, 0) \\
\hline
\text{Theoretical coverage} & 0.8 & 0.8 & 0.95 & 0.95 & 0.99
                            & 0.99 \\
\hline
{\bf \text{Empirical coverage}} & {\bf 0.826} & {\bf 0.826} & {\bf 0.933} & {\bf 0.964} & {\bf 0.974} & {\bf 0.985} \\
  \hline
{ \text{Ave. conf. set size}} & 5 & 57 & 12 & 155 & 31
          & 295 \\
  \hline
\end{tabular}
\end{center}
\caption{Empirical coverage of our confidence set for the set of $K=2$ root
  nodes. We report the average over 200 trials. Graph has $n=700$ nodes
  and $m= 1,000$ edges in all cases.}
\label{tab:coverage2}
\end{table}

\begin{figure}[htp]
  \centering
  \includegraphics[scale=.5]{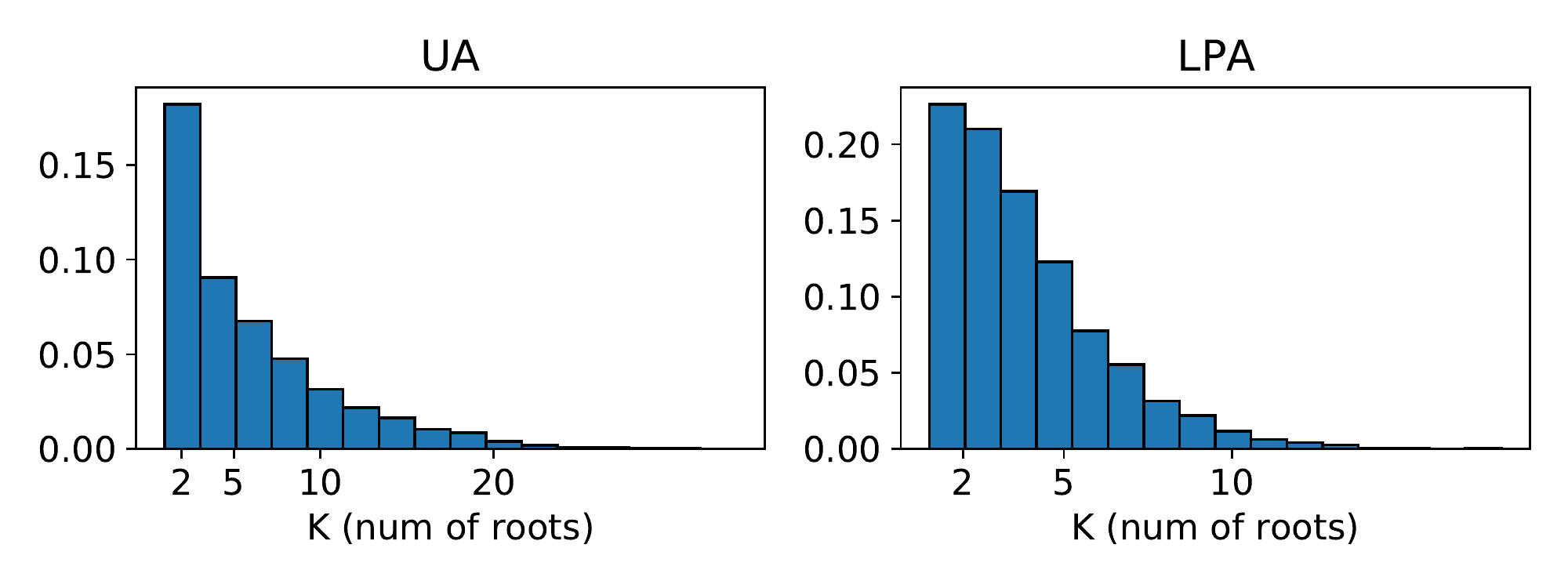}
  \caption{Posterior distribution over $K$ averaged across 20
    independent trials. \textbf{Left:} 
    networks have two latent UA trees. \textbf{Right:} networks have
    two latent LPA trees.}
  \label{fig:randomK}
\end{figure}

\noindent \textbf{Posterior on $K$ in the random $K$ roots setting:}
In our last simulation experiment, we generate PAPER graphs with
$K=2$ roots but perform posterior inference using the
$\text{PAPER}(\alpha, \beta, \alpha_0, \theta)$ model and study
resulting posterior distribution over the number of roots $K$. We consider two different settings of parameters:
$\alpha=0, \beta=1$ (LPA) and $\alpha=1, \beta=0$ (UA). We generate $\bm{G}_n^*$ according to the
$\text{PAPER}(\alpha, \beta, K, \theta)$ model with $n=700$ nodes,
$m=1,000$ edges, and $K=2$. We report the posterior distribution over
$K$, averaged over 20 independent trials, in
Figure~\ref{fig:randomK}. We observe that, in both cases, the mode of the posterior
distribution over $K$ is 2, which is the true number of
roots. However, the distributions exhibits high variance, which could
be due to the fact that the two true latent trees may have
significantly different sizes. 

\subsection{Single root analysis on real data}

We now apply the single root PAPER model on real world networks. In
a few cases (Section~\ref{sec:flu}), we can ascertain from domain knowledge that the network originated
from a single root node but more often, we use the single root model
to identify important nodes and subgraphs (Section~\ref{sec:single-root-subgraph}).

\subsubsection{Flu transmission network}
\label{sec:flu}

\begin{figure}
  \centering
  \includegraphics[scale=.23]{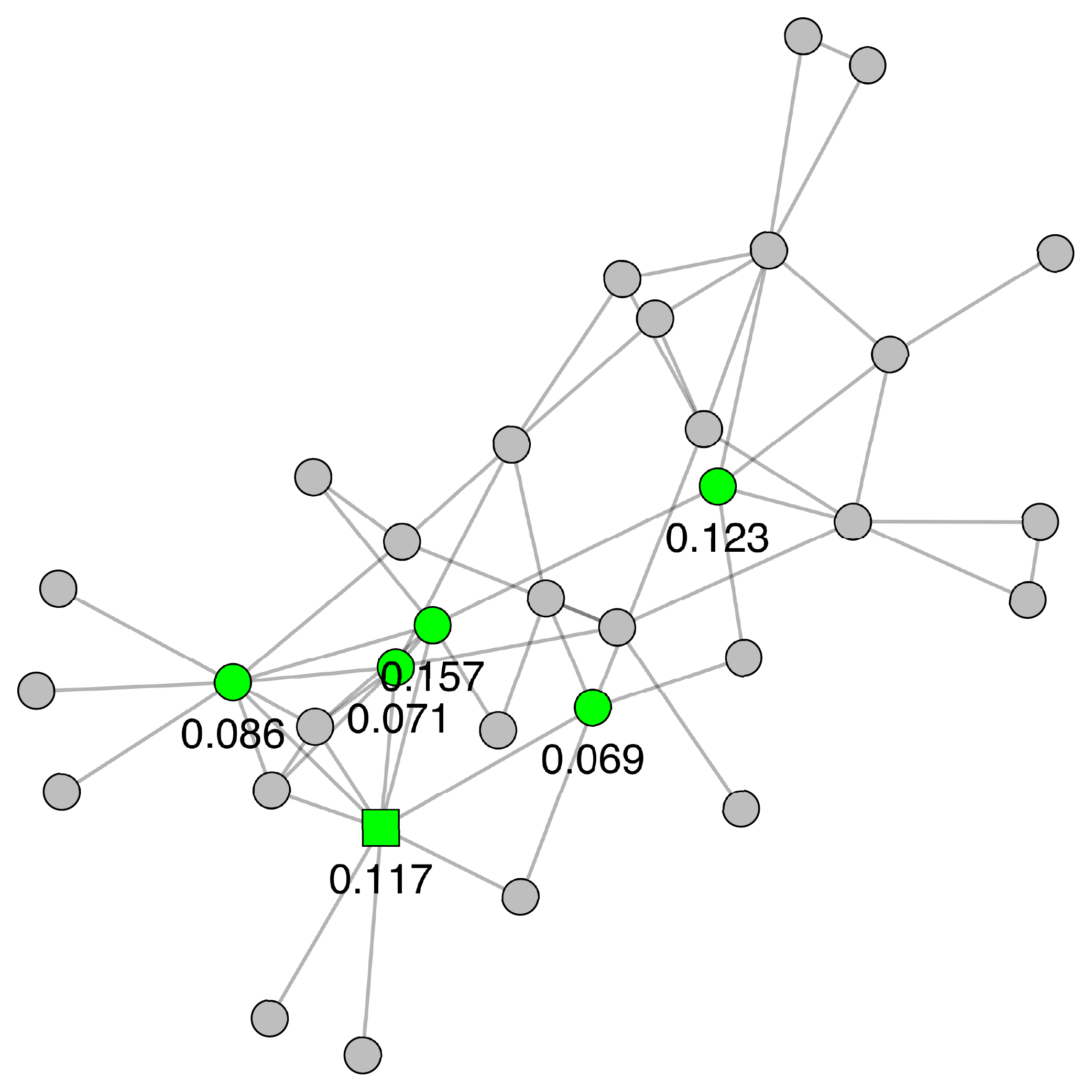}
  \includegraphics[scale=.23]{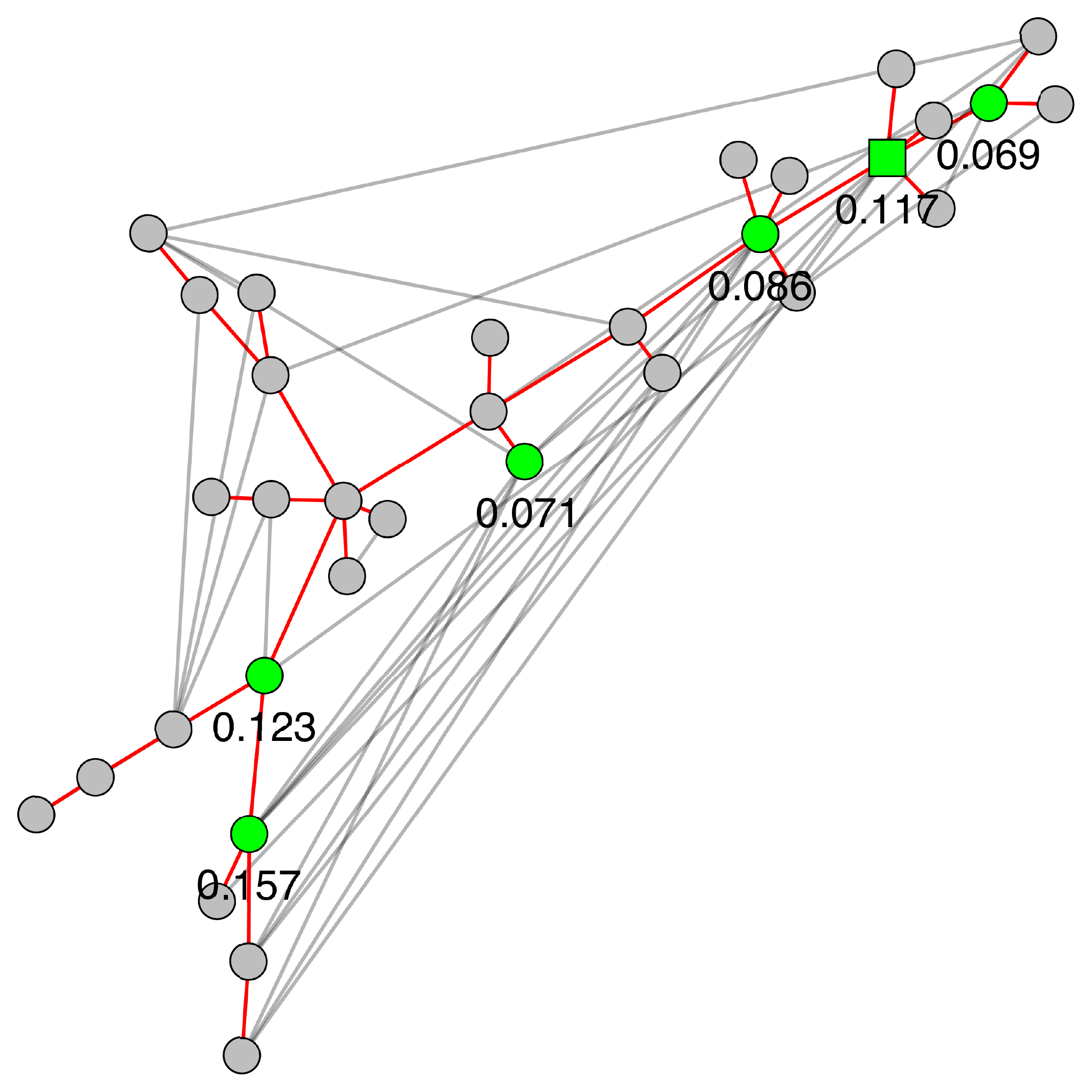}
  \includegraphics[scale=.23]{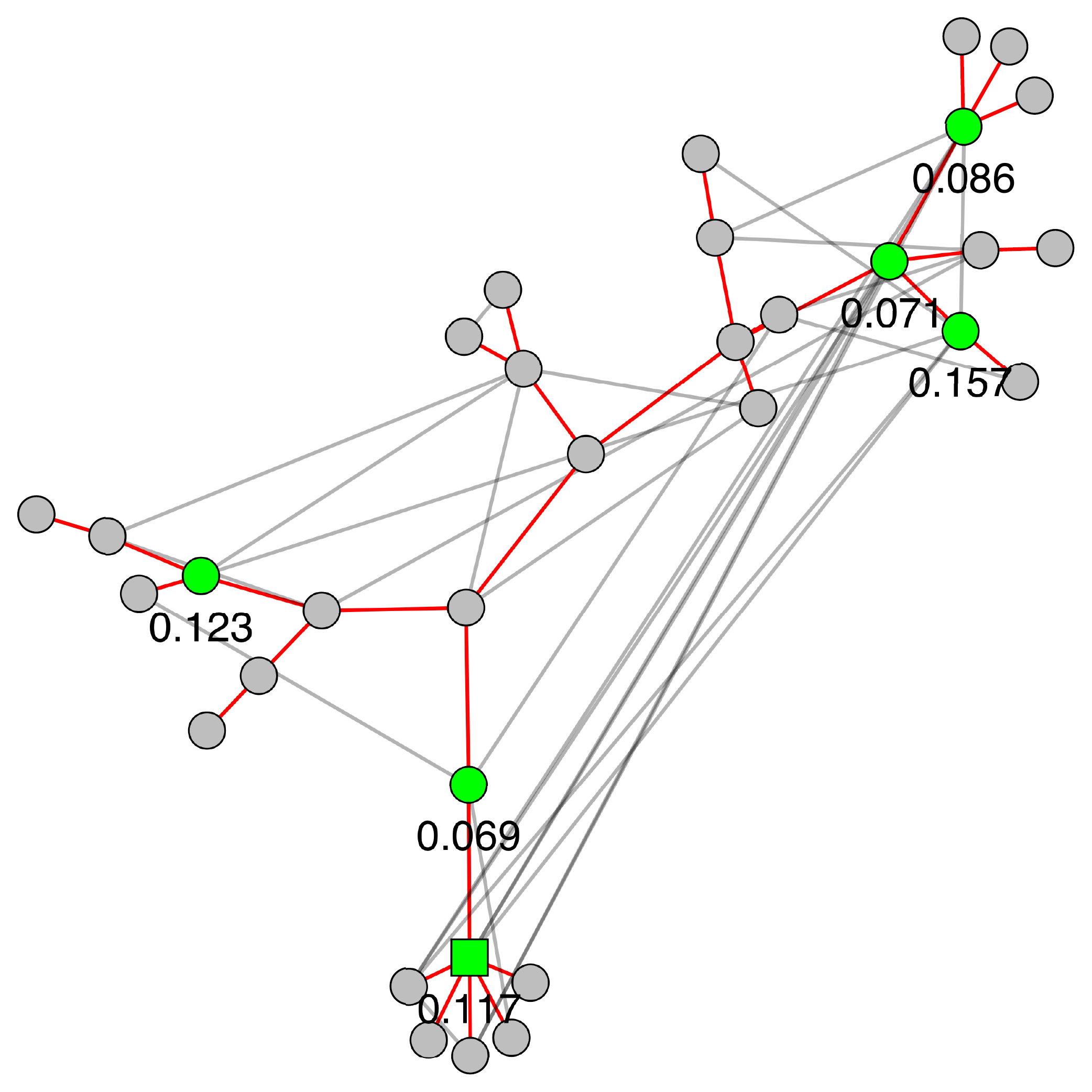}
  \caption{\textbf{Left:} contact network among 32 students in a flu
    outbreak. \textbf{Center and right:} two examples of the latent
    tree generated by the Gibbs sampler.}
  \label{fig:flu}
\end{figure}

We analyze a person-to-person contact network
among 32 students in a London classroom during a flu outbreak
\citep{hens2012robust}. We extract the data from Figure 3 in
\cite{hens2012robust} and illustrate the network in the left
sub-figure of Figure~\ref{fig:flu}. Public health investigation revealed that the
outbreak originated from a single student, which is the true
patient zero and shown as the orange node in
Figure~\ref{fig:flu}. We apply the PAPER model with a single root to
this network.  We estimate that $\beta = 1$ and $\alpha = 53.06$ using
the method described in Section~\ref{sec:parameter-estimation} and
compute the $60\%, 80\%, 95\%$, and $99\%$ confidence sets. All the
confidence sets contain the true patient zero and their sizes are as followed:
\begin{align*}
 \text{ $60\%$: 6 nodes } \quad   \text{ $80\%$: 10
                                        nodes } \quad
  \text{ $95\%$: 19 nodes } \quad  \text{ $99\%$: 27 nodes.}
\end{align*}

We provide the approximate posterior root probabilities of the top 7
nodes in Figure~\ref{fig:flu}. The true patient zero has a posterior root probability of
$0.11$ is the node with the 3rd highest posterior root probability. In
the center and right sub-figure of Figure~\ref{fig:flu}, we also show
two of the latent trees $\tilde{\bm{T}}_n$ that were generated by the
Gibbs sampler. 

\subsubsection{Visualizing central subgraphs}
\label{sec:single-root-subgraph}

Large scale real graphs are difficult to visualize but one can often
learn salient structural properties of a graph by visualizing a
smaller subgraph that contains the most important nodes. In this section,
we apply the single root PAPER model on four large networks and, for each graph, display the
subgraph that comprises the 200 nodes with the highest posterior root
probability. We see that the result reveals striking differences
between the different graphs. Unfortunately, we do not have the node
labels on any of these four graphs and can only make qualitative
interpretations of the results.\\

\noindent \textbf{MathSciNet collaboration network:} We first consider
a collaboration network of research publications from MathSciNet, which is publicly
available in the Network Repository \citep{networkrepository} at the
link \texttt{http://networkrepository.com/ca-MathSciNet.php}. This
network has $n=332,689$ nodes and $m=820,644$ edges, with a maximum degree of
$496$. Using the method described in
Section~\ref{sec:parameter-estimation}, we estimate $\beta = 1$ and
$\alpha = 0$. The sizes of confidence sets are:
\begin{align*}
 \text{ $60\%$: 3 nodes } \quad   \text{ $80\%$: 6
                                        nodes } \quad
  \text{ $95\%$: 21 nodes } \quad  \text{ $99\%$: 112 nodes.}
\end{align*}

\begin{figure}[htp]
\centering
	\begin{subfigure}[b]{0.35\textwidth}
		\centering
		{\includegraphics[scale=.2]{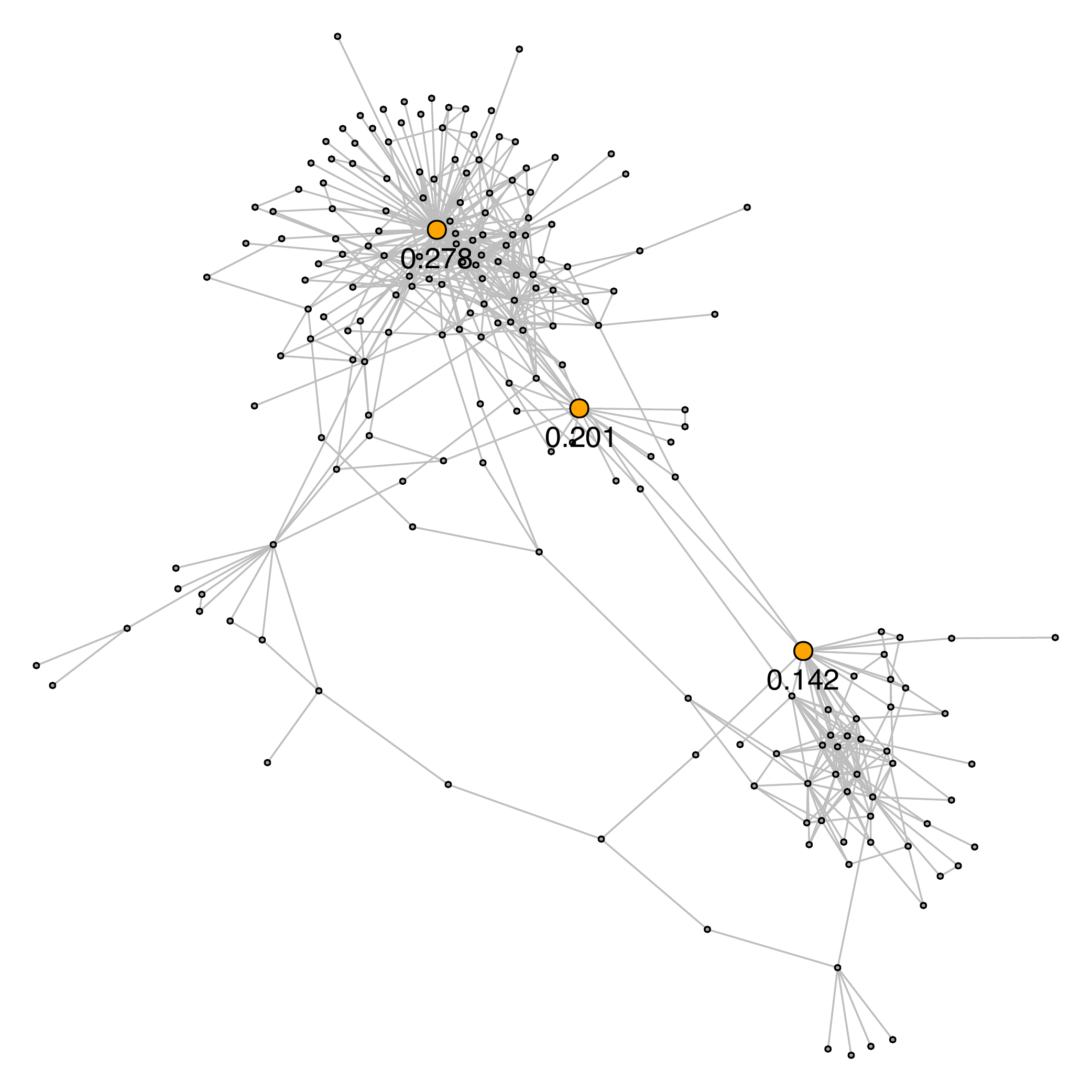}}
		\caption{MathSciNet subgraph}
                \label{fig:mathscinet}
	\end{subfigure}
	\begin{subfigure}[b]{0.35\textwidth}
		\centering
		{\includegraphics[scale=.2]{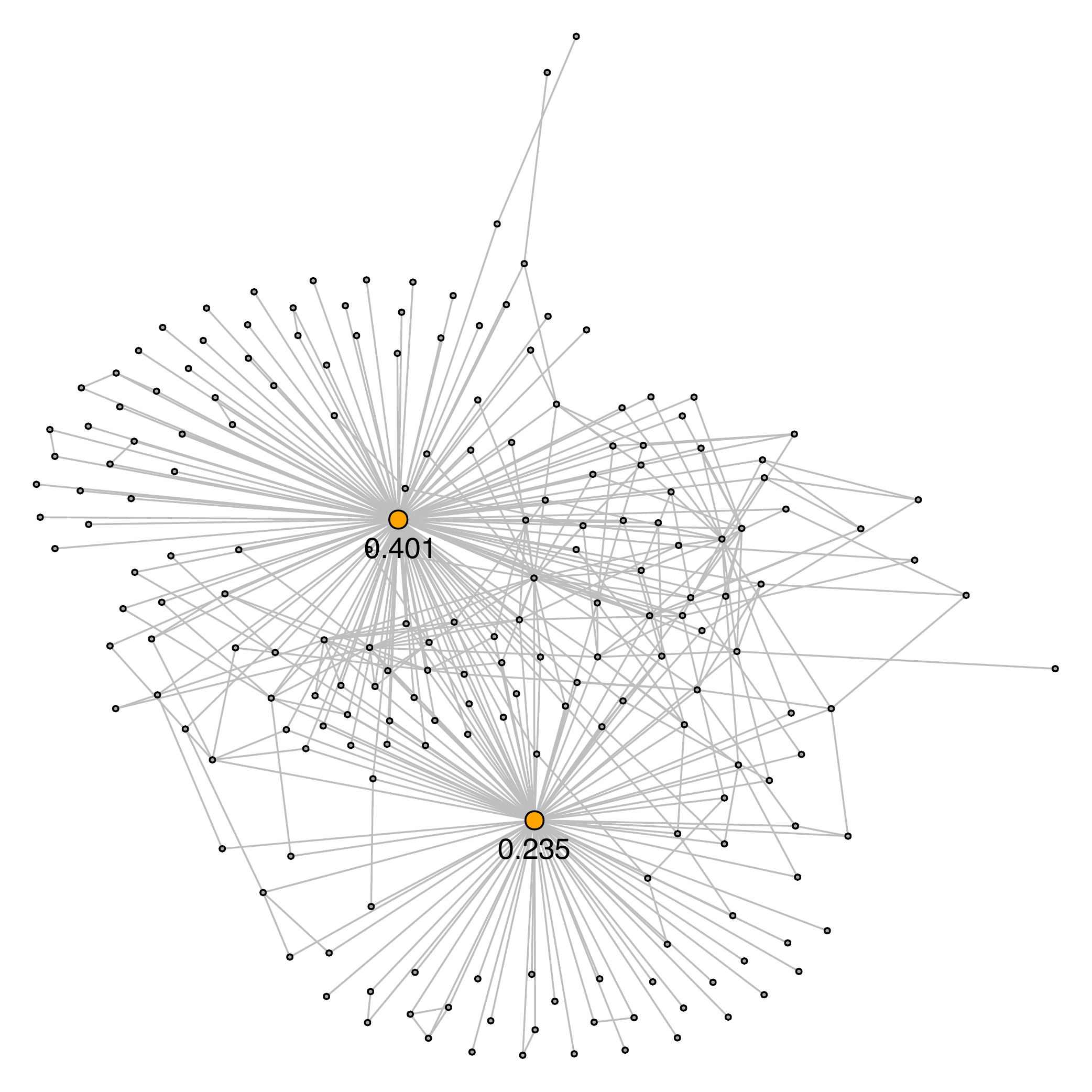}}
		\caption{Notre Dame subgraph}
                \label{fig:notredame}
         \end{subfigure}
\caption{Subgraph of the 200 nodes with highest posterior root
  probabilities.}
\end{figure}

We display the subgraph containing the 200 nodes with the highest
posterior root probability in Figure~\ref{fig:mathscinet}. We observe
that the subgraph reveals a cluster structure that may represent
the different academic disciplines. \\

\noindent \textbf{University of Notre Dame website network:} We study
a network of hyperlinks between webpages of University of Notre
Dame \citep{albert1999diameter}, which is publicly available at the
website 
\texttt{https://snap.stanford.edu/data/web-NotreDame.html}. This
network has $n=325,729$ nodes and $m=1,090,108$ edges, with a maximum degree
of 10,721. Using the method described in
Section~\ref{sec:parameter-estimation}, we estimate $\beta = 1$ and
$\alpha = 0$. The sizes of confidence sets are:
\begin{align*}
 \text{ $60\%$: 2 nodes }  \quad  \text{ $80\%$: 21
                                        nodes } \quad
  \text{ $95\%$: 524 nodes } \quad  \text{ $99\%$: 3498 nodes }.
\end{align*}

We observe
that the central subgraph (shown in Figure~\ref{fig:notredame})
reveals two hub nodes with many sparsely connected ``spokes''. \\

\subsection{Community recovery with the fixed $K$ model}
\label{sec:fixed-K-real-data}

In this section, we show that we can use the PAPER model
with multiple roots for community recovery on real world networks. To
estimate the community membership from the posterior samples, we use
a greedy matching procedure. To be precise, our Gibbs sampler outputs a sequence of forests $\tilde{\bm{f}}_n^{(1)}, \ldots, \tilde{\bm{f}}_n^{(J)}$ where
$J$ is the number of Monte Carlo samples. Each forest
$\tilde{\bm{f}}_n^{(j)}$ contains $K$ component trees which we denote
$\tilde{\bm{t}}^{(1, j)}, \tilde{\bm{t}}^{(2, j)}, \ldots,
\tilde{\bm{t}}^{(K, j)}$. We write $Q_k^{(j)}(\cdot) := \Pb( \Pi_1 = \cdot \,|\, \tilde{\bm{T}} =
\tilde{\bm{t}}^{(k, j)})$ as the posterior root distribution of the $k$-th tree of the $j$-th Monte
Carlo sample. Since the tree labels may switch from sample to sample,
we use the following matching procedure: we maintain $K$ distributions $Q_1(\cdot), Q_2(\cdot), \ldots, Q_K(\cdot)$ and
initially set $Q_k = Q_k^{(1)}$ for all $k \in [K]$. Then, for
$j=2,3,\ldots, J$, we use the Hungarian algorithm to compute a one-to-one matching $\sigma \,:\, [K]
\rightarrow [K]$ that minimizes the overall total variation distance
\[
\sum_{k = 1}^K \text{TV}(Q_k^{(j)}, Q_{\sigma(k)}).
\]
Once we compute the matching, we then update $Q_{\sigma(k)} \leftarrow \frac{j-1}{j} Q_{\sigma(k)} + \frac{1}{j} Q_k^{(j)}$.

In this way, we interpret $Q_1, \ldots, Q_K$ as the average posterior root
distributions for the $K$ trees across all the Monte Carlo samples and
using the matching, we may also compute the posterior probability $\Pb(
\text{ $u$ in tree $k$ } \,|\, \tilde{\bm{G}}_n = \tilde{\bm{g}}_n)$,
which allows us to perform community detection -- we put node $u$ in
cluster $k$ if 
$\Pb( \text{ $u$ in tree $k$ } \,|\, \tilde{\bm{G}}_n =
\tilde{\bm{g}}_n) \geq \Pb( \text{ $u$ in tree $k'$ } \,|\, \tilde{\bm{G}}_n =
\tilde{\bm{g}}_n)$ for all $k' \neq k$. We use the greedy matching
procedure for computational efficiency -- slower but more principles
approaches are studied by e.g. \cite{wade2018bayesian}. 

\subsubsection{Karate club network}

\begin{figure}[htp]
  \centering
  \includegraphics[scale=.22]{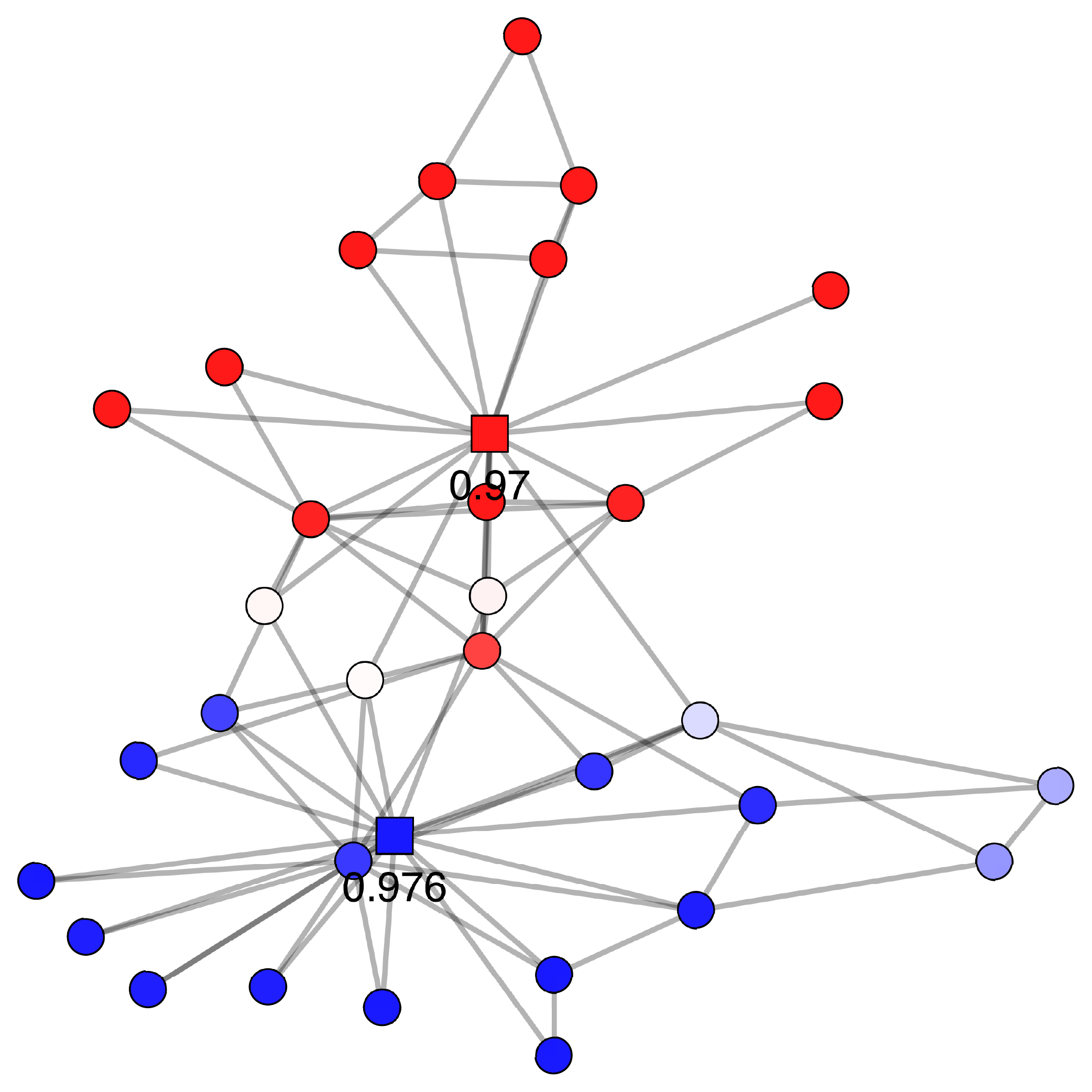}
  \includegraphics[scale=.22]{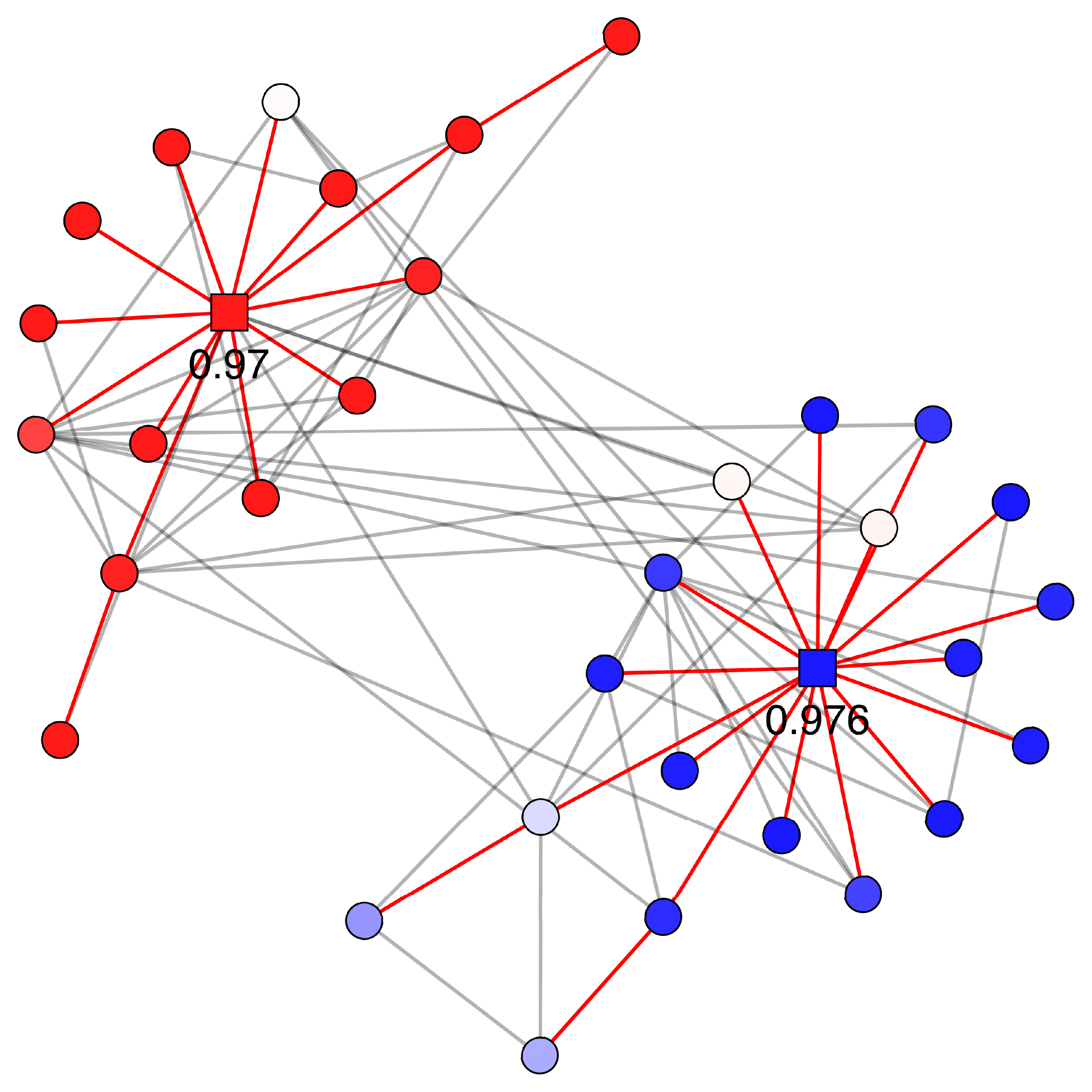}
  \includegraphics[scale=.22]{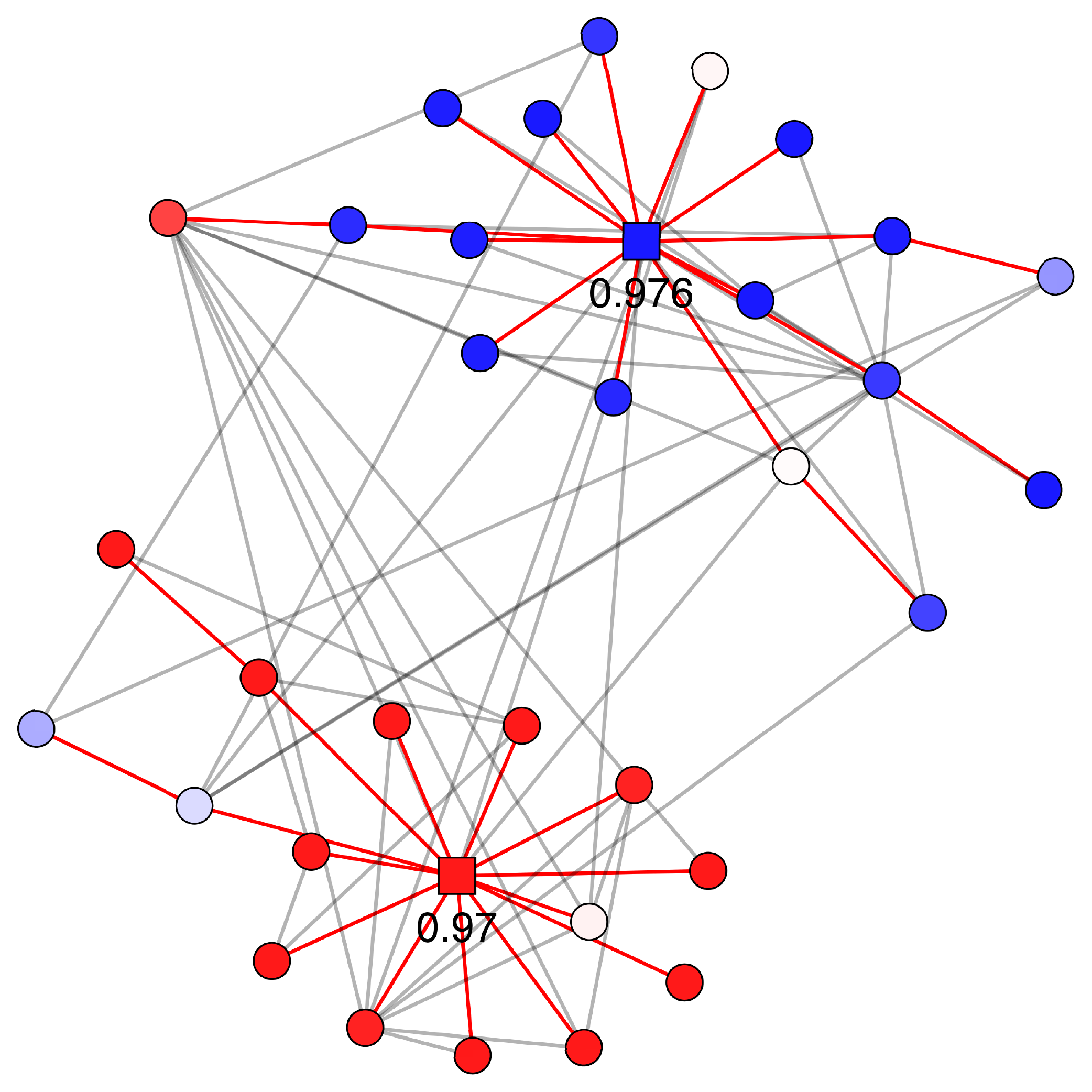}
  \caption{\textbf{Left:} karate club network where node color reflects
    community membership probability. \textbf{Center and right:} two
    examples of the latent forest generated by the
    Gibbs sampler.}
  \label{fig:karate}
\end{figure}

We apply the PAPER model to Zachary's karate club network
\cite{zachary1977information}, which is publicly available at
\texttt{http://www-personal.umich.edu/~mejn/netdata/}.
The karate club network has $n=34$ nodes and $m=76$ edges, where two
individuals share an edge if they socialize with each other. The network has two ground truth communities, one led by the instructor and one led by the
administrator (shown as rectangular nodes in
Figure~\ref{fig:karate}. These two communities later split into two
separate clubs. In this case, we apply the PAPER model with $K=2$
roots. For every node $u$, we consider the community membership
probability $\Pb( \text{$u$ in tree 1} \,|\,
\tilde{\bm{G}}_n)$ and assign $u$ to community $1$ if and only if this
value is greater than $0.5$. We show the result in in
Figure~\ref{fig:karate}, where each node has a color that reflects its
community membership probability.

We correctly cluster
all but one node, which matches the performance of degree-corrected
SBM \cite{karrer2011stochastic, amini2013} (DCSBM)--the current the state of the art
model for community detection. The node that we misclassify has a
posterior probability $\Pb( \text{$u$ in tree 1} \,|\,
\tilde{\bm{G}}_n) = 0.47$, indicating that the model is indeed unsure
of whether it belong in community 1 or 2. We note that the PAPER model requires only 3 parameters whereas the DCSBM for this network requires 38 parameters because each node
has a degree
correction parameter. SBM without degree correction performs badly \cite{karrer2011stochastic}. 

\subsubsection{Political blogs network}

Next, we analyze a political blogs network~\citep{adamic2005political}
that is frequently used as a benchmark for network clustering
algorithms; the full network is publicly available at the website 
\texttt{http://www-personal.umich.edu/~mejn/netdata/}.
This network contains $m=16,714$ edges between $n=1,222$
blogs, where two blogs are connected if one contains a link to the
other. For simplicity, we treat the network as undirected.

\begin{figure}
  \centering
  \includegraphics[scale=.33]{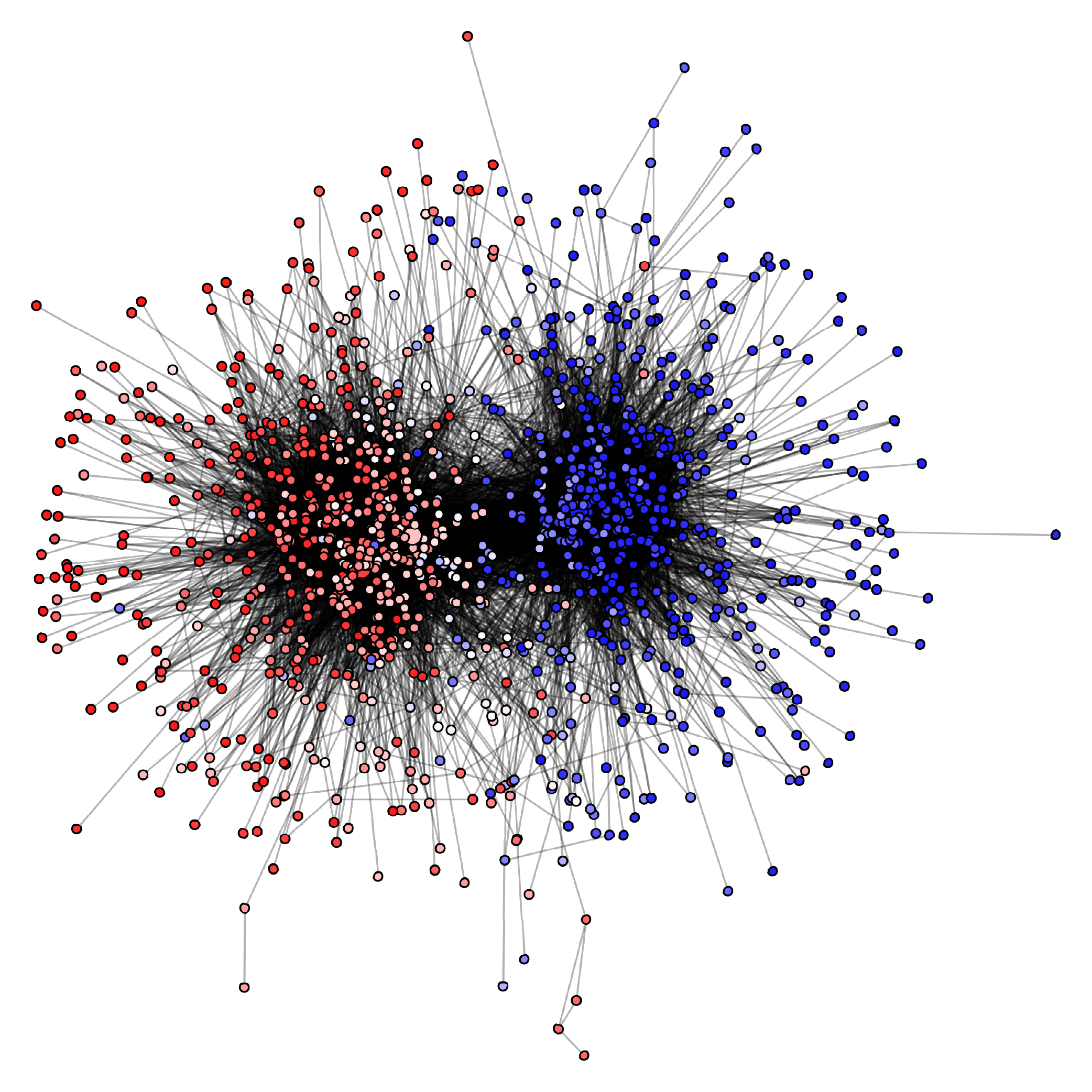}
  \includegraphics[scale=.33]{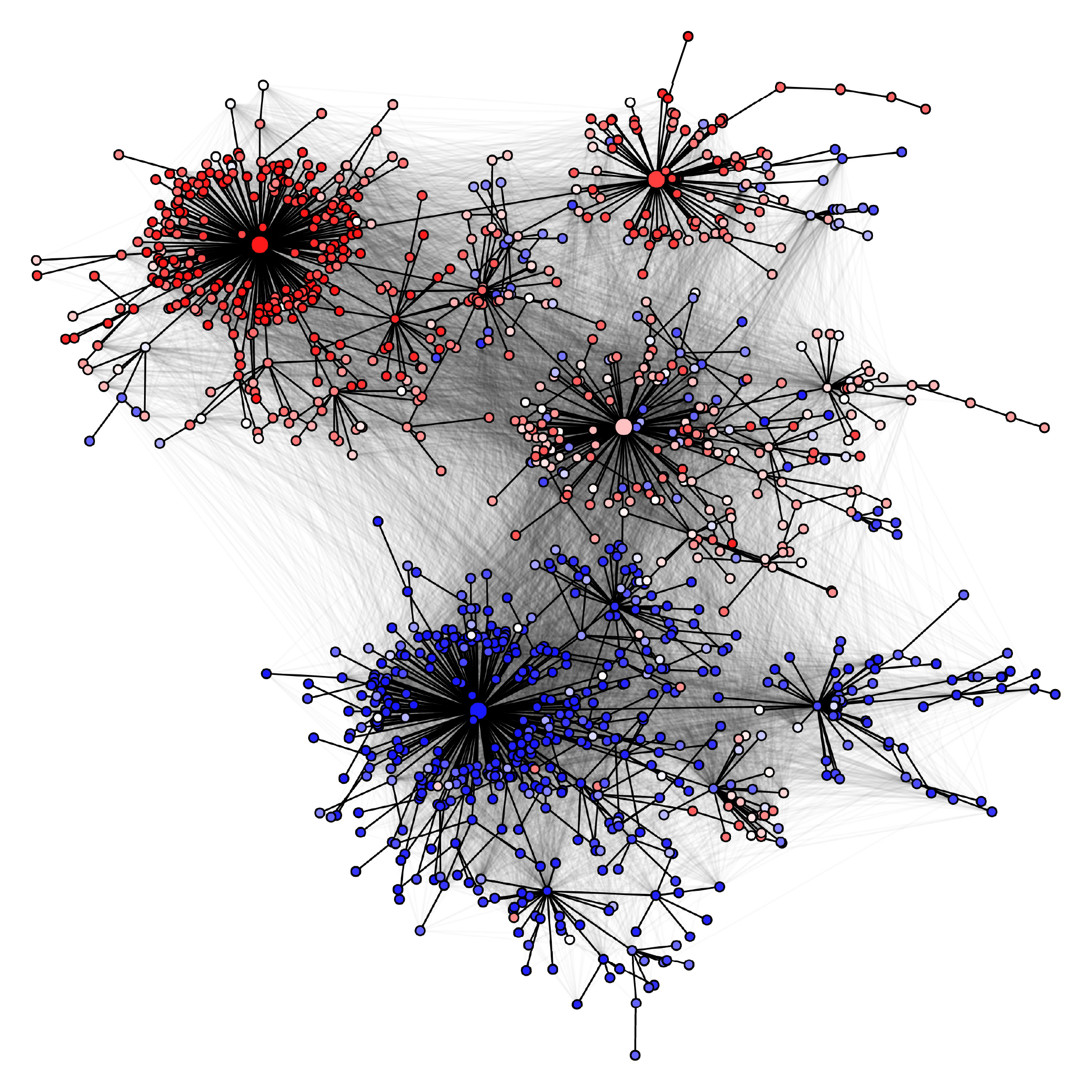}
  \caption{\textbf{Left:} political blog network where node color reflects
    community membership probability. \textbf{Right:} one example of a
    forest generated by the Gibbs sampler. The 5 nodes with the larger marker comprise the $95\%$ confidence
  set for the roots. }
  \label{fig:polblog}
\end{figure}

The network again has two ground truth communities, one that comprise
of left-leaning blogs and one that comprises of right-leaning blogs. 
We again apply the PAPER model with $K=2$
roots and for every node $u$, we compute the community membership
probability $\Pb( \text{$u$ in tree 1} \,|\,
\tilde{\bm{G}}_n)$ and assign $u$ to community $1$ if and only if this
value is greater than $0.5$. We show the result in in
Figure~\ref{fig:polblog}, where each node has a color that reflects its
community membership probability.

Our overall misclustering error rate is \textbf{9.1\%}, which is
high compared to current state of the art approaches; for example, the
SCORE method \citep{jin2015fast} attains
an error rate of about 5\%. However, we compute the misclustering
error rate with respect to only the top 400 nodes with the highest
posterior root probabilities, which can be interpreted as the most
important nodes in the graph, our misclustering error rate drops to
\textbf{3.5\%}. This confirms our intuition that the PAPER model, when
used for clustering, is more reliable for central nodes than for
peripheral nodes.

\subsection{Community discovery with the random $K$ model}
\label{sec:random-K-real-data}
For networks with an unknown number of small and possibly overlapping
communities, 
the random $K$ model $\text{PAPER}(\alpha, \beta,
\alpha_0, \theta)$ can be useful for discovering complex community structures. To extract
community information from the posterior samples, we again use a greedy
matching procedure. To be precise, in the random $K$ setting, our proposed Gibbs sampler outputs a sequence of
forests $\tilde{\bm{f}}_n^{(1)}, \ldots, \tilde{\bm{f}}_n^{(J)}$ where
$J$ is the number of Monte Carlo samples. We write each forest
$\tilde{\bm{f}}_n^{(j)}$, for $j \in [J]$, as a collection of trees $\{
\tilde{\bm{t}}^{(1, j)}, \ldots, \tilde{\bm{t}}^{(K_j, j)}\}$ where
$K_j$ is the number of trees in $\tilde{\bm{f}}_n^{(j)}$. For $j \in
[J]$ and $k \in [K_j]$, we write $Q_k^{(j)}(\cdot) = \Pb( \Pi_1 = \cdot \,|\, \tilde{\bm{T}} =
    \tilde{\bm{t}}^{(k, j)})$ as the posterior root distribution of
    the $k$-th tree in the $j$-th Monte Carlo sample. To summarize the output in an interpretable way, we do the following: 
\begin{enumerate}
  \item We initialize $K_{\text{all}} = \max_{j \in [J]} K_j$ and 
    $Q_k = Q_k^{(1)}$ for $k = 1,2,\ldots, K_1$. For $k = K_1 + 1,
    \ldots, K_{\text{all}}$, we initialize $Q_k(\cdot) = 0$. 
  \item For $j = 2, 3, \ldots, J$, we match
    $\{ Q_1, \ldots, Q_{K_{\text{all}}} \}$ with $\{Q_1^{(j)}, \ldots,
    Q_{K_j}^{(j)}\}$ by computing a one-to-one matching $\sigma \,:\, [K_j]
    \rightarrow [K_{\text{all}}]$ that minimizes
    \[
      \sum_{k = 1}^{K_j} \text{TV}(Q_k^{(j)},
      Q_{\sigma(k)}). 
    \]
    For every $k \in [K_j]$, if the total variation distance between the $k$-th pair of the
    matching is too large, that is $\text{TV}( Q_k^{(j)},
    Q_{\sigma(k)}) > 0.75$, then we create a new  set $K_{\text{all}} \leftarrow K_{\text{all}} + 1$
    and set $Q_{K_{\text{all}}+1} \leftarrow Q_k^{(j)}$; otherwise, we perform the
    update $Q_{\sigma(k)} \leftarrow \frac{j-1}{j} Q_{\sigma(k)} + \frac{1}{j} Q_{k}^{(j)}$.
    \item We output $\{Q_1, \ldots, Q_{K_{\text{all}}} \}$ as the discovered clusters, represented as posterior root probability distributions.
\end{enumerate}

\noindent For all of our experiments, we only include trees that contain at
least $1\%$ of the total number of nodes. For each discovered cluster $Q_\ell$ for $\ell \in [K_{\text{all}}]$, we also compute $\rho_{Q_\ell}$ as the number of Monte Carlo iteration $j \in [J]$ where we match $Q_\ell$ with $Q_k^{(j)}$, i.e. $\sigma(k) = \ell$, and update $Q_{\ell}$. We then compute $\frac{\rho_{Q_\ell}}{J}$ as the posterior frequency of cluster $Q_{\ell}$.

\begin{figure}
  \centering
  \includegraphics[scale=0.4]{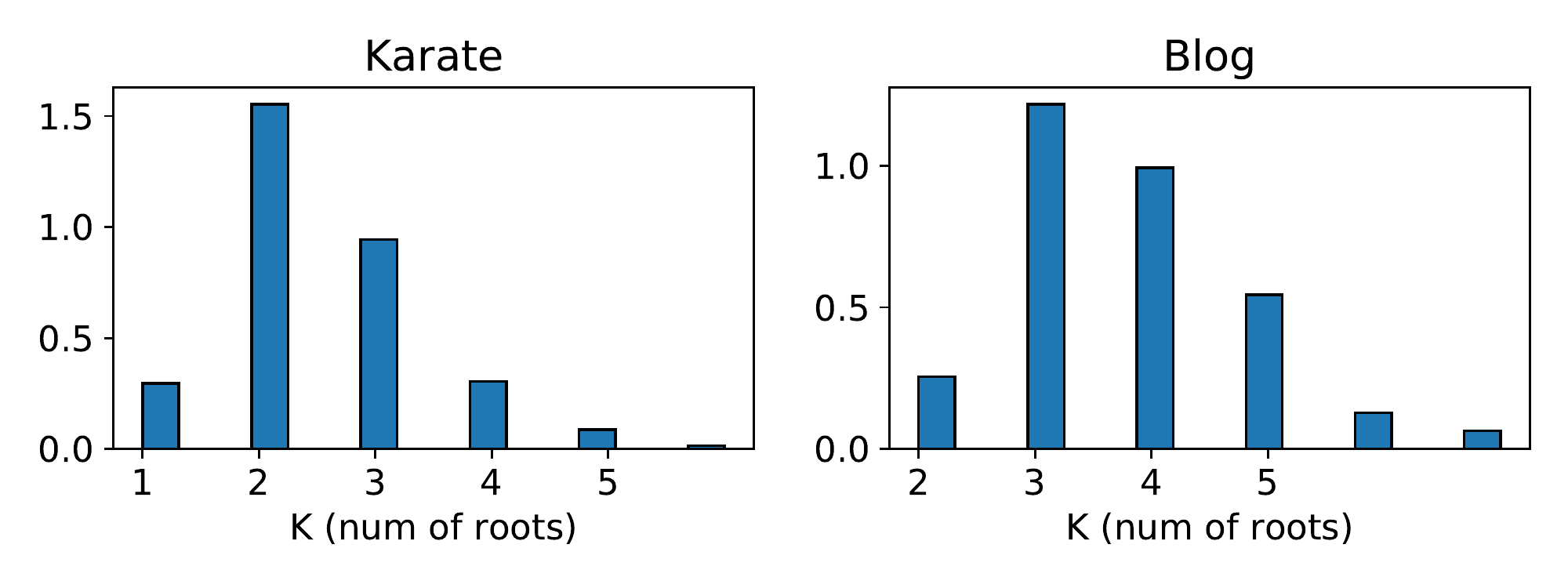}
  \caption{Posterior over $K$ using the random $K$ roots model on the karate club network (\textbf{left})
    and the political blog network (\textbf{right}).}
  \label{fig:randomKreal}
\end{figure}

In order to check that the random $K$ model is reasonable, we first
apply it to the karate club and the political blog networks, which we know
contain two underlying clusters, and analyze
the resulting posterior distribution over the number of cluster-trees
$K$. We provide the results for the karate club network in the left
part of Figure~\ref{fig:randomKreal}, in which we see that the
posterior mode is at $K=2$. For the political blog network, the Gibbs
sampler tends to produce a few large clusters and many tiny clusters
of fewer than 10 nodes. Therefore, to compute the posterior over $K$,
we count only clusters that have at least 12 nodes (1\% of the total
number of nodes) and give the results in the right part of
Figure~\ref{fig:randomKreal}. The posterior mode in this case is
$K=3$, which is reasonably close to the ground truth. 

We also analyze an air route network \citep{guimera2005worldwide} of $n=3,618$ airports and $m=14,142$
edges where two airports share an edge if there is a regularly
scheduled flight between them. We remove the direction
of the edges and treat the network as undirected. The dataset is
publicly available at
\texttt{http://seeslab.info/downloads/air-transportation-networks/}. Using the random $K$ model, we discover a large central cluster containing major airports around the world and various small clusters that correspond to more remote regions such as airports on Pacific and Polynesian islands, airports in Alaska, and airports in the Canadian Northwest Territories. For sake of brevity, we defer the detailed results to Section~\ref{sec:airroute} of the Appendix. 

{\color{black}

\subsection{Analysis of statistician co-authorship network}
\label{sec:coauthor}

We now apply PAPER models to perform an extensive analysis of a statistician co-authorship network constructed by \cite{ji2016coauthorship}. In this network, each node corresponds to a statistician and two nodes $u$ and $v$ have an edge between them if they have co-authored 1 or more papers in either \emph{Journal of Royal Statistical Society: Series B}, \emph{Journal of the American Statistical Association}, \emph{Annals of Statistics}, or \emph{Biometrika} from 2002 to 2013. We consider only the largest connected component which has $n=2263$ nodes and $m=4388$ edges. \cite{ji2016coauthorship} in their manuscript (Section 4.3) refers to this network as "Coauthorship Network (B)". We emphasize that since the data reflect only coauthorship in 4 journals in the period 2002-2013, the results that we produce cannot be used to compare researchers---we use this network only to illustrate PAPER models in a setting where we can more easily assess whether the output is meaningful or not. \\

\begin{figure}[htp]
\centering
\includegraphics[scale=.36]{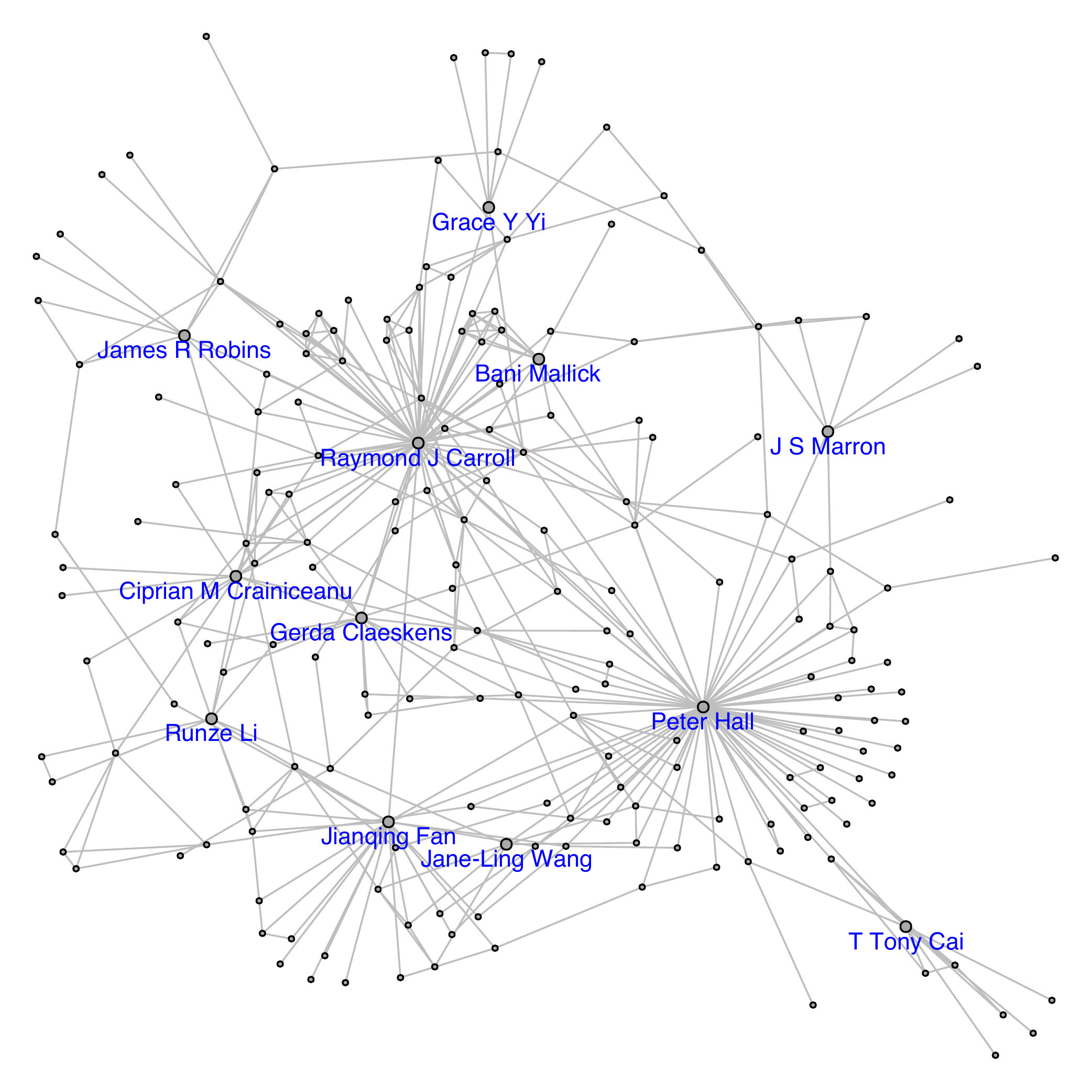}
\caption{Subgraph of the co-authorship graph comprising the 200 nodes with the highest posterior root probabilities. We label the 12 nodes with the highest root probabilities.}
\label{fig:stats_sm}
\end{figure}

\textbf{Single root analysis: } We first use the single root $\text{PAPER}(\alpha, \beta, \theta)$ model where we estimate $\alpha = 0$, $\beta=1$ using the EM algorithm described in Section~\ref{sec:parameter-estimation}. We find that the following 4 nodes have the highest posterior root probabilities: (1) Raymond Carroll with root probability 0.32, (2) Peter Hall with root probability 0.26, (3) Jianqing Fan with root probability 0.086, and (4) James Robins with root probability 0.048. The root probability ranking align closely with betweenness centrality ranking, in which Raymond Carroll, Peter Hall, and Jianqing Fan are also the top 3 most central nodes; see Table 2 of \cite{ji2016coauthorship}. Both the root probability ranking and the betweenness ranking differ significantly from degree ranking. We also display the subgraph of the 200 nodes with the highest posterior root probabilities in Figure~\ref{fig:stats_sm} where we labeled the top 12 nodes with the highest root probabilities. \\

\textbf{Community detection with random $K$ roots model: } Using our inference algorithm and the greedy matching procedure in~\ref{sec:random-K-real-data}, we compute clusters $\{Q_1, \ldots, Q_{K_{\text{all}}}\}$ where we find about $K_{\text{all}} \approx 40$ significant clusters. We order the clusters by their posterior frequencies and display the top 9 clusters in Figure~\ref{fig:coauthor_cluster1}, along with labels that we curated; we display the nodes in the cluster as word clouds in which the word size is proportional to the posterior root probabilities. We display 18 additional clusters in Section~\ref{sec:coauthor_appendix} of the Appendix. We note that the clusters can overlap since they are constructed from a sequence of posterior samples by matching; see the first paragraph of Section~\ref{sec:random-K-real-data}.  
}

\begin{figure}[ht]
\begin{subfigure}{0.32\textwidth}
\centering
\includegraphics[scale=.24, trim=1in 2in 1in 1in, clip]{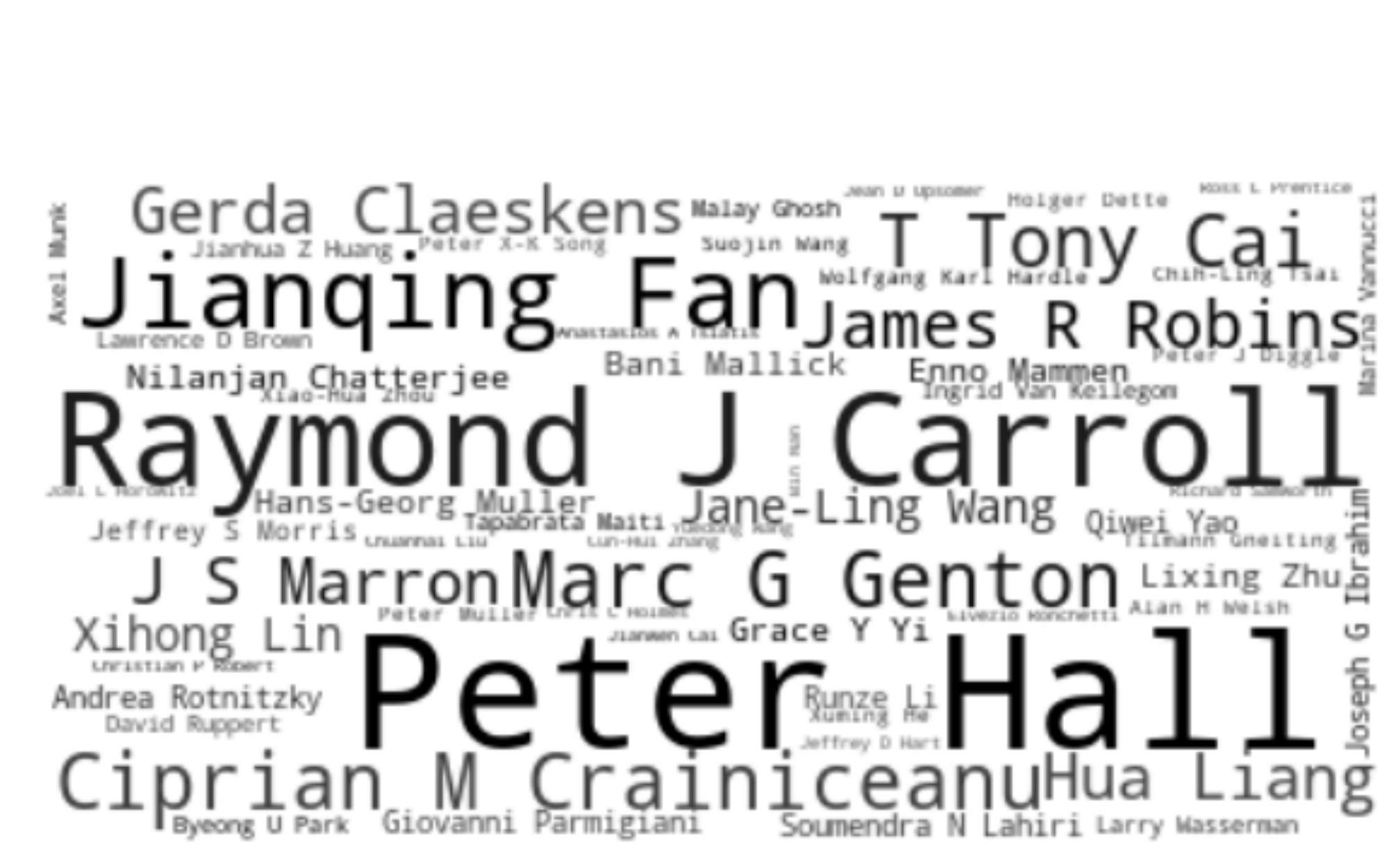}
\caption{Central super-cluster}
\end{subfigure}
\begin{subfigure}{0.32\textwidth}
\centering
\includegraphics[scale=.24, trim=1in 2in 1in 1in, clip]{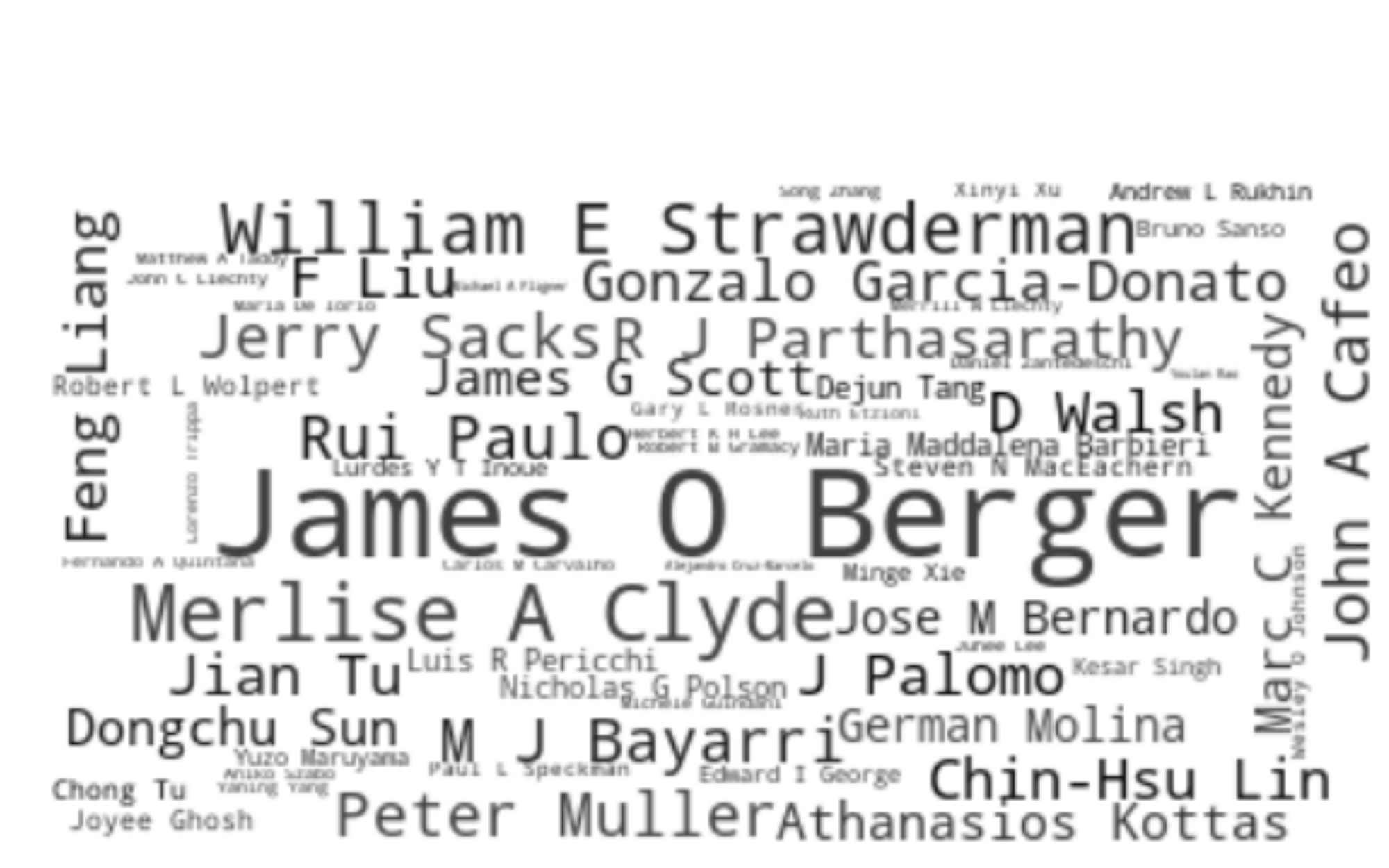}
\caption{Bayesian}
\label{fig:bayes1}
\end{subfigure}
\begin{subfigure}{0.32\textwidth}
\centering
\includegraphics[scale=.24, trim=1in 2in 1in 1in, clip]{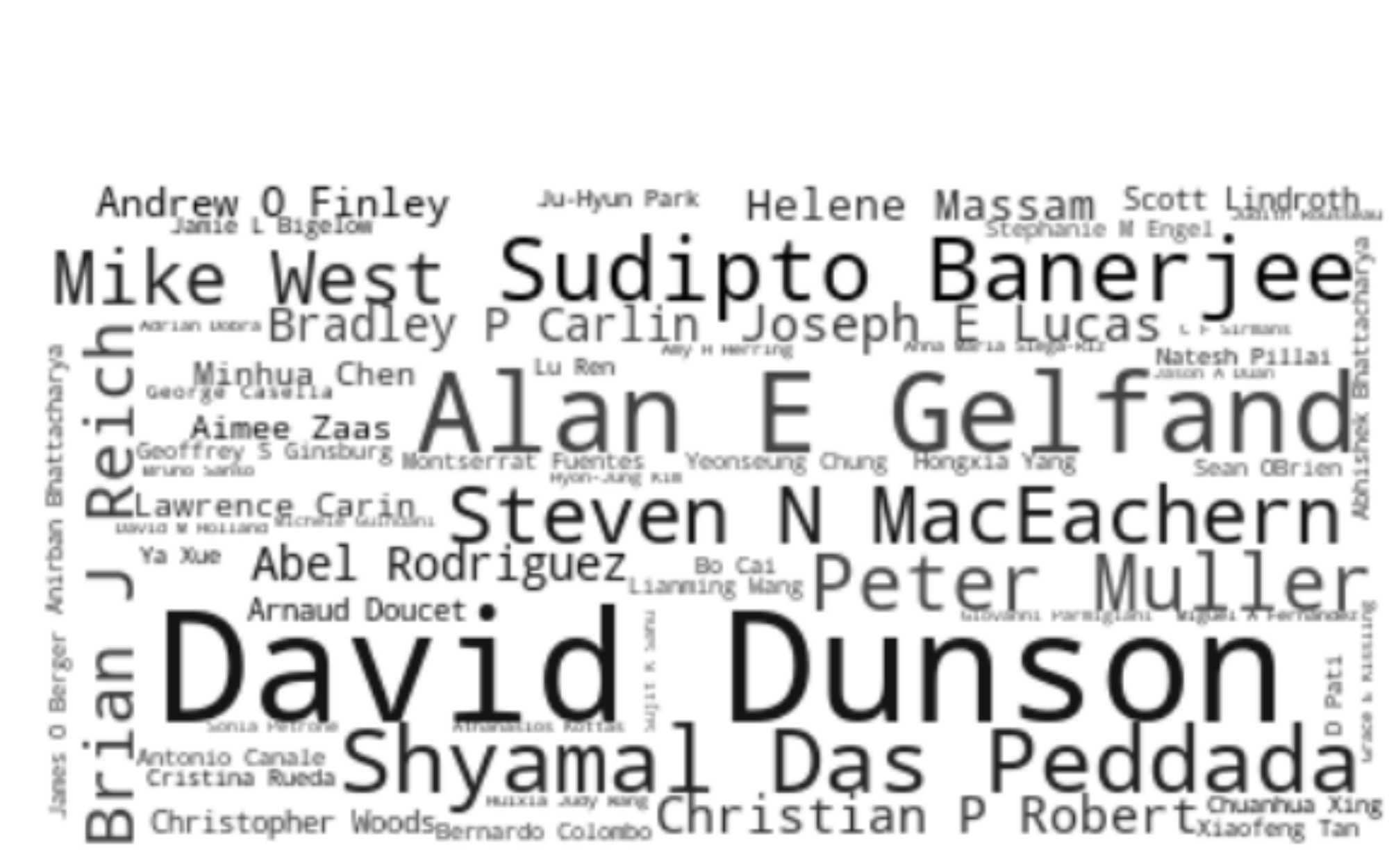}
\caption{Bayesian}
\label{fig:bayes2}
\end{subfigure}
\begin{subfigure}{0.32\textwidth}
\centering
\includegraphics[scale=.24, trim=1in 2in 1in 1in, clip]{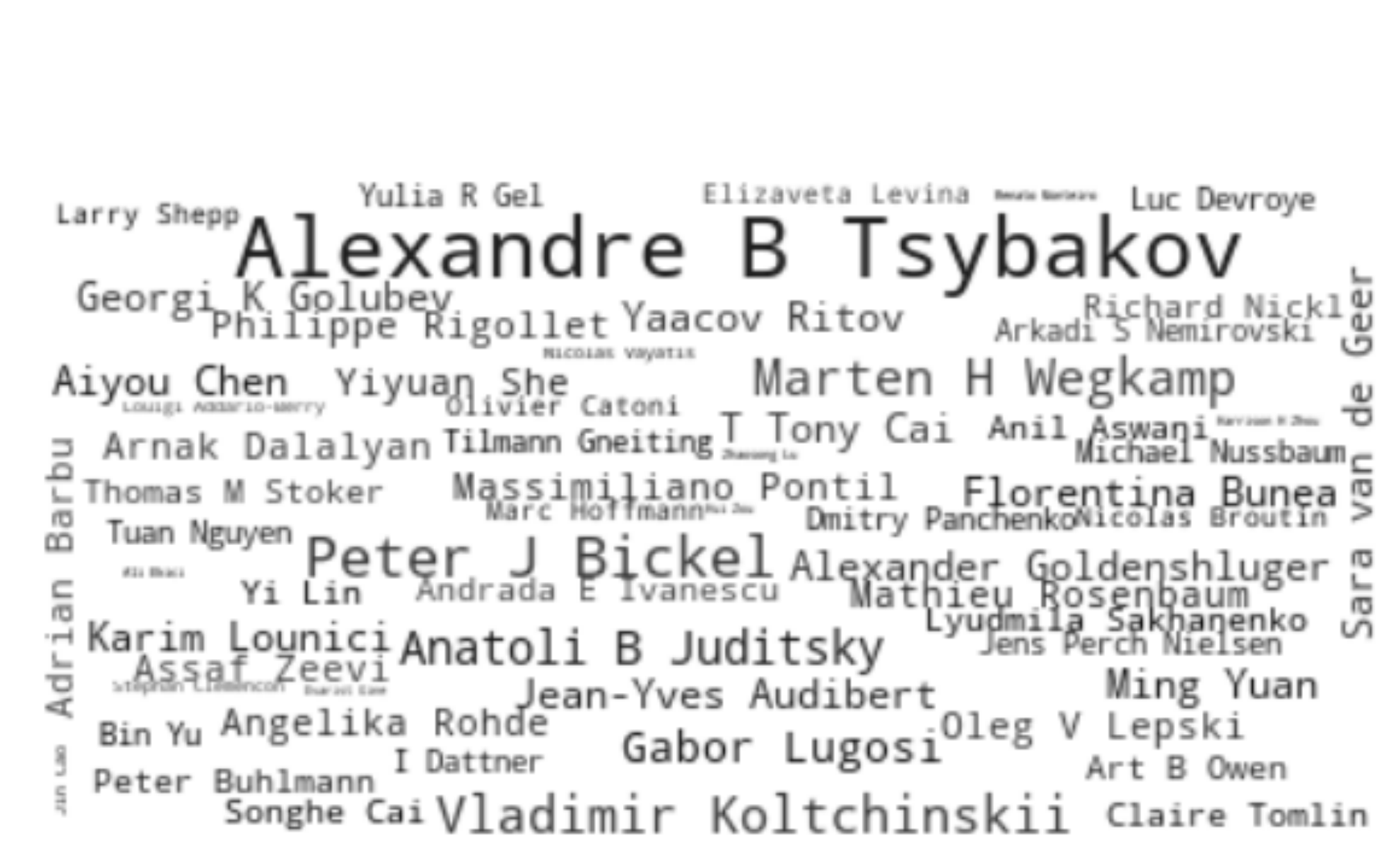}
\caption{Theory}
\end{subfigure}
\begin{subfigure}{0.32\textwidth}
\centering
\includegraphics[scale=.24, trim=1in 2in 1in 1in, clip]{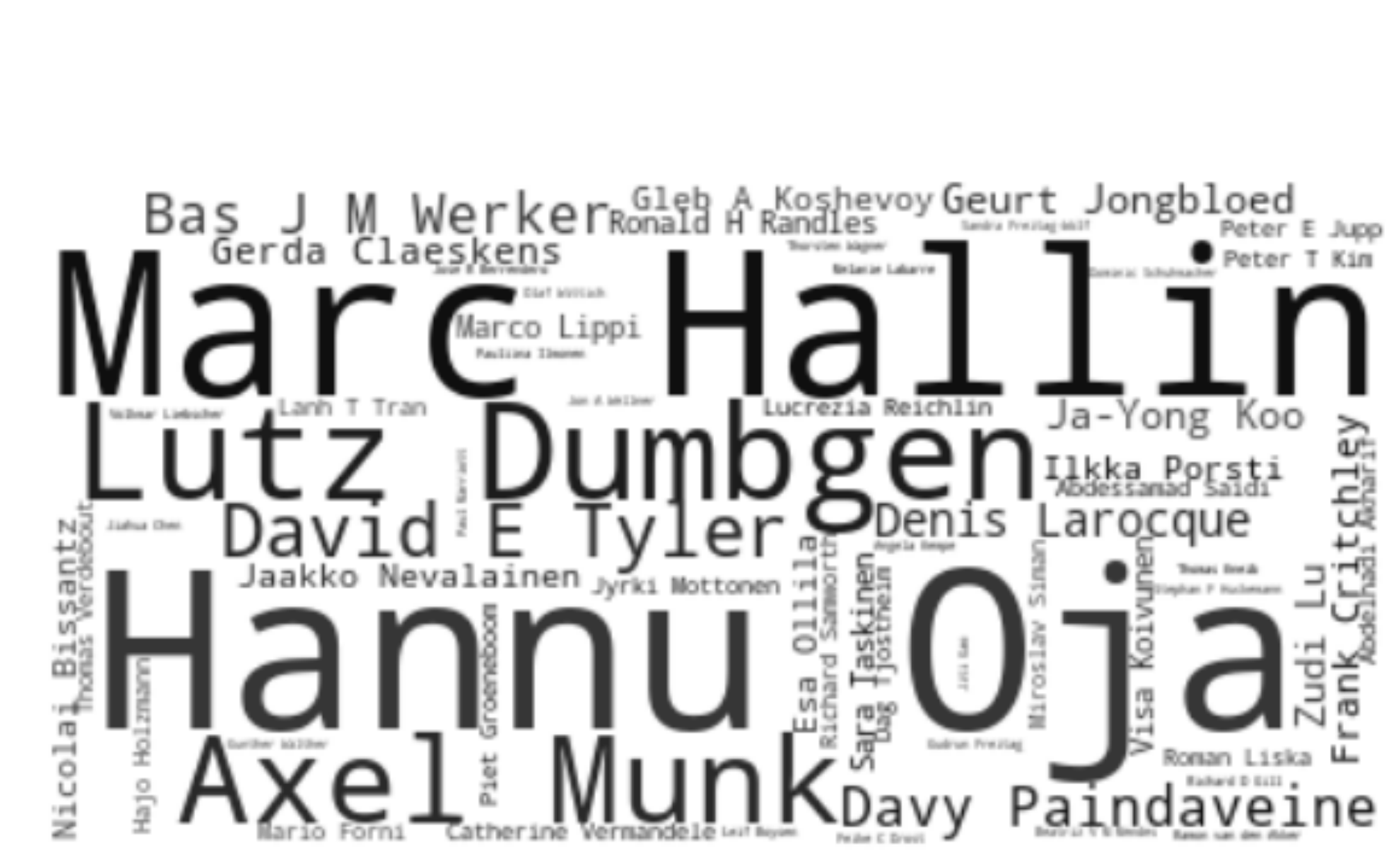}
\caption{Multivariate Analysis}
\end{subfigure}
\begin{subfigure}{0.32\textwidth}
\centering
\includegraphics[scale=.24, trim=1in 2in 1in 1in, clip]{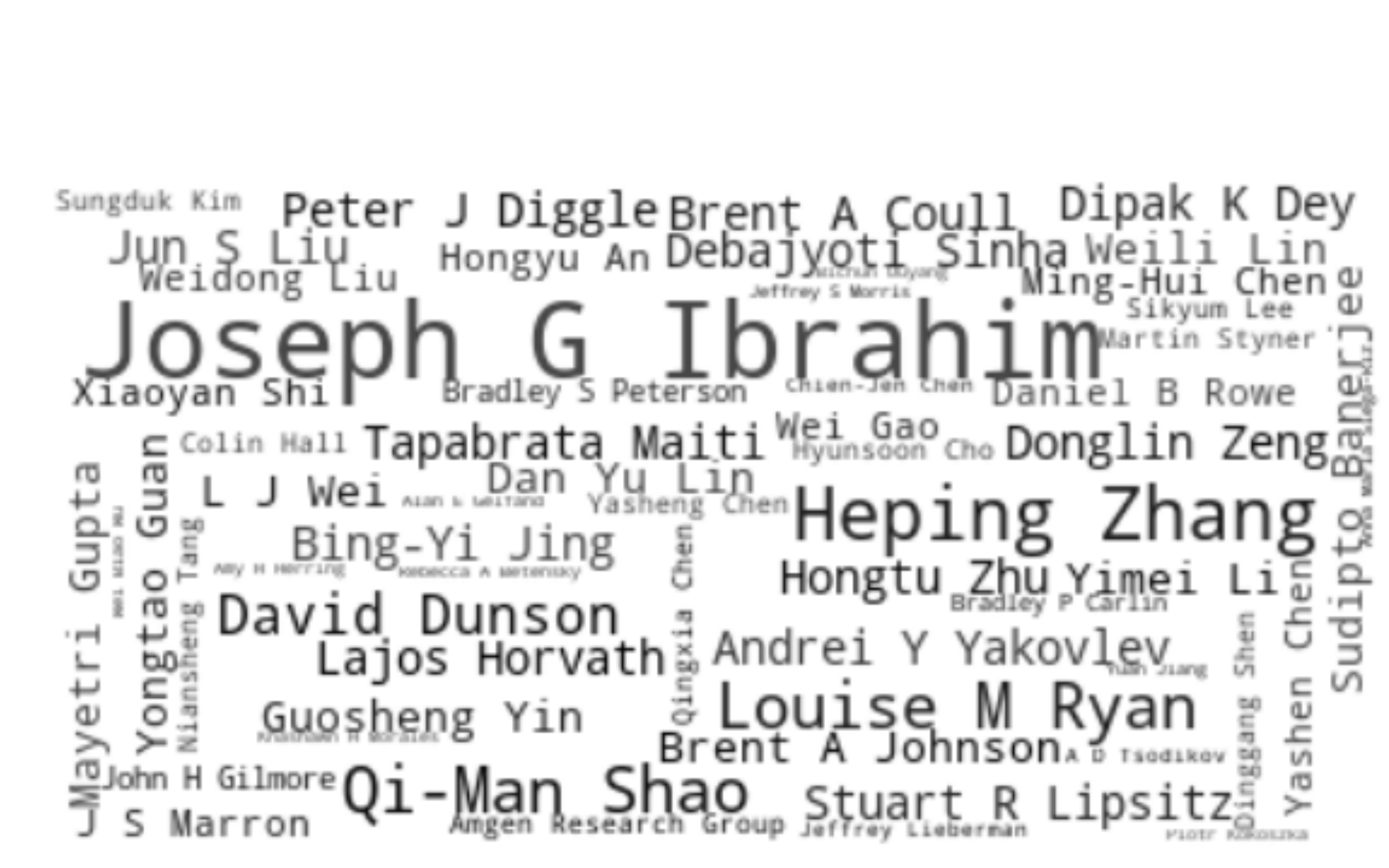}
\caption{Biostat}
\label{fig:biostat1}
\end{subfigure}
\begin{subfigure}{0.32\textwidth}
\centering
\includegraphics[scale=.24, trim=1in 2in 1in 1in, clip]{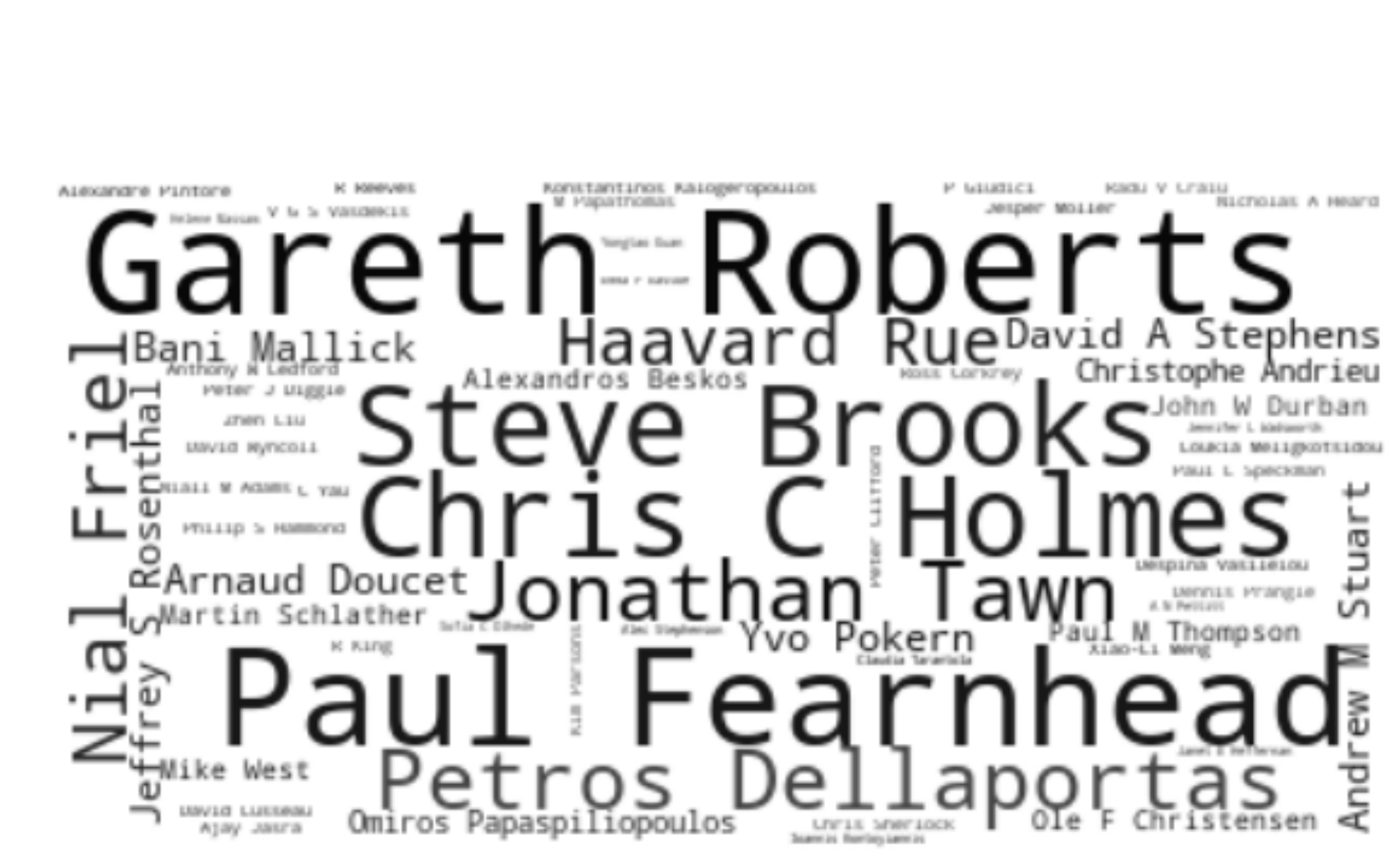}
\caption{Computation/UK}
\end{subfigure}
\begin{subfigure}{0.32\textwidth}
\centering
\includegraphics[scale=.24, trim=1in 2in 1in 1in, clip]{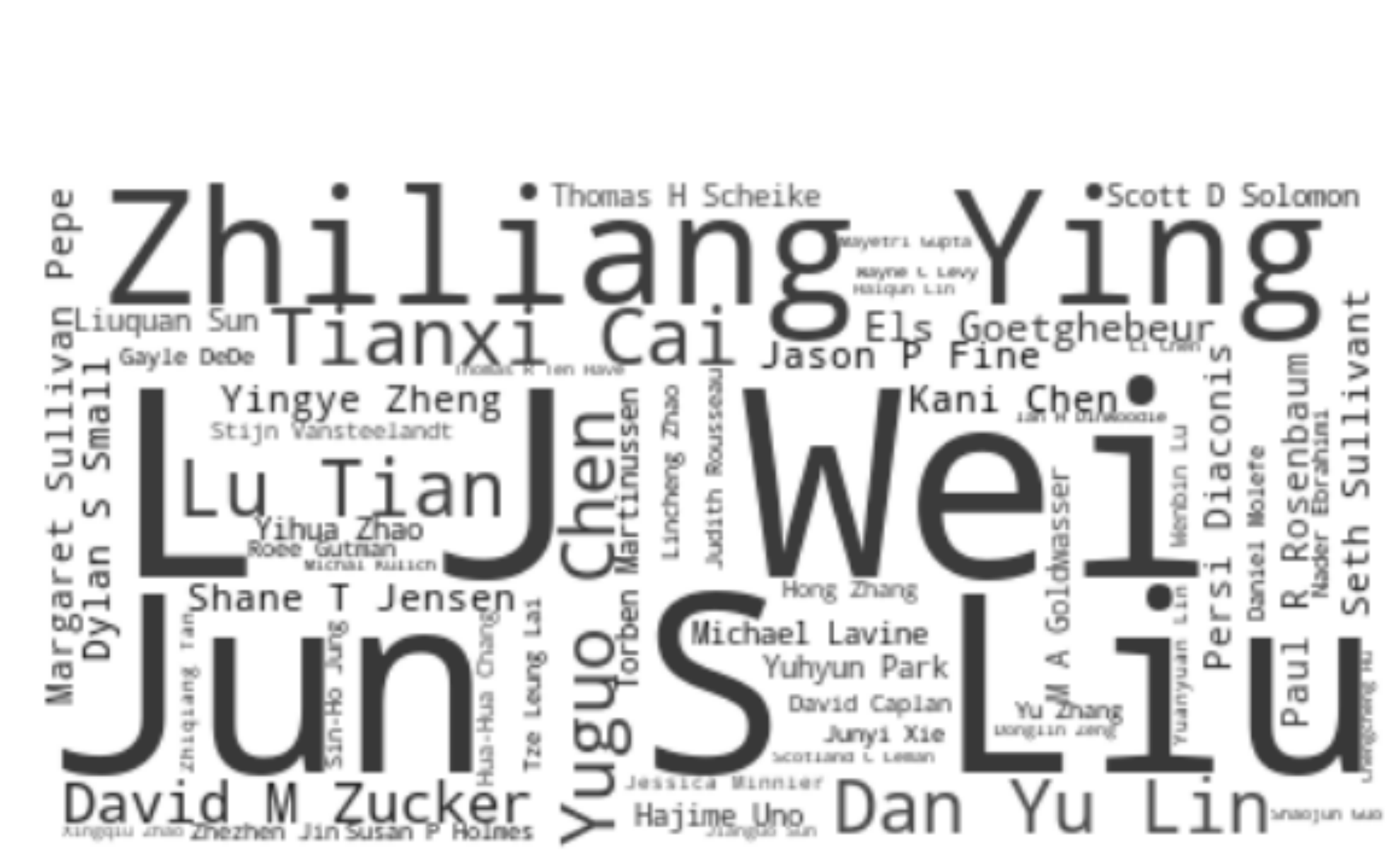}
\caption{Biostat}
\label{fig:biostat2}
\end{subfigure}
\begin{subfigure}{0.32\textwidth}
\centering
\includegraphics[scale=.24, trim=1in 2in 1in 1in, clip]{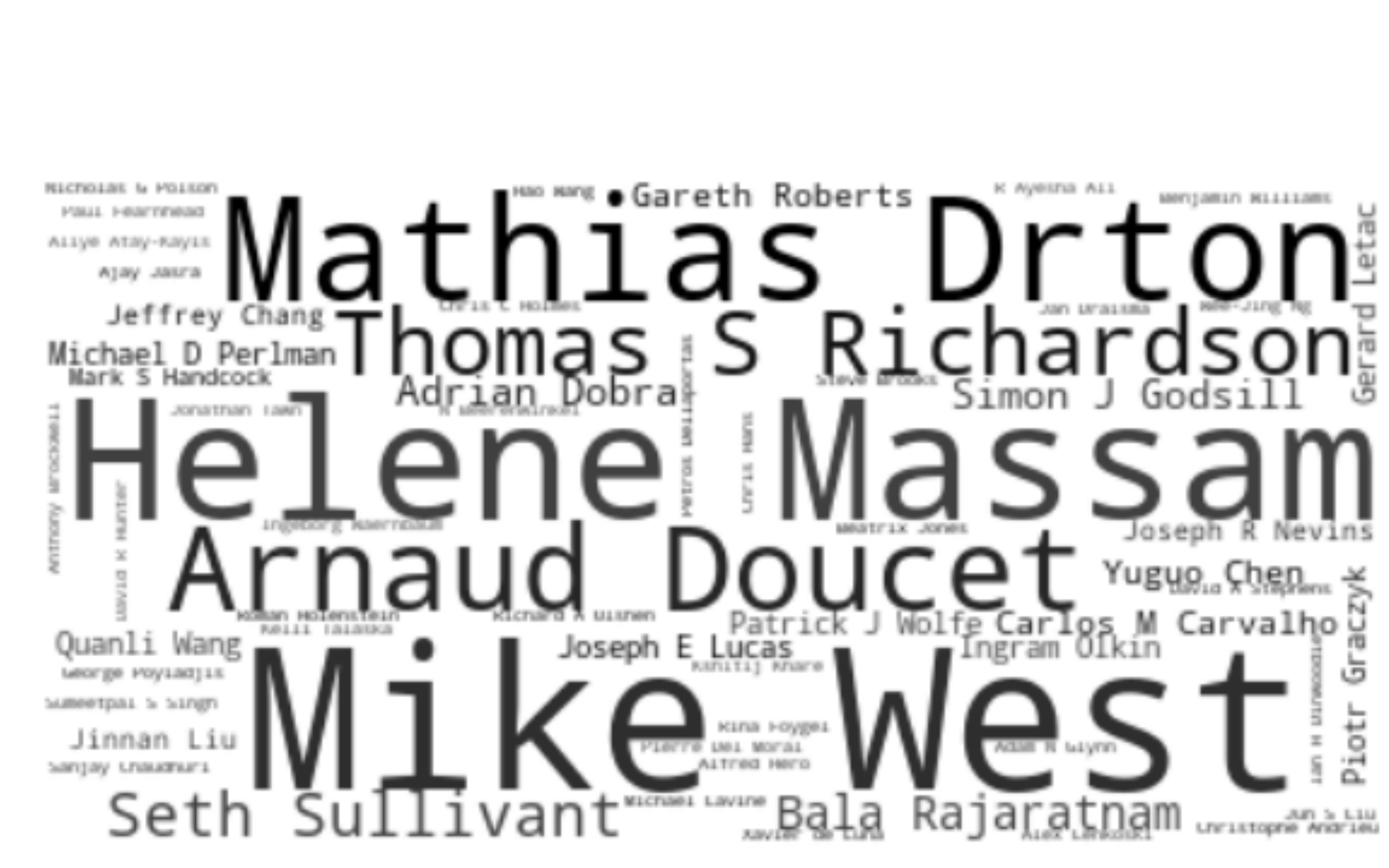}
\caption{Graphical Models}
\label{fig:bayes3}
\end{subfigure}
\caption{Nine of the clusters that most frequently appear in the posterior samples. Word sizes are proportional to the posterior root probability with respect to the cluster.}
\label{fig:coauthor_cluster1}
\end{figure}

\cite{ji2016coauthorship} on the same network uses scree plot to conclude that there are $K=3$ clusters, which are shown in Figures 9, 10, and 11 in their paper. They refer to the three clusters as a "high-dimensional" supercluster, a "biostatistics" cluster, and a "Bayes" cluster. We find a giant supercluster, but we also find a large number of smaller clusters which accurately reflect actual research communities in statistics. For example, we find the same "Bayes" cluster in \cite{ji2016coauthorship} (see Figure~\ref{fig:bayes1}), but we also discover other Bayesian clusters such as ones shown in Figure~\ref{fig:bayes2}. Similarly, we find the "biostat" community in \cite{ji2016coauthorship} (see Figure~\ref{fig:biostat1}) but we find other biostat clusters as well such as the one shown in Figure~\ref{fig:biostat2} and the one centered on Jason Fine and Michael Korsorok in Figure~\ref{fig:coauthor_cluster3} in the Appendix. In addition, we find many other meaningful communities, such as the experimental design community or the high-dimensional statistics community shown in Figure~\ref{fig:coauthor_cluster2}, or the survey and theory community in Figure~\ref{fig:coauthor_cluster3} in the Appendix. We believe that PAPER model gives highly coherent clusters for this network because the network itself is locally tree-like, as shown in two cluster subgraphs that we display in Figure~\ref{fig:coauthor_cluster2}. 

\begin{figure}[ht]
\centering
\begin{subfigure}{0.9\textwidth}
\centering
\includegraphics[scale=.35, trim=2cm 3cm 2cm 4cm, clip]{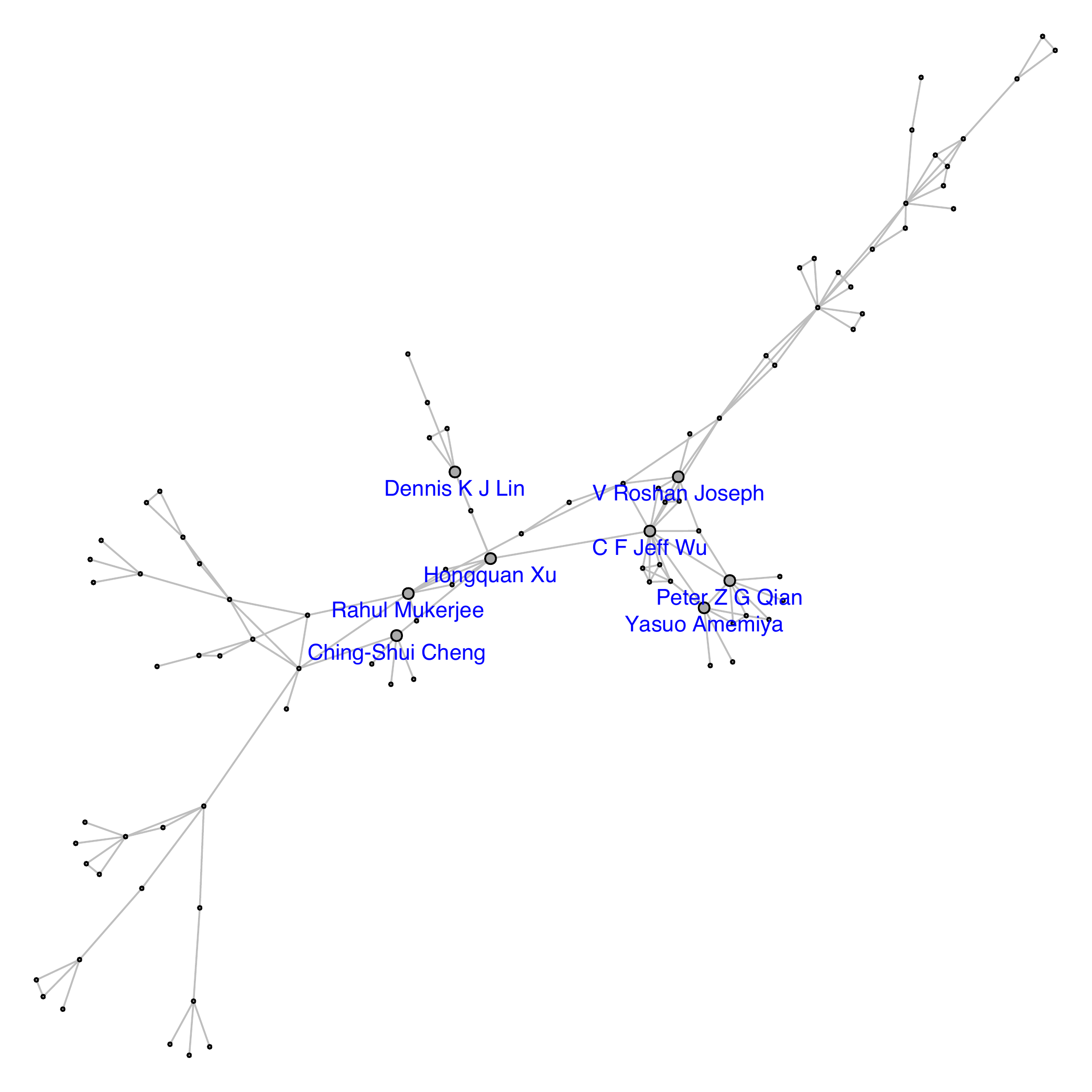}
\includegraphics[scale=.3, trim=1in 1in 1in 2in, clip]{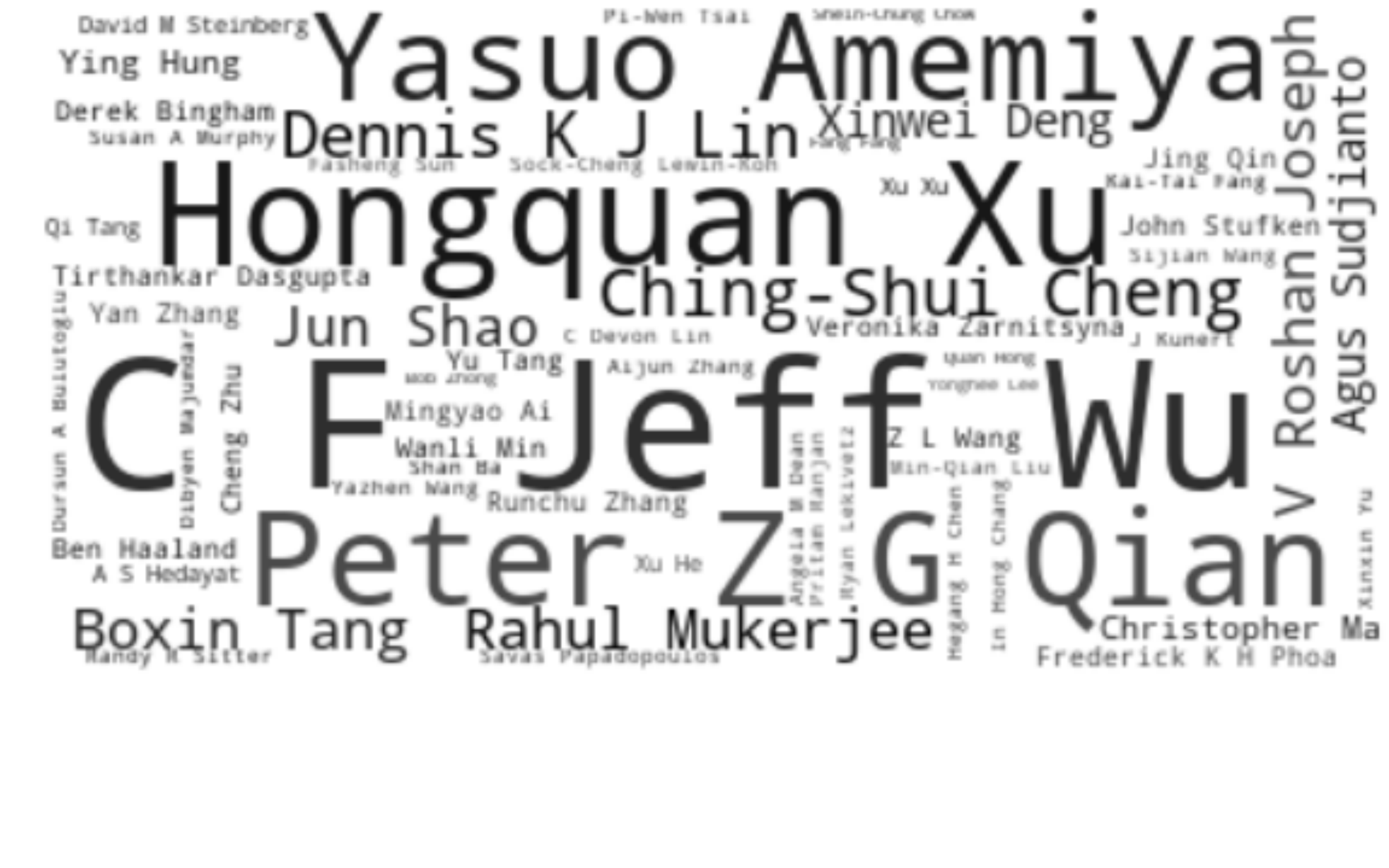}
\caption{Experimental design community}
\end{subfigure}
\begin{subfigure}{0.9\textwidth}
\centering
\includegraphics[scale=.35, trim=2cm 3cm 2cm 4cm, clip]{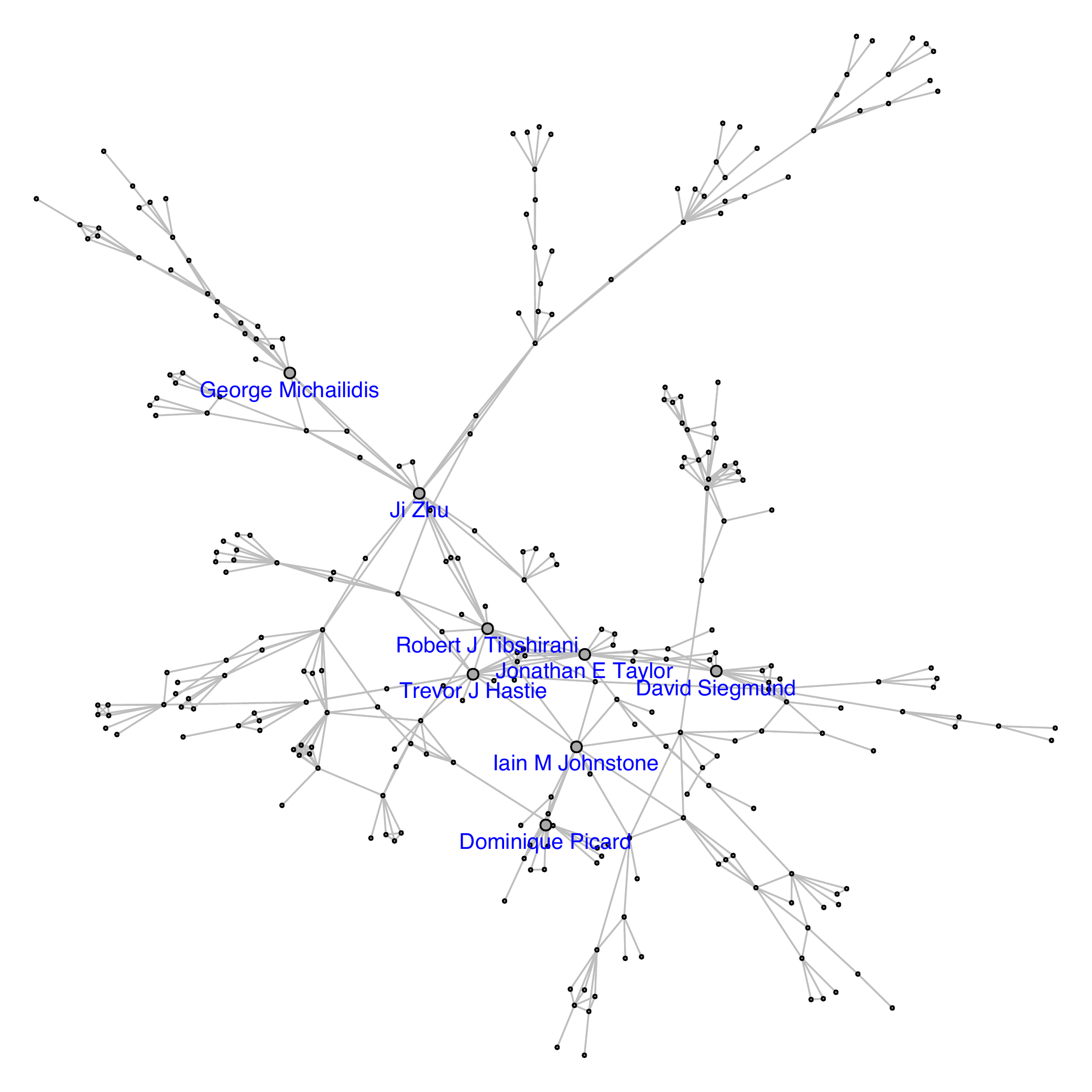}
\includegraphics[scale=.3, trim=1in 1in 1in 2in, clip]{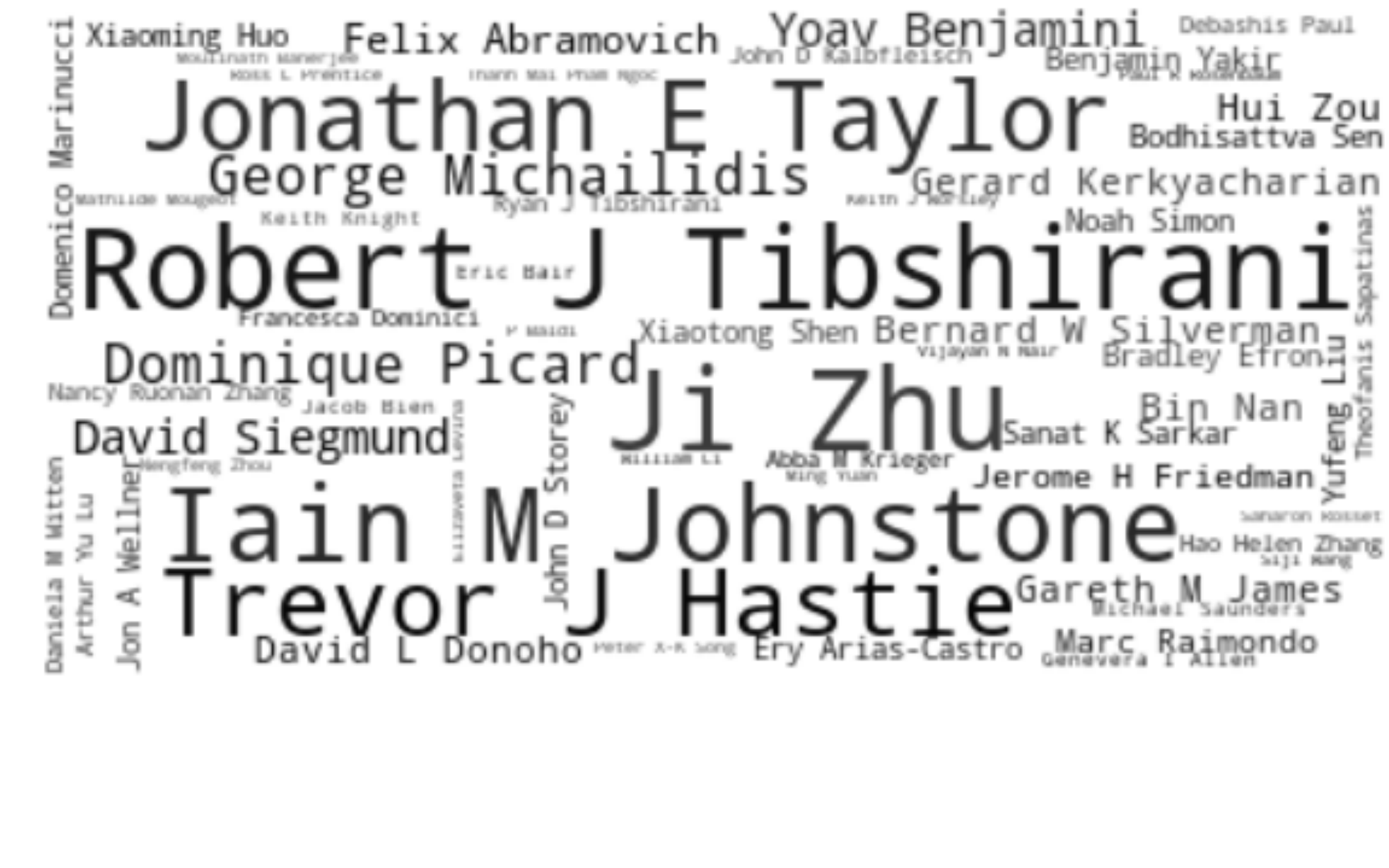}
\caption{High-dimensional statistics community}
\end{subfigure}
\caption{Two additional clusters along with the subgraphs that correspond to the clusters. In the subgraph, we label the 8 nodes with the highest posterior root probability with respect to that cluster. We observe that the subgraphs are tree-like.}
\label{fig:coauthor_cluster2}
\end{figure}

\section{Discussion}
\label{sec:discussion}
In this paper, we present the PAPER model for networks with
underlying formation processes and formalize the problem of root
inference. We extend the PAPER model to the setting of multiple
roots to reflect the growth of multiple communities. There are a number
of important open questions from modeling, theoretical,
and algorithmic perspectives.

From a modeling perspective, an interesting direction is to
suppose that the graph start not as singleton nodes but as a small subgraph. The goal then is to infer the seed-graph
instead of the root node (c.f. \cite{devroye2018discovery}). Model extensions such as the PAPER-SBM mixture described in Remark~\ref{rem:imbalance} are also interesting; in these models, a subtle question is to what extend we have to estimate the parameters of the noise model well in order to recover the root nodes of the latent forest.

There are many open theoretical questions related to PAPER model and
root inference. For instance, in Conjecture~\ref{con:size}, we
hypothesize that the size of the optimal confidence set for the root
node is of a constant order if so long as the noise level is below a
certain threshold. If the noise level is above the threshold, then
every confidence set has size that diverges with $n$. The lower bound
of this conjecture seems especially difficult and may require new
techniques. Another interesting theoretical question is the analysis
of community recovery using the PAPER model with multiple
roots. Intuitively, we expect be able to correctly cluster the early
nodes since they tend to have more central positions in the final
graph. The late arriving nodes on the other hand would be more
peripheral and difficult to cluster.

Algorithmically, we observe that the Gibbs sampler that we derived in
Section~\ref{sec:algorithm} converges very quickly in practice (see Section~\ref{sec:diagnostics}). It
would be interesting to study its mixing time, especially how the
mixing time depends on the noise level. 

\subsection*{Acknowledgement} 

This work is supported by the U.S. National Science Foundation DMS grant \#2113671. The second
author is grateful to Justin Khim for insightful discussions in the
early stage of the work and to Rong Chen for helpful feedback and comments. {\color{black} The authors would like to thank anonymous referees for insightful comments which helped improve the paper.}

\subsection*{List of Figures}

\bibliographystyle{dcu}
\bibliography{refs}

\clearpage

\setcounter{section}{0}
\setcounter{equation}{0}
\setcounter{theorem}{0}
\def\theequation{S\arabic{section}.\arabic{equation}}
\def\thesection{S\arabic{section}}
\def\thetheorem{S\arabic{theorem}}

\begin{center}
\Large{Supplementary material to ``Root and community inference on latent network growth
  processes using noisy attachment models''} \\ \vspace{0.2in}
\large{Harry Crane and Min Xu}
\end{center}

\section{Supplement for Section~\ref{sec:model}}

\subsection{Model Likelihood}
\label{sec:model-likelihood}

We first give the likelihood of any time
labeled tree under the $\text{APA}(\alpha, \beta)$ model. Define, for any integer $k \geq 1$,
  \begin{align*}
      \psi_{\alpha, \beta}(k) &:= \left\{
  \begin{array}{cc}
    \prod_{j=1}^{k-1} (\beta j + \alpha)  & \text{ if $k \geq 2$, } \\
    1 & \text{ if $k = 1$. }
  \end{array} \right.
  \end{align*}

  \begin{proposition}
    \label{prop:apa-prob1}
Let $\bm{T}_n \sim \text{APA}(\alpha, \beta)$. Then, for any
time labeled tree $\bm{t}_n$, we have that
\begin{align}
    \Pb( \bm{T}_n = \bm{t}_n ) = L_{\alpha, \beta}( \bm{t}_n) := \frac{ \prod_{v \in
    [n]} \psi_{\alpha, \beta}( D_{\bm{t}_n}(v) ) }
{ \prod_{t=3}^n (2(t-2)\beta + (t-1)\alpha) }. \label{eq:apa-prob1}
    \end{align}
  \end{proposition}
The fact that the likelihood depends on the tree
  $\bm{t}_n$ only through its degree distribution
  $D_{\bm{t}_n}(\cdot)$ remains true in the
  multiple roots setting except that the likelihood also depends on
  the root nodes. One complication with the multiple roots setting is that we give
  each root node an imaginary self-loop. To deal with this, we first define $\psi^r_{\alpha, \beta}(k) := \prod_{j=2}^{k+1} (\beta j + \alpha)$.
  
  \begin{proposition}
\label{prop:apa-prob2}
Let $\bm{F}_n \sim \text{APA}(\alpha, \beta, K)$. Then, for any
time labeled forest $\bm{f}_n$, we have that
\begin{align}
    \Pb( \bm{F}_n = \bm{f}_n ) = L_{\alpha, \beta, K}( \bm{f}_n) :=
  \frac{ \prod_{v \in \pi_{1:K}} \psi_{\alpha, \beta}^r(D_{\bm{f}_n}(v))
  \prod_{v \notin \pi_{1:K}} \psi_{\alpha, \beta}( D_{\bm{f}_n}(v) ) }
{ \prod_{t=K+1}^n (2(t-1)\beta + (t-1)\alpha) }. \label{eq:apa-prob2}
\end{align}
\end{proposition}

In the random $K$ setting, the likelihood is very similar except that
the set of root nodes is not necessarily $\pi_{1:K}$. 
\begin{proposition}
  \label{prop:apa-prob3}
Let $\bm{F}_n \sim \text{APA}(\alpha, \beta, \alpha_0)$. Then, for any
time labeled forest $\bm{f}_n$ with $K$ component trees, we have that
\begin{align}
    \Pb( \bm{F}_n = \bm{f}_n ) = L_{\alpha, \beta, S}( \bm{f}_n) :=
  \frac{ \prod_{v \in S} \psi_{\alpha, \beta}^r(D_{\bm{f}_n}(v))
  \prod_{v \notin S} \psi_{\alpha, \beta}( D_{\bm{f}_n}(v) ) }
{ \prod_{t=K+1}^n (2(t-1)\beta + (t-1)\alpha) }. \label{eq:apa-prob3}
\end{align}
where $S$ is the set of root nodes of $\bm{f}_n$, that is, a node is
in $S$ if and only if it has the earliest arrival time in its
component tree. 
\end{proposition}

Under the PAPER model, the complete data likelihood is also simple
owing to the fact that any non-forest edge of the random graph
$\bm{G}_n$ is Erd\H{o}s--R\'{e}nyi and any forest with $K$ component trees has
exactly $n-K$ edges. Therefore, for a time labeled graph
$\bm{g}_n$ with $m$ edges and a time labeled sub-forest $\bm{f}_n$, we
have that, under the PAPER model and conditional on $\bm{G}_n$ having $m$ edges,
\[
\Pb( \bm{G}_n = \bm{g}_n, \bm{F}_n = \bm{f}_n ) = \binom{ n(n-1)/2 - (n - K)}{m -
  (n-K)}^{-1} \Pb( \bm{F}_n = \bm{f}_n).
\]
We do not observe the forest of course. This is one of the main
hurdles that we address in Section~\ref{sec:algorithm}.

{\color{black}
\subsection{Poisson attachment approximation}
\label{sec:poisson_approx}

\begin{proposition}
\label{prop:poisson_approx}
Let $t \in \mathbb{N}$, and let $q_1, \ldots, q_t \in [0, 1]$ satisfy $\sum_{j=1}^t q_j \leq 1$, and let $\theta > 0$ such that $\theta q_j \leq 1$ for all $i \in [t]$. Let $X$ and $Y$ denote two \textbf{random subsets} of $[t]$, where $X$ is generated by adding each $j \in [t]$ independently with probability $\theta q_j$, and where $Y$ is generated by first drawing $M \sim \text{Poisson}(\theta)$ and then repeating $M$ times the procedure where we randomly choose $j \in [t]$ with probability $q_j$ and with replacement and add $j$ to $Y$. Let $P^{(X)}$ and $P^{(Y)}$ denote the distribution of $X$ and $Y$ respectively. Then, we have that
\begin{align*}
d_{\text{TV}}( P^{(X)}, P^{(Y)}) \leq \theta^2 \max_{ j \in [t]} q_j.
\end{align*}
\end{proposition}

\begin{proof}
For $j = 1, \ldots, n$, define $X_j = \mathbbm{1}\{ j \in X \}$ and $Y_j$ as the number of copies of element $j$ in $Y$. Direct calculation then shows that $X_1, \ldots, X_t$ are independent where $X_j \sim \text{Ber}(\theta q_j)$ and that $Y_1, \ldots, Y_t$ are independent where $Y_j \sim \text{Poisson}(\theta q_j)$. 

Therefore, by a coupling argument (see e.g. Example 2 in Chapter 10.1 of \cite{pollard2002user}) and the fact that $\sum_{j=1}^t q_j \leq 1$, we have 
\begin{align*}
d_{\text{TV}}( P^{(X)}, P^{(Y)}) 
&\leq \mathbb{P}( X \neq Y) = \sum_{j=1}^t \mathbb{P}(X_j \neq Y_j) \\
&\leq \sum_{j=1}^t \theta^2 q_j^2 \leq \theta^2 \max_j q_j,
\end{align*}
as desired. 
\end{proof}

If we apply Proposition~\ref{prop:poisson_approx} to the independent Bernoulli noise model described in Section~\ref{sec:seq_noise}, where $X$ and $Y$ denote the random set of edges added under the Bernoulli noise model and the Poisson noise model respectively and where $q_j = \frac{ \tbeta D_{\bm{T}_{t-1}}(j) + \talpha}{2(t-2)\tbeta + (t-1)\talpha}$, then we may use the fact that $\max_{j \in [t]} q_j = O_p( \frac{1}{\sqrt{t}})$ (see e.g. Section 8.7 in \cite{van2016random}) to see that the two noise models are approximately equivalent for large $t$.
}

\section{Supplement for Section~\ref{sec:methodology}}
\label{sec:methodology-supp}

Recall that for an alphabetically labeled tree $\tilde{\bm{t}}_n$, we
define the $\text{hist}(\tilde{\bm{t}}_n)$ as the set of all label
ordering $\pi \in \text{Bi}([n], \mathcal{U}_n)$ such that $\pi^{-1}
\tilde{\bm{t}}_n$ is a time labeled tree that has a positive
probability over the APA model (Definition~\ref{def:apa1}). For a node
$u$, we also
define $\text{hist}(u, \tilde{\bm{t}}_n)$ as all $\pi \in
\text{hist}(\tilde{\bm{t}}_n)$ such that $\pi_1 =
u$ and $h(u, \tilde{\bm{t}}_n) = |\text{hist}(u, \tilde{\bm{t}}_n)|$. \cite{shah2011rumors} derives an $O(n)$ runtime algorithm that
computes the whole collection $\{ h(u, \tilde{\bm{t}}_n) \}_{u \in
  \mathcal{U}_n}$, which is shown as Algorithm~\ref{alg:count-history}.

\begin{algorithm}
\caption{Computing $\{ h(u, \tilde{\mathbf{t}}_n)\}_{u \in \mathcal{U}_n}$ \citep{shah2011rumors}}
\label{alg:count-history}
\textbf{Input:} a labeled tree $\tilde{\mathbf{t}}_n$. \\
\textbf{Output:} $h(u, \tilde{\mathbf{t}}_n)$ for all nodes $u \in \mathcal{U}_n$.
\begin{algorithmic}
\State Arbitrarily select root $u_0 \in \mathcal{U}_n$.
\For{$u \in \mathcal{U}_n$}
 \State Compute and store $n^{(u_0)}_u := | \tilde{\mathbf{t}}^{(u_0)}_u |$.
 \EndFor
 \State Compute $h(u_0, \tilde{\mathbf{t}}_n) = n! \prod_{u \in
   \mathcal{U}_n} \frac{1}{ | \tilde{\bm{t}}^{(u_0)}_u |}$.
 \State Set $\mathcal{S} = \{ \text{Children}(u_0) \}$.
\While{$\mathcal{S}$ is not empty}
 \State Remove an arbitrary node $u \in \mathcal{S}$.
 \State Compute $h(u, \tilde{\mathbf{t}}_n) = h(\text{pa}(u), \tilde{\mathbf{t}}_n) \frac{ n_u^{(u_0)}}{n- n_u^{(u_0)}}$
 \State Add $\text{Children}(u)$ to $\mathcal{S}$
 \EndWhile
\end{algorithmic}
\end{algorithm}

\subsection{Equivalence to maximum likelihood}
\label{sec:equivalence-mle}

Before deriving the likelihood formally, it is useful to have the following
standard definitions. For two labeled graphs $\bm{g}, \bm{g}'$, we say
that $\bm{g} \sim \bm{g}'$ if there exists $\rho \in
\text{Bi}(V(\bm{g}), V(\bm{g}'))$ such that $\rho \bm{g} =
\bm{g}'$. In this case, we say that $\bm{g}$ and $\bm{g}'$ are
isomorphic, or that they have the same shape, or that they are
equivalent as unlabeled graphs. The $\sim$ relationship defines
equivalence classes on the set of all labeled graphs, which we refer
to as the \emph{unlabeled shape} or just \emph{shape} for short. We write
\begin{align*}
I(\bm{g}, \bm{g'}) := \{ \rho \in \text{Bi}(V(\bm{g}), V(\bm{g}'))
  \,:\, \rho \bm{g} = \bm{g}'\}.
\end{align*}
Note that $I(\bm{g}, \bm{g})$ is the set of automorphisms of the graph
$\bm{g}$. To represent an unlabeled shape, we write
$\text{sh}(\bm{g})$ where $\bm{g}$ an arbitrary representative element
in the equivalence class.

Similarly, given a node $u \in V(\bm{g})$ and $u' \in V(\bm{g}')$, we
say that $(\bm{g}, u) \sim_0 (\bm{g}', u')$ if there exists $\rho \in
\text{Bi}(V(\bm{g}), V(\bm{g}'))$ such that $\rho \bm{g} =
\bm{g}'$ and $\rho(u) = u'$. In this case, we say that $(\bm{g}, u)$
and $(\bm{g}', u')$ have the same \emph{rooted shape}. The $\sim_0$
relationship defines an equivalence class on the pairs $(\bm{g}, u)$. We write 
\begin{align*}
I(\bm{g}, u, \bm{g'}, u') := \{ \rho \in \text{Bi}(V(\bm{g}), V(\bm{g}'))
  \,:\, \rho \bm{g} = \bm{g}', \rho(u) = u'\}.
\end{align*}

We have the following facts:
\begin{enumerate}
  \item $I(\bm{g}, \bm{g}')$ is non-empty if and only if $\bm{g},
    \bm{g}'$ have the same shape. Moreover, the cardinality of
    $I(\bm{g}, \bm{g}')$ depends only on that shape. For instance,
    $|I(\bm{g}, \bm{g}')| = |I(\bm{g}, \bm{g})|$ if the former is non-zero. In discrete
    mathematics, this cardinality is referred to as the size of the automorphism
    group of $\bm{g}$.
  \item $I(\bm{g}, u, \bm{g}', u')$ is non-empty if and only if
    $(\bm{g}, u), 
    (\bm{g}', u')$ have the same shape. Moreover, the cardinality of
    $I(\bm{g}, u, \bm{g}', u')$ depends only on that shape.
\end{enumerate}

Now, for a labeled graph $\bm{g}$ and a node $u \in V(\bm{g})$, we
define
\begin{align*}
\text{Eq}(u, \bm{g}) = \{ u' \in \bm{g} \,:\, (\bm{g}, u) \sim_0
  (\bm{g}, u') \}. 
\end{align*}
Nodes in $\text{Eq}(u, \bm{g})$ are indistinguishable from node $u$
once the node labels are removed. 

\begin{figure}[htp]
\centering
\includegraphics[scale=.4]{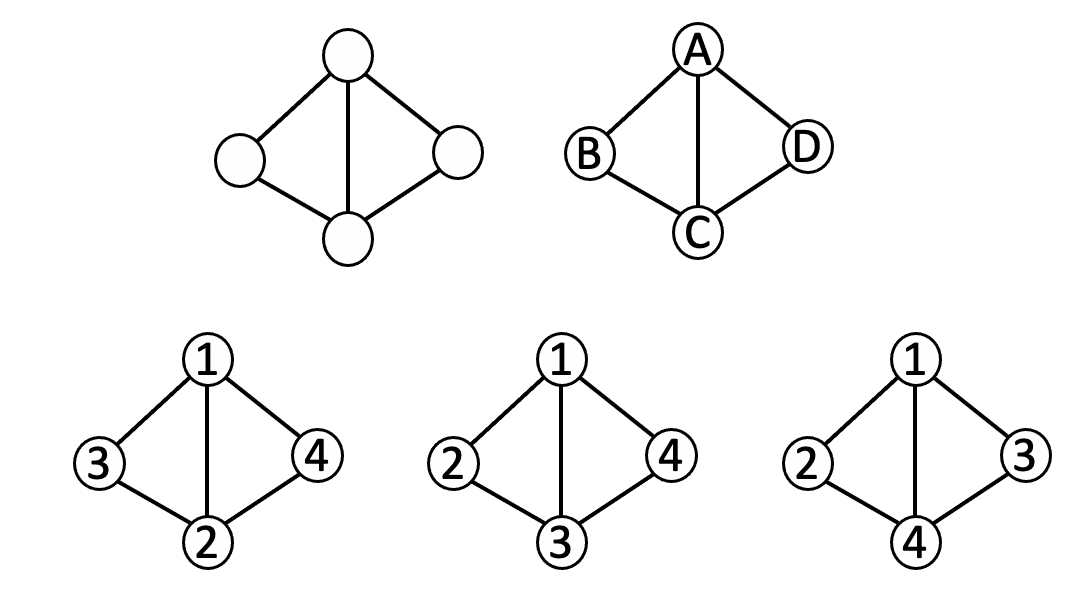}
\caption{Viewing the top right graph as $\tilde{\bm{g}}$ and the
  bottom graphs as $\bm{g}^1, \bm{g}^2, \bm{g}^3$, we have
  $\text{Eq}(A, \tilde{\bm{g}}) = \{A, C\}$ and $\mathcal{L}(A,
  \tilde{\bm{g}}_n) = \frac{1}{2} \bigl\{ \Pb(\bm{G}_n = \bm{g}^1) +
  \Pb(\bm{G}_n = \bm{g}^2)+ \Pb(\bm{G}_n = \bm{g}^3) \bigr\}$.}
\label{fig:likelihood}
\end{figure}

On observing an unlabeled graph $\text{sh}(\tilde{\bm{g}}_n)$, the
likelihood of a node $u$ being the root node is therefore
\begin{align*}
\mathcal{L}(u, \tilde{\bm{g}}_n) := \frac{1}{ |\text{Eq}(u,
  \tilde{\bm{g}}_n)|} \sum_{ \bm{g}_n \text{ time labeled}} \Pb(
  \bm{G}_n = \bm{g}_n) \mathbbm{1}\{ (\bm{g}_n, 1) \sim_0
  (\tilde{\bm{g}}_n, u) \},
\end{align*}
where $\bm{G}_n$ has the $\text{PAPER}(\alpha, \beta, \theta)$
distribution. It is straightforward to check that $\mathcal{L}(u,
\tilde{\bm{g}}_n)$ depends only on the unlabeled shape of
$(\tilde{\bm{g}}_n, u)$. We give a concrete example of the likelihood
in Figure~\ref{fig:likelihood}.

\begin{theorem}
\label{thm:equivalence-mle}
  For any alphabetically labeled graph $\tilde{\bm{g}}_n$, we have
  \[
\Pb( \Pi_1 = u \,|\, \tilde{\bm{G}}_n = \tilde{\bm{g}}_n) =
\frac{\mathcal{L}(u, \tilde{\bm{g}}_n)}{\sum_{v \in \mathcal{U}_n}
  \mathcal{L}(v, \tilde{\bm{g}}_n)}.
  \]
\end{theorem}

\begin{proof}
We have that
  \begin{align*}
    \Pb( \Pi_1 = u \,|\, \tilde{\bm{G}}_n = \tilde{\bm{g}}_n)
    &\propto \sum_{\pi \in \text{Bi}([n], \mathcal{U}_n),\, \pi_1 = u}
      \Pb( \tilde{\bm{G}}_n = \tilde{\bm{g}}_n \,|\, \Pi = \pi) \frac{1}{n!} \\
    &\propto \sum_{\pi \in \text{Bi}([n], \mathcal{U}_n),\, \pi_1 = u}
      \Pb( \bm{G}_n = \pi^{-1} \tilde{\bm{g}}_n) \\
    &= \sum_{\bm{g}_n \text{ time labeled}}
      \sum_{\stackrel{\pi \in \text{Bi}([n], \mathcal{U}_n),}{\pi_1 = u, \pi
      \bm{g}_n = \tilde{\bm{g}}_n}} \Pb( \bm{G}_n = \bm{g}_n) \\
    &=  \sum_{\bm{g}_n \text{ time labeled}} |I(\bm{g}_n, 1,
      \tilde{\bm{g}}_n, u)|  \Pb( \bm{G}_n = \bm{g}_n) \\
    &= 
      \frac{| I(\tilde{\bm{g}}_n, \tilde{\bm{g}}_n)|}{|\text{Eq}(u,
      \tilde{\bm{g}}_n)|} \sum_{\bm{g}_n \text{ time labeled}} 
      \Pb(
      \bm{G}_n = \bm{g}_n) \mathbbm{1}\{ (\bm{g}_n, 1) \sim_0
      (\tilde{\bm{g}}_n, u) \},
  \end{align*}
where the second equality follows by the definition of $I(\bm{g}, 1,
\tilde{\bm{g}}_n, u)$ and the final equality follows by
Lemma~\ref{lem:isomorphism}. The desired conclusion immediately
follows. 
  
\end{proof}

\begin{lemma}
  \label{lem:isomorphism}
For any labeled graphs $\bm{g}, \bm{g}'$ and nodes $u \in V(\bm{g})$,
$u' \in V(\bm{g}')$, if $(\bm{g}, u) \sim_0 (\bm{g}', u')$, then 
\[
|I( \bm{g}, \bm{g'})| = |I(\bm{g}, u, \bm{g}', u')| |\text{Eq}(u, \bm{g})|.
\]
\end{lemma}

\begin{proof}
  Suppose $|I(\bm{g}, u, \bm{g}', u')| > 0$. We note for any node $v
  \in V(\bm{g})$, we have that
  $|I(\bm{g}, v, \bm{g}', u')|$ is either zero or equal to $|I(\bm{g},
  u, \bm{g}', u')|$. Moreover, it is non-zero if and only if $v \in
  \text{Eq}(u, \bm{g})$. 
  
  Therefore, using the fact that $I(\bm{g}, \bm{g}') = \cup_{v \in
    V(\bm{g})} I(\bm{g}, v,
    \bm{g}', u')$, we have
  \begin{align*}
    |I( \bm{g}, \bm{g'})| = \sum_{v \in V(\bm{g})} | I(\bm{g}, v,
    \bm{g}', u')| = |\text{Eq}(u, \bm{g}_n)| | I(\bm{g}, u, \bm{g}', u')|,
  \end{align*}
  as desired. 
\end{proof}

\section{Supplement for Section~\ref{sec:algorithm}  }

\subsection{Parameter estimation for the PAPER model via EM}
\label{sec:parameter-estimation}

The PAPER models are parametrized by $\alpha, \beta$ which control the
attachment mechanism, by $\theta$ which is the noise level, and by
either $K$ or $\alpha_0$ in the multiple roots setting. We discuss some ways to select the number of trees $K$ in the
fixed $K$ root setting and ways
to estimate $\alpha_0$ in the random $K$ roots setting in Section~\ref{sec:practical} of the
appendix. 

In this section therefore, we consider only the estimation of the parameters $\alpha$ and
$\beta$. We assume that $\beta > 0$, in which case, without loss of
generality, we may assume $\beta = 1$ so that we only need to estimate
$\alpha$. We note that assuming $\beta > 0$ does not exclude uniform
attachment if we allow $\alpha = \infty$. We first consider the single root setting. For any tree
$\tilde{\bm{t}}_n$, by Proposition~\ref{prop:apa-prob1} that
\begin{align}
  \Pb_\alpha(
  \tilde{\bm{T}}_n = \tilde{\bm{t}}_n)
  &=
     \sum_{\pi \in \text{hist}(\tilde{\bm{t}}_n)}
    \Pb_\alpha(\bm{T}_n = \pi^{-1} \tilde{\bm{t}} \,|\, \Pi=\pi)
  \Pb(\Pi = \pi) \nonumber \\
  &= h(\tilde{\bm{t}}_n) \frac{ \prod_{v \in \mathcal{U}_n}
    \prod_{j=1}^{D_{\tilde{\bm{t}}_n}(v)-1 } (j + \alpha)}
    { \prod_{k=3}^n (2 (k-2)  + (k-1)\alpha ) } \frac{1}{n!}. \label{eq:likelihood1}
\end{align}

Therefore, keeping only terms that depend on $\alpha$, we have that
the log-likelihood is
\begin{align*}
  \ell(\alpha ; \tilde{\bm{T}}_n)
  &= \sum_{v \in \mathcal{U}_n}
  \sum_{j=1}^{\infty} \log (j + \alpha) \mathbbm{1}\{ j <
  D_{\tilde{\bm{T}}_n}(v)\} - \sum_{k=3}^n \log \bigl( 2(k-2) +
  (k-1)\alpha \bigr) \\
  &= \sum_{j=1}^\infty \log(j+\alpha) W_{\tilde{\bm{T}}_n}(j) -
    \sum_{k=3}^n \log \bigl( 2(k-2) + (k-1)\alpha \bigr), 
\end{align*}
where we define $W_{\tilde{\bm{T}}_n}(j) := | \{ v \in \mathcal{U}_n \,:\,
D_{\tilde{\bm{T}}_n}(v) > j \} |$. We note that, in this case, the
log-likelihood of $\alpha$ depends on the tree $\tilde{\bm{T}}_n$ only
through its degree sequence. 

In the PAPER model where $\tilde{\bm{G}}_n =
\tilde{\bm{T}}_n + \tilde{\bm{R}}_n$, for every node $v \in
\mathcal{U}_n$, we have that
$D_{\tilde{\bm{G}}_n}(v) = D_{\tilde{\bm{T}}_n}(v) +
D_{\tilde{\bm{R}}_n}(v)$ where the tree degree
$D_{\tilde{\bm{T}}_n}(v)$ is now latent. We propose an approximate
EM algorithm in this setting. 

The complete data log-likelihood in this case is
\begin{align*}
\ell( \alpha ; D_{\tilde{\bm{G}}_n}, D_{\tilde{\bm{T}}_n}) = \sum_{j=1}^\infty
  \log (j + \alpha)
  \sum_v \mathbbm{1} \{ j < D_{\tilde{\bm{T}}_n}(v) \} -
  \sum_{k=s}^n \log\bigl( 2(k-2) + (k-1)\alpha \bigr).
\end{align*}

For a given value $\alpha'$, the EM update is then to maximize
\begin{align}
  M(\alpha | \alpha') &:= \mathbb{E}_{\alpha'}
  \biggl[ \sum_{j=1}^\infty
  \log (j + \alpha)
  \sum_v \mathbbm{1} \{ j < D_{\tilde{\bm{T}}_n}(v) \}
  \,\bigg|\, \tilde{\bm{G}}_n \biggr]
  -
  \sum_{k=s}^n \log\bigl( 2(k-2) + (k-1)\alpha \bigr) \nonumber \\
  &= \sum_{j=1}^\infty \log( j + \alpha)  \sum_v \mathbb{P}_{\alpha'} \biggl\{
    j < D_{\tilde{\bm{T}}_n}(v) \,\bigg|\, \tilde{\bm{G}}_n \biggr\}
    - \sum_{k=s}^n \log\bigl( 2(k-2) + (k-1)\alpha \bigr). \label{eq:EMupdate}
\end{align}

The conditional probability term $\Pb_{\alpha'}( j <
D_{\tilde{\bm{T}}_n}(v) \,|\, \tilde{\bm{G}}_n)$ can be computed by
Gibbs sampling, but we can significantly reduce the computation
time by approximating
$\Pb_{\alpha'}( j < D_{\tilde{\bm{T}}_n}(v) \,|\, 
\tilde{\bm{G}}_n)$ with $\Pb_{\alpha'}( j < D_{\tilde{\bm{T}}_n}(v) \,|\,
  D_{\tilde{\bm{G}}_n}(v) )$, which ignores the mild dependence
  between the degrees of all the nodes. To further improve the quality of the approximation, we observe that
 \[
   \sum_{j=1}^\infty \sum_v \Pb_{\alpha'}\bigl( j < D_{\tilde{\bm{T}}_n}(v)
 \,\big|\, \tilde{\bm{G}}_n  \bigr) = \sum_v (D_{\tilde{\bm{T}}_n}(v)-1) =
 n-2
\]
while the sums of the approximate conditional probabilities $\sum_{j=1}^\infty \sum_v \Pb_{\alpha'}( j < D_{\tilde{\bm{T}}_n}(v) \,|\,
  D_{\tilde{\bm{G}}_n}(v) )$ may be different. Thus, we normalize
  $\Pb_{\alpha'}( j < D_{\tilde{\bm{T}}_n}(v) \,|\,
  D_{\tilde{\bm{G}}_n}(v))$ by defining $\tilde{W}_{\tilde{\bm{G}}_n}(j) = (n-2) \frac{ \Pb_{\alpha'}( j < D_{\tilde{\bm{T}}_n}(v) \,|\,
  D_{\tilde{\bm{G}}_n}(v) )}{ \sum_{j=1}^\infty \Pb_{\alpha'}( j < D_{\tilde{\bm{T}}_n}(v) \,|\,
  D_{\tilde{\bm{G}}_n}(v) )}$ so that $\sum_{j=1}^\infty
\tilde{W}_{\tilde{\bm{G}}_n}(j) = n-2$ and, instead of
maximizing~\eqref{eq:EMupdate}, we update
  \begin{align}
    \tilde{M}(\alpha | \alpha') := \sum_{j=1}^\infty \log( j + \alpha)
    \tilde{W}_{\tilde{\bm{G}}_n}(j)
    - \sum_{k=s}^n \log\bigl( 2(k-2) + (k-1)\alpha \bigr). \label{eq:EMupdate2}
  \end{align}
In practice, we find that the normalization significant improves the
quality of the approximation. 

  To compute $\tilde{W}_{\tilde{\bm{G}}_n}$, we have by Bayes rule that for any $k \in [n]$ and $s \leq k$, 
\begin{align}
\mathbb{P}_{\alpha'}( D_{\tilde{\bm{T}}_n}(v) = s \,|\, D_{\tilde{\bm{G}}_n}(v)
  = k )
  &= \frac{ \mathbb{P}_{\alpha'}( D_{\tilde{\bm{T}}_n}(v) = s,
    D_{\tilde{\bm{R}}_n}(v) = k - s)}
    { \sum_{t = 1}^k \mathbb{P}_{\alpha'}( D_{\tilde{\bm{T}}_n}(v) = t,
    D_{\tilde{\bm{R}}_n}(v) = k - t)
    } \\
  &= \frac{ P_{\text{Bin}(n-s,\theta)}(k-s) \Pb_{\alpha'}(
    D_{\tilde{\bm{T}}_n}(v) = s)}
    { \sum_{t=1}^k
    P_{\text{Bin}(n-t,\theta)}(k-t) \Pb_{\alpha'}(
    D_{\tilde{\bm{T}}_n}(v) = t)
    }, \label{eq:degree-asymp}
\end{align}

where $P_{\text{Bin}(n-s,\theta)}(\cdot)$ denotes the probability of a
binomial distribution with $n-s$ trials and success probability
$\theta$. The exact distribution of the degree  
$D_{\tilde{\bm{T}}_n}(v)$ of a node $v$ under the $\text{APA}_{\alpha',
  1}$ is intractable but we can approximate it by its limiting distribution
\begin{align*}
  P_{\alpha'}(s) := (2+\alpha') \frac{\Gamma(s + \alpha')
    \Gamma(3 + 2\alpha')}{ \Gamma(s + 3 + 2\alpha') \Gamma(1 + \alpha')}
  = \frac{2+\alpha'}{3+2\alpha'} \prod_{j=1}^{s-1} \frac{j + \alpha'}{j+3+2\alpha'}.
\end{align*}
By \citet[Theorem 8.2][]{van2016random}, we have that, for any node $v$, 
\begin{align*}
\sup_{s \in \mathbb{N}} \bigl| \mathbb{P}_{\alpha'}\bigl( D_{\tilde{\bm{T}}_n}(v) = s \bigr)
   - P_{\alpha'}(s) \bigr| \leq C_{\alpha'} \sqrt{ \frac{\log n}{n}} 
\end{align*}
with probability that tends to 1 as $n \rightarrow \infty$. Therefore,
we may replace $\mathbb{P}_{\alpha'}\bigl( D_{\tilde{\bm{T}}_n}(v) = s
\bigr)$ with $P_{\alpha'}(s)$ in~\eqref{eq:degree-asymp} to obtain a
tractable approximation which is accurate in the limit. 
 
To summarize, our estimation procedure generates a sequence $\alpha^j$
where $\alpha^j$ maximizes $\tilde{M}(\cdot \,|\, \alpha^{j-1})$ and
where $\tilde{M}$ is computed
using~\eqref{eq:degree-asymp}. Although we approximate $M(\cdot \,|\,
\cdot)$ by $\tilde{M}(\cdot \,|\, \cdot)$ and approximate the
distribution of the random degree $D_{\tilde{\bm{T}}_n}(v)$ by its
asymptotic limit, we find empirically that the resulting procedure always
converges and performs well. We test
the estimation procedure on simulated PAPER graphs of $n=3,000$ nodes
and $m=15,000$ edges and report the estimation performance in
Table~\ref{tab:estimatealpha}. We find that the estimator is biased upwards
when $\alpha$ is large, which is possibly because the
likelihood~\eqref{eq:likelihood1} is much less sensitive to a change in
$\alpha$ when $\alpha$ is large than when $\alpha$ is
small. In our simulation studies (Section~\ref{sec:simulation}), we
show that the confidence sets constructed with the estimated
parameters still attain their nominal coverage so that estimation
error does not significantly impact the inference quality.

\begin{table}[htp]
\begin{center}
\begin{tabular}{|c|c|c|c|c|c|}
\hline
\text{True $\alpha$} & 0 & 1 & 3 & 6 & $\infty$ (UA) \\
\hline
\text{Estimated $\alpha$} &  0.03 (0.04) & 1.04 (0.2) &
                                                                   3.3 (1.34) &
 10.7 (13.57) & 85.4 (20.9) \\
\hline
\end{tabular}
\end{center}
\caption{Mean and standard deviation of the estimated $\alpha$ computed on 200 independent trials
  on graphs with $n=3,000$ nodes and $m=15,000$ edges.}
\label{tab:estimatealpha}
\end{table}

We use the same estimator in the fixed $K > 1$ setting and the
variable $K$ setting. In these cases, the log-likelihood is slightly
different because the root nodes have imaginary self-loop
edges. However, if the number of root nodes is small, the
log-likelihood is virtually identical.

\subsection{Derivation of root sampling probability~\eqref{eq:multi-tree-root} for the fixed $K$ and random $K$ setting}
\label{sec:rootderivation}

Let $\tilde{\bm{f}}_n$ be an alphabetically labeled forest with
component trees $\tilde{\bm{t}}^1, \ldots, \tilde{\bm{t}}^K$. For a
specific tree $\tilde{\bm{t}}^k$ and a node $u \in
V(\tilde{\bm{t}}^k) \subset \mathcal{U}_n$, we compute the
probability, under the $\text{PAPER}(\alpha, \beta, K, \theta)$ model
and label randomization, that $u^k$ is the first node of
$\tilde{\bm{t}}^k$ given $\tilde{\bm{F}}_n = \tilde{\bm{f}}_n$.

To formally derive this, denote the $K$ random component trees of the random forest
$\tilde{\bm{F}}_n$ by $\tilde{\bm{T}}^1, \ldots, \tilde{\bm{T}}^K$,
and define $\Pi^k$ as the random latent \emph{relative} ordering of the
nodes in the $k$-th random component tree $\tilde{\bm{T}}^k$. In other
words, $\Pi^k$ takes value in $\text{Bi}( [n^k], V(\tilde{\bm{T}}^k))$
(where $n^k = | V(\tilde{\bm{T}}^k)|$) and $\Pi^k_t = v$ implies that
$v$ is the $t$-th node, among the nodes of $\tilde{\bm{T}}^k$, to
arrive in $\tilde{\bm{T}}^k$.

Then, we have that, for any $u \in V(\tilde{\bm{t}}^k)$,
\begin{align*}
  \Pb( \Pi^k_1 = u \,|\, \tilde{\bm{T}}^k = \tilde{\bm{t}}^k)
  &= \sum_{\pi^k \in \text{hist}(u, \tilde{\bm{t}}^k)} 
    \Pb( \Pi^k = \pi^k \,|\, \tilde{\bm{T}}^k =
    \tilde{\bm{t}}^k)\\
  &\propto \sum_{\pi^k \in \text{hist}(u, \tilde{\bm{t}}^k)}
    \Pb( \tilde{\bm{T}}^k =
    \tilde{\bm{t}}^k \,|\,  \Pi^k = \pi^k )\\
  &\propto h(u, \tilde{\bm{t}}^k) \prod_{j=2}^{D_{\tilde{\bm{t}}^k}(u)
    + 1} (\beta j + \alpha) 
    \prod_{v \neq u, v \in V(\tilde{\bm{t}}^k)} \prod_{j=1}^{D_{\tilde{\bm{t}}^k}(v)-1} (\beta j
    + \alpha) \\
  &= h(u, \tilde{\bm{t}}^k) ( \beta D_{\tilde{\bm{t}}^k}(u) +
    \beta + \alpha) (\beta D_{\tilde{\bm{t}}^k}(u) + \alpha)
    \prod_{v \in V(\tilde{\bm{t}}^k)} \prod_{j=1}^{D_{\tilde{\bm{t}}^k}(v)-1} (\beta j
    + \alpha)\\
  &\propto h(u, \tilde{\bm{t}}^k) ( \beta D_{\tilde{\bm{t}}^k}(u) +
    \beta + \alpha) (\beta D_{\tilde{\bm{t}}^k}(u) + \alpha),
\end{align*}
  where the third proportionality (equality up to multiplicative
  factor that is constant with respect to $u$) follows from
  Proposition~\ref{prop:apa-prob2}. Formula~\eqref{eq:multi-tree-root}
  thus follows. 

\subsection{Collapsed Gibbs sampler}
\label{sec:collapsed-gibbs}

We give an alternative Gibbs sampler in which we sample only a set of
root nodes instead of sampling an entire history $\pi$. More
precisely, we alternate between the following two stages:
\begin{enumerate}
\item[(A)] We fix the forest $\tilde{\bm{f}}$ and sample a set of root nodes $\tilde{s}$ with probability
\begin{align}
\Pb( \tilde{S} = \tilde{s} \,|\, \tilde{\bm{F}}_n =
  \tilde{\bm{f}}_n, \tilde{\bm{G}}_n = \tilde{\bm{g}}) 
  \propto \Pb(  \tilde{S} = \tilde{s} \,|\, \tilde{\bm{F}}_n =
  \tilde{\bm{f}}_n),
\end{align}
where $\tilde{s}$ comprise of a single node from each of the component
trees of $\tilde{\bm{f}}_n$.
\item[(B)] We fix the root set $\tilde{s}$ and generate a new forest
  $\tilde{\bm{f}}_n$ by iteratively sampling a new parent for each of
  the nodes. 
\end{enumerate}

To sample the root set for the first stage of the Gibbs sampler, we write $\tilde{\bm{t}}^{1},
\ldots, \tilde{\bm{t}}^{K}$ as the $K$ disjoint trees of the fixed
forest $\tilde{\bm{f}}_n$. Then, to generate the root set $\tilde{s}$,
we generate, for each tree $\tilde{\bm{t}}^k$, the root node $u^k$
with probability~\eqref{eq:multi-tree-root}. 

For the second stage of the Gibbs sampler, we place the nodes in some
arbitrary order and for each node $u$, we generate a parent
$\tilde{u}$, which could be equal to the old parent, according to the distribution
\begin{align}
\Pb\bigl\{ pa(u) = \tilde{u} \,|\, \{ pa(v) \}_{v\neq u}, \tilde{S}=\tilde{s},
  \tilde{\bm{G}}_n = \tilde{\bm{g}}_n \bigr\}. \label{eq:cond-pa2}
\end{align}
The action of generating a new parent is equivalent to replacing the
edge between $u$ and its old parent with a new one between $u$ and
$\tilde{u}$. Because we do not condition on the ordering $\Pi$, the
new parent $\tilde{u}$ can be any node in the network connected to $u$
that is not a descendant of $u$--that is, we only require that
$\tilde{u}$ is not in the subtree $\tilde{\bm{t}}^{(\tilde{s})}_u$ of
node $u$, where we view $\tilde{s}$ as the roots for the whole
forest. 

\begin{figure}
  \centering
  \includegraphics[scale=0.3]{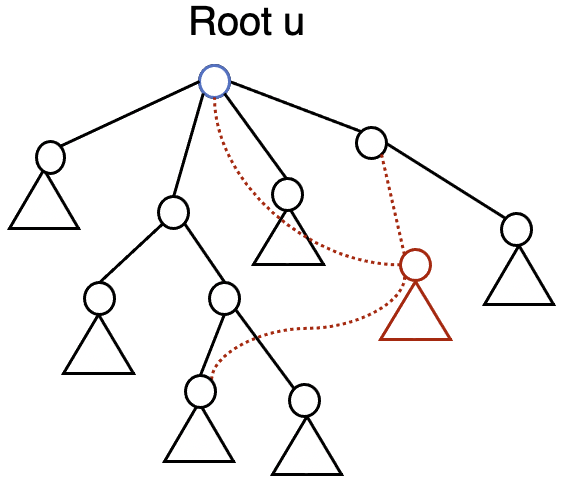}\hspace{0.2in}
  \includegraphics[scale=0.3]{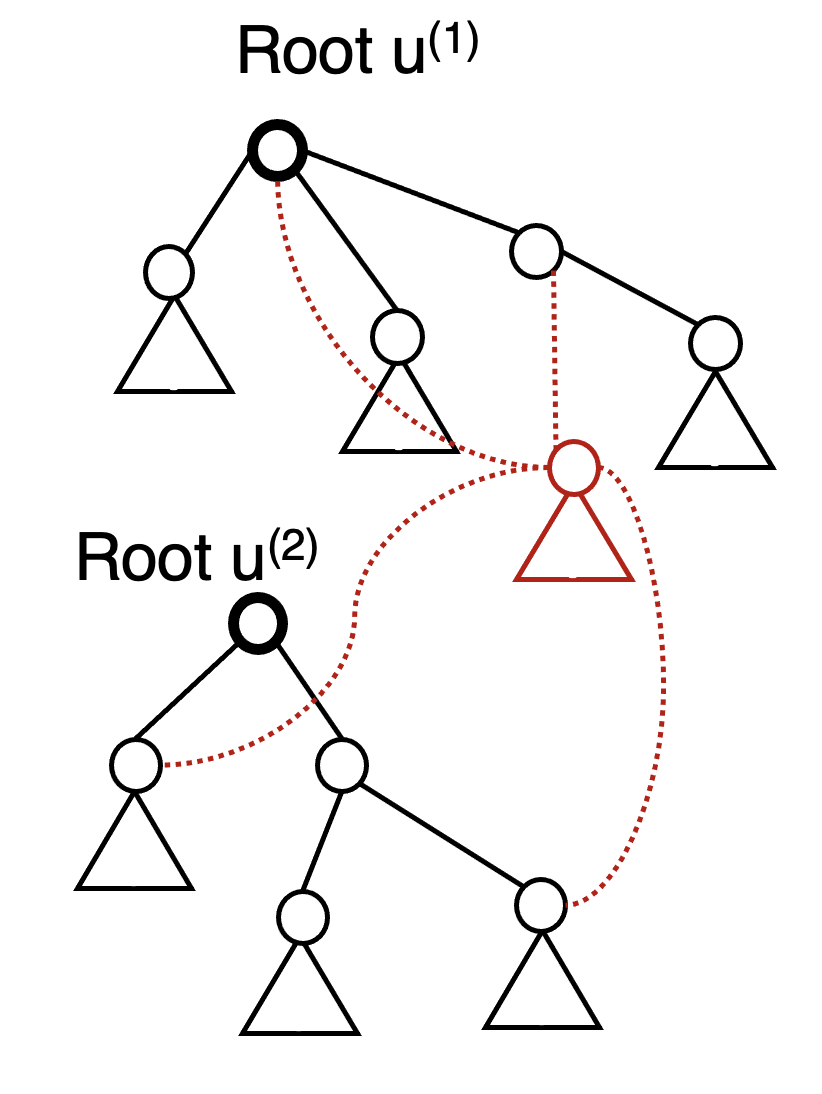}
  \caption{Selecting a new parent for a node. \textbf{Left:}
  the single root setting. \textbf{Right:} the multiple roots setting.}
  \label{fig:graft}
\end{figure}

Another way to think of the second stage is that we take the subtree
$\tilde{\bm{t}}^{(\tilde{s})}_u$ and \emph{graft} it onto another part
of the forest. In the multiple roots setting, a subtree may be
transferred from one component tree to another. In the random $K$
setting, two disjoint subtrees may be merged into a single tree or, a
subtree may be split and forms a new component. See
Figure~\ref{fig:graft} for a visual illustration.

In contrast with~\eqref{eq:cond_pa}, we do not condition on $\Pi$ and
must therefore sum over all histories when computing~\eqref{eq:cond-pa2}:
\begin{align*}
&\Pb( \tilde{\bm{F}}_n = \tilde{\bm{f}}_n \,|\, \tilde{\bm{G}}_n =
  \tilde{\bm{g}}_n, \tilde{S} = \tilde{s}) \\
  &\qquad \propto \Pb( \tilde{\bm{G}}_n = \tilde{\bm{g}}_n \,|\,
    \tilde{\bm{F}}_n = \tilde{\bm{f}}_n)
    \Pb( \tilde{\bm{F}}_n = \tilde{\bm{f}}_n, \tilde{S} = \tilde{s}) \\
  &\qquad = \binom{n(n-1)/2 - n + K}{m - n + k}^{-1} \sum_{\pi \in
    \text{hist}(\tilde{s}, \tilde{\bm{f}}_n)}
    \Pb( \tilde{\bm{F}}_n = \tilde{\bm{f}}_n, \Pi= \pi)\\
  &\qquad \propto \prod_{k=1}^K \frac{m-n+k}{n(n-1)/2 - n +k} \sum_{\pi \in
    \text{hist}(\tilde{s}, \tilde{\bm{f}}_n)}
    \Pb( \bm{F}_n = \pi^{-1} \tilde{\bm{f}}_n \,|\, \Pi= \pi) \\
  &\qquad \propto \prod_{k=1}^K \frac{m-n+k}{n(n-1)/2 - n +k}
    h(\tilde{s}, \tilde{\bm{f}}_n)
    \left\{
    \begin{array}{cc}
      \tilde{L}_{\alpha, \beta}(D_{\tilde{\bm{f}}_n})
      & \text{ if single root }
      \\
      \tilde{L}_{\alpha, \beta, K}(\tilde{s}, D_{\tilde{\bm{f}}_n})
      &  \text{ if fixed $K$ roots}\\
      \tilde{L}_{\alpha, \beta, \alpha_0}(\tilde{s},
      D_{\tilde{\bm{f}}_n})
      & \text{ if random $K$}
    \end{array}
     \right.
\end{align*}
where we have that
\begin{align*}
  \tilde{L}_{\alpha, \beta}(D_{\tilde{\bm{f}}_n})
  &= \prod_v
    \prod_{j=1}^{D_{\tilde{\bm{f}}_n}(v)-1} \beta j + \alpha \\
  \tilde{L}_{\alpha, \beta, K}(\tilde{s}, D_{\tilde{\bm{f}}_n})
  &= \prod_{v \in \tilde{s}} \prod_{j=2}^{D_{\tilde{\bm{f}}_n}(v)+1} (\beta j
    + \alpha) \prod_{v \notin \tilde{s}}
    \prod_{j=1}^{D_{\tilde{\bm{f}}_n}(v) -1} (\beta j
    + \alpha)\\
    \tilde{L}_{\alpha, \beta, \alpha_0}(\tilde{s}, D_{\tilde{\bm{f}}_n})
  &= \alpha_0^K \prod_{v \in \tilde{s}}
    \prod_{j=2}^{D_{\tilde{\bm{f}}_n}(v) +1} (\beta j
    + \alpha) \prod_{v \notin \tilde{s}}
    \prod_{j=1}^{D_{\tilde{\bm{f}}_n}(v) -1} (\beta j
    + \alpha).
\end{align*}

We may characterize the count of the history as follows:
\begin{align*}
  h(\tilde{s}, \tilde{\bm{f}}_n) =
  \left\{
  \begin{array}{cc}
    n! \prod_v \frac{1}{|\tilde{\bm{t}}_v^{(\tilde{s})}|}
    & \text{if single root }\\
    (n-K)! \prod_{v \notin \tilde{s}}
    \frac{1}{|\tilde{\bm{t}}_v^{(\tilde{s})}| }
    & \text{if fixed $K$ roots} \\
     (n-1)! \prod_{v}
    \frac{1}{|\tilde{\bm{t}}_v^{(\tilde{s})}|}
    & \text{if random $K$ roots}
      \end{array} \right.
\end{align*}

We summarize the resulting procedure in
Algorithm~\ref{alg:collapse-gibbs1}
and~\ref{alg:collapse-gibbs2}. These are similar to
Algorithm~\ref{alg:draw-forest1} and~\ref{alg:draw-forest2} except
that we take into account how the choice of the graft affects the size
of the history of the resulting forest. 

\begin{algorithm}
  \caption{Collapsed Gibbs sampler for fixed $K$ or single root settings}
  \label{alg:collapse-gibbs1}
  \textbf{Input:} labeled forest $\tilde{\bm{f}}_n$, a set of $K$ root
  nodes $\tilde{s}$.\\
  \textbf{Effect:} Modifies $\tilde{\bm{f}}_n$ in place.
  \begin{algorithmic}
    \For{ each node $u \in \mathcal{U}_n$}:
    \State if $u \in \tilde{s}$, continue.
    \State Remove the edge $(u, p(u))$ from $\tilde{\bm{f}}_n$. 
    \State Generate a node $w \in N_{\tilde{\bm{g}}_n} \backslash
    V(\tilde{\bm{t}}_u^{(\tilde{s})})$ with probability proportional to
    \[
      w \mapsto \prod_{v \in A_{\tilde{\bm{f}}_n}(w), v\notin \tilde{s}} 
      \frac{| \tilde{\bm{t}}_v^{(\tilde{s})}|}
      {|\tilde{\bm{t}}_v^{(\tilde{s})}| + |
        \tilde{\bm{t}}_u^{(\tilde{s})}|}
      (\beta D_{\tilde{\bm{f}}_n}(w) + \underbrace{2\beta \mathbbm{1}\{ w \in
      \tilde{s} \}}_{\text{ only for $K > 1$}} + \alpha),
    \]
    \State $\;\;$ where $A_{\tbm{f}_n}(w)$ is the set of ancestors (parent, parent of parent, etc) of $w$ \emph{including} $w$ itself.
    \State Add edge $(u, w)$ to $\tilde{\bm{f}}_n$.
    \EndFor
  \end{algorithmic}
\end{algorithm}

\begin{algorithm}
  \caption{Collapsed Gibbs Sampler for the random $K$ setting}
  \label{alg:collapse-gibbs2}
 \textbf{Input:} labeled forest $\tilde{\bm{f}}_n$, a set of root
  nodes $\tilde{s}$.\\
  \textbf{Effect:} Modifies $\tilde{\bm{f}}_n$ and $\tilde{s}$ in place.
  \begin{algorithmic}
    \For{each node $u \in \mathcal{U}_n$}:
    \State If $u \in \tilde{s}$ and $|\tilde{s}| = 1$, continue.
    \State If $u \in \tilde{s}$ and $|\tilde{s}| > 1$, set $\tilde{s} = \tilde{s}
    \backslash \{u\}$; else, remove the edge $(u, p(u))$ from $\tilde{\bm{f}}_n$. 
    \State Generate $w \in \{ \emptyset \} \cup \bigl(N_{\tilde{\bm{g}}_n} \backslash
    V(\tilde{\bm{t}}_u^{(\tilde{s})}) \bigr)$ with probability proportional to
    \[
      \left\{
        \begin{array}{lc}
      w \mapsto \prod_{v \in A_{\tilde{\bm{f}}_n}(w)} 
      \frac{| \tilde{\bm{t}}_v^{(\tilde{s})}|}
      {|\tilde{\bm{t}}_v^{(\tilde{s})}| + |
        \tilde{\bm{t}}_u^{(\tilde{s})}|}
      (\beta D_{\tilde{\bm{f}}_n}(w) + 2\beta \mathbbm{1}\{ w \in
      \tilde{s} \} + \alpha), & \text{ if $w \in N_{\tilde{\bm{g}}_n} \backslash
                             V(\tilde{\bm{t}}_u^{(\tilde{s})})$}\\
          w \mapsto \alpha_0 \frac{m - n + |\tilde{s}|}{n(n-1)/2 - n +
          |\tilde{s}|} & \text{ if $w = \emptyset$},
        \end{array}
        \right. 
    \]
    \State $\;\;$ where $A_{\tbm{f}_n}(w)$ is the set of ancestors (parent, parent of parent, etc) of $w$ \emph{including} $w$ itself.
    \State If $w \in N_{\tilde{\bm{g}}_n} \backslash
    V(\tilde{\bm{t}}_u^{(\tilde{s})})$, add edge $(u, w)$ to
    $\tilde{\bm{f}}_n$. Else, if $w = \emptyset$, let $\tilde{s} = \tilde{s} \cup \{ w \}$. 
    \EndFor
  \end{algorithmic}
\end{algorithm}

\subsection{Practical details on the Gibbs sampler}
\label{sec:practical}

\noindent \textbf{Convergence criterion:} We use a simple convergence
criterion where we run two chains simultaneously and keep track of the
resulting posterior root distributions, which we denote $Q^{A}$ and
$Q^{B}$ for the two chains. We continue the chain until the distance
(we use Hellinger distance or total variation distance in all the
experiments) between $Q^{A}$ and $Q^{B}$ is smaller than some
threshold $\tau$. We find that $\tau = 0.1$ suffices to generate
accurate confidence sets for the root node in the single root
setting. However, in the multiple roots setting, we require $\tau =
0.01$ or smaller. We observe in our experiments that the UA setting
($\alpha=1, \beta=0$)
requires far more iterations to converge than the LPA model
($\alpha=0, \beta=1$). It is important to note that the chains $A$ and $B$ are initialized with uniformly random spanning trees and uniformly random orderings on those trees so that the initialization is guaranteed to be overdispersed.

\textbf{Estimating $K$ in the fixed $K$ roots setting:} one way to
select $K$ is by maximum likelihood. For $K = 1, 2, 3, \ldots$, let $\tilde{\bm{G}}_n$ be
distributed according to $\text{PAPER}(\alpha, \beta, K, \theta)$ and
let 
\begin{align*}
  \mathcal{L}(K) &:= \Pb( \tilde{\bm{G}}_n = \tilde{\bm{g}}_n) \\
  &= \sum_{\tilde{\bm{f}}_n \in \mathcal{F}_K(\tilde{\bm{g}}_n), \pi}
    \Pb( \tilde{\bm{G}}_n = \tilde{\bm{g}}_n \,|\, \tilde{\bm{F}}_n =
    \tilde{\bm{f}}_n, \Pi = \pi) \Pb( \tilde{\bm{F}}_n =
    \tilde{\bm{f}}_n, \Pi = \pi) \\
  &= \binom{ n(n-1)/2 - (n-K)}{m - (n-K)}^{-1} \sum_{\tilde{\bm{f}}_n
    \in \mathcal{F}_K(\tilde{\bm{g}}_n), \pi}
    \Pb( \bm{F}_n = \pi^{-1} \tilde{\bm{f}}_n) \frac{1}{n!}. 
\end{align*}

Using the Gibbs sampler, we would then evaluate $\mathcal{L}(K)$ for
all $K \in [n]$. This however would be computationally
intensive. We therefore recommend the random $K$ model in settings
where $K$ is unknown and potentially large. \\

\noindent \textbf{Estimating $\alpha_0$ in the random $K$ roots
  setting:} We estimate $\alpha_0$ by adding one more step in the
Gibbs sampler where, after we generate a new forest and potentially a
new $K$, we sample $\alpha_0$ from the posterior distribution $\Pb(
\alpha_0 \,|\, K )$. To that end, we use an $\text{Exponential}(\lambda)$ prior on
$\alpha_0$ (we use $\lambda = 0.1$ yielding a variance of 100 in all
experiments) and follow \cite{west1992hyperparameter} to generate
posterior samples from $\Pb( \alpha_0 \,|\, K)$. We find that the
resulting estimate is insensitive to the choice of the hyperparameter $\lambda$
and performs well in practice. \\

{\color{black}    
\subsection{Details for algorithm under sequential noise models}
\label{sec:seq_noise_detail}

Let $\tbm{g}_n$ be an alphabetically labeled graph and $\tbm{t}_n$ be a spanning tree of $\tbm{g}_n$. Write $\tilde{\bm{r}}_n = \tilde{\bm{g}}_n \backslash \tilde{\bm{t}}_n$ as the subgraph of $\tbm{g}_n$ that comprises of the noise edges. We then have
\begin{align*}
  \mathbb{P}( \tbm{G}_n = \tbm{g}_n \,|\, \Pi=\pi, \tbm{T}_n = \tbm{t}_n)
  &= \prod_{j=3}^{n-1} \prod_{k=j+1}^n Q_{jk}, \\
\text{where }\, Q_{jk} &\equiv Q_{jk}(\pi, \tbm{t}_n) = 
    \biggl\{ \frac{\theta ( \tbeta D_{\tbm{t}_{k-1}}(\pi_j) + \talpha
    )}{2(k-2)\tbeta + (k-1)\talpha} \biggr\}^{\mathbbm{1}\{(\pi_j, \pi_k) \in
    \tilde{\bm{r}}_n\}} \\
  &\qquad 
    \biggl\{ \frac{ 2(k-2)\tbeta + (k-1)\talpha -
    \theta ( \tbeta D_{\tbm{t}_{k-1}}(\pi_j) + \talpha) }{
    2(k-2)\tbeta + (k-1)\talpha } \biggr\}^{\mathbbm{1}\{(\pi_j, \pi_k) \notin
    \tilde{\bm{g}}_n \}}.
\end{align*}

In some cases, it is convenient to refer to nodes through alphabetical labels
$\mathcal{U}_n$. Let $u, v \in \mathcal{U}_n$ be a pair of nodes and suppose $\pi_u^{-1} < \pi_v^{-1}$; we write
\begin{align}
Q_{uv} &\equiv Q_{uv}(\pi, \tbm{t}_n) := \biggl\{ \frac{\theta ( \tbeta D_{\tbm{t}_{\pi^{-1}_v-1}}(u) + \talpha
    )}{2(\pi^{-1}_v -2)\tbeta + (\pi^{-1}_v -1)\talpha} \biggr\}^{\mathbbm{1}\{(u, v) \in
    \tilde{\bm{r}}_n\}} \\
&\quad    \biggl\{ \frac{ 2(\pi_v^{-1} -2)\tbeta + (\pi_v^{-1} -1)\talpha -
    \theta ( \tbeta D_{\tbm{t}_{\pi_v^{-1} -1}}(u) + \talpha) }{
    2(\pi_v^{-1}-2)\tbeta + (\pi_v^{-1}-1)\talpha } \biggr\}^{\mathbbm{1}\{(u, v) \notin
    \tilde{\bm{g}}_n \}}. \label{eq:Q_defn}
\end{align}

For simplicity, we leave implicit the dependence of $Q_{uv}$ on $\pi$, $\tbm{g}_n$, and $\tbm{t}_n$. 

\subsubsection{Preliminary calculations}
\label{sec:seq_noise_preliminary}

To simplify notation, for two positive integers $j < k$, we write $[j, k] := \{j, j+1, \ldots, k\}$, $[j, k) := \{j, j+1, \ldots, k-1 \}$, and $(j, k] := \{j+1, j+2,\ldots, k\}$.

We first describe a fast algorithm to compute, for a particular node $u$ and a time interval $[j, k]$ where $\pi^{-1}_u < j$, the quantity 
\begin{align}
\prod_{t \in [j, k]} Q_{u, \pi_t}, 
\label{eq:first_Q_term}
\end{align}
which can be interpreted as the part of the noise likelihood associated with node
$u$ on a time interval $[j, k]$. We first observe that

\begin{align*}
  \prod_{t \in [j, k]} Q_{u, \pi_t}
  &= \prod_{t \in [j, k]} \biggl\{ \frac{\theta ( \tbeta D_{\tbm{t}_{t-1}}(u) + \talpha
    )}{2(t -2)\tbeta + (t -1)\talpha} \biggr\}^{\mathbbm{1}\{(u, \pi_t) \in
    \tilde{\bm{r}}_n\}} \\
 &\qquad  \biggl\{ \frac{ 2(t -2)\tbeta + (t -1)\talpha -
    \theta ( \tbeta D_{\tbm{t}_{t -1}}(u) + \talpha) }{
    2(t-2)\tbeta + (t-1)\talpha } \biggr\}^{\mathbbm{1}\{(u, \pi_t) \notin
   \tilde{\bm{g}}_n \}} .
\end{align*}

We extract the term $C_1 := \prod_{t \in [j, k]} \frac{1}{2(t-2)\tbeta +
  (t-1)\talpha}$ to obtain
\begin{align*}
  \prod_{t \in [j, k]} Q_{u, \pi_t}
  &= C_1  \prod_{t \in [j, k]}
    \bigl\{ \theta ( \tbeta D_{\bm{t}_{t-1}}(u) + \talpha) \bigr\}^{\mathbbm{1}\{(u, \pi_t)
    \in \tilde{\bm{r}}_n \}}  \bigl\{ 2(t -2)\tbeta + (t -1)\talpha -
    \theta ( \tbeta D_{\bm{t}_{t -1}}(u) + \talpha) \bigr\}^{\mathbbm{1}\{(u, \pi_t) \notin
    \tilde{\bm{g}}_n\}} \\
  &\qquad \qquad \bigl\{ 2(t-2)\tbeta + (t-1)\talpha \bigr\}^{\mathbbm{1}\{(u,
    \pi_t) \in \tilde{\bm{t}}_n\}}.
\end{align*}

We divide the time interval $[j, k]$ into sub-intervals in which
$D_{\bm{t}_{t-1}}(u)$ is constant. To that end, define $j = t_0 < t_1 < \ldots <
t_M = k+1$ such that
\begin{align*}
  D_{\tbm{t}_{t-1}}(u) &= d_0 \text{ for all $t \in [t_0, t_1)$, } \\
  D_{\tbm{t}_{t-1}}(u) &= d_1 = d_0 + 1 \text{ for all $t \in [t_1, t_2)$, } \\
  \ldots \\
  D_{\tbm{t}_{t-1}}(u) &= d_{M-1} = d_0 + M - 1 \text{ for all $t \in [t_{M-1}, t_M)$.}
\end{align*}

Then, we have that
\begin{align*}
 \prod_{t \in [j, k]} Q_{u, \pi_t}
  &= C_1 \prod_{\ell=0}^{M-1}   \prod_{t \in [t_\ell, t_{\ell+1})}
    \bigl\{ \theta ( \tbeta d_{\ell} + \talpha) \bigr\}^{\mathbbm{1}\{(u, \pi_t)
    \in \tilde{\bm{r}}_n \}} \\
  &\qquad \qquad \bigl\{ 2(t -2)\tbeta + (t -1)\talpha -
    \theta ( \tbeta d_{\ell} + \talpha) \bigr\}^{\mathbbm{1}\{(u, \pi_t) \notin
    \tilde{\bm{g}}_n\}}
    \bigl\{ 2(t-2)\tbeta + (t-1)\talpha \bigr\}^{\mathbbm{1}\{(u,
    \pi_t) \in \tilde{\bm{r}}_n\}}  \\
  &= C_1 \prod_{\ell=0}^{M-1}   \prod_{t \in [t_\ell, t_{\ell+1})}
    \biggl\{ \frac{\theta ( \tbeta d_{\ell} + \talpha)}
    { 2(t-2)\tbeta + (t-1)\talpha - \theta( \tbeta d_\ell + \talpha) }
    \biggr\}^{\mathbbm{1}\{(u, \pi_t)
    \in \tilde{\bm{r}}_n \}} \\
  &\qquad \qquad
    \biggl\{ \frac{ 2(t-2)\tbeta + (t-1)\talpha }
    { 2(t-2)\tbeta + (t-1)\talpha - \theta( \tbeta d_\ell + \talpha) }
    \biggr\}^{\mathbbm{1}\{(u, \pi_t) \in \tilde{\bm{t}}_n\}}
    \bigl\{
     2(t-2)\tbeta + (t-1)\talpha - \theta( \tbeta d_\ell + \talpha)
    \bigr\}.
\end{align*}

To simplify, we observe that $2(t-2) \tbeta + (t-1)\talpha = (2\tbeta
+ \talpha)t - (4 \tbeta + \talpha)$ and hence,
\begin{align*}
  \prod_{t \in [t_\ell, t_{\ell+1})}
  &\bigl\{ (2 \tbeta + \talpha)t - (4
  \tbeta + \talpha) - \theta( \tbeta d_\ell + \talpha) \bigr\} \\
  &= (2 \tbeta + \talpha)^{t_{\ell+1} - t_{\ell}}
    \frac{\Gamma( t_{\ell+1} - \frac{(4 \tbeta + \talpha) + \theta(
    \tbeta d_\ell + \talpha)}{2 \tbeta + \talpha})}
    { \Gamma( t_{\ell} - \frac{(4 \tbeta + \talpha) + \theta(
    \tbeta d_\ell + \talpha)}{2 \tbeta + \talpha})}.
\end{align*}

Therefore, we may re-write $\prod_{t \in [j, k]} Q_{u, \pi_t}$ as
follows:
\begin{align*}
  &\prod_{t \in [j, k]} Q_{u, \pi_t}
  = C_1 \prod_{\ell=0}^{M-1} (2 \tbeta + \talpha)^{t_{\ell+1} - t_{\ell}}
    \frac{\Gamma( t_{\ell+1} - \frac{(4 \tbeta + \talpha) + \theta(
    \tbeta d_\ell + \talpha)}{2 \tbeta + \talpha})}
    { \Gamma( t_{\ell} - \frac{(4 \tbeta + \talpha) + \theta(
  \tbeta d_\ell + \talpha)}{2 \tbeta + \talpha})} \\
  &\quad \prod_{t \in [t_\ell, t_{\ell+1})}
  \biggl\{ \frac{\theta ( \tbeta d_{\ell} + \talpha)}
    { 2(t-2)\tbeta + (t-1)\talpha - \theta( \tbeta d_\ell + \talpha) }
    \biggr\}^{\mathbbm{1}\{(u, \pi_t)
    \in \tilde{\bm{r}}_n \}}
     \biggl\{ \frac{ 2(t-2)\tbeta + (t-1)\talpha }
    { 2(t-2)\tbeta + (t-1)\talpha - \theta( \tbeta d_\ell + \talpha) }
    \biggr\}^{\mathbbm{1}\{(u, \pi_t) \in \tilde{\bm{t}}_n\}}.
\end{align*}

The quantities $\{ t_\ell, d_\ell\}_{\ell=0}^{M-1}\}$ can be readily computed by iterating through the neighbors of $u$ in $\tbm{g}_n$. Therefore, this entire expression can be computed in time at most $O( D_{\tbm{g}_n}(u))$, as $M \leq D_{\tbm{g}_n}(u)$. This concludes the description of the algorithm for computing~\eqref{eq:first_Q_term}. 

Now, suppose $u$ is a node such that $\pi_u^{-1} \leq k$ and that $D_{\bm{T}_{t-1}}(u) = 1$ for all $t \in [\pi_u^{-1} + 1, k]$. We now give an efficient method to compute
\begin{align}
\prod_{t=1}^k Q_{u, \pi_t}. 
\label{eq:second_Q_term}
\end{align}
This is the part of the noise likelihood associated with node $u$ on the time interval $[1, k]$. We have that
\begin{align}
\prod_{t=1}^k Q_{u, \pi_t} = \underbrace{\prod_{t=1}^{\pi_u^{-1}-1} Q_{u, \pi_t}}_{\text{first term}}
 \underbrace{ \prod_{t=\pi_u^{-1}+1}^k Q_{u, \pi_t}}_{\text{second term}}.
\label{eq:second_Q_term_decomp}
\end{align}
    
To compute the first term of~\eqref{eq:second_Q_term_decomp}, we have
\begin{align*}
   \prod_{t=1}^{\pi_u^{-1}-1} Q_{u, \pi_t} 
  &= \prod_{t=1}^{\pi_u^{-1} - 1}
    \biggl\{ \frac{\theta ( \tbeta D_{\bm{t}_{\pi_u^{-1}-1}}(\pi_t) + \talpha
    )}{2(\pi_u^{-1} -2)\tbeta + (\pi_u^{-1} -1)\talpha} \biggr\}^{\mathbbm{1}\{(u, \pi_t) \in
    \tilde{\bm{r}}_n\}} \\
&\qquad    \biggl\{ \frac{ 2(\pi_u^{-1}-2)\tbeta + (\pi_u^{-1}-1)\talpha -
    \theta ( \tbeta D_{\bm{t}_{\pi_u^{-1}-1}}(\pi_t) + \talpha) }{
    2(\pi_u^{-1}-2)\tbeta + (\pi_u^{-1}-1)\talpha } \biggr\}^{\mathbbm{1}\{(u, \pi_t) \notin
    \tilde{\bm{g}}_n \}}.
\end{align*}

Define $C_2 = \prod_{t=1}^{\pi_u^{-1}-1}
\frac{2(\pi_u^{-1}-2)\tbeta + (\pi_u^{-1}-1)\talpha - \theta(\tbeta
  D_{\bm{t}_{\pi_u^{-1}-1}}(\pi_t) + \talpha) }{2(\pi_u^{-1}-2)\tbeta +
  (\pi_u^{-1}-1)\talpha}$. Then,
\begin{align*}
  \prod_{t=1}^{\pi_u^{-1}-1} Q_{u, \pi_t}
  &= C_2 \biggl\{
  \frac{ 2(\pi_u^{-1} - 2)\tbeta + (\pi_u^{-1}-1)\talpha}
  {2(\pi_u^{-1} - 2)\tbeta + (\pi_u^{-1}-1)\talpha -
  \theta ( \tbeta D_{\bm{t}_{\pi_u^{-1}-1}}(\text{pa}(u)) + \talpha)} \biggr\}
  \\
  &\qquad \prod_{t=1}^{\pi_u^{-1}-1} \biggl\{
    \frac{\theta( \tbeta D_{\bm{t}_{\pi_u^{-1}-1}}(\pi_t) + \talpha)}
    {2(\pi_u^{-1}-2)\tbeta + (\pi_u^{-1} - 1)\talpha -
    \theta (\tbeta D_{\bm{t}_{\pi_u^{-1}-1}}(\pi_t) + \talpha) }
    \biggr\}^{\mathbbm{1}\{(\pi_t, u) \in \tilde{\bm{r}}_n\}}.
\end{align*}
Since it takes at most $O(D_{\tbm{g}_n}(\text{pa}(u)))$ time to compute $D_{\bm{T}_{\pi_u^{-1} - 1}}(\text{pa}(u))$, we see that the above expression, excluding $C_2$, can be computed in time at most $O( D_{\tbm{g}_n}(\text{pa}(u)) \vee D_{\tbm{g}_n}(u))$. We do not need to compute the $C_2$ term in practice as we care only about ratios of likelihoods.

For the second term of~\eqref{eq:second_Q_term_decomp}, we have that
\begin{align*}
  \prod_{t=\pi_u^{-1}+1}^{k} Q_{u, \pi_t}
  &= \prod_{t=\pi_u^{-1}+1}^k
    \bigg\{
    \frac{ \theta (\tbeta D_{\bm{t}_{t-1}}(u) + \talpha)}
    {2(t-2)\tbeta + (t-1)\talpha }
    \biggr\}^{\mathbbm{1}\{(u, \pi_t) \in \tilde{\bm{r}}_n\}} \\
  &\qquad \bigg\{
    \frac{2(t-2)\tbeta + (t-1)\talpha -  \theta (\tbeta D_{\bm{t}_{t-1}}(u) + \talpha)}
    {2(t-2)\tbeta + (t-1)\talpha }
    \biggr\}^{\mathbbm{1}\{(u, \pi_t) \notin \tilde{\bm{g}}_n\}}.
\end{align*}

Since we assume that $(\pi_t, u)$ is not a tree edge for every $t =
\pi_u^{-1}+1, \ldots, k$, we have that $D_{\bm{t}_{t-1}}(u) = 1$ for all $t \in
(\pi_u^{-1}, k]$ and thus,
\begin{align*}
\prod_{t=\pi_u^{-1}+1}^{k} Q_{u, \pi_t}
  &= \prod_{t=\pi_u^{-1}+1}^k
    \bigg\{
    \frac{ \theta (\tbeta + \talpha)}
    {2(t-2)\tbeta + (t-1)\talpha }
    \biggr\}^{\mathbbm{1}\{(u, \pi_t) \in \tilde{\bm{r}}_n\}} \\
  &\qquad \bigg\{
    \frac{2(t-2)\tbeta + (t-1)\talpha -  \theta (\tbeta + \talpha)}
    {2(t-2)\tbeta + (t-1)\talpha }
    \biggr\}^{\mathbbm{1}\{(u, \pi_t) \notin \tilde{\bm{g}}_n\}}.
\end{align*}

Define $C_3 = \prod_{t=\pi_u^{-1} + 1}^k \bigg\{
    \frac{2(t-2)\tbeta + (t-1)\talpha -  \theta (\tbeta + \talpha)}
    {2(t-2)\tbeta + (t-1)\talpha }
    \biggr\}$, we then have
\begin{align*}
  \prod_{t=\pi_u^{-1}+1}^k Q_{u, \pi_t}
  &= C_3 \prod_{t=\pi_u^{-1}+1}^k \bigg\{
    \frac{ \theta (\tbeta + \talpha)}
    {2(t-2)\tbeta + (t-1)\talpha - \theta(\tbeta + \talpha)}
    \biggr\}^{\mathbbm{1}\{(u, \pi_t) \in \tilde{\bm{r}}_n\}}.
\end{align*}    
    
\subsubsection{Calculation for transposition sampling}
\label{sec:seq_noise_transposition}

In this section, we provide an efficient way to compute the acceptance probability in the Metropolis--Hastings algorithm for updating our sample of $\pi$. For clarity, we write $Q_{jk}(\pi) \equiv Q_{jk}(\pi, \tbm{t}_n)$ to highlight the dependence of $Q_{jk}$ on $\pi$. We first state a Lemma that gives an easy way to check if a proposed $\pi^*$ is a valid history with respect to a given tree $\tbm{t}_n$.

\begin{lemma}
  \label{lem:swap_valid}
Let $\pi \in \text{hist}(\pi_1, \tilde{\bm{t}}_n)$. Let $\pi^*$ be equal to $\pi$ except that nodes $u$ and $v$, neither equal to $\pi_1$, are swapped. Assume without loss of generality that $\pi^{-1}_{u} < \pi^{-1}_{v}$. Then, $\pi^* \in \text{hist}(\pi_1, \tilde{\bm{t}}_n)$ if and only 
\begin{enumerate}
\item[(1)] For any child $w$ of $u$, we have $\pi^{-1}_w > \pi^{-1}_{v}$ and 
\item[(2)] the parent $\text{pa}(v)$ satisfies $\pi^{-1}_{\text{pa}(v)} < \pi^{-1}_{v}$.
\end{enumerate}
\end{lemma}

\begin{proof}
If $\pi^*$ is in $\text{hist}(\pi_1, \tbm{t}_n)$, it is clear that it must satisfy the two conditions (1) and (2).

Now assume conditions (1) and (2), we aim to show that $\pi^* \in \text{hist}(\pi_1, \tbm{t}_n)$. Since $\pi$ is a valid history, condition (1) implies that $v$ cannot be a descendant (e.g. child, grand-child, etc) of $u$. Moreover, (2) implies that all ancestors of $v$ have a $\pi$-position earlier than $u$. Therefore, it follows that swapping $u$ and $v$ yields a valid history $\pi^*$. The lemma follows as desired.
\end{proof}

We choose a pair $u = \pi_j$ and $v = \pi_k$ and define a new $\pi^*$
equal to $\pi$ except that
\begin{align*}
\pi^*_j = v, \quad \pi^*_k = u.
\end{align*}
Suppose $\pi^*$ satisfies the conditions of Lemma~\ref{lem:swap_valid}
so that $\pi^* \in \text{hist}(\pi_1, \tilde{\bm{t}}_n)$.

For a pair of nodes $x,y \in \mathcal{U}_n$, recall the definition of $Q_{x,y}(\pi)$ from~\eqref{eq:Q_defn}, where we now explicitly state the dependence of $Q_{x,y}$ on $\pi$. We have that
\begin{align*}
  &\frac{\mathbb{P}( \tbm{G}_n = \tbm{g}_n \,|\, \Pi = \pi^*, \tbm{T}_n =
  \tbm{t}_n) \mathbb{P}( \Pi = \pi^* \,|\, \tbm{T}_n = \tbm{t}_n)}
  {\mathbb{P}( \tbm{G}_n = \tbm{g}_n \,|\, \Pi = \pi, \tbm{T}_n =
  \tbm{t}_n) \mathbb{P}( \Pi = \pi \,|\, \tbm{T}_n = \tbm{t}_n)} 
  = \prod_{(x,y)} \frac{Q_{xy}(\pi^*) }{Q_{xy}(\pi)}.
\end{align*}

We claim that $\frac{Q_{xy}(\pi^*)}{Q_{xy}(\pi)} = 1$ for all $x,y$
that satisfy one of the following three conditions:
\begin{enumerate}
\item both $x, y \notin \{ u, v, \text{pa}(u), \text{pa}(v) \}$;
\item $x \in \{ u, v, \text{pa}(u), \text{pa}(v) \}$ and $\pi_y^{-1} >
  k$;
\item $x \in \{ \text{pa}(u), \text{pa}(v) \}$ and $\pi_y^{-1} < j$.
\end{enumerate}

This follows from the definition of $Q_{xy}(\pi)$. Therefore, we have that
\begin{align}
  \prod_{(x,y)} \frac{Q_{xy}(\pi^*) }{Q_{xy}(\pi)}
  &=
    \prod_{\substack{y \,:\, \pi_y^{-1} \leq k, \\ y\notin
  \{\text{pa}(u), \text{pa}(v)\}}}
  \frac{Q_{uy}(\pi^*)
    }{Q_{uy}(\pi)}
  \prod_{\substack{y \,:\, \pi_y^{-1} \leq k, \\ y \notin
  \{\text{pa}(u), \text{pa}(v), v \}}} \frac{Q_{vy}(\pi^*) 
  }{Q_{vy}(\pi)} \nonumber \\
  & \prod_{y \,:\, \pi_y^{-1} \in [j, k]}
    \frac{Q_{\text{pa}(u),y}(\pi^*)}
    {Q_{\text{pa}(u),y}(\pi)}
    \prod_{y \,:\, \pi_y^{-1} \in [j, k]}
    \frac{Q_{\text{pa}(v),y}(\pi^*)}
    {Q_{\text{pa}(v),y}(\pi)}. 
    \label{eq:mh_accept_prob}
\end{align}

The first two terms on the RHS of~\eqref{eq:mh_accept_prob} are of the form~\eqref{eq:second_Q_term}. The second two terms of the RHS of~\eqref{eq:mh_accept_prob} are of the form~\eqref{eq:first_Q_term}. Therefore, the whole expression~\eqref{eq:mh_accept_prob} can be computed in time at most $O( D_{\tbm{g}_n}(u) \vee D_{\tbm{g}_n}(v) \vee D_{\tbm{g}_n}(\text{pa}(u)) \vee D_{\tbm{g}_n}(\text{pa}(v)))$.

\subsubsection{Calculations for tree sampling}
\label{sec:seq_noise_tree_sampling}

\noindent \textbf{For seq-PAPER model without deletion of tree edges:}

For clarity, we write $Q_{jk}(\tbm{t}_n) \equiv Q_{jk}(\pi, \tbm{t}_n)$ to highlight the dependence of $Q_{jk}$ on $\tbm{t}_n$. For convenience, let us define
\begin{align}
F(\tbm{t}_n) &:= \mathbb{P}( \tbm{G}_n = \tbm{g}_n \,|\, \Pi = \pi, \tbm{T}_n =
  \tbm{t}_n) \mathbb{P}( \tbm{T}_n = \tbm{t}_n \,|\, \Pi = \pi) \\
  &= \prod_{x,y} Q_{xy}(\tbm{t}_n)
    \frac{\prod_{v \in \mathcal{U}_n} 
    \prod_{j=1}^{D_{\tbm{t}_n}(v)-1} (\beta j + \alpha)}
    { \prod_{t=3}^n 2(t-2)\beta + (t-1)\alpha}.
    \label{eq:F_defn}
\end{align}

We iterate $t=2,3,\ldots,n$ and sample a new parent for
$\pi_t$ among the candidate set $\pi_{1:(t-1)} \cap
N_{\tbm{g}_n}(\pi_t)$. For each $w \in \pi_{1:(t-1)} \cap
N_{\tbm{g}_n}(\pi_t)$, define $\tbm{t}^{(\cdot, \pi_t)}_n$ as the disconnected graph that results from removing the edge $(\text{pa}(\pi_t), \pi_t)$ from $\tbm{t}_n$, and define $\tbm{t}^{(w, \pi_t)}_n$ as the tree that
results from adding the edge
$(w, \pi_t)$ to $\tbm{t}_n^{(\cdot, \pi_t)}$.

For $t=1,2,\ldots, n$, we then sample a new parent for $\pi_t$ by removing $(\text{pa}(\pi_t), \pi_t)$ and then randomly choosing $w \in \pi_{1:(t-1)} \cap N_{\tbm{g}_n}(\pi_t)$ with probability
\begin{align}
\frac{F( \tbm{t}_n^{(w, \pi_t)} )}{\sum_{u \in \pi_{1:(t-1)} \cap N_{\tbm{g}_n}(\pi_t)} F(\tbm{t}_n^{(u, \pi_t)}) }.
\end{align}

Calculating $F(\tbm{t}_n^{(w, \pi_t)})$ naively according to~\eqref{eq:F_defn} takes time $O(n^2)$. We give a faster algorithm here. 

We start by noting that if \textbf{(1)} $x, y \notin \pi_{1:(t-1)} \cap
N_{\tbm{g}_n}(\pi_t)$ or \textbf{(2)} $x \in \pi_{1:(t-1)}\cap N_{\tbm{g}_n}(\pi_t)$ and
$y$ is such that $\pi_y^{-1} < t$, then the tree degree of $x$ at time $\pi_y^{-1}-1$ (or the tree degree of $y$ at time $\pi_x^{-1}-1$) is the same under both $\tbm{t}_n$ and $\tbm{t}^{(w, \pi_t)}_n$ for any $w$ and hence, $Q_{xy}( \tbm{t}^{(w, \pi_t)}_n) = Q_{xy}(\tbm{t}_n)$. Therefore, we have that
\begin{align}
F(\tbm{t}^{(w, \pi_t)}_n) = C \bigl\{\beta (D_{\tbm{t}^{(w, \pi_t)}_n}(w)-1)
  + \alpha\bigr\} \prod_{u \in \pi_{1:(t-1)} \cap
  N_{\tbm{g}_n}(\pi_t) } \, \prod_{y \,:\, \pi_y^{-1} \geq t}
  Q_{uy}(\tbm{t}^{(w, \pi_t)}_n) , 
  \label{eq:F_simple}
\end{align}
where $C$ is a term that does not depend on $w$; more precisely, we
have that
\begin{align*}
C &=\biggl\{ \prod_{x,y \notin \pi_{1:(t-1)}\cap
  N_{\tbm{g}_n}(\pi_t)} Q_{xy}(\tbm{t}_n) \biggr\} \biggl\{ \prod_{u
  \in \pi_{1:(t-1)}\cap N_{\tbm{g}_n}(\pi_t)} \, \prod_{y \,:\,
    \pi_y^{-1} < t} Q_{uy}(\tbm{t}_n) \biggr\} \\
  & \biggl\{
    \frac{
    \prod_{v \in \pi_{1:(t-1)} \cap N_{\tbm{g}_n}(\pi_t)}
    \prod_{j=1}^{D_{\tbm{t}^{(\cdot, \pi_t)}_n}(v)-1} (\beta j + \alpha)
    \prod_{v \notin \pi_{1:(t-1)} \cap N_{\tbm{g}_n}(\pi_t)}
    \prod_{j=1}^{D_{\tbm{t}_n}(v)-1} (\beta j + \alpha)
    }
    {\prod_{t=3}^n 2(t-2)\beta + (t-1)\alpha}
    \biggr\}.
\end{align*}
We make one further
simplication. Since $Q_{uy}(\tbm{t}^{(w, \pi_t)})$ depends on the tree $\tbm{t}^{(w, \pi_t)}_n$ only
through its degree sequence across time,
we observe that, for an arbitrary fixed $u \in
\pi_{1:(t-1)} \cap N_{\tbm{g}_n}(\pi_t)$, the quantity $\prod_{y \,:\, \pi_y^{-1}
  \geq t} Q_{uy}(\tbm{t}^{(w, \pi_t)}_n)$ depends on $w$ only through the binary
value of whether $w
= u$ or $w \neq u$. Therefore, for any $u \in \pi_{1:(t-1)}\cap N_{\tbm{g}_n}(\pi_t)$, we write
\begin{align}
B(u) &= \prod_{y \,:\, \pi_y^{-1}
       \geq t} Q_{uy}(\tbm{t}^{(w, \pi_t)}_n) \quad \text{ for any $w \neq u$}
  \nonumber \\
  A(u) &= \prod_{y \,:\, \pi_y^{-1}
       \geq t} Q_{uy}(\tbm{t}^{(u, \pi_t)}_n).
       \label{eq:AB_defn}
\end{align}

Then, by defining $C' = \prod_{u \in \pi_{1:(t-1)} \cap
  N_{\tbm{g}_n}(\pi_t)} B(u)$, we have that
\begin{align*}
F(\tbm{t}^{(w, \pi_t)}_n) = C \cdot C' \cdot \frac{A(w)}{B(w)} \bigl\{\beta (D_{\tbm{t}^{(w, \pi_t)}_n}(w)-1)
  + \alpha\bigr\}.
\end{align*}

The terms $A(w), B(w)$ are of the form~\eqref{eq:first_Q_term} and can thus be computed in time proportional to the degree $D_{\tbm{g}_n}(w)$. Therefore, the whole term $F(\tbm{t}_n^{(w, \pi_t)})$ can be, up to constants $C, C'$ which do not depend on $w$, computed in time $O( D_{\tbm{g}_n}(w) )$.\\

\noindent \textbf{For $\text{seq-PAPER}^*$ model with potential deletion of tree edges:}

With deletion noise, we must incorporate the likelihood of tree edge removal into~\eqref{eq:F_defn}. We denote $E(\tbm{t}_n)$ and $E(\tbm{g}_n)$ as the sets of edges of $\tbm{t}_n$ and $\tbm{g}_n$ respectively and define
\begin{align}
F(\tbm{t}_n) &:= \mathbb{P}( \tbm{G}_n = \tbm{g}_n \,|\, \Pi = \pi, \tbm{T}_n =
  \tbm{t}_n) \mathbb{P}( \tbm{T}_n = \tbm{t}_n \,|\, \Pi = \pi) \\
  &= \prod_{x,y} Q_{xy}(\tbm{t}_n)
    \frac{\prod_{v \in \mathcal{U}_n} 
    \prod_{j=1}^{D_{\tbm{t}_n}(v)-1} (\beta j + \alpha)}
    { \prod_{t=3}^n 2(t-2)\beta + (t-1)\alpha} (1- \eta)^{ | E(\tbm{t}_n) \cap E(\tbm{g}_n) |} \eta^{ | E(\tbm{t}_n) \backslash E(\tbm{g}_n) |}.
\end{align}

Define $\tbm{t}^{(\cdot, \pi_t)}_n$ as the disconnected graph that results from removing $(\text{pa}(\pi_t), \pi_t)$ just as in the discussion following~\eqref{eq:F_defn} and, for $w \in \pi_{1:(t-1)}$, define $\tbm{t}_n^{(w, \pi_t)}$ as the tree that results from adding $(w, \pi_t)$. Note that we do not require $w \in N_{\tbm{g}_n}(\pi_t)$, i.e. $(w, \pi_t)$ need not be an edge in $\tbm{g}_n$, and hence, $\tbm{t}_n^{(w, \pi_t)}$ may not be a subgraph of $\tbm{g}_n$. 

Following the same derivation as~\eqref{eq:F_simple}, we have that
\begin{align}
F(\tbm{t}^{(w, \pi_t)}_n) &= C \bigl\{\beta (D_{\tbm{t}^{(w, \pi_t)}_n}(w)-1)
  + \alpha\bigr\} \nonumber \\
  &\quad 
  \eta^{ \mathbbm{1}\{ (w, \pi_t) \notin \tbm{g}_n\}} 
  (1-\eta)^{\mathbbm{1}\{ (w, \pi_t) \in \tbm{g}_n\}}
  \prod_{u \in \pi_{1:(t-1)}} \, \prod_{y \,:\, \pi_y^{-1} \geq t}
  Q_{uy}(\tbm{t}^{(w, \pi_t)}_n) , 
  \label{eq:F_simple2}
\end{align}
where $C$ is a term that does not depend on $w$.

Defining $A(\cdot)$ and $B(\cdot)$ as in~\eqref{eq:AB_defn}, we then have
\begin{align}
F(\tbm{t}^{(w, \pi_t)}_n) = C \cdot C' \cdot  \frac{A(w)}{B(w)} \bigl\{\beta (D_{\tbm{t}^{(w, \pi_t)}_n}(w)-1)
  + \alpha \bigr\} \cdot \eta^{ \mathbbm{1}\{ (w, \pi_t) \in \tbm{g}_n\} } (1 - \eta)^{ \mathbbm{1}\{ (w, \pi_t) \notin \tbm{g}_n \}}
  \label{eq:F_final2}
\end{align}

Since $A(w)$ and $B(w)$ can be computed in time $O( D_{\tbm{g}_n})$, we have that $F( \tbm{t}_n^{(w, \pi_t)})$ can be computed in time $O( D_{\tbm{g}_n} )$ as well.

The overall procedure is then to sample $w \in \pi_{1:(t-1)}$ with probability proportional to~\eqref{eq:F_final2} and replacing the edge $(\text{pa}(\pi_t), \pi_t)$ with $(w, \pi_t)$ in the tree $\tbm{t}_n$. 

\subsubsection{Parameter sampling for the seq-PAPER model}
\label{sec:seq_param}

Although it may be possible to derive an EM algorithm to estimate the parameters $\alpha, \beta, \theta, \talpha, \tbeta$ in the seq-PAPER model, we propose to take a full Bayesian approach where we impose a prior and sample the parameters after sampling the ordering $\pi$ and the tree $\tbm{\pi}_n$ in the Gibbs sampler. 

As in Section~\ref{sec:parameter-estimation}, we assume that $\beta, \tbeta > 0$ so that we may assume without loss of generality that $\beta = \tbeta = 1$ and only estimate $\alpha$ and $\talpha$. We propose to use an $\text{Exponential}(\lambda)$ prior for $\alpha$, $\theta$, and $\talpha$ with $\lambda = 0.1$. Conditional on the ordering $\pi$ and the tree $\tbm{t}_n$, the likelihood for $\alpha$ is the same as that of $\ell(\alpha ; \tbm{T}_n)$ in Section~\ref{sec:parameter-estimation}; the likelihood for $\talpha$ and $\theta$ is, writing $\tbm{r}_n = \tbm{g}_n \backslash \tbm{t}_n$, 
\[
\ell(\talpha ; \tbm{g}_n, \tbm{t}_n, \pi) = \prod_{j=3}^{n-1} \prod_{k=j+1}^n Q_{jk}
\]
where $Q_{jk} = \bigl\{ \frac{ \theta (D_{\tbm{t}_{k-1}}(\pi_j) + \talpha)}{2(k-2) + (k-1)\talpha} \bigr\}^{\mathbbm{1}\{ (\pi_j, \pi_k) \in \tbm{r}_n \} } 
\bigl\{ 1 - \frac{ \theta (D_{\tbm{t}_{k-1}}(\pi_j) + \talpha)}{2(k-2) + (k-1)\talpha} \bigr\}^{\mathbbm{1}\{ (\pi_j, \pi_k) \notin \tbm{g}_n \} } $ is the contribution to the likelihood from the pair $(\pi_j, \pi_k)$. 

Conditionally on $\tbm{t}_n$ and $\pi$, it is still intractable to directly sampling $\alpha, \theta, \talpha$ so we propose Metropolis updates where we generate the new proposal either by adding a draw from $\text{Unif}[-\delta, \delta]$ or by multiplying with log-normal $e^{Z}$ for $Z \sim N(0, \delta)$, where the ratio of proposal probabilities is easy to compute with a Jacobian adjustment.

}

\section{Proof of results in Section~\ref{sec:theory}}
\label{sec:proof}
We first give the proof of the optimality lemma for $B_{\epsilon}(\cdot)$.

\begin{proof} (of Lemma~\ref{lem:comparison}) \\
  Fix $\epsilon, \delta \in (0,1)$ and suppose that $C_{\delta
    \epsilon}(\cdot)$ is a labeling-equivariant (see
  Remark~\ref{rem:label-equivariance}) confidence set for the root
  node with asymptotic coverage $1-\delta \epsilon$, that is, there
  exists a sequence $\mu_n \rightarrow 0$ such that $\mathbb{P}(
  \rho_1 \in C_{\delta
    \epsilon}(\mathbf{G}^*_n)) \geq 1 - \delta \epsilon - \mu_n$.

Let $\Lambda$ be a random permutation drawn uniformly from
$\text{Bi}(\mathcal{U}_n, \mathcal{U}_n)$ and write $\Pi = \Lambda
\circ \rho$ so that
$\tilde{\bm{G}}_n := \Lambda \bm{G}^*_n = \Pi \bm{G}_n$ is the randomly labeled graph. Then, there exists a real-valued sequence $\mu_n \rightarrow 0$ such that
\begin{align}
&\mathbb{P} \bigl\{ \Pi_1 \in C_{\delta \epsilon}(\tilde{\mathbf{G}}_n) \bigr\} \nonumber \\
  &= \sum_{\pi \in \text{Bi}([n], \mathcal{U}_n)} \mathbb{P} \bigl\{
    \pi_1 \in C_{\delta \epsilon}( \pi
    \mathbf{G}_n) \,|\, \Pi = \pi \bigr\} \mathbb{P}( \Pi = \pi) \nonumber \\
&= \mathbb{P}( \rho_1 \in C_{\delta \epsilon}(\rho \mathbf{G}_n) ) \nonumber \\
&= \mathbb{P}( \rho_1 \in C_{\delta \epsilon}(\mathbf{G}^*_n)) \geq 1 - \delta \epsilon + \mu_n, \label{eq:random-coverage}
\end{align}
where the penultimate equality follows from the labeling-equivariance of $C_{\delta \epsilon}(\cdot)$.

For any labeled graph $\tilde{\mathbf{g}}_n$, we have from
definition~\eqref{eq:posterior_rootset_credset} that
$B_{\epsilon}(\tilde{\mathbf{g}}_n)$ is the smallest
labeling-equivariant subset of $\mathcal{U}_n$ such that $\mathbb{P}(
\Pi_1 \in B_{\epsilon}(\tilde{\mathbf{t}}_n) \,|\, \tilde{\mathbf{G}}_n = \tilde{\mathbf{g}}_n ) \geq 1- \epsilon$. Then, if $|B_\epsilon(\tilde{\mathbf{g}}_n)| > |C_{\delta \epsilon}(\tilde{\mathbf{g}}_n)|$, then it must be that $\mathbb{P}( \Pi_1 \in C_{\delta \epsilon}(\tilde{\mathbf{G}}_n) \,|\, \tilde{\mathbf{G}}_n = \tilde{\mathbf{g}}_n ) < 1- \epsilon$.

Therefore, we have from~\eqref{eq:random-coverage} that
\begin{align*}
1 - \delta \epsilon + \mu_n &\leq \mathbb{P}(\Pi_1 \in C_{\delta \epsilon}(\tilde{\mathbf{G}}_n) ) \\
&= \sum_{\tilde{\mathbf{g}}_n}
\mathbb{P}( \Pi_1 \in C_{\delta \epsilon}(\tilde{\mathbf{G}}_n) \,|\, \tilde{\mathbf{G}}_n = \tilde{\mathbf{g}}_n ) \mathbb{P}(\tilde{\mathbf{G}}_n = \tilde{\mathbf{g}}_n ) \\
&\leq \mathbb{P}\bigl\{ |B_\epsilon(\tilde{\mathbf{G}}_n)| \leq |C_{\delta \epsilon}(\tilde{\mathbf{G}}_n)| \bigr\} + (1 - \epsilon) \mathbb{P}\bigl\{ |B_\epsilon(\tilde{\mathbf{G}}_n)| > |C_{\delta \epsilon}(\tilde{\mathbf{G}}_n)| \bigr\}.
\end{align*}

We then obtain by algebra that
\[
\mathbb{P}\bigl\{ |B_\epsilon(\tilde{\mathbf{G}}_n)| > |C_{\delta
  \epsilon}(\tilde{\mathbf{G}}_n)| \bigr\} \leq \delta + \mu_n / \epsilon,
\]
which yields the desired conclusion.
\end{proof}

\subsection{Proof of results in LPA setting}
\label{sec:lpa-proof}

Next, we give the proof of all statements regarding the LPA setting.

\begin{proof} (of Theorem~\ref{thm:pa-conf-size})\\

  Since $\bm{G}_n = \bm{T}_n + \bm{R}_n$ for a linear preferential attachment
  tree $\bm{T}_n$ and an Erd\H{o}s--R\'{e}nyi graph $\bm{R}_n$, we have that
  $D_{\bm{G}_n} = D_{\bm{T}_n} + D_{\bm{R}_n}$.

  By \cite{pekoz2014joint}, we have that, for any $q > 2$,
\[
\frac{1}{\sqrt{n}} (D_{\bm{T}_n}(1), D_{\bm{T}_n}(2), \ldots, D_{\bm{T}_n}(n), 0, 0, \ldots) \stackrel{d}{\rightarrow} (Y_1, Y_2, Y_3, \ldots ),
\]
in distribution with respect to the $\ell_q$ metric where $(Y_1, Y_2,
\ldots)$ is a random sequence satisfying $\sum_{j=1}^\infty \mathbb{E} Y_j^q <
\infty$ and each random variable $Y_j$ has a density with respect to
the Lebesgue measure.

We first claim that, for any $q > \frac{1}{\delta}$, if $\theta \leq n^{-\frac{1}{2} - \delta}$, then
\[
\frac{1}{\sqrt{n}} (D_{\bm{R}_n}(1), D_{\bm{R}_n}(2), \ldots,
D_{\bm{R}_n}(n), 0, 0 \dots) \rightarrow
(0, 0, 0 \ldots)
\]
in $\ell_q$ metric. Indeed, we have
\begin{align*}
\mathbb{E} & \| n^{-1/2} (D_{\bm{R}_n}(1), D_{\bm{R}_n}(2), \ldots, D_{\bm{R}_n}(n), 0, 0, \ldots ) \|^q_q 
= n^{ -\frac{q}{2}} \sum_{k=1}^n \mathbb{E} D_{\bm{R}_n}(k)^q \\
&\qquad \stackrel{(a)}{\leq} n^{1 - \frac{q}{2}} \mathbb{E} \bigl( \text{Bin}(n-1, \theta)^q \bigr) 
 \stackrel{(b)}{\leq} n^{1 - \frac{q}{2}} ((2 \theta n)^q + C_q) \\
&\qquad \leq 2^q n^{1 - \frac{q}{2}} n^{\frac{q}{2} - q \delta} + C_q n^{1 - \frac{q}{2}} =  2^q n^{1 - q \delta} + C_q n^{1 - \frac{q}{2}},
\end{align*}
where the inequality $(a)$ follows since $D_{\bm{R}_n}(k)$ is Binomial
with $n - D_{\bm{T}_n}(k)$ trials and hence stochastically dominated
by $\text{Bin}(n-1, \theta)$ and where inequality $(b)$ follows from
Lemma~\ref{lem:binomial-moment-bound}.

Since $q > 2 \vee 1/\delta$ by assumption, we have that 
\[
\limsup_{n \rightarrow \infty} \mathbb{E}  \| n^{-1/2} (D_{\bm{R}_n}(1), D_{\bm{R}_n}(2), \ldots, D_{\bm{R}_n}(n), 0, 0, \ldots ) \|^q_q = 0
\]
and thus $\frac{1}{\sqrt{n}} (D_{\bm{R}_n}(1), D_{\bm{R}_n}(2), \ldots, D_{\bm{R}_n}(n), 0, 0 \dots) \rightarrow (0, 0, \ldots )$ in distribution. 

Since $D_{\bm{G}_n}(k) = D_{\bm{T}_n}(k) + D_{\bm{R}_n}(k)$ for all $k \in [n]$, we have by Slutsky's lemma that
\[
 \frac{1}{\sqrt{n}} (D_{\bm{G}_n}(1), D_{\bm{G}_n}(2), \ldots, D_{\bm{G}_n}(n), 0, 0, \ldots) \stackrel{d}{\rightarrow} (Y_1, Y_2, Y_3, \ldots ).
\]

We claim that, for any $\epsilon \in (0, 1)$, there exists $L_\epsilon \in
\mathbb{N}$ such that $\mathbb{P}( Y_1 \leq L_\epsilon\mhyphen\max(
\{Y_n\} ) ) \leq \epsilon$. To see this, recall that $Y_1$ has a
density $q(\cdot)$ on $[0, \infty)$  with respect to the Lebesgue measure and, fixing some $q > 2$, that $\mathbb{E} Y_j^q
\rightarrow 0$ as $j \rightarrow \infty$. Therefore, choosing any
$\delta > 0$ such that $\Pb( Y_1 \leq \delta) \leq \frac{\epsilon}{2}$
and $L_\epsilon$ such that $\mathbb{E} Y_{L_\epsilon}^q \leq
\frac{\epsilon}{2} \delta^j$, we have by Markov's inequality that
 \begin{align*}
    \mathbb{P}( Y_1 \leq Y_{L_\epsilon})
    &\leq \int_0^\infty \mathbb{P}( Y_{L_\epsilon}  > t) q(t) \, dt \\
    &\leq \int_0^\delta q(t) \, dt + \int_{\delta}^\infty \mathbb{P}(
    Y_{L_\epsilon}  > t) q(t) \,dt\\
  &\leq \mathbb{P}( Y_1 \leq \delta) + \mathbb{P}(
    Y_{L_\epsilon}  > \delta) \int_{\delta}^\infty  q(t) \, dt \\
  &\leq \frac{\epsilon}{2} + \frac{ \mathbb{E} Y_{L_{\epsilon}}^q}{ \delta^q}
    \leq \epsilon. 
\end{align*}
Since $L_\epsilon\mhyphen\max(\cdot)$
function on sequences is continuous with respect to $\ell_q$, we have
by continuous mapping theorem and Portmanteau lemma that
\[
\limsup_{n \rightarrow \infty} \mathbb{P}\bigl\{ D_{\bm{G}_n}(1) \leq
L_\epsilon \mhyphen\max(D_{\bm{G}_n}) \bigr\} \leq \Pb\bigl\{ Y_1 \leq
L_\epsilon\mhyphen\max(\{Y_n\}) \bigr\} \leq \epsilon.
\]
This proves the first conclusion of Theorem~\ref{thm:pa-conf-size}. 

To obtain the second conclusion, note that $C_\epsilon(\bm{G}^*_n) := \bigl\{
1\mhyphen\max(D_{\tilde{\bm{G}}_n}),
2\mhyphen\max(D_{\tilde{\bm{G}}_n}), \ldots,
L_\epsilon\mhyphen\max(D_{\tilde{\bm{G}}_n}) \bigr\}$ is a
labeling-equivariant confidence set for the root at asymptotical level
$1-\epsilon$. The second conclusion follows from
Lemma~\ref{lem:comparison}. 

\end{proof}

\begin{lemma}
\label{lem:binomial-moment-bound}
Let $X$ be a random variable with $\text{Bin}(n, \theta)$
distribution. For any $q \geq 1$, $\theta \in [0,1]$ and any $n \in \mathbb{N}$, we have that
\[
\mathbb{E} X^q \leq (2 \theta n)^q + C_q,
\]
where $C_q > 0$ is a constant that depends only on $q$.
\end{lemma}

\begin{proof}

Write $X$ as a random variable with the $\text{Bin}(n, \theta)$ distribution. Then,
\begin{align}
\mathbb{E} X^q &= \int_0^\infty \mathbb{P}( X^q \geq t ) dt \nonumber \\
&\leq (2\theta n)^q + \int_{(2\theta n)^q}^\infty \mathbb{P}( X^q \geq t) dt. \label{eqn:init_binomial_moment}
\end{align}

We note that $\text{Var}X \leq \theta n$. By Bernstein's inequality, we have that for all $t \geq (2\theta n)^q$,
\begin{align*}
\mathbb{P}( X^q \geq t) &= 
\mathbb{P}( X - \theta n \geq t^{1/q} - \theta n) \\
&\leq \exp \biggl( - \frac{1}{2} \frac{(t^{1/q} - \theta n)^2}{(t^{1/q} - \theta n) + \theta n} \biggr) \\
&\leq \exp ( - \frac{1}{8} t^{1/q} ).
\end{align*}

Therefore, we may bound the second term of~\eqref{eqn:init_binomial_moment} as 
\begin{align*}
\int_{(2\theta n)^q}^\infty \mathbb{P}( X^q \geq t) dt &\leq \int_{(2\theta n)^q}^\infty e^{ - \frac{t^{1/q}}{8} } dt \\
&\leq \int_0^\infty q s^{q-1} e^{- \frac{s}{8} } ds.
\end{align*}

\end{proof}

\subsection{Proof of results in UA setting}
\label{sec:ua-proof}

\begin{proof} (of Theorem~\ref{thm:ua-conf-size})\\

Let $\bm{T}_n$ be a random recursive tree with the UA
distribution. Let $s \in [n]$ be a node with arrival time $s$ and
assume that $s \geq n^{\eta}$. For any integer $i
\geq 1$, we define the random variable
\[
Z^{(s)}_i := \left\{ \begin{array}{cc}
    1 & \text{ if node $i+1$ is attached to node $1$}\\
    -1 & \text{if node $i+1$ is attached to node $s$} \\
    0 & \text{else}
    \end{array} \right.
\]

We note then that $\{ Z^{(s)} \}_{i=1}^n$ are independent. If $i \geq
s$, then $\mathbb{E} Z^{(s)}_i = 0$ and $\text{Var} Z^{(s)}_i =
\frac{2}{i}$, and if $i < s$, then we cannot attach to node $s$ and
hence, $\mathbb{E} Z^{(s)}_i = \frac{1}{i}$ and $\text{Var} Z^{(s)}_i
\leq \frac{1}{i}$. Define $Z^{(s)} = \sum_{i=1}^n Z_i^{(s)}$ so that
\[
Z^{(s)} = D_{\bm{T}_n}(1) - D_{\bm{T}_n}(s).
\]

Then, we have that
\begin{align*}
\mathbb{E} Z^{(s)} &= \sum_{i=1}^n \mathbb{E} Z^{(s)}_i = \sum_{i=1}^s \frac{1}{i} \geq (1 + \mu_1) \log s \\
\text{Var} Z^{(s)} &= \sum_{i=1}^n \text{Var} Z^{(s)}_i \leq \sum_{i=2}^s \frac{1}{i} + \sum_{i=s+1}^n \frac{2}{i} \leq 
(1+\mu_2) \bigl\{ \log s +  2 \bigl(\log n - \log s\bigr) \bigr\} .
\end{align*}
where we use $\mu_1, \mu_2$ to represent terms that are $o(1)$ as $n
\rightarrow \infty$. Therefore, we obtain that
\begin{align*}
  \mathbb{E} \bigl( D_{\bm{G}_n}(1) - D_{\bm{G}_n}(s) \bigr) =
  \mathbb{E} Z^{(s)} + \mathbb{E} \bigl( D_{\bm{R}_n}(1) - D_{\bm{R}_n}(s) \bigr)
 \leq (1 + \mu_1) \log s,
\end{align*}
where the inequality follows since $D_{\bm{R}_n}(s)$ has the
$\text{Bin}( n - D_{\bm{T}_n}(s), \theta)$ distribution; since
$D_{\bm{T}_n}(1)$ stochastically dominates $D_{\bm{T}_n}(s)$, we have
that $D_{\bm{R}_n}(s)$ stochastically dominates $D_{\bm{R}_n}(1)$. We
also have the following bound on the variance of $D_{\bm{G}_n}(1) -
D_{\bm{G}_n}(s)$:
\begin{align*}
\text{Var} \bigl( D_{\bm{G}_n}(1) - D_{\bm{G}_n}(s) \bigr) 
&= \text{Var} \biggl( \sum_{i=1}^n Z^{(s)}_i + D_{\bm{R}_n}(1) - D_{\bm{R}_n}(s) \biggr) \\
&\leq \mathbb{E} \, \text{Var} \biggl( \sum_{i=1}^n Z^{(s)}_i + D_{\bm{R}_n}(1) - D_{\bm{R}_n}(s) \bigg| D_{\bm{R}_n}(1), D_{\bm{R}_n}(s) \biggr) \\
&\qquad \qquad + \text{Var} \, \mathbb{E} \biggl[ \sum_{i=1}^n Z^{(s)}_i + D_{\bm{R}_n}(1) - D_{\bm{R}}(s) \bigg| D_{\bm{R}_n}(s), D_{\bm{R}_n}(1) \biggr]   \\
&\leq (1 + \mu_2)\{ \log s + 2(\log n - \log s)\} + 2n\theta \\
&\leq (1 + \mu_3) (2 - \eta) \log n.
\end{align*}

Hence, we have by Proposition~\ref{prop:bennett} that
\begin{align*}
\mathbb{P}( D_{\bm{G}_n}(s) \geq D_{\bm{G}_n(1)} ) 
&= \mathbb{P}\biggl( \sum_{i=1}^n Z^{(s)}_i + D_{\bm{R}_n}(1) - D_{\bm{R}_n}(s) \leq 0 \biggr)\\
  &\leq \mathbb{P}\biggl( \sum_{i=1}^n Z^{(s)}_i + D_{\bm{R}_n}(1) -
    D_{\bm{R}_n}(s) -
    \mathbb{E}\bigl[ Z^{(s)}+D_{\bm{R}_n}(1) - D_{\bm{R}_n}(s)\bigr] \leq - (1+\mu_1) \log s \biggr) \\
  &\leq 2 \exp\biggl( - (1 + \mu_3)(2 - \eta) \log n \cdot h\biggl( \frac{(1 + \mu_1) \eta \log n}{(1 + \mu_3)(2-\eta) \log n} \biggr) \biggr) \\
&\leq 2  (1 + \mu_4) n^{ - (2-\eta) h ( \frac{\eta}{2-\eta} )}.
\end{align*}

Therefore, we have
\begin{align*}
  \mathbb{P}\bigl( |
  & \{ s \geq n^\eta \,:\, D_{\bm{G}_n}(s) >
    D_{\bm{G}_n}(1)\}| \leq 2 \epsilon^{-1} n^{1 - (2-\eta) h(\frac{\eta}{2-\eta}) } \bigr) \\
 &\leq \epsilon n^{- 1 + (2-\eta) h(\frac{\eta}{2-\eta}) } \mathbb{E} | \{ s \geq n^\eta \,:\, D_{\bm{G}_n}(s) >  D_{\bm{G}_n}(1)\}| \\
  &\leq \epsilon n^{- 1 + (2-\eta) h(\frac{\eta}{2-\eta}) }
    \sum_{s =
    \lfloor n^{\eta}\rfloor }^n \mathbb{P}( D_{\bm{G}_n}(s) \geq D_{\bm{G}_n}(1)) \\
&\leq \epsilon (1 + \mu_4).
\end{align*}

Hence, we have that with probability at least $1 - (1+\mu_4)\epsilon$,
\[
D_{\bm{G}_n}(1) \geq L_{\eta, n, \epsilon}\mhyphen\max(D_{\bm{G}_n}).
\]

By optimizing $\eta$, we have that for some $\gamma < 0.8$ and
universal constant $C > 0$, with probability at least $1 - (1
+ \mu_4)\epsilon$,
\[
D_{\bm{G}_n}(1) \geq \frac{C}{\epsilon} n^\gamma\mhyphen\max(D_{\bm{G}_n}).
\]
Therefore, we may form a level $1-\epsilon$ asymptotically valid
confidence set for the root node by taking the $\frac{C}{\epsilon} n^\gamma$ nodes with
the highest degree in the observed alphabetically labeled graph
$\bm{G}^*_n$. The second claim of the theorem follows directly from Lemma~\ref{lem:comparison}.
\end{proof}

The next concentration inequality is standard. 

\begin{proposition}
\label{prop:bennett}
  (Bennett's inequality)\\
Let $X_1, \ldots X_n$ be independent random variables such that $|X_i|\leq b$. Let $V \geq \sum_{i=1}^n \text{Var} (X_i)$. Then, for any $t \geq 0$,
\[
\mathbb{P}\biggl( \biggl| \sum_{i=1}^n X_i - \mathbb{E}X_i \biggr| > t \biggr) \leq 2 \exp\biggl( - \frac{V}{b^2} h\biggl( \frac{bt}{V} \biggr) \biggr),
\]
where $h(z) = (1+z) \log(1+z) - z$.
\end{proposition}

\section{Supplement for Section~\ref{sec:empirical}}

\subsection{Additional results for central subgraph visualization}
\label{sec:central_subgraph_appendix}

\noindent \textbf{Enron email network:} This dataset consists of email exchanges
between members of the Enron corporation shortly before its
bankruptcy and the network is publicly available at the website 
\texttt{https://snap.stanford.edu/data/email-Enron.html}
(c.f. \cite{leskovec2009community}) for more details on the
network). This network has $n=33,696$ nodes and $m=180,811$ edges, with a
maximum degree of 1,383. We estimate $\beta = 1$ and
$\alpha = 0$ and the sizes of confidence sets are:
\begin{align*}
\text{ $60\%$: 7 nodes }  \quad  \text{ $80\%$: 11
                                        nodes } \quad
\text{ $95\%$: 42 nodes } \quad  \text{ $99\%$: 2393 nodes }.
\end{align*}

\begin{figure}
\centering
\begin{subfigure}[b]{0.35\textwidth}
		\centering
		{\includegraphics[scale=.2]{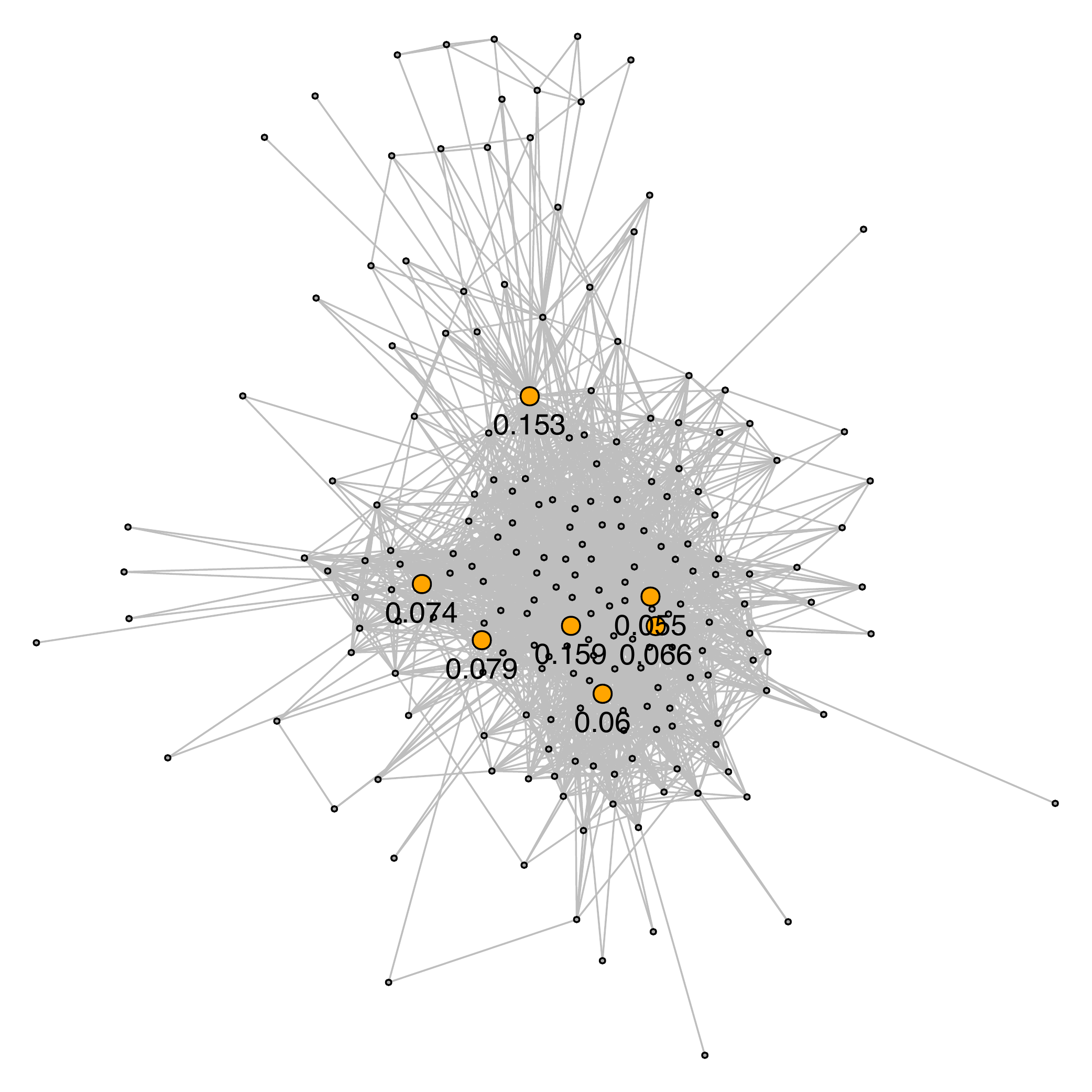}}
                \caption{Enron email subgraph}
	\label{fig:enron}
\end{subfigure}
\begin{subfigure}[b]{0.35\textwidth}
		\centering
		{\includegraphics[scale=.2]{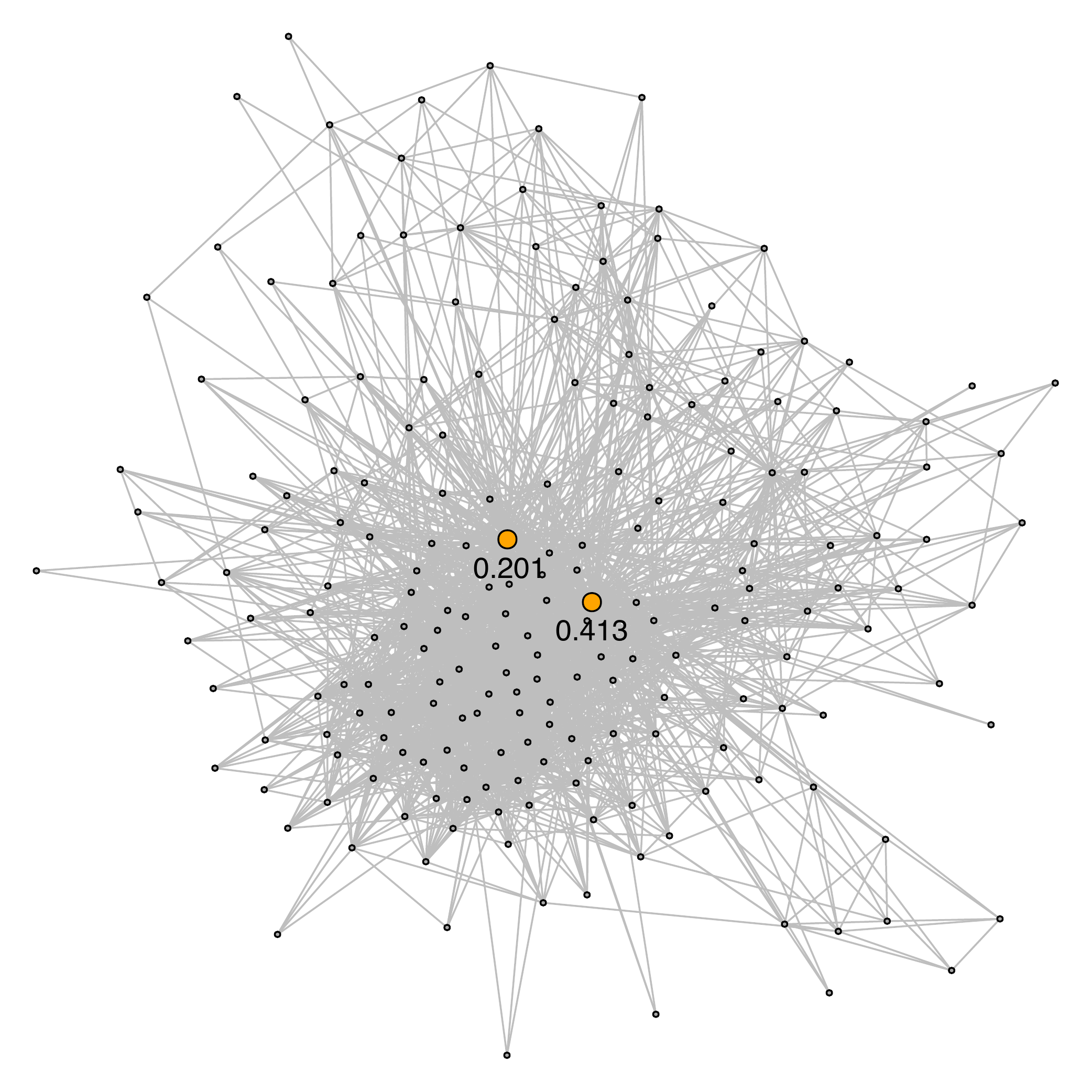}}
                \caption{Youtube subgraph}
	\label{fig:youtube}
\end{subfigure}	
\caption{Subgraph of the 200 nodes with highest posterior root
  probabilities.}
\end{figure}

The central subgraph of this network (shown in Figure~\ref{fig:enron}) exhibits a large central cluster
with many nodes that have relatively large posterior root
probabilities. These nodes may correspond to leadership personnel in
the company. \\

\noindent \textbf{Youtube social network:} This dataset consists of
friendship links between users in Youtube
\citep{mislove-2007-socialnetworks} and it is publicly available at
\texttt{https://snap.stanford.edu/data/com-Youtube.html}. This network
has $n=1,134,890$ nodes and $m=2,987,624$ edges, with a maximum degree
of 28,754. We estimate $\beta = 1$ and
$\alpha = 0$ and the sizes of confidence sets are:
\begin{align*}
\text{ $60\%$: 2 nodes }  \quad  \text{ $80\%$: 35
                                        nodes } \quad
 \text{ $95\%$: 1874 nodes } \quad  \text{ $99\%$: 16368 nodes }.
\end{align*}

The central subgraph of this network (shown in
Figure~\ref{fig:youtube}) also contains a large central cluster, which
may contain the most popular accounts on Youtube.

\subsection{Random $K$ roots analysis on air route network}
\label{sec:airroute}

We analyze an air route network \citep{guimera2005worldwide} of $n=3,618$ airports and $m=14,142$
edges where two airports share an edge if there is a regularly
scheduled flight between them. We remove the direction
of the edges and treat the network as undirected. The dataset is
publicly available at
\texttt{http://seeslab.info/downloads/air-transportation-networks/}. 

\begin{figure}[ht]
  \centering
  \includegraphics[scale=.45]{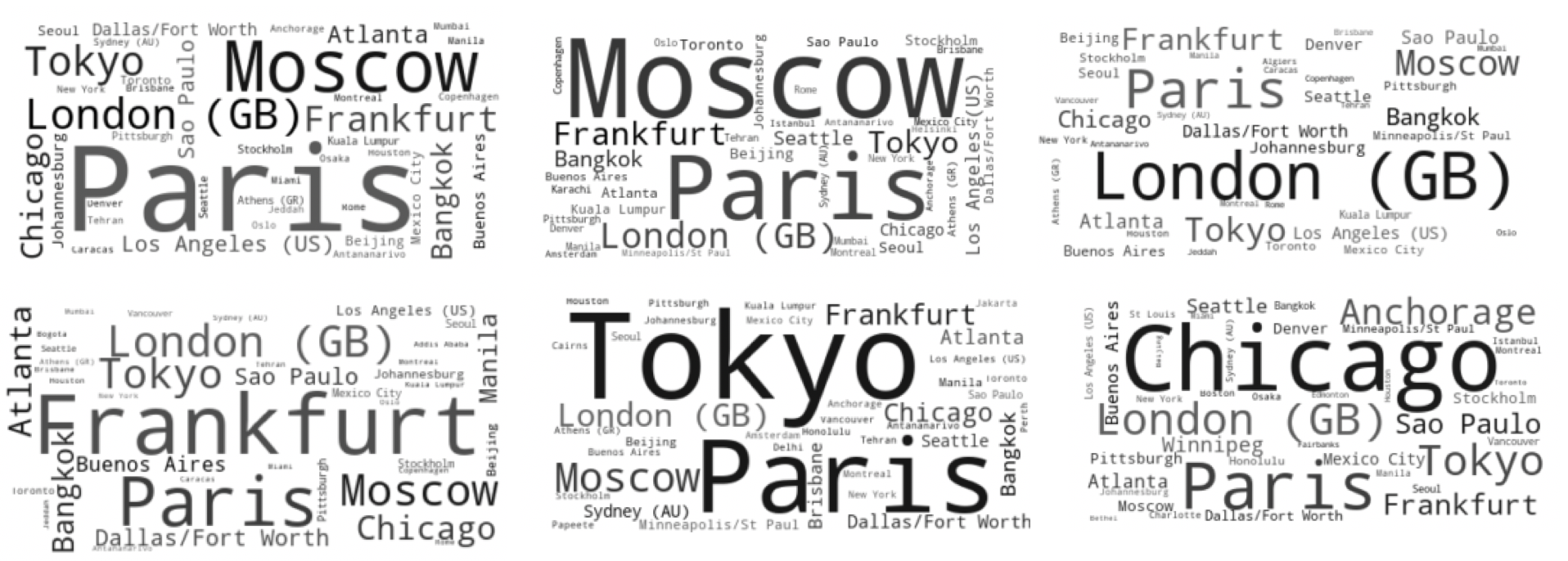}
  \includegraphics[scale=.45]{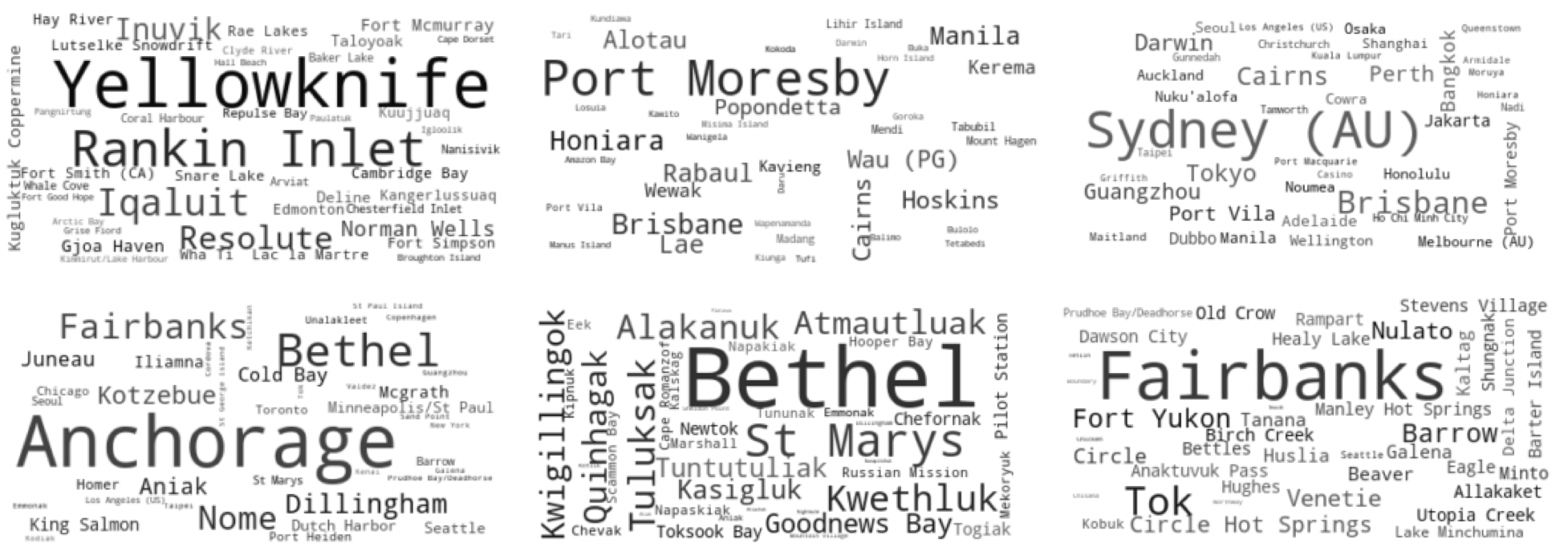}
  \caption{Top 12 community--trees on the air route network; first 6
    trees reflect the hub of major global airports centered at
    different cities; tree 7 contains remote regional airports in the Northwest
    Territories province of Canada; tree 8 contains remote regional
    airports in southeast Asian Pacific
    islands; tree 9 contains Australia/Southeast Asia airports; tree
    10 contains Alaskan airports while tree 11 and 12 contain western
    Alaskan and Northern Alaskan airports respectively.}
  \label{fig:airport}
\end{figure}

We perform our inference algorithm and display the top 12
community--trees in Figure~\ref{fig:airport}. That is, we take $\{Q_1,
\ldots, Q_{K_{\text{all}}}\}$ and display the 12 that has the largest
posterior probability of occuring. The first 6 community--trees
represent the same community, basically of all the major airports in the
world, centered at various potential root nodes (Paris, London,
Moscow, Tokyo, Chicago, Frankfurt). 

The 7th community--tree comprise
of regional airports in the remote Northwest Territories province of Canada and it is
centered at Yellowknife, which is the capital of the
province. This is not surprising because most regional airports in
Northern Canada are very small and are built only to connect remote settlements to larger
nearby cities such as Yellowknife. 

The 8th
community--tree comprise of regional airports on various Pacific and
Polynesian islands and it is centered at Port Moresby, the capital of
Papua New Guinea. The 9th community--tree is the Australia/Southeast
Asia cluster centered at Sydney. This result is sensible again because
most airports in the pacific islands are built only to connect the
small islands to larger nearby cities such as Port Moresby or Cairns. From a network respectively, these
remote airports are reachable only through a few cities such as Port
Moresby. 

The 10th to 12th community--trees comprise of airports in Alaska, many
of which are regional. The 10th community--tree is the whole Alaska
cluster centered at Anchorage while the 11th community--tree and the
12th community--tree represent, respectively, Western Alaska (centered
at Bethel, AK) and Northern Alaska (centered at Fairbanks, AK).

\subsection{Additional clusters for statistician co-authorship network}
\label{sec:coauthor_appendix}

\begin{figure}[!ht]
\begin{subfigure}{0.32\textwidth}
\centering
\includegraphics[scale=.24, trim=1in 2in 1in 1in, clip]{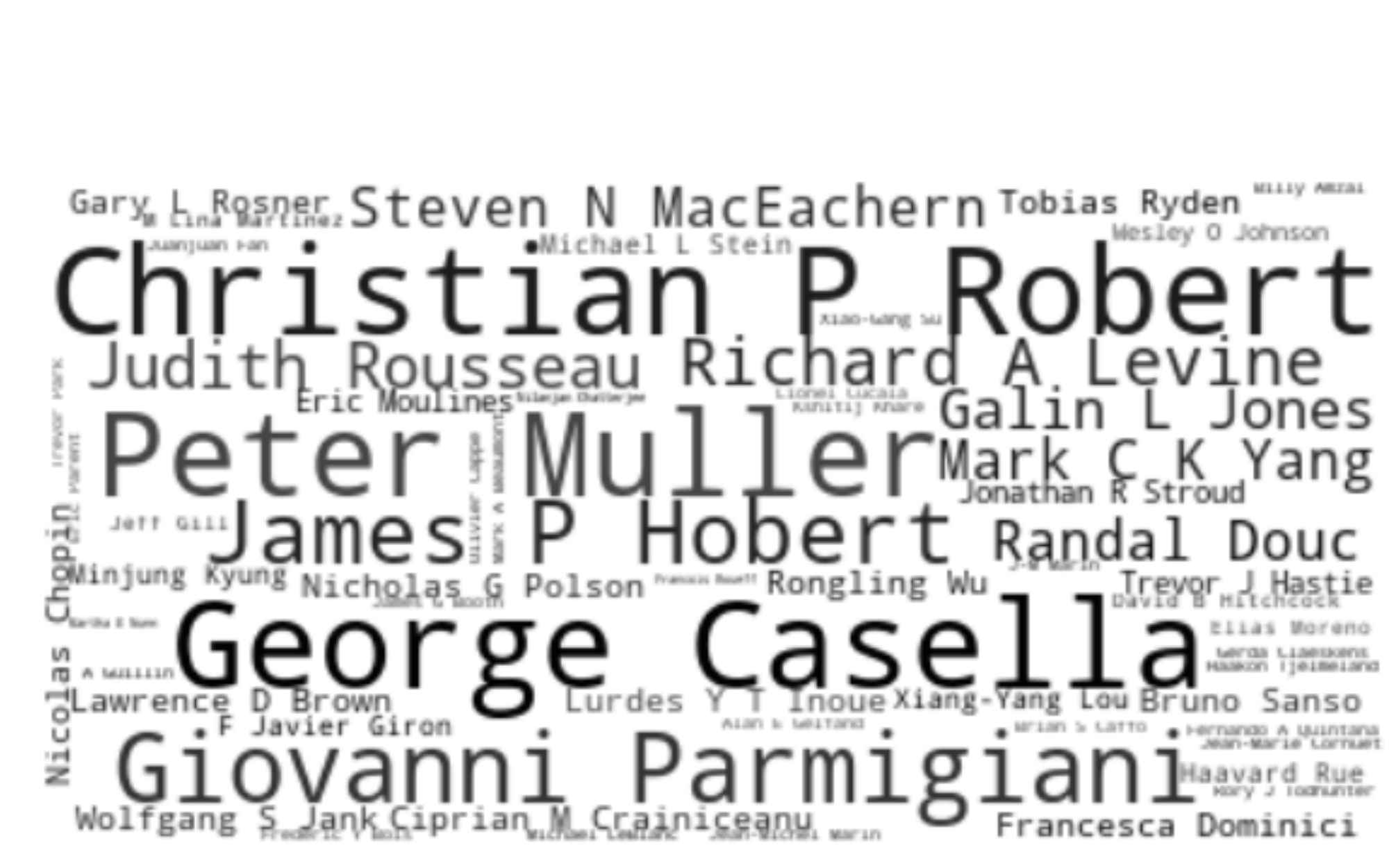}
\caption{Bayesian}
\end{subfigure}
\begin{subfigure}{0.32\textwidth}
\centering
\includegraphics[scale=.24, trim=1in 2in 1in 1in, clip]{10wordcloud.pdf}
\caption{Experimental design}
\end{subfigure}
\begin{subfigure}{0.32\textwidth}
\centering
\includegraphics[scale=.24, trim=1in 2in 1in 1in, clip]{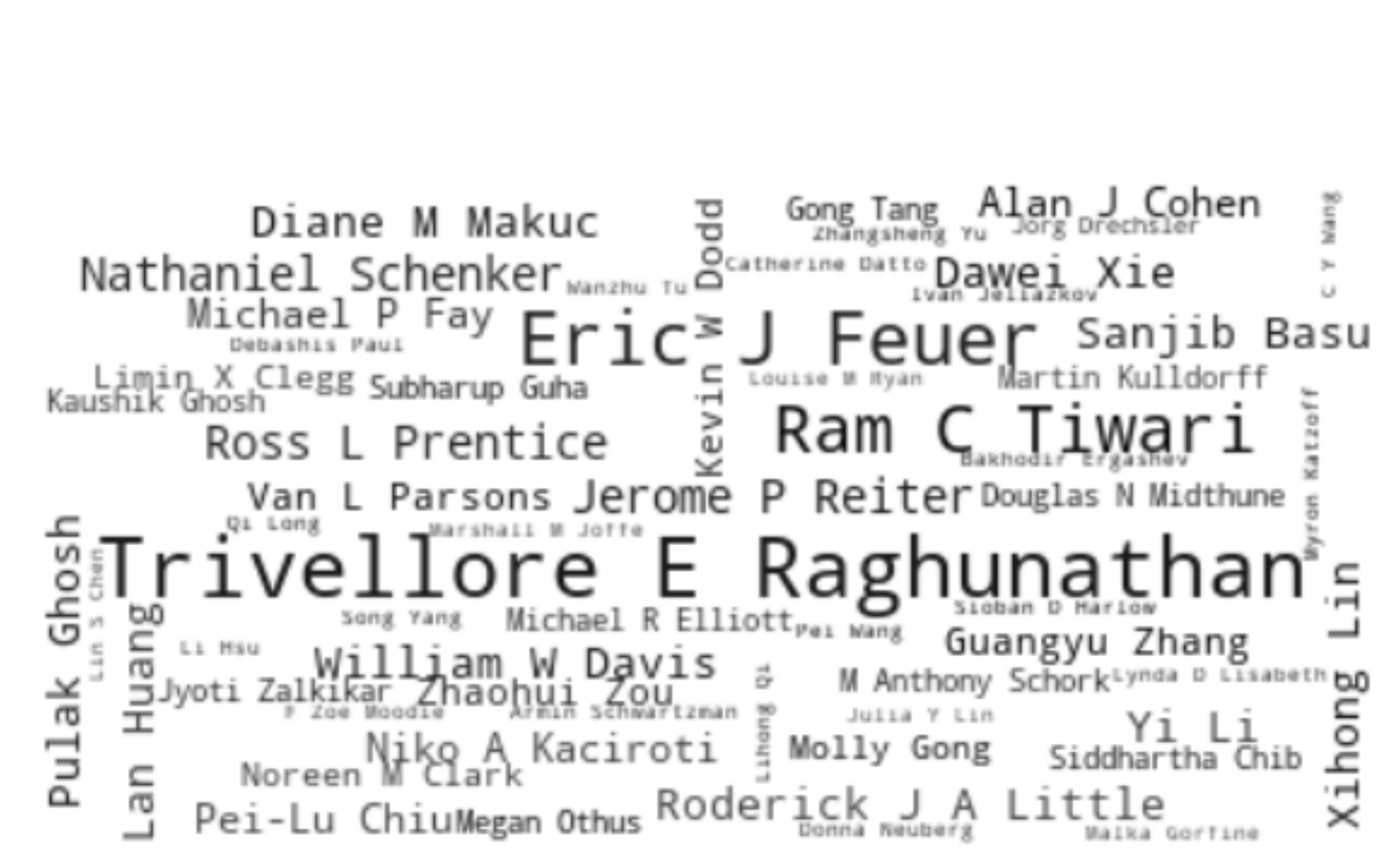}
\caption{Biostat/Survey}
\end{subfigure}
\begin{subfigure}{0.32\textwidth}
\centering
\includegraphics[scale=.24, trim=1in 2in 1in 1in, clip]{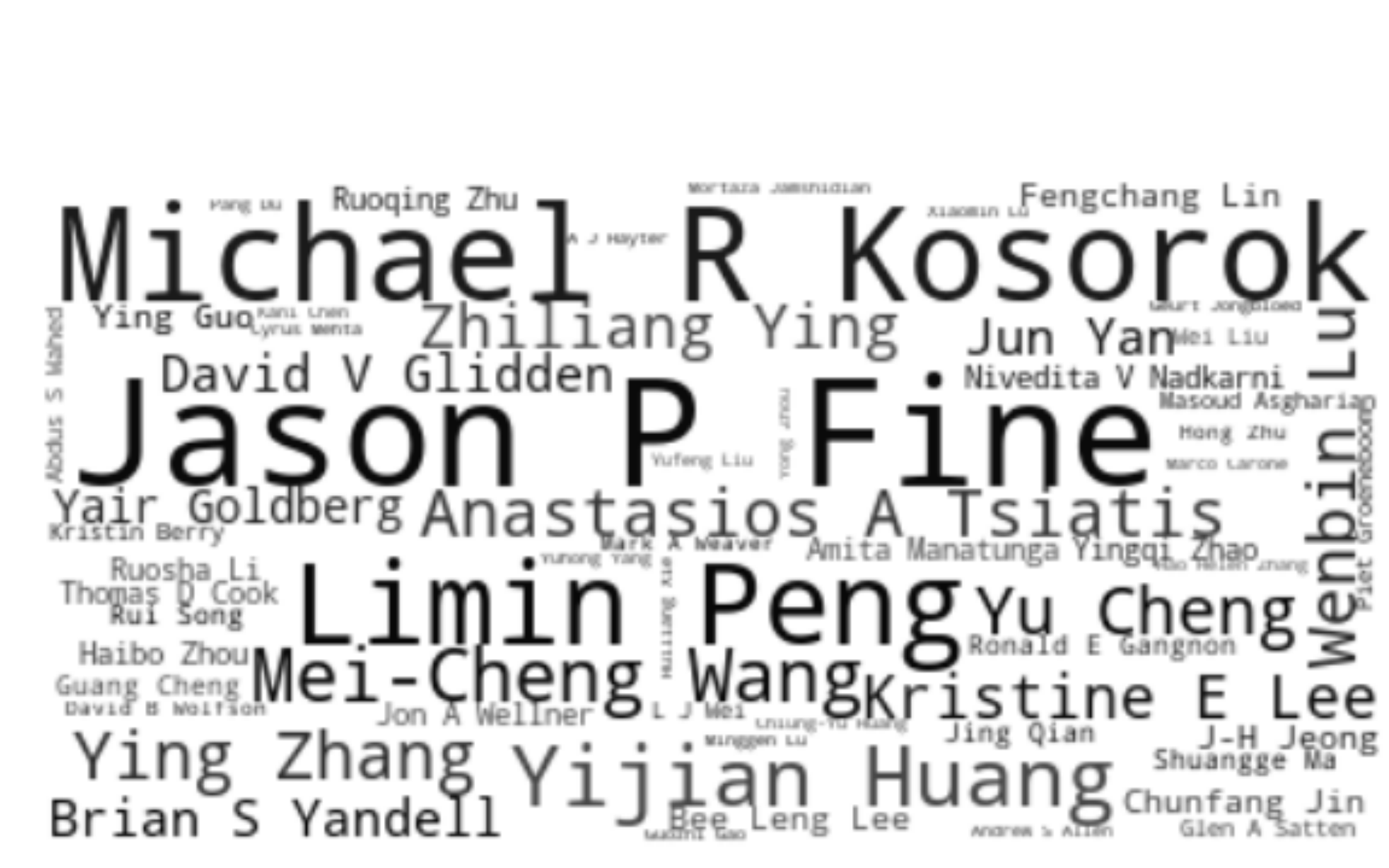}
\caption{Biostat}
\end{subfigure}
\begin{subfigure}{0.32\textwidth}
\centering
\includegraphics[scale=.24, trim=1in 2in 1in 1in, clip]{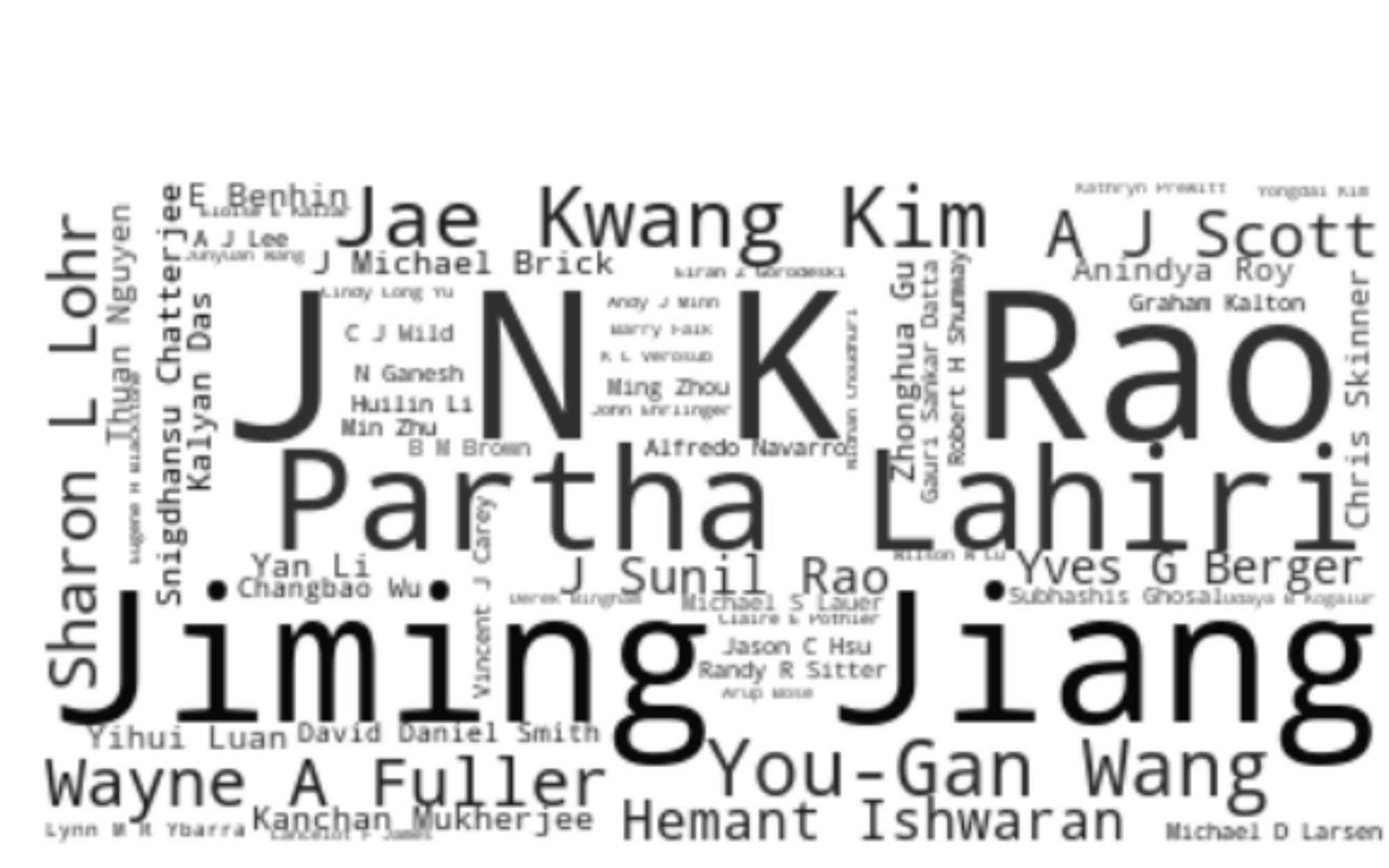}
\caption{Survey}
\end{subfigure}
\begin{subfigure}{0.32\textwidth}
\centering
\includegraphics[scale=.24, trim=1in 2in 1in 1in, clip]{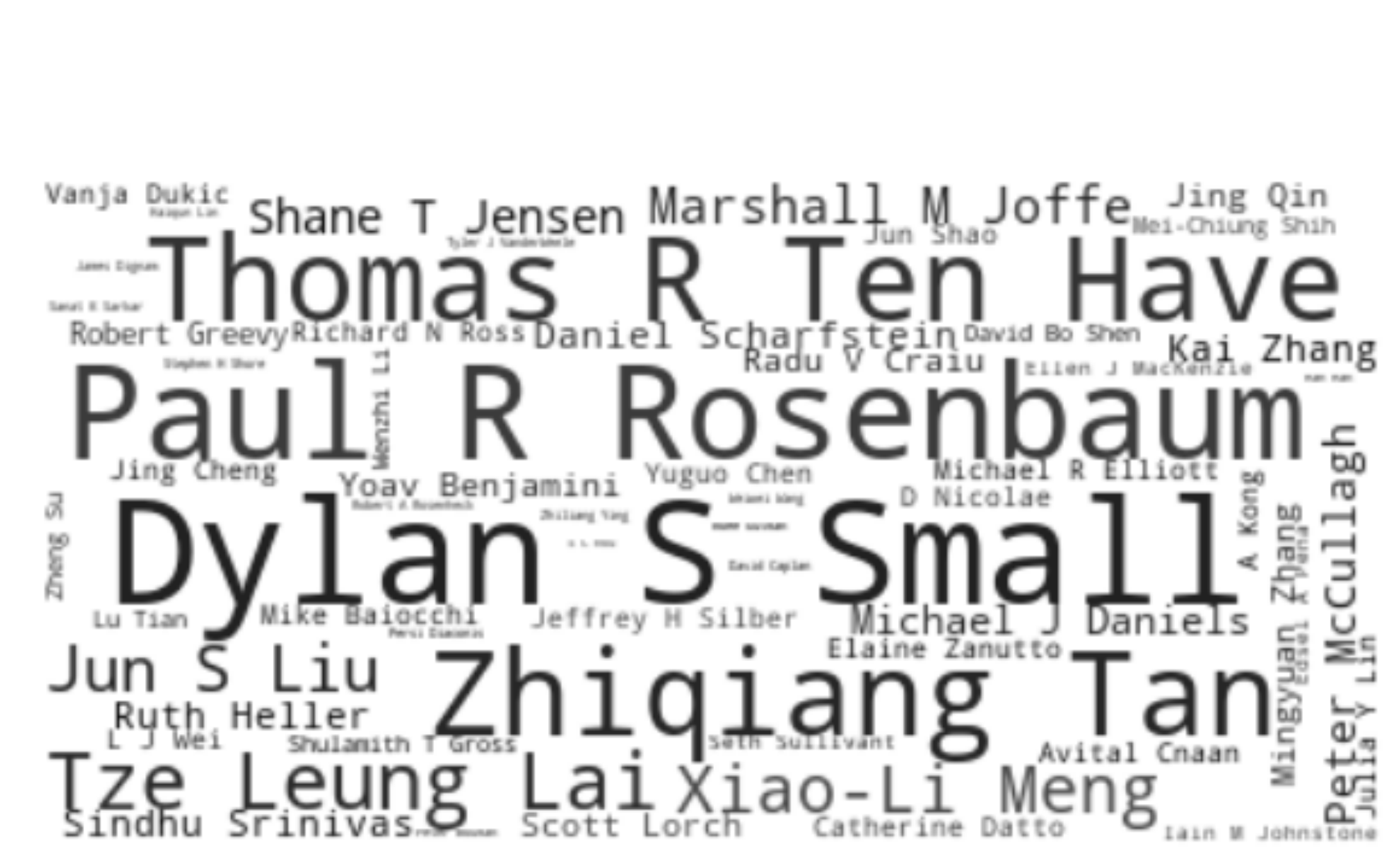}
\caption{Causal}
\end{subfigure}
\begin{subfigure}{0.32\textwidth}
\centering
\includegraphics[scale=.24, trim=1in 2in 1in 1in, clip]{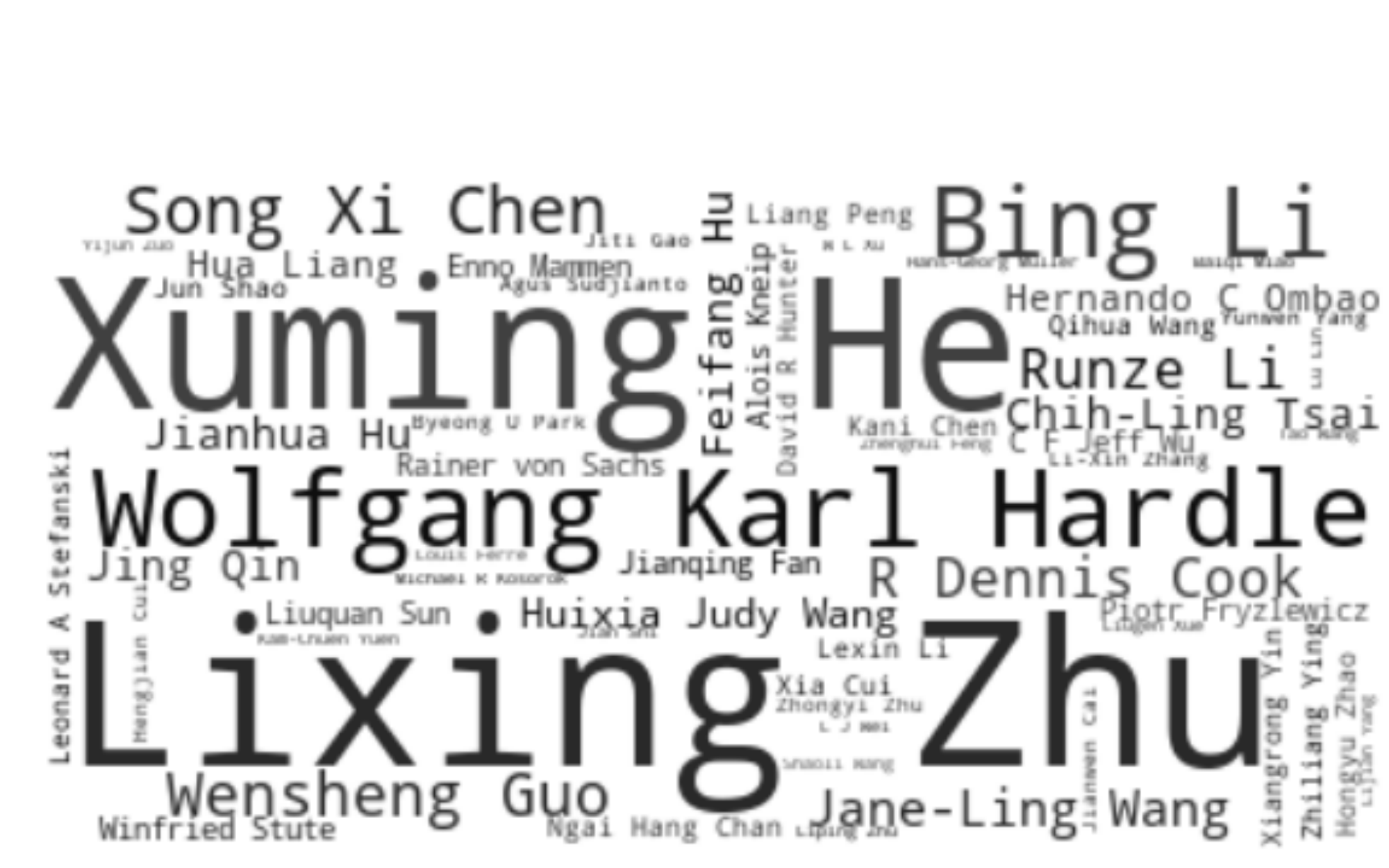}
\caption{Econometrics}
\end{subfigure}
\begin{subfigure}{0.32\textwidth}
\centering
\includegraphics[scale=.24, trim=1in 2in 1in 1in, clip]{16wordcloud.pdf}
\caption{High dimensional}
\end{subfigure}
\begin{subfigure}{0.32\textwidth}
\centering
\includegraphics[scale=.24, trim=1in 2in 1in 1in, clip]{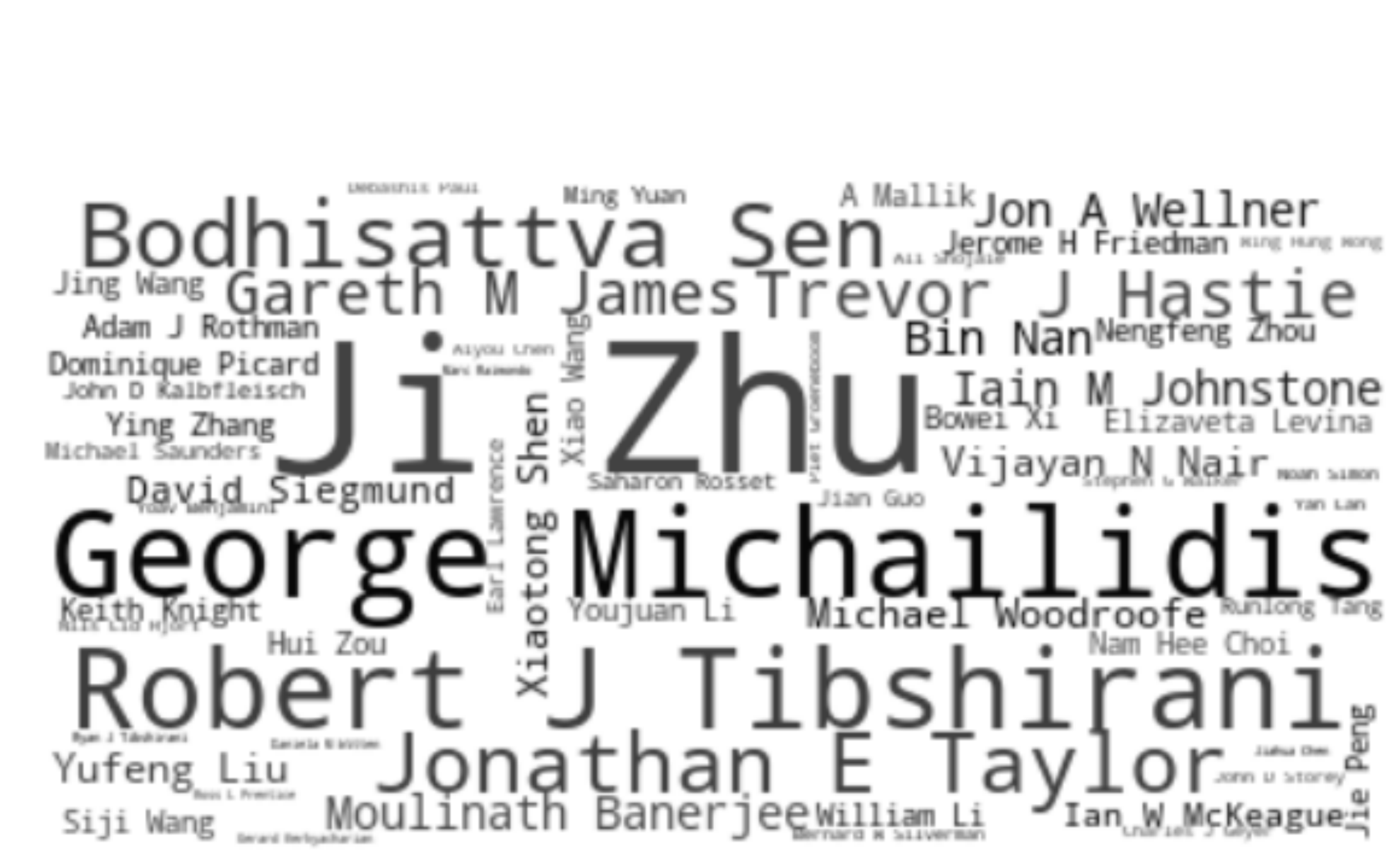}
\caption{High dimensional}
\end{subfigure}
\begin{subfigure}{0.32\textwidth}
\centering
\includegraphics[scale=.24, trim=1in 2in 1in 1in, clip]{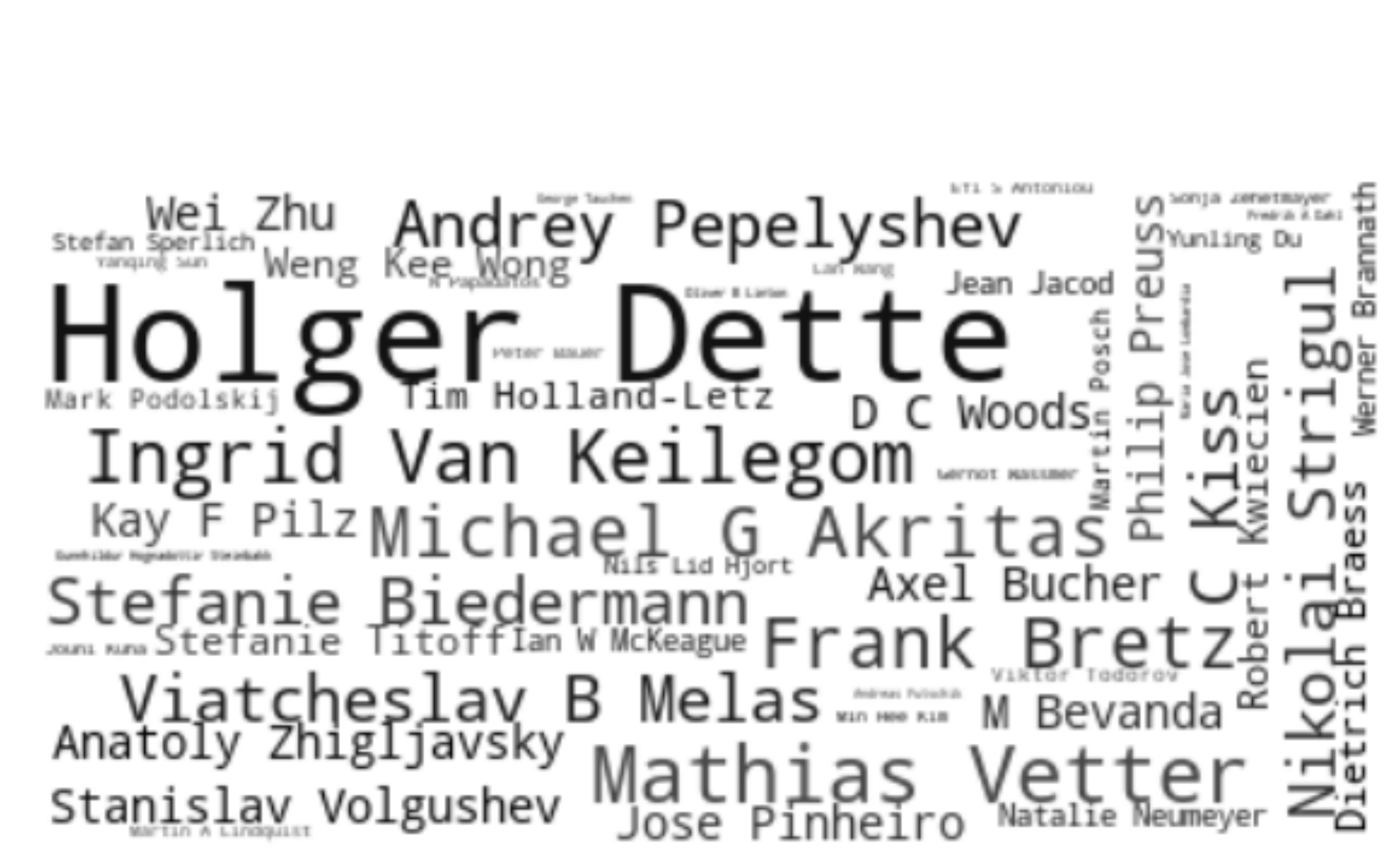}
\caption{Theory}
\end{subfigure}
\begin{subfigure}{0.32\textwidth}
\centering
\includegraphics[scale=.24, trim=1in 2in 1in 1in, clip]{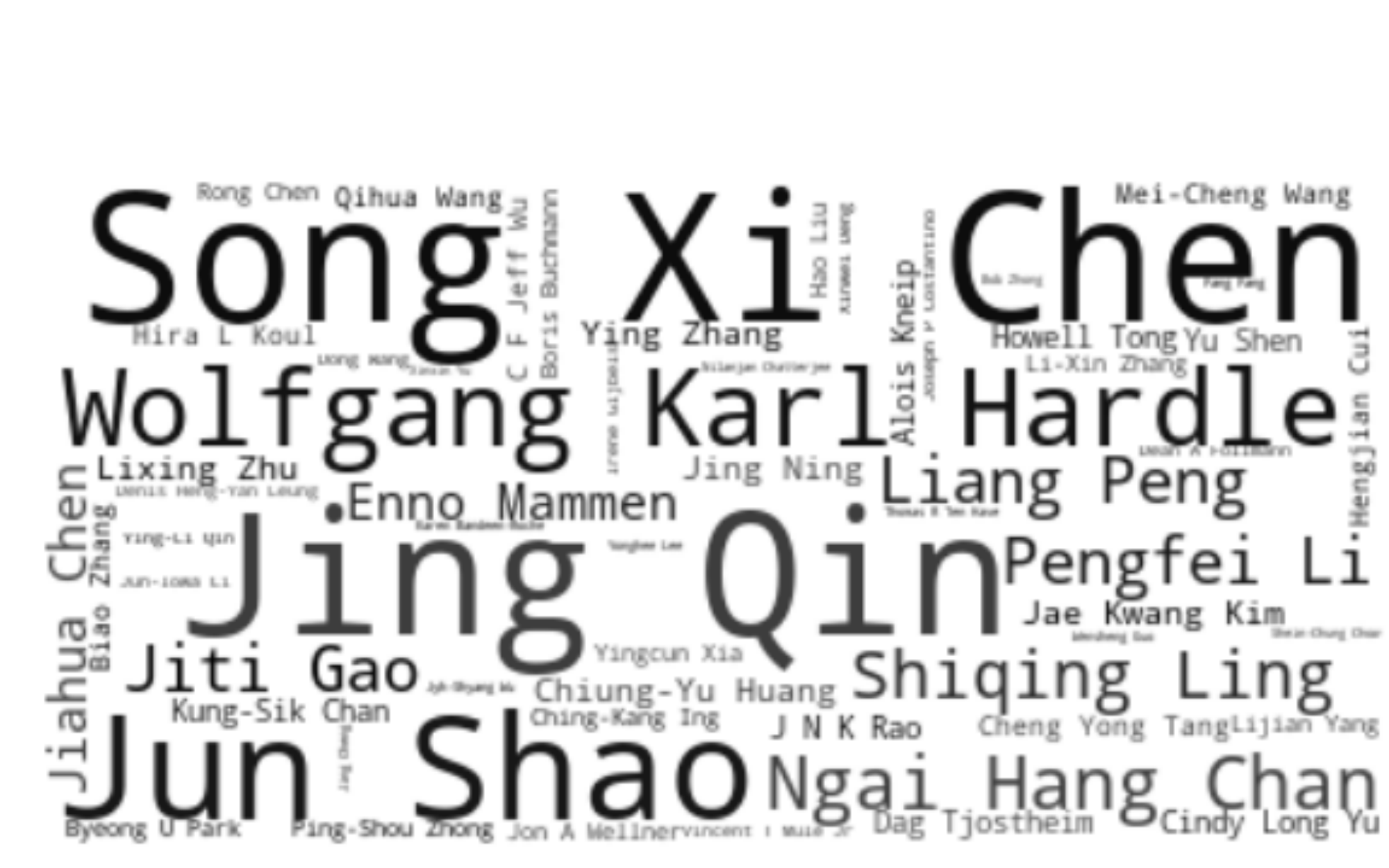}
\caption{Empirical likelihood/Inference}
\end{subfigure}
\begin{subfigure}{0.32\textwidth}
\centering
\includegraphics[scale=.24, trim=1in 2in 1in 1in, clip]{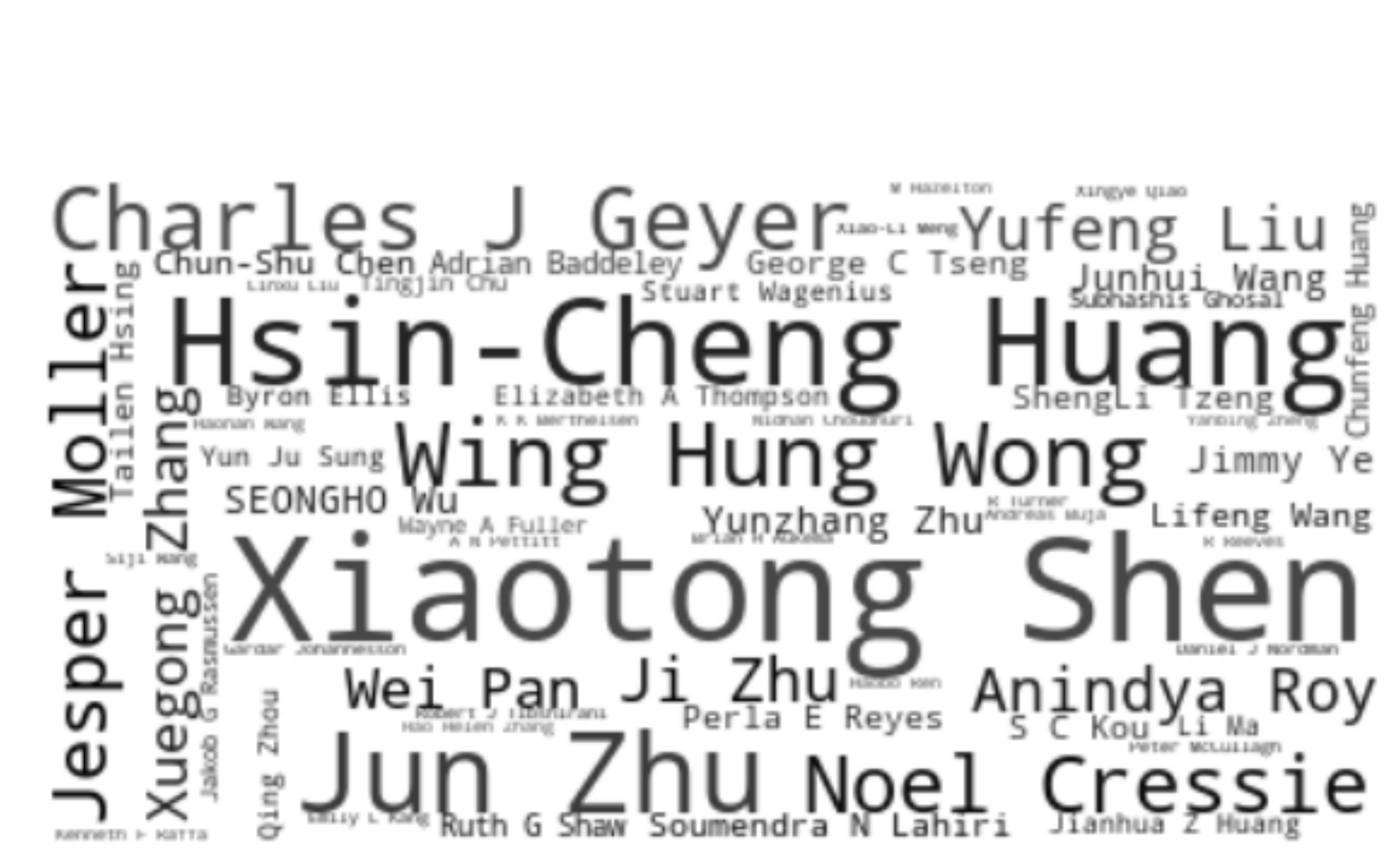}
\caption{High dim/Multivariate}
\end{subfigure}
\begin{subfigure}{0.32\textwidth}
\centering
\includegraphics[scale=.24, trim=1in 2in 1in 1in, clip]{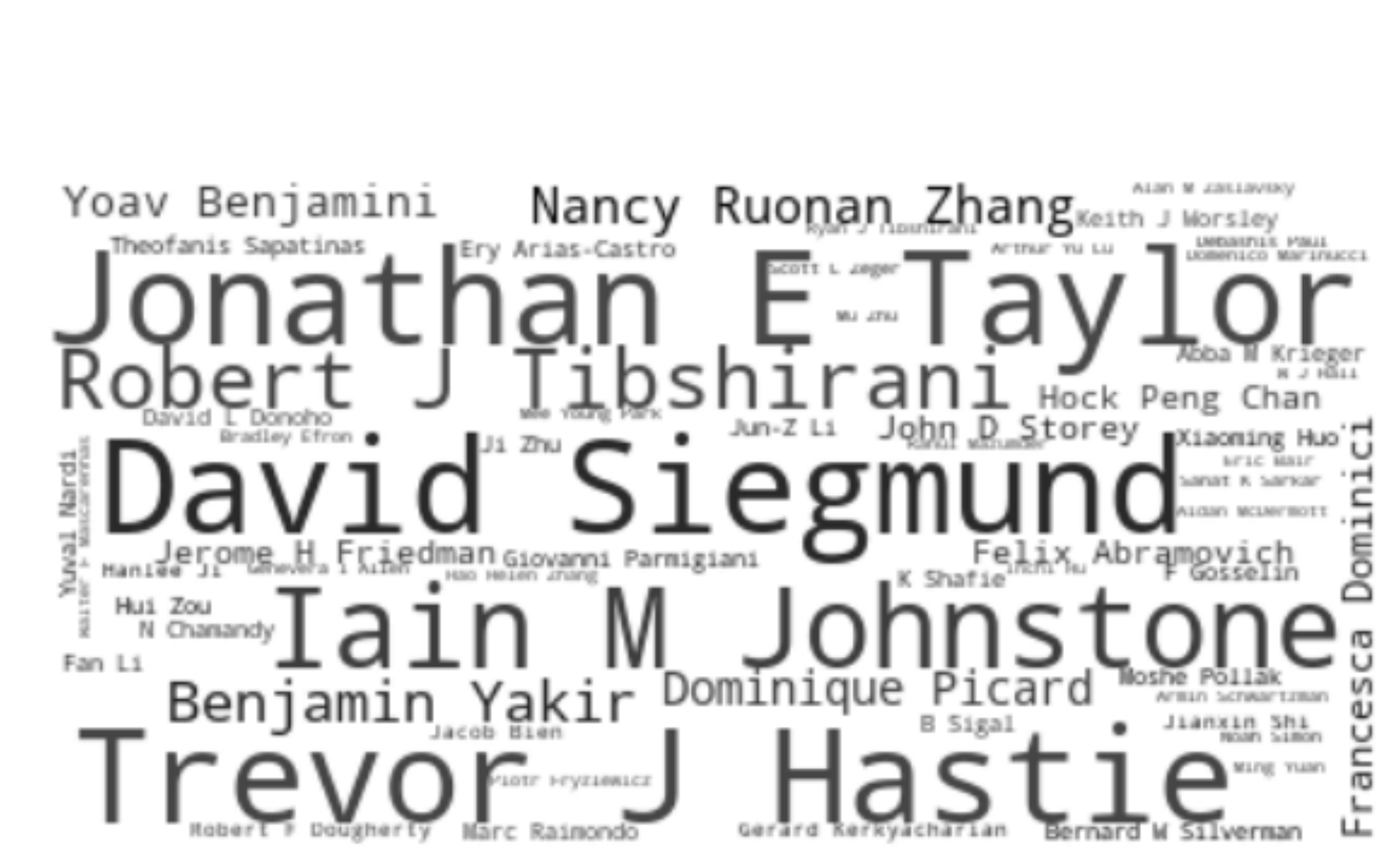}
\caption{Sequential}
\end{subfigure}
\begin{subfigure}{0.32\textwidth}
\centering
\includegraphics[scale=.24, trim=1in 2in 1in 1in, clip]{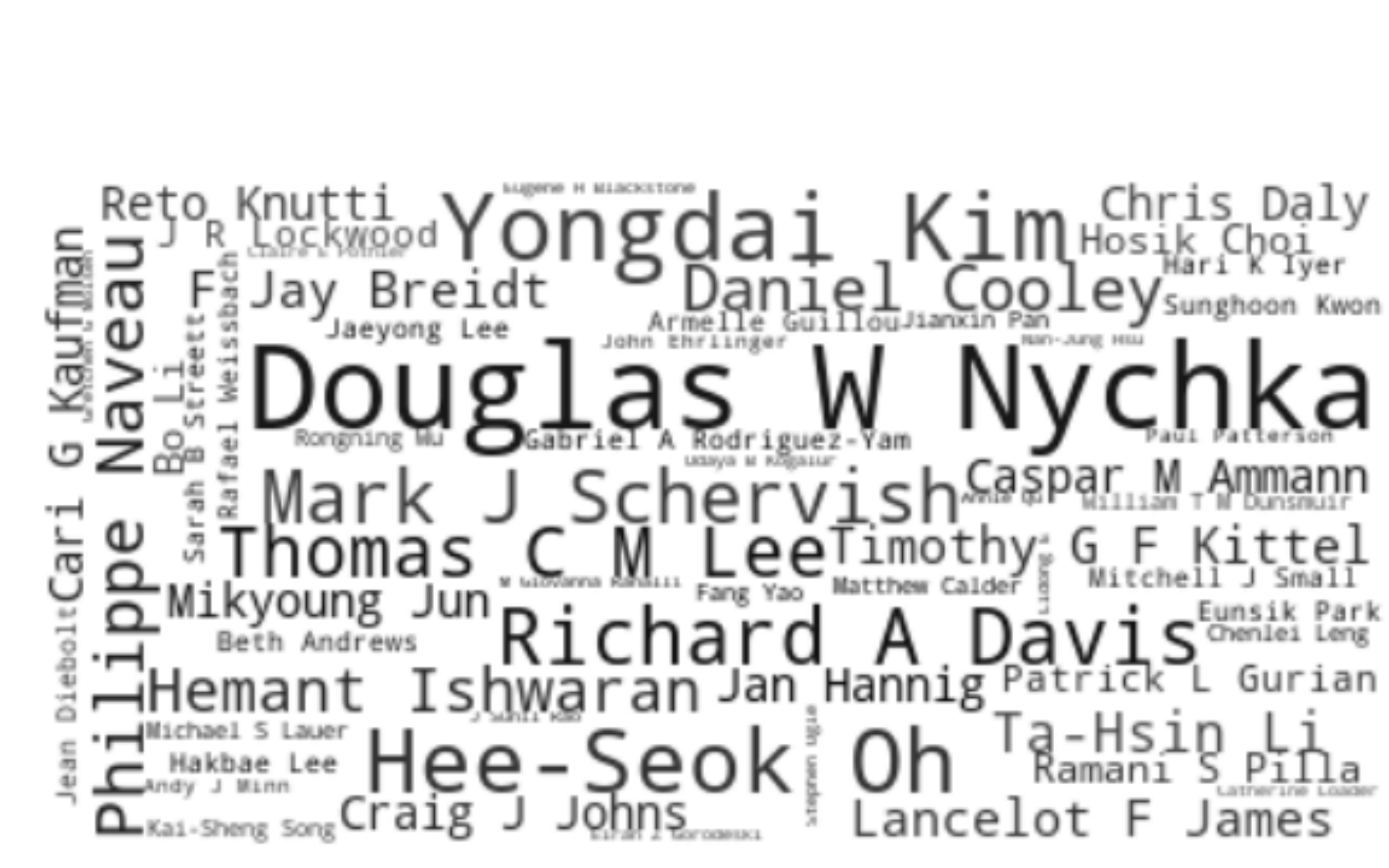}
\caption{Spatial/Image}
\end{subfigure}
\begin{subfigure}{0.32\textwidth}
\centering
\includegraphics[scale=.24, trim=1in 2in 1in 1in, clip]{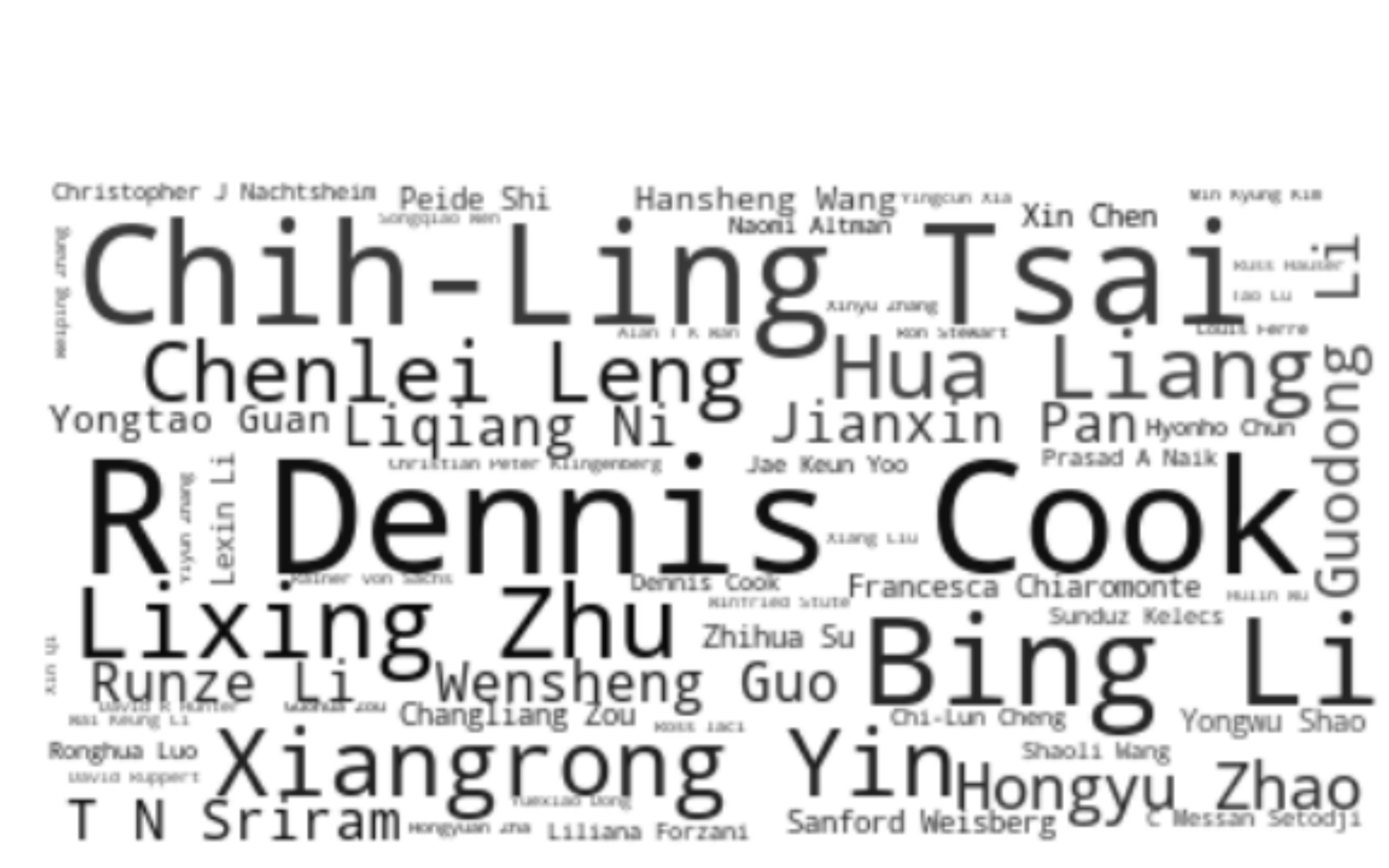}
\caption{Dimensionality reduction}
\end{subfigure}
\begin{subfigure}{0.32\textwidth}
\centering
\includegraphics[scale=.24, trim=1in 2in 1in 1in, clip]{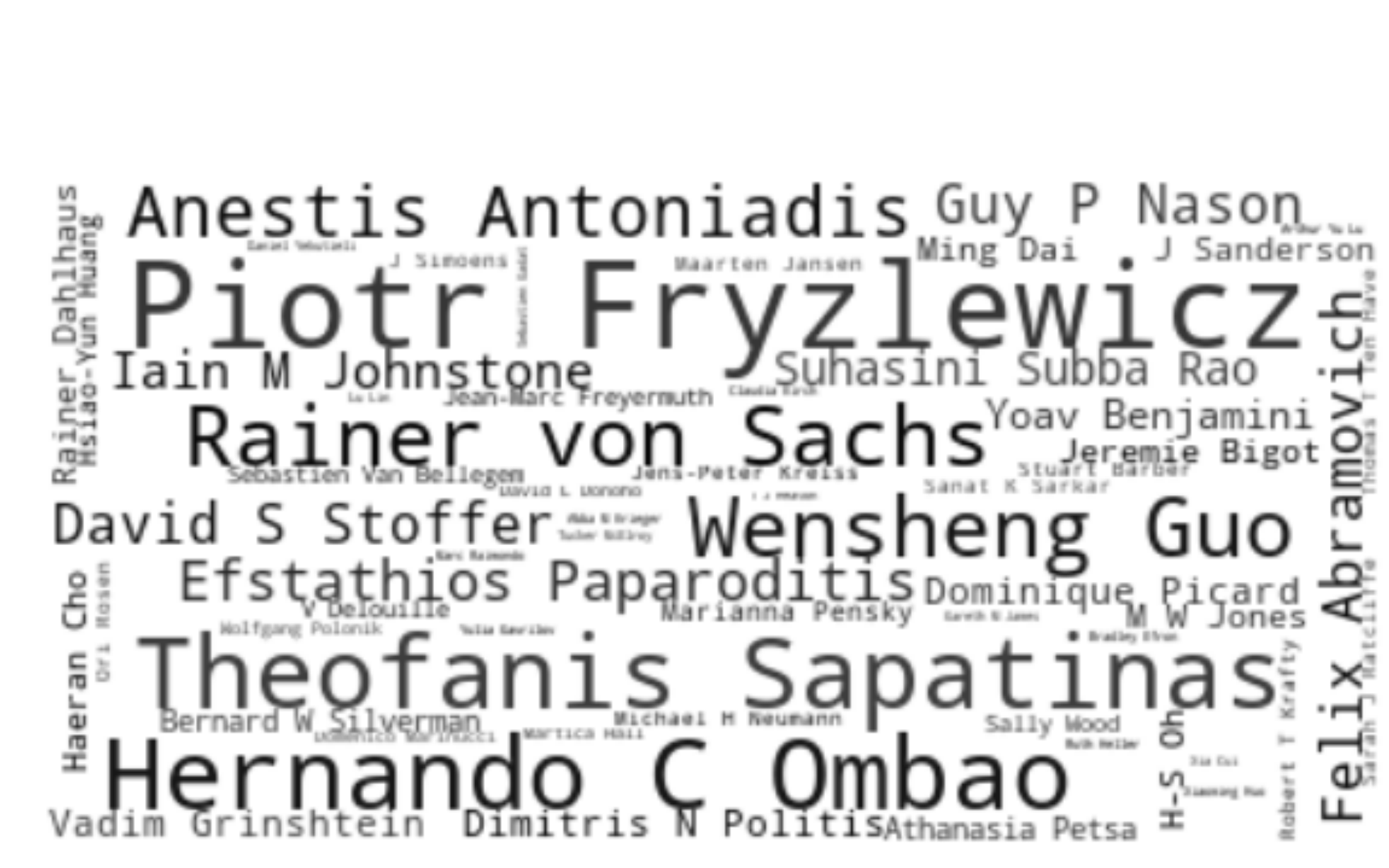}
\caption{Time series}
\end{subfigure}
\begin{subfigure}{0.32\textwidth}
\centering
\includegraphics[scale=.24, trim=1in 2in 1in 1in, clip]{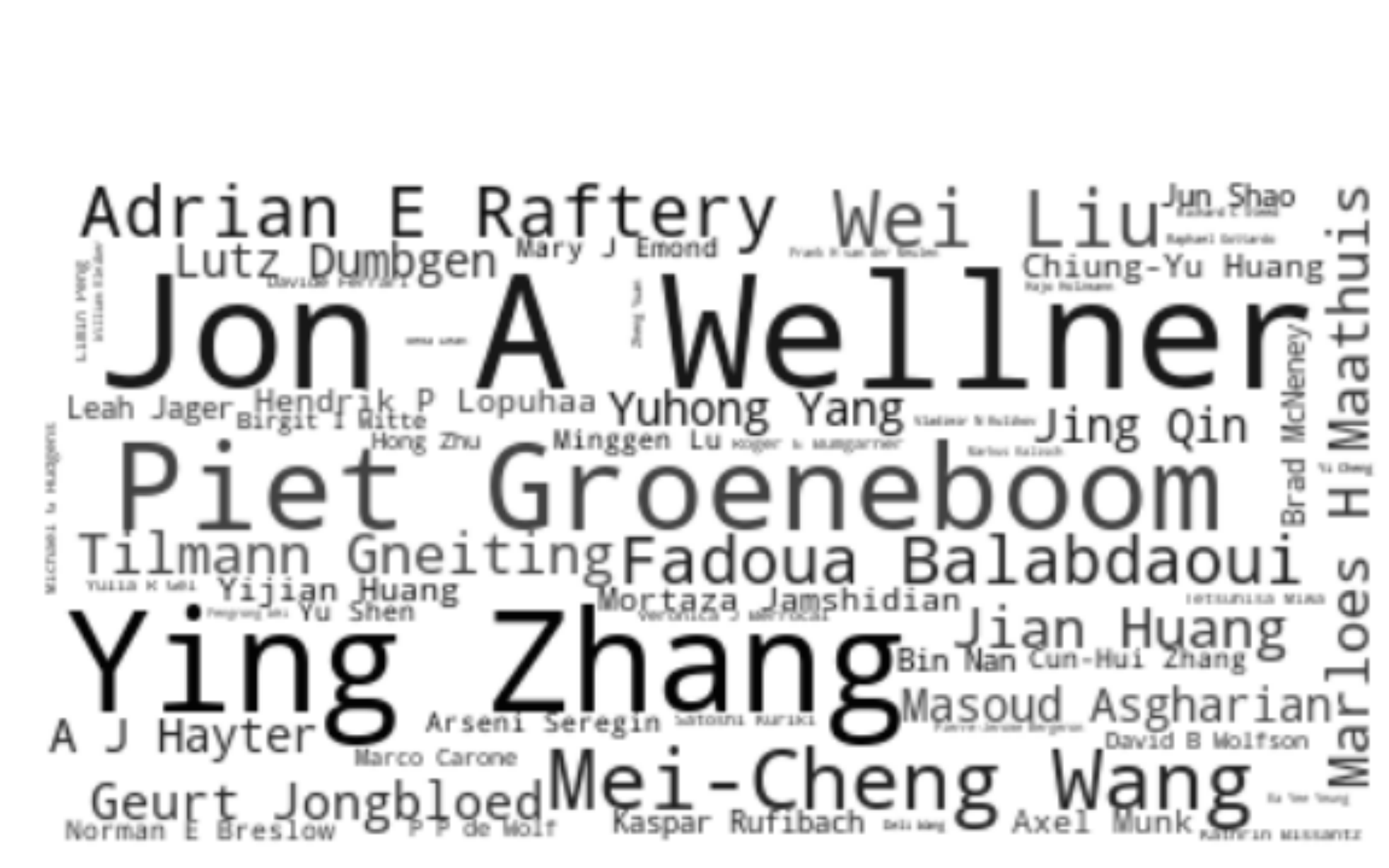}
\caption{Shape constrained}
\end{subfigure}
\begin{subfigure}{0.32\textwidth}
\centering
\includegraphics[scale=.24, trim=1in 2in 1in 1in, clip]{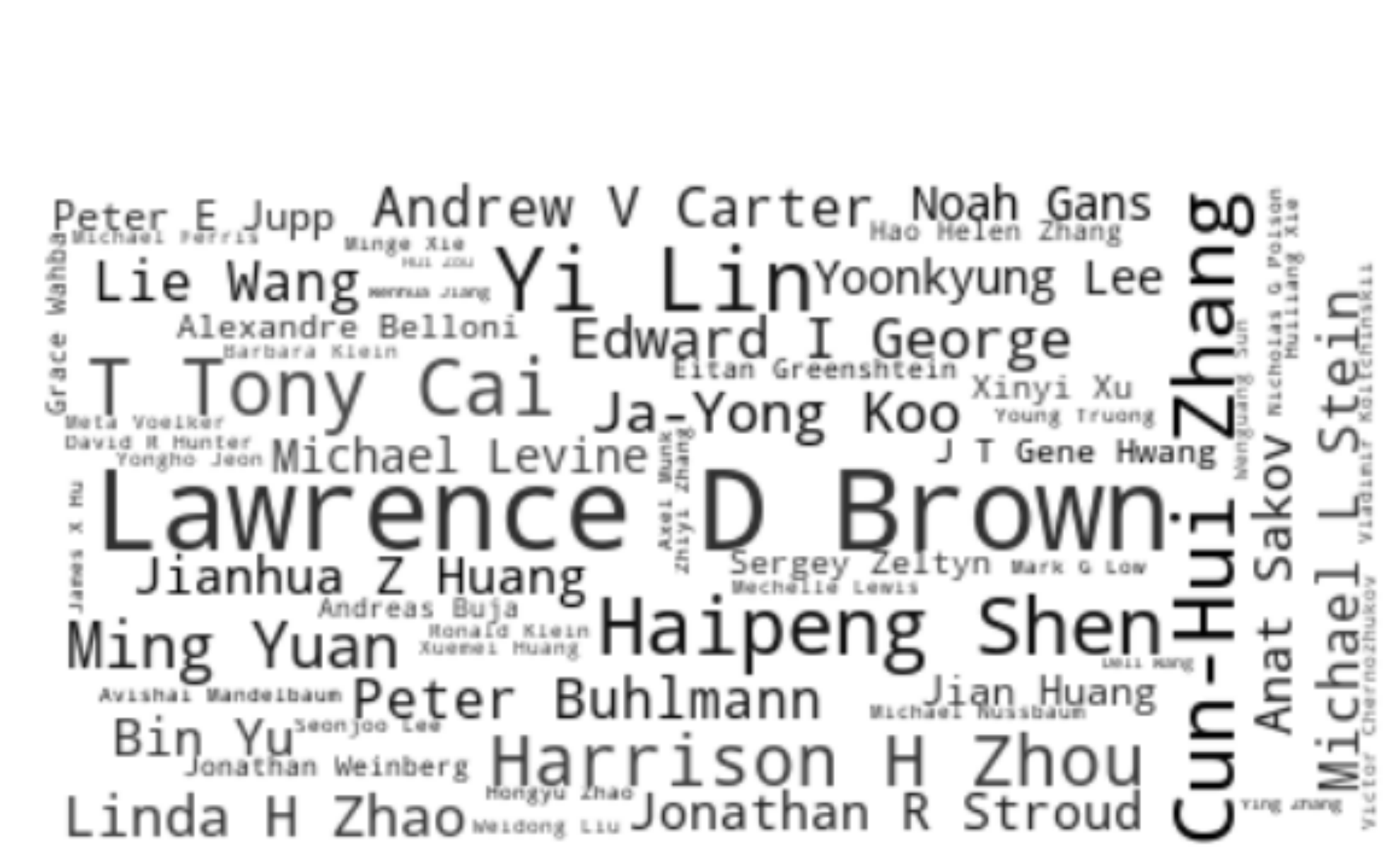}
\caption{Theory}
\end{subfigure}
\caption{Additional clusters from the statistician co-authorship network. We hand label a subset.}
\label{fig:coauthor_cluster3}
\end{figure}

In this section, we give 18 additional clusters discovered on the statistican co-authorship network in Figure~\ref{fig:coauthor_cluster3}, expanding the results given in Section~\ref{sec:coauthor} of the main paper.  

{\color{black}
\subsection{Sampler diagnostic information}
\label{sec:diagnostics}

In this section, we give detailed sampler diagnostic information of the Gibbs sampling algorithm proposed in Section~\ref{sec:algorithm}. We use a simulation setting where we generate a PAPER network with $n=2000$ nodes and $m=4000$ edges with $K=1$ and we also use the statistician co-authorship network analyzed in Section~\ref{sec:coauthor}, which has $n=2263$ nodes and $m=4388$ edges. 

Recall that our Gibbs sampler produces a sequence of samples of a spanning tree $\tilde{\bm{t}}_n^{(j)}$ and ordering $\pi^{(j)}$ for $j=1,2,\ldots, J$ where $J$ is the number of Gibbs outer iterations. We use $\tilde{\bm{t}}_n^{(j)}$ to compute the "sampled" posterior root probability $Q^{(j)}(\cdot) = \mathbb{P}(\Pi_1 = \cdot \,|\, \tilde{\bm{T}}_n = \tbm{t}_n^{(j)})$. For the simulation setting, we then construct trace plot and auto-correlation plot based on the sequence $\{ Q^{(j)}(\text{true root}) \}_{j=1}^J$. For the statistician co-authorship network, we use construct the plots based on $\{ Q^{(j)}(\text{ Raymond Carroll })\}_{j=1}^J$. Figures~\ref{fig:sim_acf},~\ref{fig:coauthor_acf},~\ref{fig:sim_trace}, and~\ref{fig:coauthor_trace} suggest that the sampler is able to converge to the stationary distribution and has no significant autocorrelation. 

\begin{figure}
\centering
\begin{subfigure}{.45\textwidth}
\centering
\includegraphics[scale=.5]{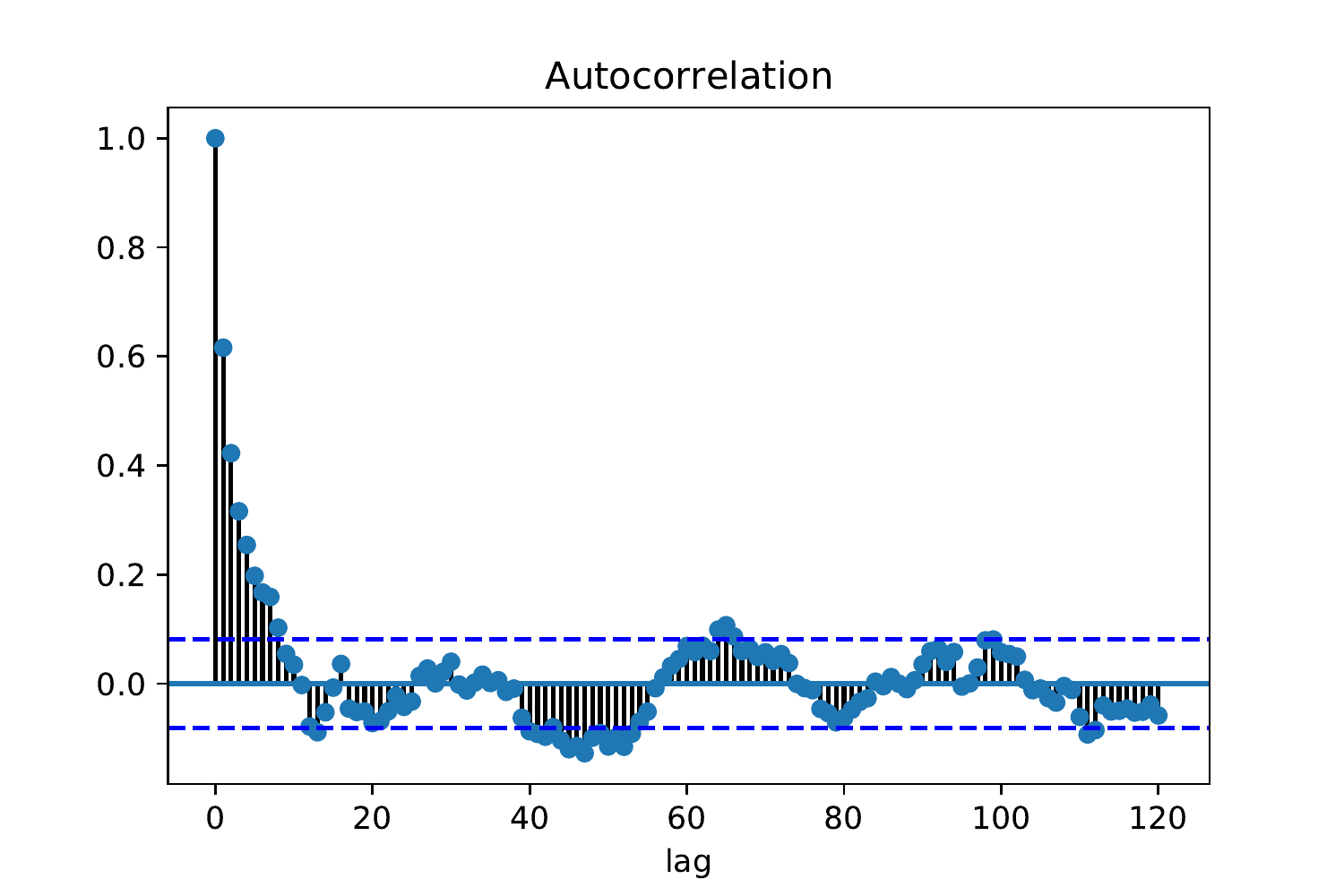}
\caption{ACF plot for a simulated network}
\label{fig:sim_acf}
\end{subfigure}
\begin{subfigure}{.45\textwidth}
\centering
\includegraphics[scale=.5]{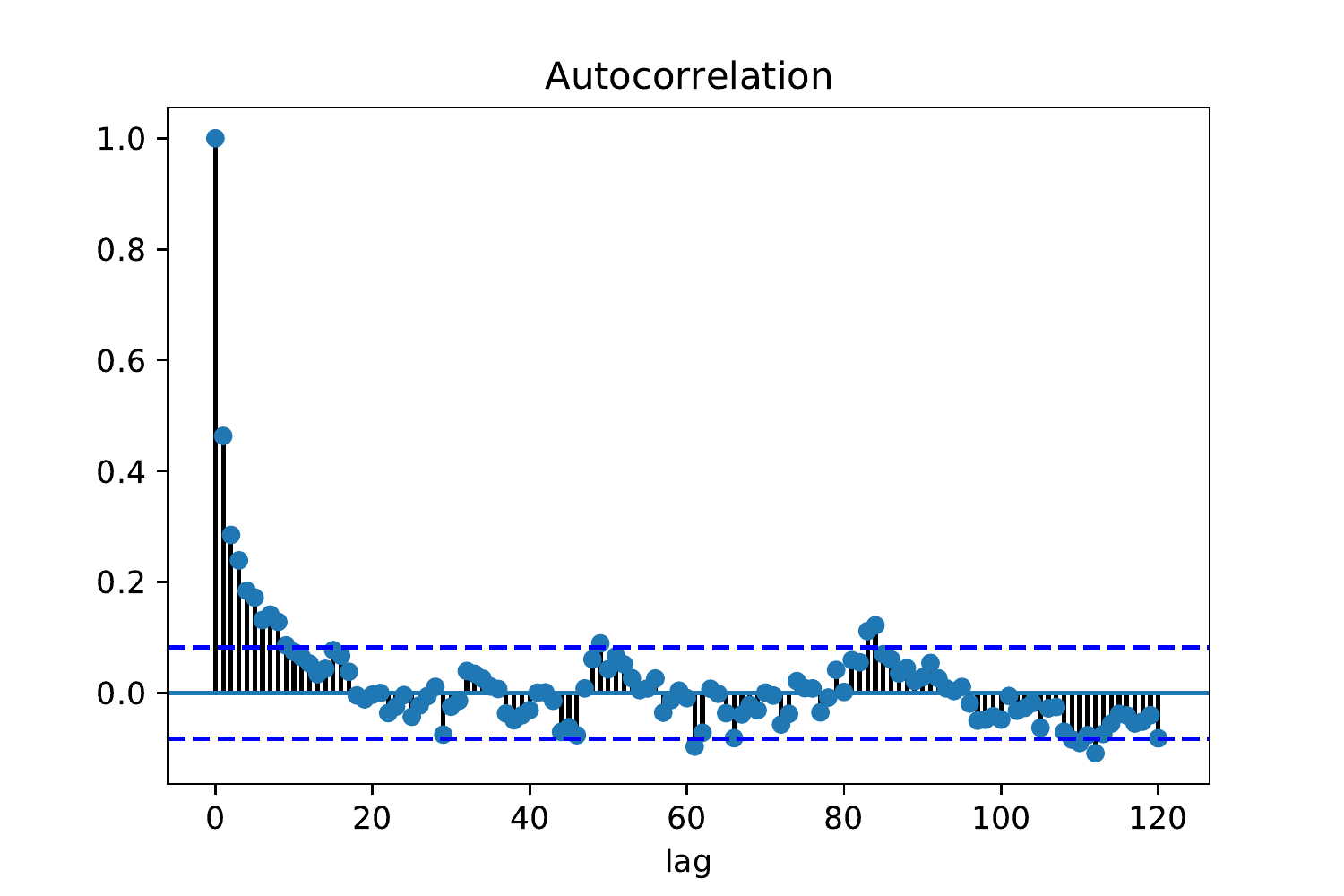}
\caption{ACF plot for the co-authorship network}
\label{fig:coauthor_acf}
\end{subfigure}
\begin{subfigure}{.45\textwidth}
\centering
\includegraphics[scale=.5]{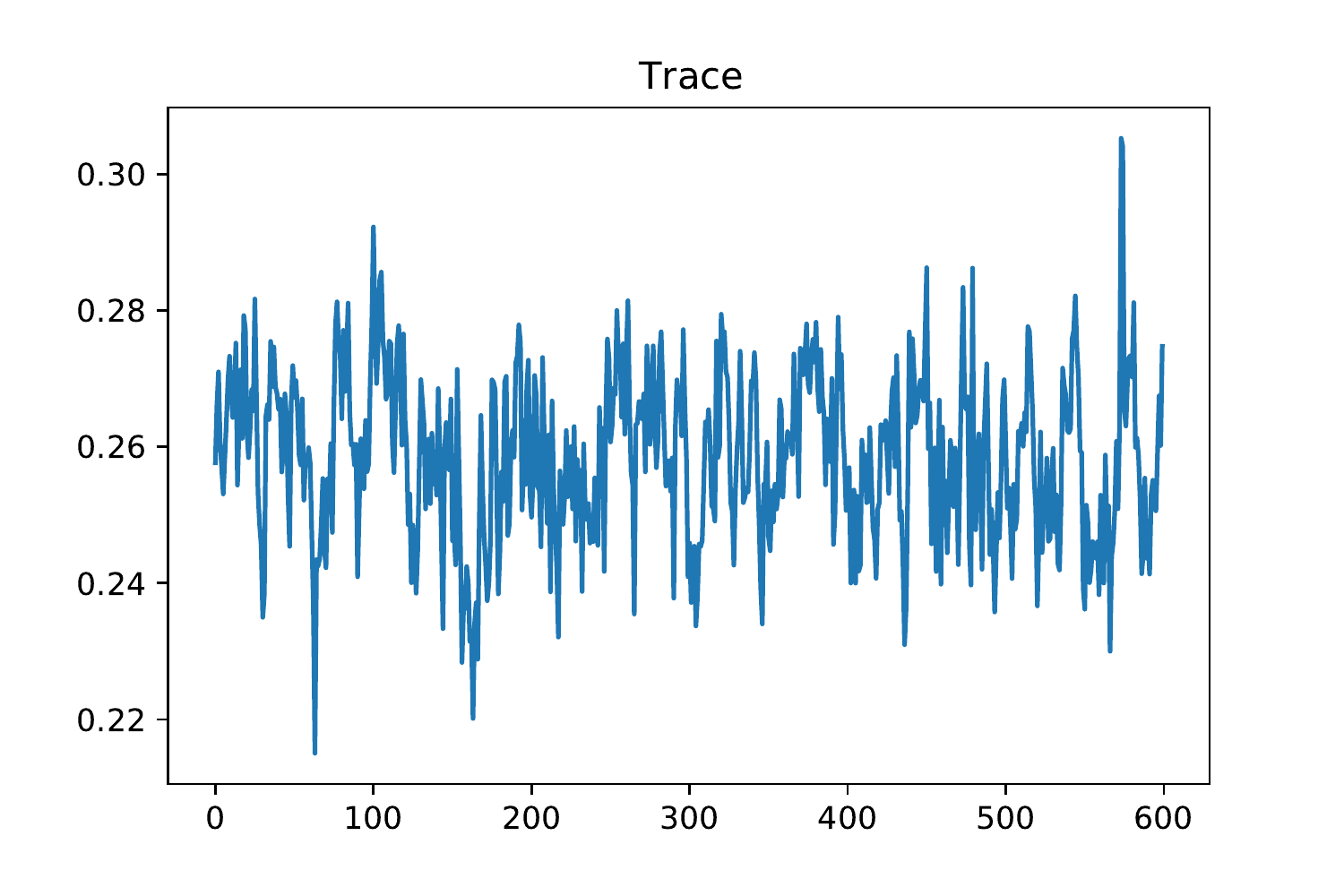}
\caption{Trace plot for a simulated network}
\label{fig:sim_trace}
\end{subfigure}
\begin{subfigure}{.45\textwidth}
\centering
\includegraphics[scale=.5]{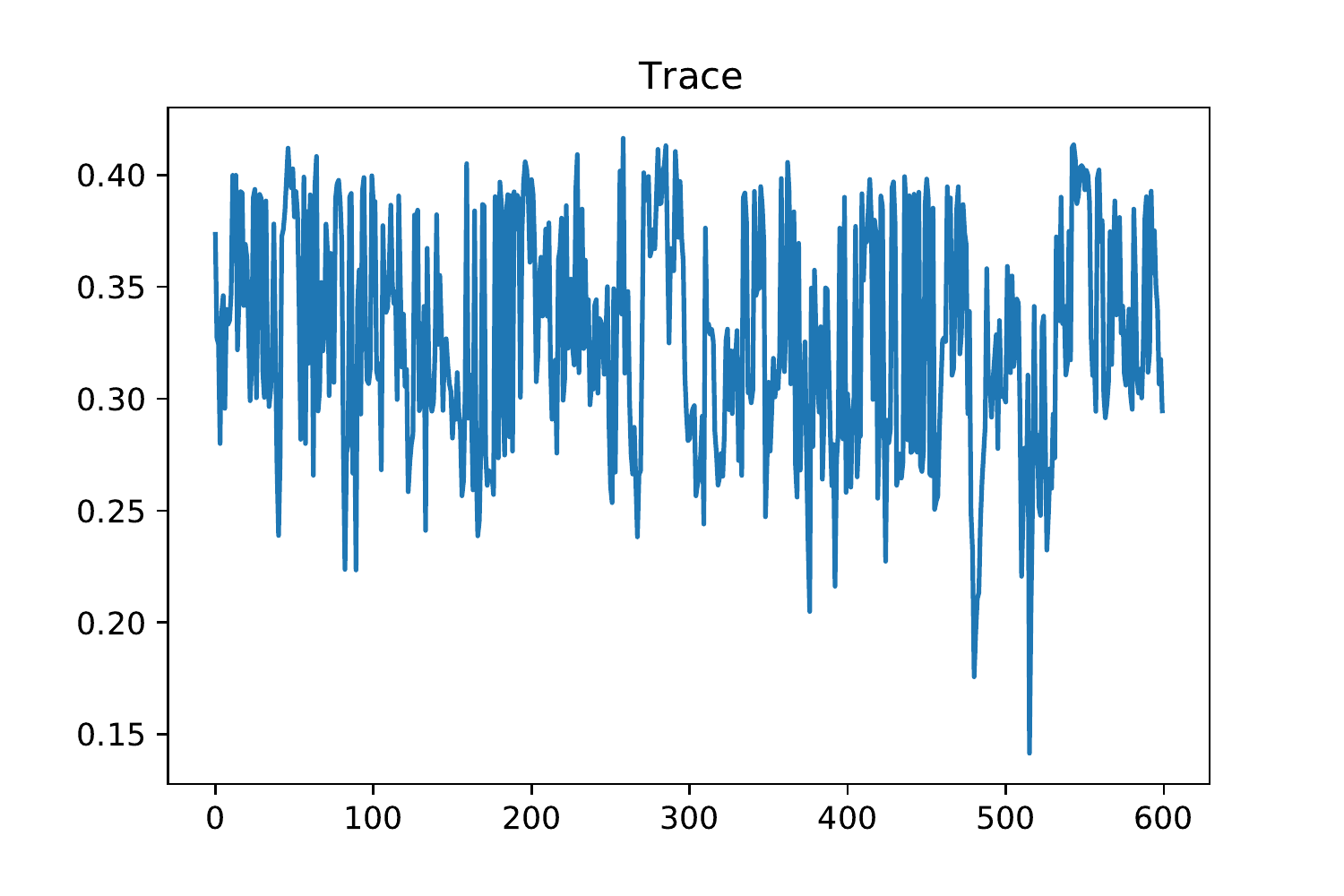}
\caption{Trace plot for the co-authorship network}
\label{fig:coauthor_trace}
\end{subfigure}
\caption{}
\end{figure}

As described in Section~\ref{sec:practical}, to assess convergence, we run two parallel chains $A$ and $B$ with corresponding posterior root probability estimates $Q^{A (1:J)}(\cdot) = \frac{1}{J} \sum_{j=1}^J Q^{A(j)}(\cdot)$ and $Q^{B (1:J)}(\cdot) = \frac{1}{J} \sum_{j=1}^J Q^{A(j)}(\cdot)$. We then compute the Hellinger distance $d_{\text{H}}(Q^{A (1:J)}, Q^{B (1:J)})$ and increase $J$ until the distance is small enough. In Figures~\ref{fig:sim_dist} and~\ref{fig:coauthor_dist}, we show that $d_{\text{H}}(Q^{A (1:J)}, Q^{B (1:J)})$ indeed converges to 0 quickly as $J$ increases. We emphasize that the chains $A$ and $B$ are initialized with a uniformly random spanning tree and a uniformly random ordering on that tree so that the initialization is guaranteed to be overdispersed. 

\begin{figure}
\centering
\begin{subfigure}{.43\textwidth}
\centering
\includegraphics[scale=.42]{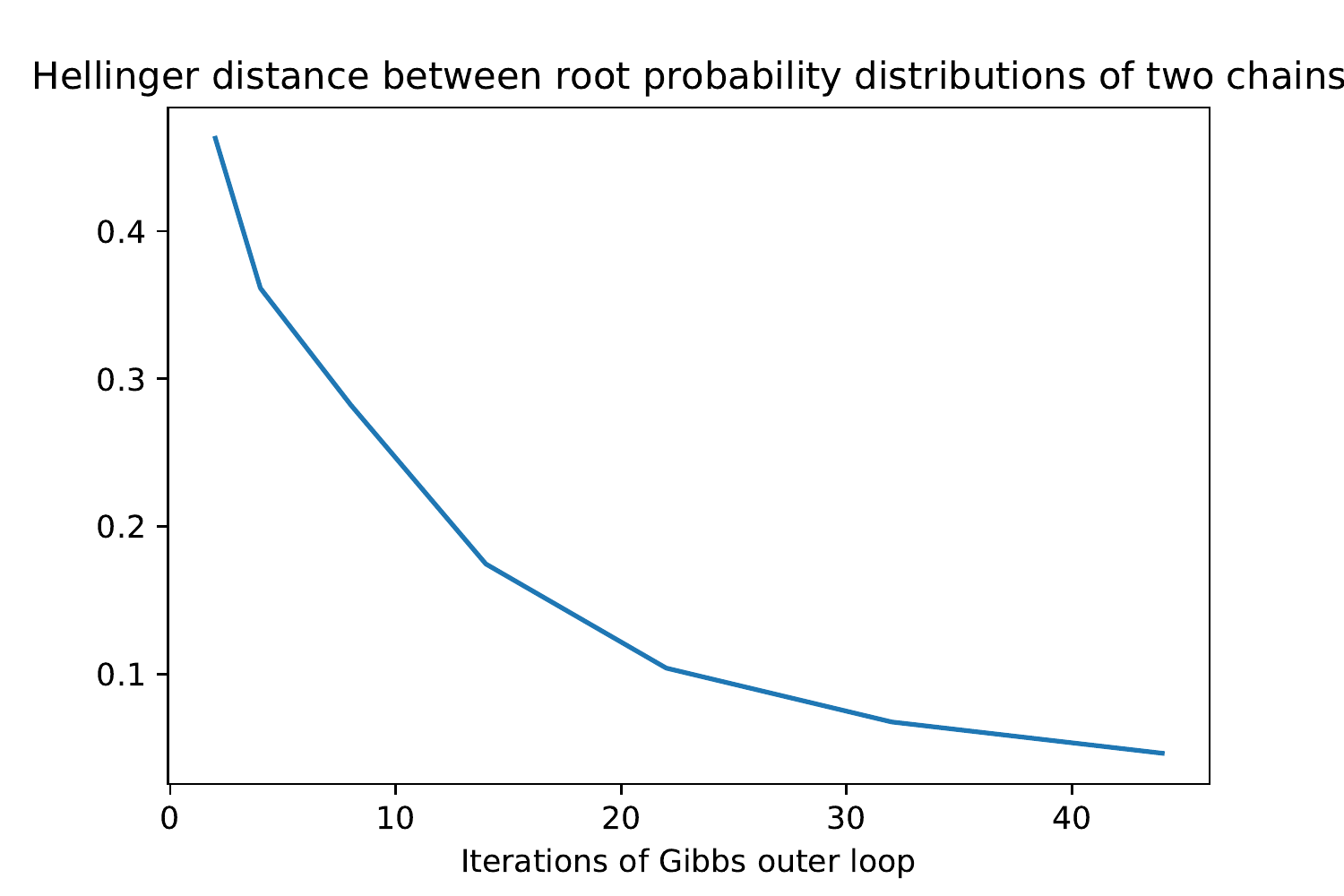}
\caption{Hellinger distance between the root probability distributions of two chains on a simulated network.}
\label{fig:sim_dist}
\end{subfigure}
\begin{subfigure}{.43\textwidth}
\centering
\includegraphics[scale=.42]{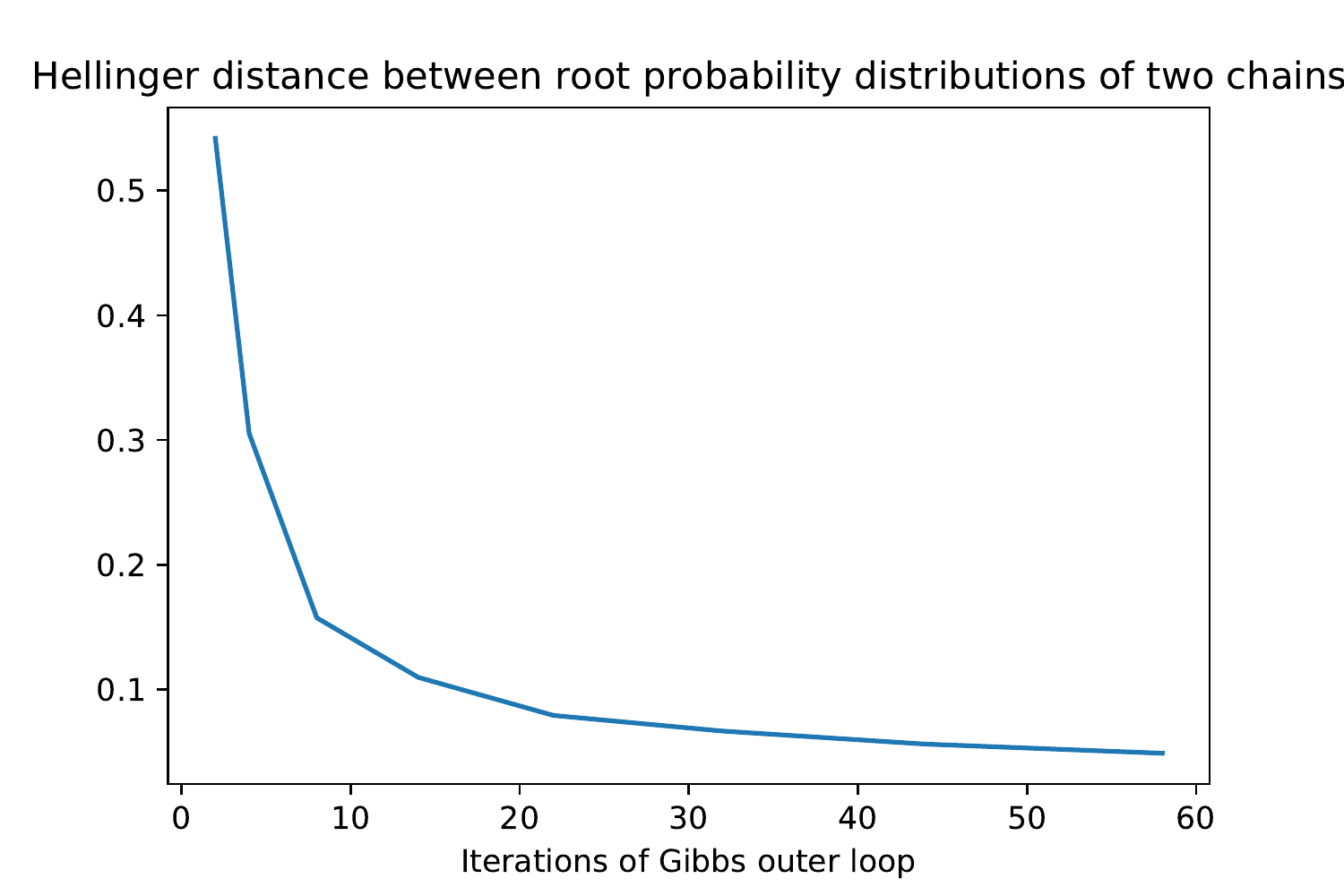}
\caption{Hellinger distance between the root probability distributions of two chains on the co-authorship network.}
\label{fig:coauthor_dist}
\end{subfigure}
\caption{}
\end{figure}

}

\end{document}